\documentclass{pasj00}
\usepackage{bm}

\def\commenta{$^*$}
\def\commentb{$^\dagger$}
\def\commentc{$^\ddagger$}

\def\inpress{in press}
\def\inprep{in preparation}

\def\arxiv#1{ (arXiv astro-ph/#1)}

\DeclareAbbreviation\AAHam{Astron. Abh. Hamburg. Sternw.}
\DeclareAbbreviation\AARv{Astron. Astrophys. Rev.}
\DeclareAbbreviation\AAS{American Astron. Soc. Meeting Abstracts}
\DeclareAbbreviation\an{Astron. Nachr.}
\DeclareAbbreviation\AcA{Acta Astron.}
\DeclareAbbreviation\Afz{Astrofizika}
\DeclareAbbreviation\AnTok{Tokyo Astron. Obs. Annals, Sec. Ser.}
\DeclareAbbreviation\Ap{Astrophysics}
\DeclareAbbreviation\ARep{Astron. Rep.}
\DeclareAbbreviation\ATel{Astron. Telegram}
\DeclareAbbreviation\ATsir{Astron. Tsirk.}
\DeclareAbbreviation\AcApS{Acta Astrophys. Sinica}
\DeclareAbbreviation\AstL{Astron. Lett.}
\DeclareAbbreviation\BaltA{Baltic Astron.}
\DeclareAbbreviation\BASI{Bull. Astron. Soc. India}
\DeclareAbbreviation\BeSN{Be Newslett.}
\DeclareAbbreviation\BHarO{Harvard Coll. Obs. Bull.}
\DeclareAbbreviation\CBET{Cent. Bur. Electron. Telegrams}
\DeclareAbbreviation\ChJAA{Chinese J. of Astron. and Astrophys.}
\DeclareAbbreviation\CoSka{Contributions of the Astronomical Observatory Skalnat\'e Pleso}
\DeclareAbbreviation\GCN{GRB Coord. Netw. Circ.}
\DeclareAbbreviation\ibvs{IBVS}
\DeclareAbbreviation\JAD{J. Astron. Data}
\DeclareAbbreviation\JAVSO{J. American Assoc. Variable Star Obs.}
\DeclareAbbreviation\JBAA{J. Br. Astron. Assoc.}
\DeclareAbbreviation\JPSJ{J. Phys. Soc. Japan}
\DeclareAbbreviation\LowOB{Lowell Obs. Bull.}
\DeclareAbbreviation\MitVS{Mitteil. Ver\"{a}nderl. Sterne}
\DeclareAbbreviation\MmSAI{Mem. Soc. Astron. Ital.}
\DeclareAbbreviation\Msngr{Messenger}
\DeclareAbbreviation\NewA{New Astron.}
\DeclareAbbreviation\NewAR{New Astron. Rev.}
\DeclareAbbreviation\OAP{Odessa Astron. Publ.}
\DeclareAbbreviation\Obs{Observatory}
\DeclareAbbreviation\OEJV{Open Eur. J. on Variable Stars}
\DeclareAbbreviation\PASA{Publ. Astron. Soc. Australia}
\DeclareAbbreviation\PCCP{Phys. Chem. Chem. Phys.}
\DeclareAbbreviation\PAZh{Pis'ma AZh}
\DeclareAbbreviation\PhR{Phys. Rep.}
\DeclareAbbreviation\PVSS{Publ. Variable Stars Sect. R. Astron. Soc. New Zealand}
\DeclareAbbreviation\PZ{Perem. Zvezdy}
\DeclareAbbreviation\PZP{Perem. Zvezdy, Prilozh.}
\DeclareAbbreviation\QJRAS{QJRAS}
\DeclareAbbreviation\RMxAA{Rev. Mexicana Astron. Astrof.}
\DeclareAbbreviation\RvMA{Reviews of Modern Astron.}
\DeclareAbbreviation\SASS{Society for Astronom. Sciences Ann. Symp.}
\DeclareAbbreviation\Sci{Science}
\DeclareAbbreviation\SPIE{SPIE Proc.}
\DeclareAbbreviation\SvA{Soviet Astronomy}
\DeclareAbbreviation\SvAL{Soviet Astronomy Letters}
\DeclareAbbreviation\VeSon{Ver\"{o}ff. Sternw. Sonneberg}
\DeclareAbbreviation\VSOLJBul{VSOLJ Variable Star Bull.}
\DeclareAbbreviation\yCat{VizieR Online Data Catalog}
\DeclareAbbreviation\ZA{Z. Astrophys.}

\def\PublisherCambridge{Cambridge: Cambridge University Press}

\newcounter{author}
\def\authorcount#1#2{\refstepcounter{author}\label{#1}
                     \altaffiltext{\ref{#1}}{#2}}

\begin{document}
\SetRunningHead{T. Kato et al.}{Period Variations in SU UMa-Type Dwarf Novae}

\Received{201X/XX/XX}%{yyyy/mm/dd}
\Accepted{201X/XX/XX}%{yyyy/mm/dd}

\title{Survey of Period Variations of Superhumps in SU UMa-Type Dwarf Novae.
    II: The Second Year (2009--2010)}

\author{Taichi~\textsc{Kato},\altaffilmark{\ref{affil:Kyoto}*}
        Hiroyuki~\textsc{Maehara},\altaffilmark{\ref{affil:HidaKwasan}}
        Makoto~\textsc{Uemura},\altaffilmark{\ref{affil:Uemura}}
%% 21
        Arne~\textsc{Henden},\altaffilmark{\ref{affil:AAVSO}}
%% 18
        Enrique~de~\textsc{Miguel},\altaffilmark{\ref{affil:Miguel}}$^,$\altaffilmark{\ref{affil:Miguel2}}
        Ian~\textsc{Miller},\altaffilmark{\ref{affil:Miller}}
%% 15
        Pavol~A.~\textsc{Dubovsky},\altaffilmark{\ref{affil:Dubovsky}}
        Igor~\textsc{Kudzej},\altaffilmark{\ref{affil:Dubovsky}}
%% 12
        Seiichiro~\textsc{Kiyota},\altaffilmark{\ref{affil:Kis}}
        Franz-Josef~\textsc{Hambsch},\altaffilmark{\ref{affil:GEOS}}$^,$\altaffilmark{\ref{affil:BAV}}$^,$\altaffilmark{\ref{affil:Hambsch}}
%% 10
        Kenji~\textsc{Tanabe},\altaffilmark{\ref{affil:OUS}}
        Kazuyoshi~\textsc{Imamura},\altaffilmark{\ref{affil:OUS}}
        Nanae~\textsc{Kunitomi},\altaffilmark{\ref{affil:OUS}}
        Ryosuke~\textsc{Takagi},\altaffilmark{\ref{affil:OUS}}
        Mikiha~\textsc{Nose},\altaffilmark{\ref{affil:OUS}}
        Hidehiko~\textsc{Akazawa},\altaffilmark{\ref{affil:OUS}}
        Gianluca~\textsc{Masi},\altaffilmark{\ref{affil:Masi}} 
%% 9
        Shinichi~\textsc{Nakagawa},\altaffilmark{\ref{affil:OKU}}
        Eriko~\textsc{Iino},\altaffilmark{\ref{affil:OKU}}
        Ryo~\textsc{Noguchi},\altaffilmark{\ref{affil:OKU}}
        Katsura~\textsc{Matsumoto},\altaffilmark{\ref{affil:OKU}}
        Daichi~\textsc{Fujii},\altaffilmark{\ref{affil:OKU}}
        Hiroshi~\textsc{Kobayashi},\altaffilmark{\ref{affil:OKU}}
        Kazuyuki~\textsc{Ogura},\altaffilmark{\ref{affil:OKU}}
        Sachi~\textsc{Ohtomo},\altaffilmark{\ref{affil:OKU}}
        Kousei~\textsc{Yamashita},\altaffilmark{\ref{affil:OKU}}
        Hirofumi~\textsc{Yanagisawa},\altaffilmark{\ref{affil:OKU}}
%% 6
        Hiroshi~\textsc{Itoh},\altaffilmark{\ref{affil:Ioh}}
        Greg~\textsc{Bolt},\altaffilmark{\ref{affil:Bolt}}
        Berto~\textsc{Monard},\altaffilmark{\ref{affil:Monard}}
%% 5
        Tomohito~\textsc{Ohshima},\altaffilmark{\ref{affil:Kyoto}}
        Jeremy~\textsc{Shears},\altaffilmark{\ref{affil:Shears}}
%% 4
        Javier~\textsc{Ruiz},\altaffilmark{\ref{affil:Ruiz}}
%% 3
        Akira~\textsc{Imada},\altaffilmark{\ref{affil:Imada}}
        Arto~\textsc{Oksanen},\altaffilmark{\ref{affil:Nyrola}}
        Peter~\textsc{Nelson},\altaffilmark{\ref{affil:Nelson}}
        Tomas~L.~\textsc{Gomez},\altaffilmark{\ref{affil:Gomez}}
        Bart~\textsc{Staels},\altaffilmark{\ref{affil:AAVSO}}$^,$\altaffilmark{\ref{affil:Staels}}
        David~\textsc{Boyd},\altaffilmark{\ref{affil:DavidBoyd}}$^,$\altaffilmark{\ref{affil:BAAVSS}}
%% 2
        Irina~B.~\textsc{Voloshina},\altaffilmark{\ref{affil:Voloshina}}
        Thomas~\textsc{Krajci},\altaffilmark{\ref{affil:Krajci}}
        Tim~\textsc{Crawford},\altaffilmark{\ref{affil:Crawford}}
        Chris~\textsc{Stockdale},\altaffilmark{\ref{affil:Stockdale}}
        Michael~\textsc{Richmond},\altaffilmark{\ref{affil:RIT}}
        Etienne~\textsc{Morelle},\altaffilmark{\ref{affil:Morelle}}
%% 1
        Rudolf~\textsc{Nov\'{a}k},\altaffilmark{\ref{affil:Novak}}
        Daisaku~\textsc{Nogami},\altaffilmark{\ref{affil:HidaKwasan}}
        Ryoko~\textsc{Ishioka},\altaffilmark{\ref{affil:Ishioka}}
        Steve~\textsc{Brady},\altaffilmark{\ref{affil:Brady}}
        Mike~\textsc{Simonsen},\altaffilmark{\ref{affil:Simonsen}} 
        Elena~P.~\textsc{Pavlenko},\altaffilmark{\ref{affil:Pavlenko}}
        Frederick~A.~\textsc{Ringwald},\altaffilmark{\ref{affil:Ringwald}}
        Tetsuya~\textsc{Kuramoto},\altaffilmark{\ref{affil:Kyoto}}
        Atsushi~\textsc{Miyashita},\altaffilmark{\ref{affil:Seikei}}
        Roger~D.~\textsc{Pickard},\altaffilmark{\ref{affil:BAAVSS}}$^,$\altaffilmark{\ref{affil:Pickard}}
        Tom\'{a}\v{s}~\textsc{Hynek},\altaffilmark{\ref{affil:JohannPalisa}}
        Shawn~\textsc{Dvorak},\altaffilmark{\ref{affil:Dvorak}}
%% outbursts
        Rod~\textsc{Stubbings},\altaffilmark{\ref{affil:Stubbings}}
        Eddy~\textsc{Muyllaert},\altaffilmark{\ref{affil:VVSBelgium}}
}

\authorcount{affil:Kyoto}{
     Department of Astronomy, Kyoto University, Kyoto 606-8502}
\email{$^*$tkato@kusastro.kyoto-u.ac.jp}

\authorcount{affil:HidaKwasan}{
     Kwasan and Hida Observatories, Kyoto University, Yamashina,
     Kyoto 607-8471}

\authorcount{affil:Uemura}{
     Astrophysical Science Center, Hiroshima University, Kagamiyama, 1-3-1
     Higashi-Hiroshima 739-8526}

\authorcount{affil:AAVSO}{
     American Association of Variable Star Observers, 49 Bay State Rd.,
     Cambridge, MA 02138, USA}

\authorcount{affil:Miguel}{
     Departamento de F\'isica Aplicada, Facultad de Ciencias
     Experimentales, Universidad de Huelva,
     21071 Huelva, Spain}

\authorcount{affil:Miguel2}{
     Center for Backyard Astrophysics, Observatorio del CIECEM,
     Parque Dunar, Matalasca\~nas, 21760 Almonte, Huelva, Spain}

\authorcount{affil:Miller}{
     Furzehill House, Ilston, Swansea, SA2 7LE, UK}

\authorcount{affil:Dubovsky}{
     Vihorlat Observatory, Mierova 4, Humenne, Slovakia}

\authorcount{affil:Kis}{
     Variable Star Observers League in Japan (VSOLJ), 405-1003 Matsushiro, Tsukuba, Ibaraki 305-0035}

\authorcount{affil:GEOS}{
     Groupe Europ\'een d'Observations Stellaires (GEOS),
     23 Parc de Levesville, 28300 Bailleau l'Ev\^eque, France}

\authorcount{affil:BAV}{
     Bundesdeutsche Arbeitsgemeinschaft f\"ur Ver\"anderliche Sterne
     (BAV), Munsterdamm 90, 12169 Berlin, Germany}

\authorcount{affil:Hambsch}{
     Vereniging Voor Sterrenkunde (VVS), Oude Bleken 12, 2400 Mol, Belgium}

\authorcount{affil:OUS}{
     Department of Biosphere-Geosphere System Science, Faculty of Informatics,
     Okayama University of Science, 1-1 Ridai-cho, Okayama, Okayama 700-0005}

\authorcount{affil:Masi}{
     The Virtual Telescope Project, Via Madonna del Loco 47, 03023
     Ceccano (FR), Italy}

\authorcount{affil:OKU}{
     Osaka Kyoiku University, 4-698-1 Asahigaoka, Osaka 582-8582}

\authorcount{affil:Ioh}{
     VSOLJ, 1001-105 Nishiterakata, Hachioji, Tokyo 192-0153}

\authorcount{affil:Bolt}{
     Camberwarra Drive, Craigie, Western Australia 6025, Australia}

\authorcount{affil:Monard}{
     Bronberg Observatory, Center for Backyard Astronomy Pretoria,
     PO Box 11426, Tiegerpoort 0056, South Africa}

\authorcount{affil:Shears}{
     ``Pemberton'', School Lane, Bunbury, Tarporley, Cheshire, CW6 9NR, UK}

\authorcount{affil:Ruiz}{
     Observatory of Cantabria,
     Centro de Investigaci\'on del Medio Ambiente (CIMA)
     Instituto de F\'isica de Cantabria (IFCA),
     Agrupaci\'on Astron\'omica C\'antabra (AAC),
     Ctra. de Rocamundo s/n, Valderredible, Cantabria, Spain
}

\authorcount{affil:Imada}{
     Okayama Astrophysical Observatory, National Astronomical
     Observatory of Japan, Asakuchi, Okayama 719-0232}

\authorcount{affil:Nyrola}{
     Nyrola observatory, Jyvaskylan Sirius ry, Vertaalantie
     419, FI-40270 Palokka, Finland}

\authorcount{affil:Nelson}{
     RMB 2493, Ellinbank 3820, Australia}

\authorcount{affil:Gomez}{
     ICMAT (CSIC-UAM-UC3M-UCM), Serrano 113bis, 28006 Madrid, Spain}

\authorcount{affil:Staels}{
     Center for Backyard Astrophysics (Flanders),
     American Association of Variable Star Observers (AAVSO),
     Alan Guth Observatory, Koningshofbaan 51, Hofstade, Aalst, Belgium}

\authorcount{affil:DavidBoyd}{
     Silver Lane, West Challow, Wantage, OX12 9TX, UK}

\authorcount{affil:Voloshina}{
     Sternberg Astronomical Institute, Moscow University,
     Universitetskiy prospekt 13, Moscow 119992, Russia}

\authorcount{affil:Krajci}{
     Astrokolkhoz Observatory,
     Center for Backyard Astrophysics New Mexico, PO Box 1351 Cloudcroft,
     New Mexico 83117, USA}

\authorcount{affil:Crawford}{
     Arch Cape Observatory, 79916 W. Beach Road, Arch Cape, OR 97102}

\authorcount{affil:BAAVSS}{
     The British Astronomical Association, Variable Star Section (BAA VSS),
     Burlington House, Piccadilly, London, W1J 0DU, UK}

\authorcount{affil:Stockdale}{
     8 Matta Drive, Churchill, Victoria  3842, Australia}

\authorcount{affil:RIT}{
     Physics Department, Rochester Institute of Technology, Rochester,
     New York 14623, USA}

\authorcount{affil:Morelle}{
     9 rue Vasco de GAMA, 59553 Lauwin Planque, France}

\authorcount{affil:Novak}{
     Institute of Computer Science, Faculty of Civil Engineering,
     Brno University of Technology, 602 00 Brno, Czech Republic}

\authorcount{affil:Ishioka}{
     Subaru Telescope, National Astronomical Observatory of Japan, 650
     North A'ohoku Place, Hilo, HI 96720, USA}

\authorcount{affil:Brady}{
     5 Melba Drive, Hudson, NH 03051, USA}

\authorcount{affil:Simonsen}{
     AAVSO, C. E. Scovil Observatory, 2615 S. Summers Rd., Imlay City,
     Michigan 48444, USA}

\authorcount{affil:Pavlenko}{
     Crimean Astrophysical Observatory, 98409, Nauchny, Crimea, Ukraine}

\authorcount{affil:Ringwald}{
     Department of Physics, California State University, Fresno,
     2345 East San Ramon Avenue, MS MH37, Fresno, CA 93740-8031, USA}

\authorcount{affil:Seikei}{
     Seikei Meteorological Observatory, Seikei High School}

\authorcount{affil:Pickard}{
     3 The Birches, Shobdon, Leominster, Herefordshire, HR6 9NG, UK}

\authorcount{affil:JohannPalisa}{
     Project Eridanus,
     Observatory and Planetarium of Johann Palisa, VSB -- Technical
     University Ostrava, Trida 17. listopadu 15, Ostrava -- Poruba 708 33, 
     Czech Republic}

\authorcount{affil:Dvorak}{
     Rolling Hills Observatory, 1643 Nightfall Drive,
     Clermont, FL 34711, USA}

\authorcount{affil:Stubbings}{
     Tetoora Observatory, Tetoora Road, Victoria, Australia}

\authorcount{affil:VVSBelgium}{
     Vereniging Voor Sterrenkunde (VVS),  Moffelstraat 13 3370
     Boutersem, Belgium}

%%% end:list of authors


\maketitle

\begin{abstract}
   As an extension of the project in \citet{Pdot}, we collected
times of superhump maxima for 61 SU UMa-type dwarf novae mainly observed
during the 2009--2010 season.  The newly obtained data confirmed
the basic findings reported in \citet{Pdot}: the presence of stages
A--C, as well as the predominance of positive period derivatives
during stage B in systems with superhump periods shorter than 0.07 d.
There was a systematic difference in period derivatives for systems
with superhump periods longer than 0.075 d between this study and
\citet{Pdot}.  We suggest that this difference is possibly caused by
the relative lack of frequently outbursting SU UMa-type dwarf novae
in this period regime in the present study.
We recorded a strong beat phenomenon during the 2009 superoutburst of IY UMa.
The close correlation between the beat period and superhump period
suggests that the changing angular velocity of the apsidal motion of
the elliptical disk is responsible for the variation of superhump periods.
We also described three new WZ Sge-type objects with established
early superhumps and one with likely early superhumps.
We also suggest that two systems, VX For and EL UMa, are WZ Sge-type dwarf
novae with multiple rebrightenings.  The $O-C$ variation in
OT J213806.6$+$261957 suggests that the frequent absence of rebrightenings in
very short-$P_{\rm orb}$ objects can be a result of sustained superoutburst
plateau at the epoch when usual SU UMa-type dwarf novae return to quiescence
preceding a rebrightening.
We also present a formulation for a variety of Bayesian extension to
traditional period analyses.
\end{abstract}

\newpage

\section{Introduction}

   In paper \citet{Pdot}, we surveyed period variations of superhumps
in SU UMa-type dwarf novae (for general information of SU UMa-type
dwarf novae and superhumps, see \cite{war95book}).
\citet{Pdot} indicated that evolution of superhump period ($P_{\rm SH}$)
is generally composed of three distinct stages: early evolutionary
stage with a longer superhump period (stage A), middle stage with
systematically varying periods (stage B), final stage with a shorter,
stable superhump period (C).  It was also shown that the period
derivatives ($P_{\rm dot} = \dot{P}/P$)
during stage B is correlated with $P_{\rm SH}$,
or binary mass-ratios ($q = M_2/M_1$).  Although this relation
commonly applies to classical SU UMa-type dwarf novae,
WZ Sge-type dwarf novae, a subtype of SU UMa-type dwarf novae with
very infrequent superoutbursts, tend to deviate from this picture:
they rarely show a distinct stage B--C transition, and some objects
show relatively small period derivatives, and they frequently exhibit
unusual multiple post-superoutburst rebrightenings.
The origin of these relations is not yet well understood.
We have extended the survey in order to test whether this picture is
applicable to new SU UMa-type systems and to the degree of diversity
between systems in a larger sample.
We include in this paper newly recorded objects and superoutbursts
since the publication of \citet{Pdot}.  Some of new observations have
led to revisions of analysis in \citet{Pdot}.
We also include a few past superoutburst not analyzed in the previous studies.

   The structure of the paper follows the scheme in \citet{Pdot},
in which we mostly restricted to superhump timing analysis.
We also include some more details if the paper provides the first
solid presentation of individual objects.

\section{Observation and Analysis}\label{sec:obs}

   The data were obtained under campaigns led by the VSNET Collaboration
\citep{VSNET}.  In some objects, we used archival data for published
papers, and the public data from the AAVSO International Database\footnote{
$<$http://www.aavso.org/data/download/$>$.
}.
The majority of the data were acquired
by time-resolved CCD photometry by using 30 cm-class telescopes, whose
observational details on individual objects will be presented in
future papers dealing with analysis and discussion on individual objects.
The list of outbursts and observers is summarized in table \ref{tab:outobs}.
The data analysis was performed just in the same way described
in \citet{Pdot}.  In this paper, we introduce Bayesian extension
to traditional method (Appendix), and used them if they give
significantly improved results.

   The derived $P_{\rm SH}$, $P_{\rm dot}$ and other parameters
are listed in table \ref{tab:perlist} as in the same format in
\citet{Pdot}.  The definitions of parameters $P_1, P_2, E_1, E_2$
and $P_{\rm dot}$ are the same as in \citet{Pdot}.
As in \citet{Pdot}, we present comparisons of
$O-C$ diagrams between different superoutbursts since this has been
one of the motivations of these surveys (cf. \cite{uem05tvcrv}).

\begin{table*}
\caption{List of Superoutbursts.}\label{tab:outobs}
\begin{center}
\begin{tabular}{ccccc}
\hline\hline
Subsection & Object & Year & Observers or references\commenta & ID\commentb \\
\hline
\ref{obj:kxaql}    & KX Aql     & 2010 & Mhh, AAVSO, Kis, GOT, DPV, OKU & \\
\ref{obj:nncam}    & NN Cam     & 2009 & AAVSO, deM, Mhh, JSh, IMi & \\
\ref{obj:v591cen}  & V591 Cen   & 2010 & MLF, GBo & \\
\ref{obj:zcha}     & Z Cha      & 2010 & AAVSO, Kis & \\
\ref{obj:pucma}    & PU CMa     & 2009 & OUS, Mhh, Kis & \\
\ref{obj:aqcmi}    & AQ CMi     & 2010 & GBo & \\
\ref{obj:gzcnc}    & GZ Cnc     & 2010 & OKU, OAO, Mhh, deM, OUS & \\
\ref{obj:gocom}    & GO Com     & 2010 & deM, Mhh, OKU, DPV & \\
\ref{obj:tvcrv}    & TV Crv     & 2009 & Kis & \\
\ref{obj:v337cyg}  & V337 Cyg   & 2010 & IMi, KU, OKU, DPV & \\
\ref{obj:v1113cyg} & V1113 Cyg  & 2003 & \citet{bak10v1113cyg} & \\
\ref{obj:v1454cyg} & V1454 Cyg  & 2009 & HMB, Mhh, IMi, AAVSO & \\
\ref{obj:aqeri}    & AQ Eri     & 2010 & Nel, Mhh, Kis, Ioh & \\
\ref{obj:vxfor}    & VX For     & 2009 & MLF, Nyr, Nel, AAVSO, Sto, Mhh, Kis & \\
\ref{obj:awgem}    & AW Gem     & 2010 & Hyn & \\
\ref{obj:irgem}    & IR Gem     & 2010 & deM & \\
\ref{obj:v592her}  & V592 Her   & 2010 & AAVSO, OUS, Mhh, OKU, IMi, DPV, RIT, HMB, KU, Boy, OAO, MEV & \\
\ref{obj:v660her}  & V660 Her   & 2009 & AAVSO & \\
\ref{obj:v844her}  & V844 Her   & 2009 & DPV, AAVSO & \\
                   & V844 Her   & 2010 & Mhh & \\
\ref{obj:cthya}    & CT Hya     & 2010 & Mhh & \\
\ref{obj:v699oph}  & V699 Oph   & 2010 & deM, OUS & \\
\ref{obj:v1032oph} & V1032 Oph  & 2010 & Mhh, deM, AAVSO, Kis, Rui & \\
\ref{obj:v2051oph} & V2051 Oph  & 2010 & Sto, OUS, Kis & \\
\ref{obj:efpeg}    & EF Peg     & 2009 & Mhh, IMi, OUS, AAVSO & \\
\ref{obj:v368peg}  & V368 Peg   & 2009 & CTX, AAVSO, IMi, Nyr, HMB & \\
\ref{obj:uvper}    & UV Per     & 2010 & DPV, PXR, deM, AAVSO & \\
\ref{obj:eipsc}    & EI Psc     & 2009 & GOT, Kis & \\
\ref{obj:ektra}    & EK TrA     & 2009 & Nel & \\
\ref{obj:suuma}    & SU UMa     & 2010 & DPV, SAc & \\
\ref{obj:bcuma}    & BC UMa     & 2009 & Nyr, Mhh & \\
\ref{obj:eluma}    & EL UMa     & 2010 & Mhh, Ioh, DPV, HMB & \\
\ref{obj:iyuma}    & IY UMa     & 2009 & \citet{Pdot}, DPV, AAVSO & \\
\ref{obj:ksuma}    & KS UMa     & 2010 & OUS & \\
\ref{obj:mruma}    & MR UMa     & 2010 & OUS, OKU & \\
\ref{obj:tyvul}    & TY Vul     & 2010 & MEV, AAVSO, IMi, BSt & \\
\ref{obj:j0423}    & 1RXS J0423 & 2010 & OUS, Mhh, deM & \\
\ref{obj:j0532}    & 1RXS J0532 & 2009 & DPV, Mhh & \\
\ref{obj:asas2243} & ASAS J2243 & 2009 & HMB, Kra, Mhh, JSh, GBo, IMi, deM, Ioh, AAVSO & \\
\ref{obj:lanning420} & Lanning 420 & 2010 & KU, AAVSO, deM, BXS, Mhh, HMB, Rui & \\
\ref{obj:pg0149}   & PG 0149    & 2009 & IMi, HMB, Mhh & \\
\ref{obj:j1715}    & RX J1715   & 2009 & \citet{she10j1715}, Mhh & \\
\ref{obj:j0129}    & SDSS J0129 & 2009 & JSh, Mhh, AAVSO, Boy & \\
\ref{obj:j0310}    & SDSS J0310 & 2009b & IMi & \\
\ref{obj:j0732}    & SDSS J0732 & 2010 & AAVSO, JSh & \\
\ref{obj:j0838}    & SDSS J0838 & 2010 & Mhh, deM & \\
\ref{obj:j0839}    & SDSS J0839 & 2010 & deM, JSh & \\
\ref{obj:j0903}    & SDSS J0903 & 2010 & deM, Vol, AAVSO, Mas, Mhh & \\
\hline
  \multicolumn{5}{l}{\commenta Key to observers:
AKh (Astrokolkhoz team),
Boy (D. Boyd),
BSt\commentc (B. Staels),
BXS\commentc (S. Brady),
CTX\commentc (T. Crawford),
}\\ \multicolumn{4}{l}{
deM (E. de Miguel),
DKS\commentc (S. Dvorak), 
DPV (P. Dubovsky),
GBo (G. Bolt),
GOT\commentc (T. Gomez),
HHO (Higashi-Hiroshima Obs.), 
}\\ \multicolumn{4}{l}{
Hid (Hida Obs.), 
HMB (F.-J. Hambsch),
Hyn (T. Hynek),
IMi\commentc (I. Miller),
Ioh (H. Itoh),
JSh\commentc (J. Shears),
}\\ \multicolumn{4}{l}{
Kis (S. Kiyota),
KU (Kyoto U., campus obs.),
Kra (T. Krajci),
Mas (G. Masi),
MEV\commentc (E. Morelle),
}\\ \multicolumn{4}{l}{
Mhh (H. Maehara),
MLF (B. Monard),
Nel\commentc (P. Nelson),
Nov (R. Nov\'ak),
Nyr\commentc (Nyrola and Hankasalmi Obs.),
}\\ \multicolumn{4}{l}{
OAO (Okayama Astrophys. Obs.),
OKU (Osaka Kyoiku U.),
OUS (Okayama U. of Science),
Pav (E. Pavlenko),
}\\ \multicolumn{4}{l}{
PXR\commentc (R. Pickard),
Rin (F. Ringwald),
RIT (M. Richmond),
Rui (J. Ruiz),
SAc (Seikei High School),
}\\ \multicolumn{4}{l}{
Sto (C. Stockdale),
Vol (I. Voloshina),
AAVSO (AAVSO database)
} \\
  \multicolumn{5}{l}{\commentb Original identifications or discoverers.} \\
  \multicolumn{5}{l}{\commentc Inclusive of observations from the AAVSO database.} \\
\end{tabular}
\end{center}
\end{table*}

\addtocounter{table}{-1}
\begin{table*}
\caption{List of Superoutbursts. (continied)}
\begin{center}
\begin{tabular}{ccccc}
\hline\hline
Subsection & Object & Year & Observers or references\commenta & ID\commentb \\
\hline
\ref{obj:j1152}    & SDSS J1152 & 2009 & Mhh, Kis & \\
\ref{obj:j1250}    & SDSS J1250 & 2008 & BSt & \\
                   & SDSS J1250 & 2009 & IMi, Mhh & \\
\ref{obj:j1502}    & SDSS J1502 & 2009 & AAVSO, Rin, RIT, Mas, DPV, Mhh, Rui, IMi, Ioh & \\
\ref{obj:j1610}    & SDSS J1610 & 2009 & GBo, Mhh, HMB, CTX, IMi, Boy & \\
\ref{obj:j1625}    & SDSS J1625 & 2010 & Nov, deM, RIT, BSt, Mas, HMB, Mhh, IMi & \\
\ref{obj:j1637}    & SDSS J1637 & 2004 & KU, MLF, GBo, Hid & \\
\ref{obj:j1653}    & SDSS J1653 & 2010 & OKU, DPV & \\
\ref{obj:j2048}    & SDSS J2048 & 2009 & IMi & \\
\ref{obj:j0406}    & OT J0406   & 2010 & Mhh & Itagaki \citep{yam08j0406cbet1463} \\
\ref{obj:j0506}    & OT J0506   & 2009 & Mhh, \citet{kry10j0506}, Mas & \citet{kry10j0506} \\
\ref{obj:j1026}    & OT J1026   & 2010 & DPV, IMi, Mas & Itagaki \citep{yam09j1026cbet1644} \\
\ref{obj:j1044}    & OT J1044   & 2010 & Mhh, OAO, DPV, IMi, Ioh, & CSS100217:104411$+$211307 \\
                   &            &      & HMB, Kis, Pav, deM & \\
\ref{obj:j1122}    & OT J1122   & 2010 & MLF, Mhh, AKh, Mas & CSS100603:112253$-$111037 \\
\ref{obj:j1440}    & OT J1440   & 2009 & \citet{boy10j1440}, \citet{Pdot} & CSS090530:144011$+$494734 \\
\ref{obj:j1631}    & OT J1631   & 2010 & Mhh, IMi & CSS080505:163121$+$103134 \\
\ref{obj:j1703}    & OT J1703   & 2009 & Mas & CSS090622:170344$+$090835 \\
\ref{obj:j1821}    & OT J1821   & 2010 & OKU, Mhh, deM & Itagaki (vsnet-alert 11952) \\
\ref{obj:j2138}    & OT J2138   & 2010 & Mhh, deM, OUS, AAVSO, Vol, DPV, HHO, & Yi \citep{yam10j2138cbet2273}, \\
                   &            &      & DKS, Kis, GOT, OKU, SXN, Mas, Rui, Ioh & Kaneko \citep{nak10j2138cbet2275} \\
\ref{obj:j2158}    & OT J2158   & 2010 & AKh, HMB & CSS100615:215815$+$094709 \\
\ref{obj:j2230}    & OT J2230   & 2009 & MLF, GBo, Mhh, Mas & CSS090727:223003$-$145835 \\
\ref{obj:j2344}    & OT J2344   & 2010 & MLF, Mhh, HMB, KU, Mas, deM & MLS100904:234441$-$001206 \\
\hline
\end{tabular}
\end{center}
\end{table*}

\begin{table*}
\caption{Superhump Periods and Period Derivatives}\label{tab:perlist}
\begin{center}
\begin{tabular}{cccccccccccccc}
\hline\hline
Object & Year & $P_1$ (d) & err & \multicolumn{2}{c}{$E_1$\commenta} & $P_{\rm dot}$\commentb & err\commentb & $P_2$ (d) & err & \multicolumn{2}{c}{$E_2$\commenta} & $P_{\rm orb}$ (d) & Q\commentc \\
\hline
KX Aql & 2010 & -- & -- & -- & -- & -- & -- & 0.061322 & 0.000026 & 7 & 119 & 0.06035 & B \\
NN Cam & 2009 & 0.074264 & 0.000009 & 19 & 87 & 1.1 & 1.5 & 0.073876 & 0.000045 & 87 & 134 & 0.0717 & A \\
V591 Cen & 2010 & 0.060299 & 0.000019 & 0 & 138 & 6.5 & 1.3 & -- & -- & -- & -- & -- & B \\
Z Cha & 2010 & -- & -- & -- & -- & -- & -- & 0.076977 & 0.000054 & 0 & 80 & 0.074499 & C \\
PU CMa & 2009 & 0.058090 & 0.000032 & 0 & 90 & 9.7 & 2.3 & 0.057219 & 0.000129 & 121 & 192 & 0.056694 & C \\
AQ CMi & 2010 & 0.066178 & 0.000006 & 15 & 106 & -- & -- & -- & -- & -- & -- & -- & C \\
GZ Cnc & 2010 & 0.092774 & 0.000021 & 0 & 44 & -- & -- & 0.089717 & 0.000124 & 42 & 88 & 0.08825 & C \\
GO Com & 2010 & 0.063072 & 0.000024 & 0 & 103 & 10.5 & 0.9 & 0.062660 & 0.000027 & 101 & 158 & -- & A \\
TV Crv & 2009 & 0.065058 & 0.000027 & 0 & 78 & 9.1 & 3.0 & 0.064881 & 0.000136 & 77 & 94 & 0.0629 & C \\
V337 Cyg & 2010 & 0.070326 & 0.000042 & 0 & 95 & 2.9 & 3.2 & 0.069578 & 0.000125 & 95 & 139 & -- & C \\
V1454 Cyg & 2009 & 0.057650 & 0.000039 & 0 & 95 & 10.3 & 3.5 & -- & -- & -- & -- & -- & B \\
AQ Eri & 2010 & 0.062370 & 0.000068 & 0 & 82 & 20.4 & 5.2 & -- & -- & -- & -- & 0.06094 & C \\
VX For & 2009 & 0.061327 & 0.000012 & 0 & 101 & 1.0 & 1.4 & 0.061086 & 0.000056 & 99 & 204 & -- & B \\
AW Gem & 2010 & 0.079056 & 0.000025 & 0 & 26 & -- & -- & 0.078668 & 0.000027 & 24 & 103 & 0.07621 & C \\
IR Gem & 2010 & 0.070834 & 0.000005 & 0 & 58 & -- & -- & -- & -- & -- & -- & 0.0684 & C \\
V592 Her & 2010 & 0.056607 & 0.000016 & 35 & 216 & 7.4 & 0.6 & 0.056390 & 0.000088 & 216 & 300 & 0.0561 & BE \\
V660 Her & 2009 & -- & -- & -- & -- & -- & -- & 0.080692 & 0.000089 & 0 & 61 & -- & C \\
V844 Her & 2009 & 0.055923 & 0.000021 & 0 & 111 & 9.5 & 1.7 & -- & -- & -- & -- & 0.054643 & B \\
V844 Her & 2010 & 0.055764 & 0.000004 & 0 & 54 & -- & -- & -- & -- & -- & -- & 0.054643 & CG \\
CT Hya & 2010 & 0.066505 & 0.000001 & 14 & 90 & -- & -- & -- & -- & -- & -- & -- & C \\
V699 Oph & 2010 & 0.070319 & 0.000112 & 0 & 42 & -- & -- & 0.069880 & 0.000071 & 42 & 99 & -- & B \\
V1032 Oph & 2010 & 0.085342 & 0.000047 & 0 & 106 & -- & -- & -- & -- & -- & -- & 0.081056 & CG \\
V2051 Oph & 2010 & 0.064238 & 0.000083 & 0 & 65 & -- & -- & -- & -- & -- & -- & 0.062428 & C \\
EF Peg & 2009 & 0.087347 & 0.000182 & 10 & 30 & -- & -- & 0.086784 & 0.000103 & 68 & 125 & -- & C \\
V368 Peg & 2009 & 0.070358 & 0.000040 & 0 & 99 & 9.5 & 1.8 & 0.069947 & 0.000022 & 97 & 220 & -- & B \\
UV Per & 2010 & 0.066708 & 0.000026 & 0 & 32 & 21.1 & 6.7 & -- & -- & -- & -- & 0.06489 & CG \\
EI Psc & 2009 & 0.046352 & 0.000046 & 0 & 10 & -- & -- & -- & -- & -- & -- & 0.044567 & C \\
EK TrA & 2009 & 0.064832 & 0.000006 & 29 & 139 & 0.8 & 0.6 & -- & -- & -- & -- & 0.06288 & CM \\
SU UMa & 2010 & 0.079070 & 0.000034 & 0 & 51 & -- & -- & -- & -- & -- & -- & 0.07635 & C \\
BC UMa & 2009 & 0.064553 & 0.000029 & 56 & 144 & 9.5 & 2.7 & 0.064262 & 0.000019 & 141 & 219 & 0.06261 & B \\
IY UMa & 2009 & 0.076210 & 0.000025 & 46 & 115 & 15.1 & 2.3 & 0.075729 & 0.000019 & 113 & 182 & 0.073909 & A \\
MR UMa & 2010 & -- & -- & -- & -- & -- & -- & 0.064820 & 0.000053 & 61 & 125 & -- & C \\
TY Vul & 2010 & 0.080465 & 0.000048 & 0 & 39 & 6.3 & 10.7 & 0.080075 & 0.000027 & 37 & 89 & -- & B \\
1RXS J0423 & 2010 & 0.078456 & 0.000033 & 12 & 40 & 23.1 & 7.5 & 0.077921 & 0.000075 & 79 & 142 & 0.07632 & B \\
1RXS J0532 & 2009 & 0.057136 & 0.000023 & 0 & 145 & 10.1 & 1.0 & 0.056903 & 0.000030 & 140 & 215 & 0.05620 & A \\
ASAS J2243 & 2009 & 0.069809 & 0.000014 & 0 & 101 & 6.6 & 1.0 & 0.069410 & 0.000047 & 101 & 174 & -- & A \\
Lanning 420 & 2010 & 0.061585 & 0.000022 & 0 & 66 & 5.2 & 3.6 & 0.060941 & 0.000043 & 76 & 175 & -- & B \\
PG 0149 & 2009 & 0.085096 & 0.000034 & 0 & 60 & 14.6 & 2.5 & 0.084602 & 0.000029 & 70 & 154 & 0.08242 & B \\
RX J1715 & 2009 & -- & -- & -- & -- & -- & -- & 0.070782 & 0.000083 & 0 & 48 & 0.0683 & C \\
SDSS J0129 & 2009 & 0.018050 & 0.000100 & 0 & 30 & -- & -- & -- & -- & -- & -- & -- & C \\
SDSS J0310 & 2009b & 0.067863 & 0.000029 & 0 & 31 & -- & -- & -- & -- & -- & -- & -- & C \\
SDSS J0732 & 2010 & 0.079945 & 0.000018 & 0 & 60 & 3.9 & 2.4 & 0.079300 & 0.000028 & 59 & 129 & -- & B \\
SDSS J0839 & 2010 & 0.078520 & 0.000026 & 0 & 15 & -- & -- & 0.078352 & 0.000027 & 13 & 39 & -- & C \\
SDSS J0903 & 2010 & 0.060364 & 0.000050 & 0 & 116 & 13.7 & 3.2 & 0.060073 & 0.000050 & 116 & 182 & 0.059074 & B \\
SDSS J1152 & 2009 & 0.068970 & 0.000042 & 0 & 68 & -- & -- & -- & -- & -- & -- & 0.06770 & C \\
SDSS J1250 & 2008 & 0.060330 & 0.000029 & 0 & 83 & 9.4 & 2.1 & -- & -- & -- & -- & 0.058736 & C \\
SDSS J1502 & 2009 & 0.060463 & 0.000013 & 0 & 101 & 3.7 & 1.5 & 0.060145 & 0.000019 & 95 & 135 & 0.058909 & B \\
SDSS J1610 & 2009 & 0.057820 & 0.000019 & 33 & 150 & 6.4 & 1.2 & -- & -- & -- & -- & 0.05695 & BE \\
SDSS J1625 & 2010 & 0.096054 & 0.000047 & 30 & 104 & 14.9 & 3.7 & -- & -- & -- & -- & -- & B \\
SDSS J1637 & 2004 & -- & -- & -- & -- & -- & -- & 0.069167 & 0.000033 & 0 & 44 & 0.06739 & C \\
\hline
  \multicolumn{13}{l}{\commenta Interval used for calculating the period (corresponding to $E$ in section \ref{sec:individual}).} \\
  \multicolumn{13}{l}{\commentb Unit $10^{-5}$.} \\
  \multicolumn{13}{l}{\commentc Data quality and comments. A: excellent, B: partial coverage or slightly low quality, C: insufficient coverage or}\\
  \multicolumn{13}{l}{\phantom{\commentc} observations with large scatter, G: $P_{\rm dot}$ denotes global $P_{\rm dot}$, M: observational gap in middle stage,}\\
  \multicolumn{13}{l}{\phantom{\commentc} 2: late-stage coverage, the listed period may refer to $P_2$, E: $P_{\rm orb}$ refers to the period of early superhumps.} \\
  \multicolumn{13}{l}{\phantom{\commentc} P: $P_{\rm orb}$ refers to a shorter stable periodicity recorded in outburst.} \\
\end{tabular}
\end{center}
\end{table*}

\addtocounter{table}{-1}
\begin{table*}
\caption{Superhump Periods and Period Derivatives (continued)}
\begin{center}
\begin{tabular}{cccccccccccccc}
\hline\hline
Object & Year & $P_1$ & err & \multicolumn{2}{c}{$E_1$} & $P_{\rm dot}$ & err & $P_2$ & err & \multicolumn{2}{c}{$E_2$} & $P_{\rm orb}$ & Q \\
\hline
SDSS J1653 & 2010 & 0.065221 & 0.000030 & 0 & 52 & -- & -- & 0.064961 & 0.000027 & 50 & 215 & -- & C \\
SDSS J2048 & 2009 & 0.061657 & 0.000024 & 0 & 32 & -- & -- & -- & -- & -- & -- & 0.060597 & C \\
OT J0406 & 2010 & -- & -- & -- & -- & -- & -- & 0.079959 & 0.000027 & 0 & 62 & -- & C \\
OT J0506 & 2009 & 0.069322 & 0.000053 & 23 & 85 & 18.9 & 3.5 & 0.069016 & 0.000085 & 81 & 102 & -- & B \\
OT J1026 & 2010 & 0.068732 & 0.000033 & 0 & 59 & 10.1 & 4.6 & 0.068564 & 0.000023 & 57 & 118 & -- & B \\
OT J1122 & 2010 & 0.045440 & 0.000028 & 0 & 96 & 2.5 & 4.0 & 0.045143 & 0.000131 & 93 & 132 & -- & B \\
OT J1044 & 2010 & -- & -- & -- & -- & -- & -- & 0.060236 & 0.000040 & 195 & 249 & 0.05909 & BEM \\
OT J1440 & 2009 & 0.064615 & 0.000012 & 15 & 54 & $-$5.2 & 3.2 & 0.064412 & 0.000028 & 53 & 136 & -- & B \\
OT J1631 & 2010 & 0.063945 & 0.000024 & 14 & 72 & -- & -- & -- & -- & -- & -- & -- & C2 \\
OT J1703 & 2009 & 0.060853 & 0.000023 & 0 & 51 & -- & -- & -- & -- & -- & -- & -- & C \\
OT J1821 & 2010 & -- & -- & -- & -- & -- & -- & 0.082094 & 0.000031 & 0 & 86 & -- & C \\
OT J2138 & 2010 & 0.055019 & 0.000012 & 36 & 164 & 7.9 & 0.7 & 0.054906 & 0.000006 & 267 & 521 & 0.05450 & AE \\
OT J2158 & 2010 & 0.077552 & 0.000092 & 0 & 5 & -- & -- & -- & -- & -- & -- & -- & C \\
OT J2344 & 2010 & -- & -- & -- & -- & -- & -- & 0.076711 & 0.000030 & 0 & 61 & -- & C \\
\hline
\end{tabular}
\end{center}
\end{table*}

\section{Individual Objects}\label{sec:individual}

\subsection{KX Aquilae}\label{obj:kxaql}

   KX Aql has long been known as a dwarf nova with a low outburst
frequency.  \citet{may68UG} and \citet{may73UG} reported two outbursts
in 1967 and 1972 from AAVSO observations.  \citet{gar79kxaql} examined
photographic plate archives and estimated an interval between outbursts
to be $\sim$ 1000 d.  The object underwent a very bright
($m_{\rm vis} = 11.5$) outburst in 1980 November, which was first
recorded by S. Fujino on November 9.  The long duration of the outburst
inferred from the VSOLJ and AAVSO observations is sufficient to imply
the SU UMa-type nature.  Additional short outbursts were recorded
in 1994 December\footnote{
$<$http://www.kusastro.kyoto-u.ac.jp/vsnet/Mail/1994/vsnet-obs/msg00155.html$>$.
} and 2007 May (baavss-alert 582), which later turned out to be
a normal outburst.

   \citet{tap01kxaql} first presented a spectrum of this object
clearly showing a dwarf nova at a low mass-transfer rate.
\citet{kat01hvvir} listed KX Aql as a candidate for a WZ Sge-type
dwarf nova.  \citet{tho03kxaqlftcampucmav660herdmlyr} obtained
a $P_{\rm orb}$ of 0.06035(3) d from a radial-velocity study.

   The long-waited superoutburst was finally recorded in 2010 
March by T. Gomez (cvnet-outburst 3626).  Subsequent observations
detected superhumps (vsnet-alert 11859, 11862, 11884).
The outburst was followed by a rebrightening (vsnet-alert 11895).

\begin{figure}
  \begin{center}
%    \FigureFile(88mm,110mm){kxaqlshpdm.eps}
    \FigureFile(88mm,110mm){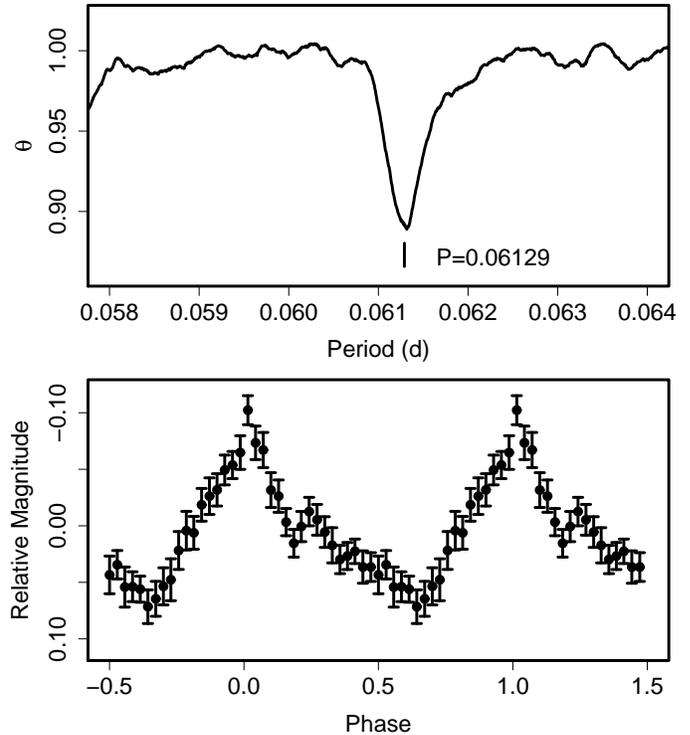}
  \end{center}
  \caption{Superhumps in KX Aql (2010) after BJD 2455266.
     (Upper): PDM analysis.
     (Lower): Phase-averaged profile.}
  \label{fig:kxaqlshpdm}
\end{figure}

   The times of superhump maxima are listed in table \ref{tab:kxaqloc2010}.
There was an apparent break in the period evolution around $E=7$.
We identified this break as a stage B--C transition since the observed
duration of the outburst after the break was only 7 d (in contrast
to $\sim$14 d for typical durations of superoutbursts of short-$P_{\rm orb}$
SU UMa-type dwarf novae) and the superhump period was almost constant
after the break.  The relatively low recorded maximum brightness
($V=13.6$) for a superoutburst also suggests that the true maximum
was missed.  The period for stage B was not determined due to the
shortness of the observed segment.
We adopted a mean superhump period of 0.06128(2) d with the PDM method
(figure \ref{fig:kxaqlshpdm}).
The recorded $\epsilon$ of 1.5 \% is slightly low for this $P_{\rm orb}$
(cf. figure 15 in \cite{Pdot}), but not as low as those of WZ Sge-type
dwarf novae with similar$P_{\rm orb}$.  This interpretation is consistent
with the presence of a single rebrightening, rather than multiple ones
in WZ Sge-type systems with similar $P_{\rm orb}$.
The scaled $q$ is 0.08 based on the refined relation in \citet{Pdot}.

\begin{table}
\caption{Superhump maxima of KX Aql (2010).}\label{tab:kxaqloc2010}
\begin{center}
\begin{tabular}{ccccc}
\hline
$E$ & max\commenta & error & $O-C$\commentb & $N$\commentc \\
\hline
0 & 55265.6366 & 0.0007 & $-$0.0110 & 145 \\
1 & 55265.6941 & 0.0018 & $-$0.0149 & 52 \\
7 & 55266.0852 & 0.0025 & 0.0077 & 56 \\
17 & 55266.6967 & 0.0028 & 0.0050 & 53 \\
26 & 55267.2439 & 0.0010 & $-$0.0005 & 183 \\
27 & 55267.3082 & 0.0005 & 0.0023 & 148 \\
38 & 55267.9827 & 0.0004 & 0.0012 & 59 \\
43 & 55268.2993 & 0.0007 & 0.0108 & 74 \\
50 & 55268.7192 & 0.0014 & 0.0008 & 43 \\
54 & 55268.9648 & 0.0003 & 0.0008 & 126 \\
55 & 55269.0288 & 0.0007 & 0.0033 & 46 \\
59 & 55269.2728 & 0.0006 & 0.0016 & 238 \\
70 & 55269.9473 & 0.0004 & 0.0006 & 53 \\
71 & 55270.0091 & 0.0008 & 0.0010 & 60 \\
75 & 55270.2611 & 0.0025 & 0.0073 & 127 \\
76 & 55270.3120 & 0.0017 & $-$0.0032 & 113 \\
82 & 55270.6850 & 0.0019 & 0.0013 & 52 \\
87 & 55270.9918 & 0.0026 & 0.0010 & 7 \\
99 & 55271.7243 & 0.0026 & $-$0.0035 & 44 \\
103 & 55271.9691 & 0.0008 & $-$0.0044 & 62 \\
119 & 55272.9490 & 0.0017 & $-$0.0072 & 16 \\
\hline
  \multicolumn{5}{l}{\commenta BJD$-$2400000.} \\
  \multicolumn{5}{l}{\commentb Against $max = 2455265.6476 + 0.061416 E$.} \\
  \multicolumn{5}{l}{\commentc Number of points used to determine the maximum.} \\
\end{tabular}
\end{center}
\end{table}

\subsection{NN Camelopardalis}\label{obj:nncam}

\begin{figure}
  \begin{center}
%    \FigureFile(88mm,90mm){nncam2009oc.eps}
    \FigureFile(88mm,90mm){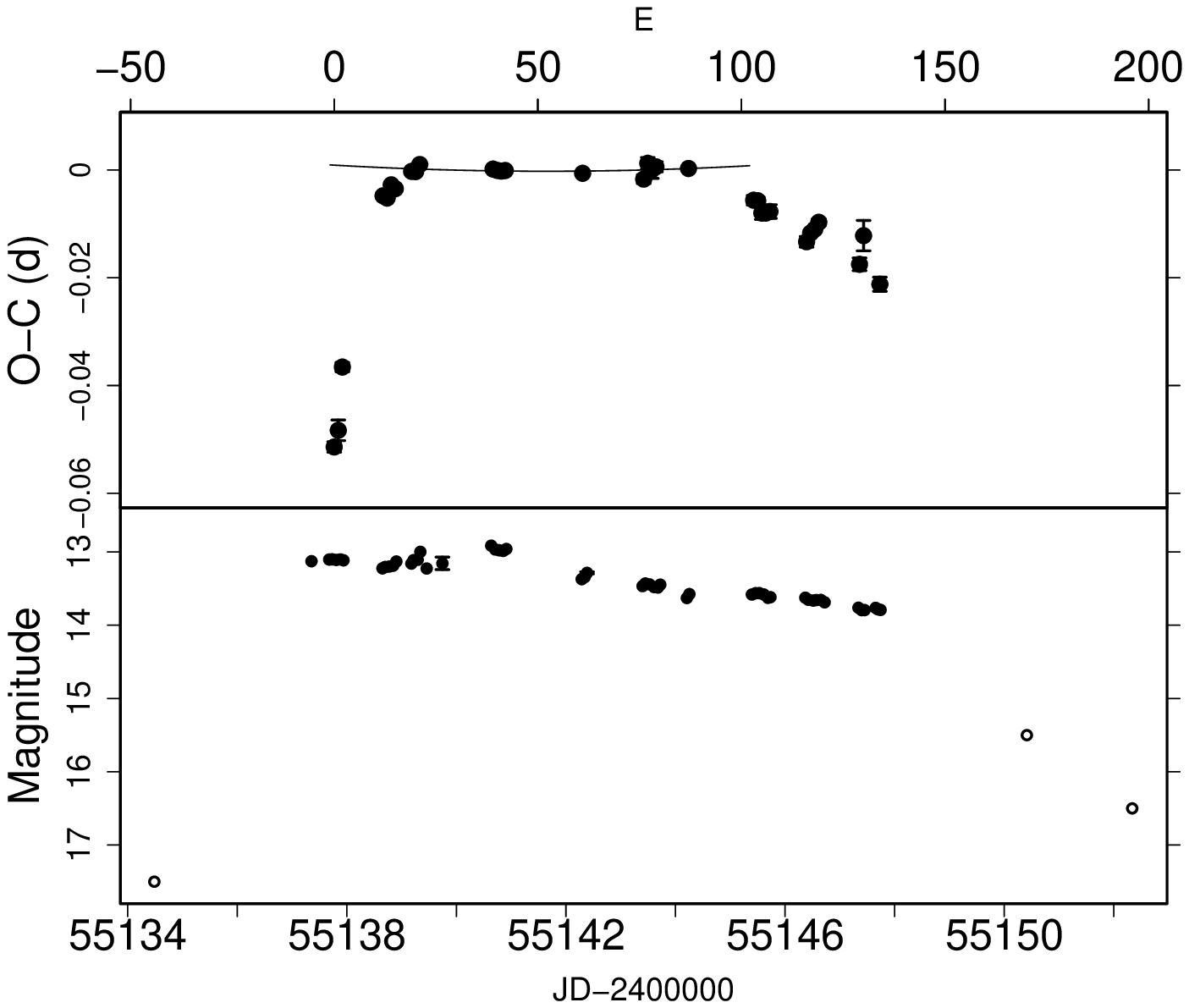}
  \end{center}
  \caption{$O-C$ of superhumps in NN Cam (2009).
  (Upper): $O-C$ diagram.  The $O-C$ values were against the mean period
  for the stage B ($19 \le E \le 87$, thin curve)
  (Lower): Light curve.  Small open circles are snapshot unfiltered
  CCD observations.  There was a rise in the light curve in accordance
  with the stage A--B transition.
  }
  \label{fig:nncam2009oc}
\end{figure}

   The superoutburst in 2009 November is the first superoutburst of
this object whose entire evolution was first observed in detail
[for general information of this object, see \citet{Pdot}].
The times of superhump maxima are listed in table \ref{tab:nncamoc2009}.
The $O-C$ diagram clearly shows all A--C stages.
There was a rise in the light curve in accordance
with the stage A--B transition (figure \ref{fig:nncam2009oc}).
The mean $P_{\rm SH}$ and $P_{\rm dot}$ for stage B ($19 \le E \le 87$)
were 0.07426(1) d and $+1.1(1.5) \times 10^{-5}$, respectively.
There was a distinct stage B--C transition around $E$ = 87.
The mean period for stage A was calculated for $E \le 15$ although
there was significant period shortening even during stage A.
\citet{she10nncam} also recently reported on two superoutbursts of
this object in 2007 and 2009.  \citet{she10nncam} did not distinguish
stages A--C during the 2009 superoutburst treated in this paper probably
due to fragmentary observational coverage.

\begin{table}
\caption{Superhump maxima of NN Cam (2009).}\label{tab:nncamoc2009}
\begin{center}
\begin{tabular}{ccccc}
\hline
$E$ & max\commenta & error & $O-C$\commentb & $N$\commentc \\
\hline
0 & 55137.7356 & 0.0010 & $-$0.0396 & 71 \\
1 & 55137.8129 & 0.0019 & $-$0.0366 & 69 \\
2 & 55137.8989 & 0.0009 & $-$0.0249 & 71 \\
12 & 55138.6733 & 0.0006 & 0.0064 & 72 \\
13 & 55138.7472 & 0.0004 & 0.0060 & 70 \\
14 & 55138.8239 & 0.0003 & 0.0084 & 67 \\
15 & 55138.8974 & 0.0003 & 0.0076 & 63 \\
19 & 55139.1977 & 0.0002 & 0.0107 & 109 \\
20 & 55139.2719 & 0.0002 & 0.0106 & 147 \\
21 & 55139.3475 & 0.0003 & 0.0119 & 90 \\
39 & 55140.6834 & 0.0004 & 0.0102 & 46 \\
40 & 55140.7574 & 0.0003 & 0.0099 & 69 \\
41 & 55140.8316 & 0.0003 & 0.0098 & 68 \\
42 & 55140.9059 & 0.0003 & 0.0098 & 63 \\
61 & 55142.3164 & 0.0003 & 0.0085 & 152 \\
76 & 55143.4293 & 0.0008 & 0.0068 & 68 \\
77 & 55143.5065 & 0.0011 & 0.0097 & 33 \\
78 & 55143.5796 & 0.0016 & 0.0084 & 25 \\
79 & 55143.6544 & 0.0009 & 0.0089 & 36 \\
87 & 55144.2482 & 0.0005 & 0.0083 & 140 \\
103 & 55145.4305 & 0.0009 & 0.0017 & 41 \\
104 & 55145.5047 & 0.0007 & 0.0015 & 39 \\
105 & 55145.5767 & 0.0012 & $-$0.0008 & 30 \\
106 & 55145.6509 & 0.0007 & $-$0.0008 & 40 \\
107 & 55145.7255 & 0.0013 & $-$0.0006 & 32 \\
116 & 55146.3882 & 0.0010 & $-$0.0066 & 65 \\
117 & 55146.4641 & 0.0006 & $-$0.0050 & 63 \\
118 & 55146.5391 & 0.0005 & $-$0.0043 & 104 \\
119 & 55146.6146 & 0.0007 & $-$0.0031 & 39 \\
129 & 55147.3495 & 0.0012 & $-$0.0113 & 33 \\
130 & 55147.4291 & 0.0028 & $-$0.0061 & 37 \\
134 & 55147.7171 & 0.0013 & $-$0.0153 & 39 \\
\hline
  \multicolumn{5}{l}{\commenta BJD$-$2400000.} \\
  \multicolumn{5}{l}{\commentb Against $max = 2455137.7752 + 0.074307 E$.} \\
  \multicolumn{5}{l}{\commentc Number of points used to determine the maximum.} \\
\end{tabular}
\end{center}
\end{table}

\subsection{V591 Centauri}\label{obj:v591cen}

   V591 Cen was discovered by \citet{hur40v591cen}
(see also \cite{wal58CVchart}).  Although the discovery observation was
already suggestive of a superoutburst, this object had long been overlooked
due to the apparent small outburst amplitude.  
The lack of CV-type signature in spectroscopic observation
by \citet{sch05CVspec} even led to a non-CV classification.
Monitoring for outbursts, however, continued because of its outbursting
nature and the likely presence of superoutbursts.
In 2006, B. Monard succeeded
in obtaining astrometry of the outbursting object, which was 3 arcsec
distant from the supposed quiescent counterpart (vsnet-alert 9158).
The true V591 Cen was invisible down to 20 mag on DSS images.

\begin{figure}
  \begin{center}
%    \FigureFile(88mm,110mm){v591censhpdm.eps}
    \FigureFile(88mm,110mm){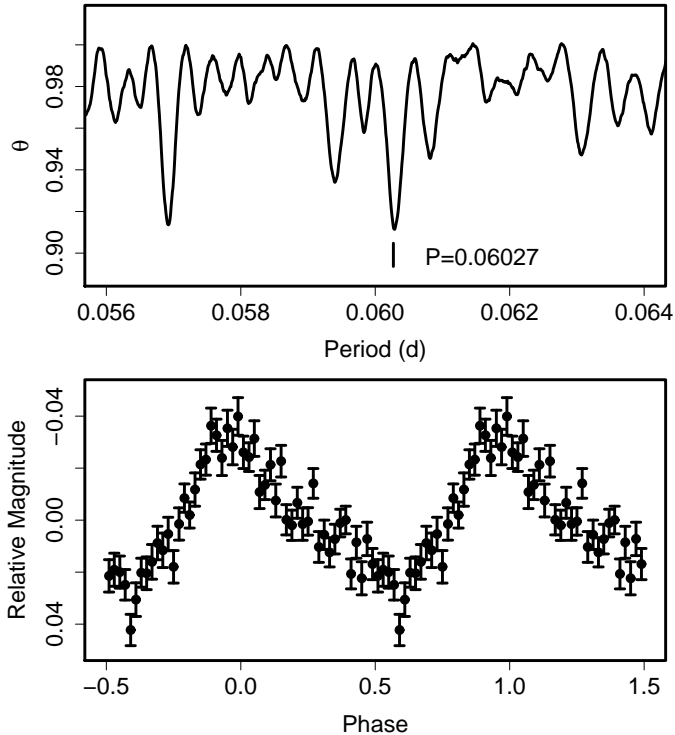}
  \end{center}
  \caption{Superhumps in V591 Cen (2010).
     (Upper): PDM analysis.  The selection of the alias was based on
     $O-C$ analysis.
     (Lower): Phase-averaged profile.}
  \label{fig:v591censhpdm}
\end{figure}

   The object again underwent a bright outburst in 2010 April
(vsnet-alert 11915), and the existence of superhumps was finally
confirmed (vsnet-alert 11919; figure \ref{fig:v591censhpdm}).
Astrometry of the outbursting object
also confirmed the identification in 2006 (vsnet-alert 11922).

   The times of superhump maxima are listed in table \ref{tab:v591cenoc2010}.
The data clearly shows a positive period derivative 
$P_{\rm dot}$ = $+6.5(1.3) \times 10^{-5}$, typical for this short
$P_{\rm SH}$.

\begin{table}
\caption{Superhump maxima of V591 Cen (2010).}\label{tab:v591cenoc2010}
\begin{center}
\begin{tabular}{ccccc}
\hline
$E$ & max\commenta & error & $O-C$\commentb & $N$\commentc \\
\hline
0 & 55296.1418 & 0.0008 & 0.0022 & 111 \\
1 & 55296.2051 & 0.0007 & 0.0052 & 109 \\
2 & 55296.2634 & 0.0011 & 0.0033 & 111 \\
17 & 55297.1647 & 0.0009 & 0.0001 & 110 \\
18 & 55297.2252 & 0.0008 & 0.0003 & 111 \\
19 & 55297.2849 & 0.0011 & $-$0.0003 & 112 \\
20 & 55297.3425 & 0.0014 & $-$0.0030 & 70 \\
86 & 55301.3271 & 0.0032 & 0.0019 & 267 \\
87 & 55301.3806 & 0.0020 & $-$0.0050 & 268 \\
88 & 55301.4362 & 0.0018 & $-$0.0097 & 267 \\
89 & 55301.4977 & 0.0017 & $-$0.0084 & 268 \\
102 & 55302.2868 & 0.0039 & $-$0.0033 & 268 \\
133 & 55304.1630 & 0.0018 & 0.0036 & 61 \\
134 & 55304.2236 & 0.0030 & 0.0040 & 103 \\
135 & 55304.2836 & 0.0015 & 0.0036 & 329 \\
136 & 55304.3447 & 0.0014 & 0.0044 & 281 \\
137 & 55304.4012 & 0.0014 & 0.0006 & 264 \\
138 & 55304.4612 & 0.0027 & 0.0004 & 73 \\
\hline
  \multicolumn{5}{l}{\commenta BJD$-$2400000.} \\
  \multicolumn{5}{l}{\commentb Against $max = 2455296.1395 + 0.060299 E$.} \\
  \multicolumn{5}{l}{\commentc Number of points used to determine the maximum.} \\
\end{tabular}
\end{center}
\end{table}

\subsection{Z Chameleontis}\label{obj:zcha}

   We analyzed AAVSO and our observations of the 2010 January superoutburst
of Z Cha.  The observation was performed during the late stage of
the superoutburst.  After removing observations within 0.10
$P_{\rm orb}$ of eclipses for the superoutburst plateau and 0.08
$P_{\rm orb}$ for later observations, we determined times of
superhump maxima (table \ref{tab:zchaoc2010}).  The times for $E \ge 91$
are secondary humps, which are likely traditional late superhumps.
The peaks of persistent ordinary superhumps were not determined because
they fell on orbital humps and around eclipses.
These superhumps ($E < 91$) most likely correspond to stage C superhumps.
The period [0.07698(5) d] determined from the superhump maxima is
close to the period [0.07681(6) d] of stage C superhumps in
1982 \citep{Pdot}.

   Although post-superoutburst observation suggests the appearance of
traditional late superhumps, this signal was not well traced due to the
strong orbital modulation and limited coverage.  A better continuous
observation is necessary to test whether dominant superhumps in the
post-superoutburst stage are stage C superhumps or traditional
late superhumps (see also a discussion on the nature of late superhumps
in \cite{ohs10qzvir}).

\begin{table}
\caption{Superhump maxima of Z Cha (2010).}\label{tab:zchaoc2010}
\begin{center}
\begin{tabular}{ccccc}
\hline
$E$ & max\commenta & error & $O-C$\commentb & $N$\commentc \\
\hline
0 & 55200.0935 & 0.0008 & 0.0032 & 152 \\
13 & 55201.0794 & 0.0003 & $-$0.0086 & 209 \\
14 & 55201.1575 & 0.0003 & $-$0.0072 & 165 \\
15 & 55201.2328 & 0.0001 & $-$0.0087 & 124 \\
27 & 55202.1711 & 0.0021 & 0.0086 & 24 \\
28 & 55202.2385 & 0.0015 & $-$0.0007 & 22 \\
65 & 55205.0878 & 0.0006 & 0.0088 & 158 \\
66 & 55205.1638 & 0.0005 & 0.0081 & 159 \\
67 & 55205.2385 & 0.0007 & 0.0060 & 94 \\
78 & 55206.0940 & 0.0007 & 0.0173 & 130 \\
79 & 55206.1667 & 0.0011 & 0.0132 & 62 \\
80 & 55206.2399 & 0.0007 & 0.0097 & 159 \\
91 & 55207.0442 & 0.0015 & $-$0.0303 & 168 \\
92 & 55207.1319 & 0.0009 & $-$0.0194 & 169 \\
\hline
  \multicolumn{5}{l}{\commenta BJD$-$2400000.} \\
  \multicolumn{5}{l}{\commentb Against $max = 2455200.0902 + 0.076750 E$.} \\
  \multicolumn{5}{l}{\commentc Number of points used to determine the maximum.} \\
\end{tabular}
\end{center}
\end{table}

\subsection{PU Canis Majoris}\label{obj:pucma}

\begin{figure}
  \begin{center}
%    \FigureFile(88mm,110mm){pucma2009prof.eps}
    \FigureFile(88mm,110mm){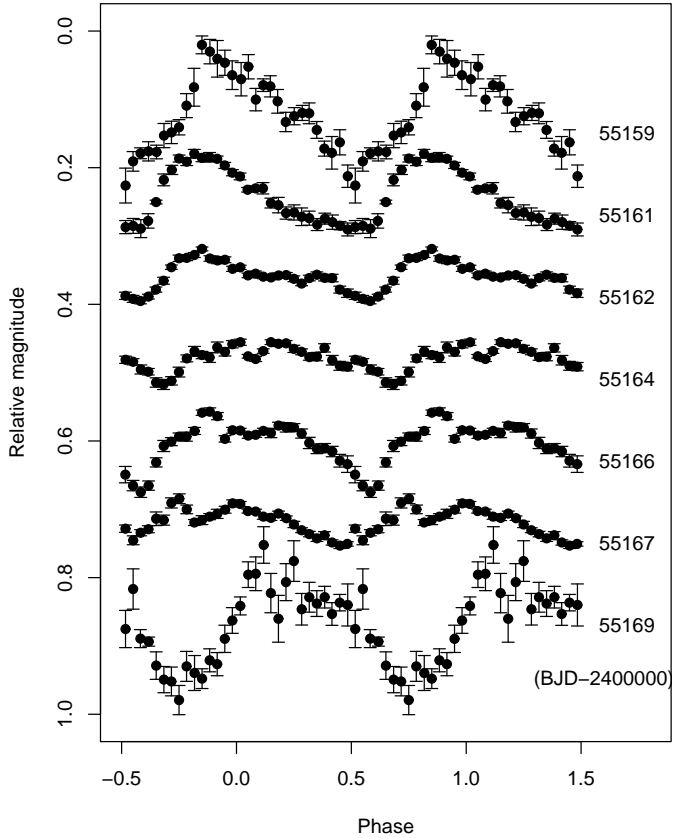}
  \end{center}
  \caption{Superhump profiles of PU CMa (2009).
     The profiles of superhumps strongly varied between nights.
     The figure was drawn against a mean period of 0.058020 d.}
  \label{fig:pucma2009prof}
\end{figure}

\begin{figure}
  \begin{center}
%    \FigureFile(88mm,70mm){pucmacomp.eps}
    \FigureFile(88mm,70mm){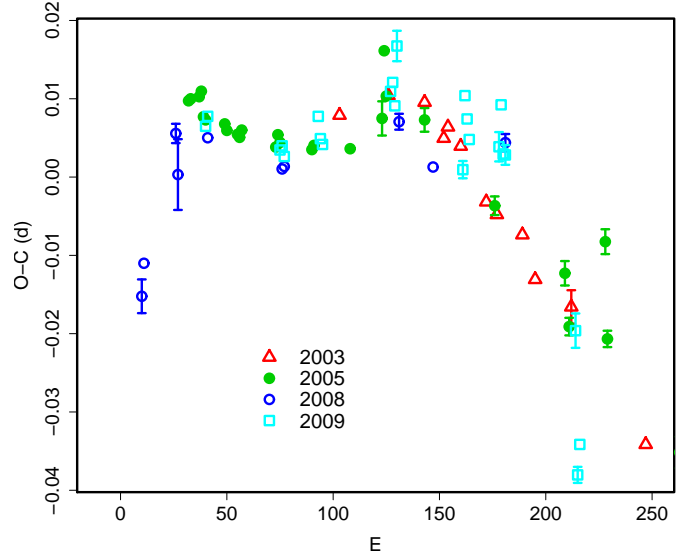}
  \end{center}
  \caption{Comparison of $O-C$ diagrams of PU CMa between different
  superoutbursts.  A period of 0.05801 d was used to draw this figure.
  Approximate cycle counts ($E$) after the appearance of the
  superhumps were used.  Since the start of the 2009 superoutburst
  was not well constrained, we shifted the $O-C$ diagrams
  to best fit the others.  The $O-C$ diagram for the 2008 superoutburst
  matched others only if $E$ was counted from the start of the main
  superoutburst, rather than from the precursor.
  }
  \label{fig:pucmacomp}
\end{figure}

   We observed the 2009 superoutburst of this object
(table \ref{tab:pucmaoc2009}).
The nightly superhump profiles (figure \ref{fig:pucma2009prof})
suggests that the superhump period first increased until
BJD 2455164, and then decreased.  The stages listed in table
\ref{tab:perlist} reflect this interpretation.
A comparison of $O-C$ diagrams between different superoutbursts
is shown in figure \ref{fig:pucmacomp}).
It would be worth noting that the $O-C$ diagram for
the 2008 superoutburst (superoutburst preceded by a precursor)
matched others only if $E$ was counted from the start of the main
superoutburst, rather than from the precursor.  This suggests that
superhumps were excited around the ignition of the main superoutburst,
rather than during the precursor.

\begin{table}
\caption{Superhump maxima of PU CMa (2009).}\label{tab:pucmaoc2009}
\begin{center}
\begin{tabular}{ccccc}
\hline
$E$ & max\commenta & error & $O-C$\commentb & $N$\commentc \\
\hline
0 & 55159.1299 & 0.0009 & $-$0.0110 & 66 \\
1 & 55159.1892 & 0.0008 & $-$0.0096 & 65 \\
35 & 55161.1572 & 0.0007 & $-$0.0083 & 53 \\
36 & 55161.2158 & 0.0003 & $-$0.0076 & 84 \\
37 & 55161.2724 & 0.0005 & $-$0.0088 & 82 \\
53 & 55162.2057 & 0.0010 & $-$0.0011 & 193 \\
54 & 55162.2609 & 0.0004 & $-$0.0038 & 186 \\
55 & 55162.3181 & 0.0005 & $-$0.0043 & 179 \\
87 & 55164.1812 & 0.0008 & 0.0076 & 85 \\
88 & 55164.2404 & 0.0008 & 0.0090 & 248 \\
89 & 55164.2954 & 0.0009 & 0.0061 & 249 \\
90 & 55164.3611 & 0.0019 & 0.0139 & 102 \\
121 & 55166.1436 & 0.0011 & 0.0032 & 150 \\
122 & 55166.2110 & 0.0007 & 0.0128 & 334 \\
123 & 55166.2661 & 0.0005 & 0.0100 & 263 \\
124 & 55166.3215 & 0.0006 & 0.0075 & 165 \\
137 & 55167.0993 & 0.0035 & 0.0334 & 128 \\
138 & 55167.1327 & 0.0019 & 0.0089 & 208 \\
139 & 55167.1960 & 0.0006 & 0.0144 & 316 \\
140 & 55167.2478 & 0.0005 & 0.0083 & 273 \\
141 & 55167.3057 & 0.0013 & 0.0083 & 168 \\
174 & 55169.1975 & 0.0022 & $-$0.0087 & 84 \\
175 & 55169.2371 & 0.0010 & $-$0.0270 & 85 \\
176 & 55169.2990 & 0.0007 & $-$0.0229 & 82 \\
192 & 55170.2172 & 0.0017 & $-$0.0304 & 78 \\
\hline
  \multicolumn{5}{l}{\commenta BJD$-$2400000.} \\
  \multicolumn{5}{l}{\commentb Against $max = 2455159.1409 + 0.057847 E$.} \\
  \multicolumn{5}{l}{\commentc Number of points used to determine the maximum.} \\
\end{tabular}
\end{center}
\end{table}

\subsection{AQ Canis Minoris}\label{obj:aqcmi}

\begin{figure}
  \begin{center}
%    \FigureFile(88mm,110mm){aqcmishpdm.eps}
    \FigureFile(88mm,110mm){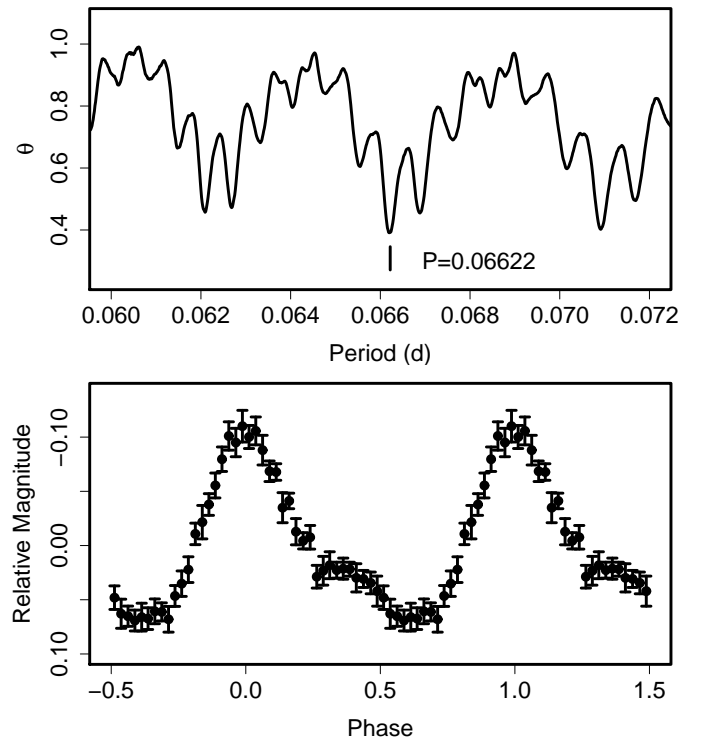}
  \end{center}
  \caption{Superhumps in AQ CMi (2010).
     (Upper): PDM analysis.  The selection of the alias is based on
     the best period reported from past observations.
     (Lower): Phase-averaged profile.}
  \label{fig:aqcmishpdm}
\end{figure}

   Although the SU UMa-type nature of this object has been well
established,\footnote{
$<$http://www.kusastro.kyoto-u.ac.jp/vsnet/DNe/aqcmi.html$>$.
}
there has unfortunately no solid publication on superhumps in this
system.  The only known superoutburst since the 1997 observation
was in 2008 September, which was too badly placed for time-series
photometry.  The superoutburst in 2010 April brought the first
chance to record superhumps since its recognition as an SU UMa-type
dwarf nova.
The times of superhump maxima are listed in table \ref{tab:aqcmioc2010}.
Although the course of the superhump evolution was not fully recorded,
there appears to have been a discontinuous period change around $E=15$.
We attributed this to stage A--B transition because the observation
recorded the early stage of the superoutburst.
The period derivative was not determined because of a gap in observation.
The mean $P_{\rm SH}$ during the entire observation was 0.06622(1) d
(figure \ref{fig:aqcmishpdm}).  This period needs to be treated with
caution because it was likely derived from different stages of superhump
evolution.

\begin{table}
\caption{Superhump maxima of AQ CMi (2010).}\label{tab:aqcmioc2010}
\begin{center}
\begin{tabular}{ccccc}
\hline
$E$ & max\commenta & error & $O-C$\commentb & $N$\commentc \\
\hline
0 & 55295.9806 & 0.0006 & $-$0.0022 & 116 \\
1 & 55296.0489 & 0.0007 & $-$0.0002 & 112 \\
15 & 55296.9775 & 0.0005 & 0.0017 & 104 \\
16 & 55297.0431 & 0.0005 & 0.0011 & 121 \\
106 & 55302.9994 & 0.0009 & $-$0.0004 & 68 \\
\hline
  \multicolumn{5}{l}{\commenta BJD$-$2400000.} \\
  \multicolumn{5}{l}{\commentb Against $max = 2455295.9828 + 0.066198 E$.} \\
  \multicolumn{5}{l}{\commentc Number of points used to determine the maximum.} \\
\end{tabular}
\end{center}
\end{table}

\subsection{GZ Cancri}\label{obj:gzcnc}

   GZ Cnc is a variable star discovered by Takamizawa (TmzV34, vsnet-obs 10504),
which later turned out to be a dwarf nova \citep{kat01gzcnc}.
\citet{kat01gzcnc} observed an outburst in 2000 February.  Based on
the relatively slow rise to an outburst maximum and the lack of periodic
modulations, they concluded that the object is likely an SS Cyg-type
dwarf nova with a long $P_{\rm orb}$.  
In 2002, \citet{kat02gzcncnsv10934} noticed an unusually increase
in the number of outburst detections, and suggested the similarity to
a proposed intermediate polar V426 Oph.
Most surprisingly, radial-velocity studies by \citet{tap03gzcnc}
yielded an $P_{\rm orb}$ of 0.08825(28) d, placing the object at the
lower edge of the period gap.  Although this $P_{\rm orb}$ was seemingly
incompatible with behavior of outbursts in this object, later observations
have detected abundant short outbursts which are compatible with
a short $P_{\rm orb}$-system.  The mystery, however, remained why
the object did not develop superhumps during its long, 2000 February
outburst.

   Later on, the object again exhibited a long outburst in 2007 December
(vsnet-alert 9783), which was unfortunately not observed for searching
superhumps.  In 2009 January, this object underwent a long, bright outburst
(vsnet-alert 10984).  Follow-up observations, however, did not detect
superhumps (vsnet-alert 11002).  GZ Cnc was then considered as a rare
object below the period gap without a signature of an SU UMa-type
dwarf nova.

\begin{figure}
  \begin{center}
%    \FigureFile(88mm,100mm){gzcnclc.eps}
    \FigureFile(88mm,100mm){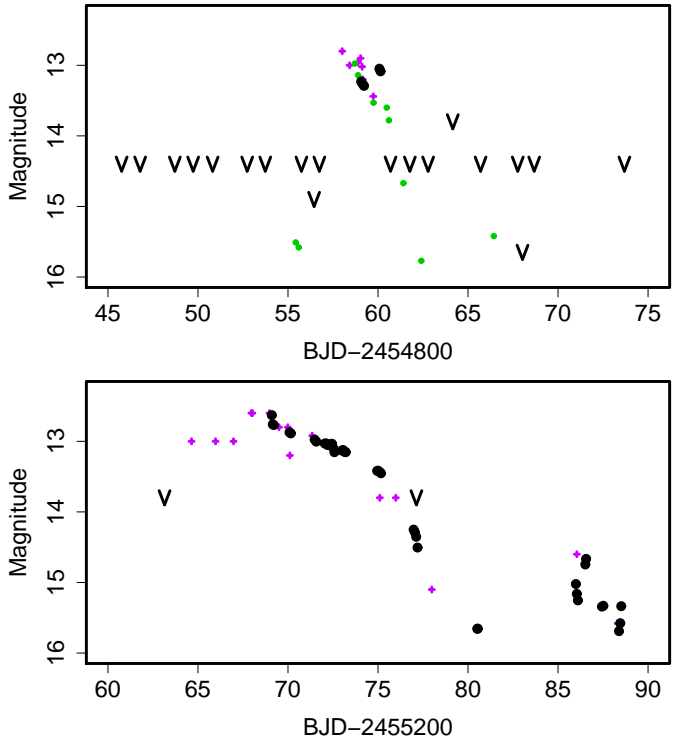}
  \end{center}
  \caption{Comparison of wide outburst (upper, 2009) and superoutburst
     (lower, 2010) of GZ Cnc.
     The filled circles, open circles, small crosses and ``v''-marks
     represent CCD observations, ASAS-3 $V$ data and visual observations,
     and upper limits, respectively.
     The outburst around BJD 2455287 is a normal outburst.}
  \label{fig:gzcnclc}
\end{figure}

   In 2010 March, the object again underwent a long, bright outburst
(vsnet-alert 11855, 11863).  Subsequent observations eventually detected
superhumps (vsnet-alert 11881, 11888, 11890, 11894).  The object is
now confirmed to be a rare object exhibiting three classes of outbursts:
normal (narrow) outbursts, long SS Cyg-type (wide) outbursts and
SU UMa-type superoutbursts (figure \ref{fig:gzcnclc}).
The mystery of the 2000 outburst could be
understood if this outburst was a ``long'' outburst failing to trigger
the tidal instability.  The only other known object having the same
property is TU Men (cf. \cite{bat89tumen}; \cite{sma00DNunsolved}),
unusual SU UMa-type dwarf nova above the period gap.
If GZ Cnc is indeed a ``borderline'' SU UMa-type dwarf nova, the object
may be analogous to BZ UMa, which also showed an usually slow rise to
a full outburst (cf. \cite{Pdot}).

\begin{figure}
  \begin{center}
%    \FigureFile(88mm,100mm){gzcncoc.eps}
    \FigureFile(88mm,100mm){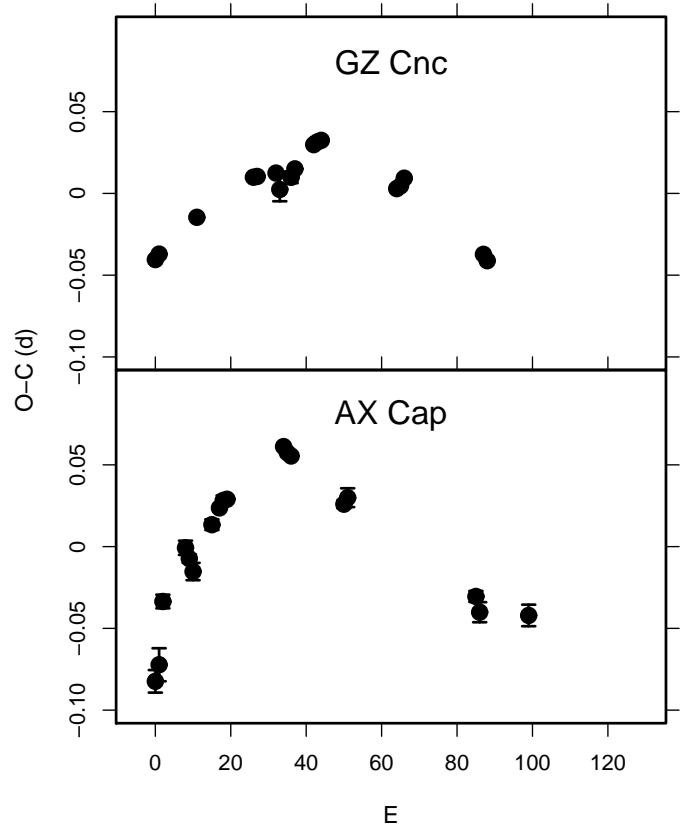}
  \end{center}
  \caption{Comparison of $O-C$ diagrams between GZ Cnc and AX Cap.}
  \label{fig:gzcncoc}
\end{figure}

   The times of superhump maxima during the 2010 superoutburst are
listed in table \ref{tab:gzcncoc2010}.  The $O-C$ diagram can be
reasonably interpreted without a phase jump only if we assume
a very large decrease in the period as in AX Cap (figure \ref{fig:gzcncoc}).
The mean $P_{\rm SH}$ for $E \le 44$ was 0.09277(2) d (PDM method),
which is equivalent to $\epsilon$ = 5.1 \%.  This fractional superhump
excess is one of the largest among SU UMa-type dwarf novae below the
period gap (cf. figure 15 in \cite{Pdot}).
The mean period 0.08972(12) d after the period decrease corresponds
to $\epsilon$ = 1.7 \%.  Since this period was fairly close to
$P_{\rm orb}$, this period decrease may have reflected the dominance
of orbital humps during the late course of the superoutburst.

\begin{figure}
  \begin{center}
%    \FigureFile(88mm,110mm){gzcncshprof.eps}
    \FigureFile(88mm,110mm){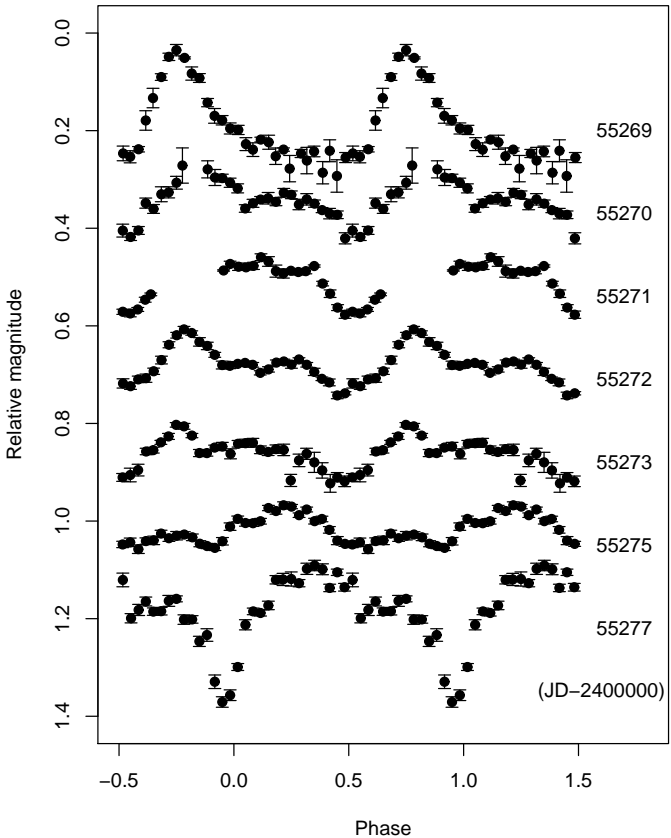}
  \end{center}
  \caption{Superhump profiles of GZ Cnc (2010).
     The figure was drawn against a mean period of 0.09277 d.
     JD 2455275 corresponds to the start of the rapid fading from
     the superoutburst plateau.}
  \label{fig:gzcncshprof}
\end{figure}

\begin{figure}
  \begin{center}
%    \FigureFile(88mm,110mm){gzcncshpdm.eps}
    \FigureFile(88mm,110mm){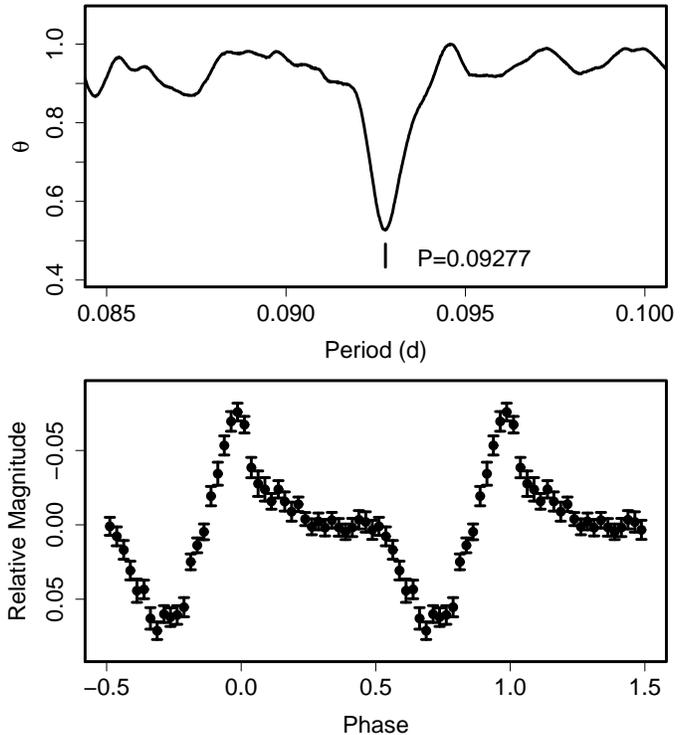}
  \end{center}
  \caption{Superhumps in GZ Cnc before JD 2455274 (2010).
     (Upper): PDM analysis.
     (Lower): Phase-averaged profile.}
  \label{fig:gzcncshpdm}
\end{figure}

\begin{table}
\caption{Superhump maxima of GZ Cnc (2010).}\label{tab:gzcncoc2010}
\begin{center}
\begin{tabular}{ccccc}
\hline
$E$ & max\commenta & error & $O-C$\commentb & $N$\commentc \\
\hline
0 & 55269.0989 & 0.0007 & $-$0.0405 & 64 \\
1 & 55269.1935 & 0.0007 & $-$0.0371 & 52 \\
11 & 55270.1286 & 0.0008 & $-$0.0146 & 175 \\
26 & 55271.5221 & 0.0007 & 0.0099 & 86 \\
27 & 55271.6139 & 0.0015 & 0.0104 & 55 \\
32 & 55272.0723 & 0.0006 & 0.0124 & 226 \\
33 & 55272.1535 & 0.0072 & 0.0024 & 48 \\
36 & 55272.4347 & 0.0036 & 0.0099 & 57 \\
37 & 55272.5312 & 0.0012 & 0.0150 & 93 \\
42 & 55273.0023 & 0.0016 & 0.0298 & 53 \\
43 & 55273.0952 & 0.0029 & 0.0315 & 74 \\
44 & 55273.1874 & 0.0016 & 0.0325 & 88 \\
64 & 55274.9832 & 0.0010 & 0.0029 & 97 \\
65 & 55275.0761 & 0.0008 & 0.0046 & 276 \\
66 & 55275.1721 & 0.0016 & 0.0093 & 160 \\
87 & 55277.0421 & 0.0006 & $-$0.0373 & 346 \\
88 & 55277.1295 & 0.0007 & $-$0.0412 & 173 \\
\hline
  \multicolumn{5}{l}{\commenta BJD$-$2400000.} \\
  \multicolumn{5}{l}{\commentb Against $max = 2455269.1393 + 0.091265 E$.} \\
  \multicolumn{5}{l}{\commentc Number of points used to determine the maximum.} \\
\end{tabular}
\end{center}
\end{table}

\subsection{GO Comae Berenices}\label{obj:gocom}

\begin{figure}
  \begin{center}
%    \FigureFile(88mm,70mm){gocomcomp2.eps}
    \FigureFile(88mm,70mm){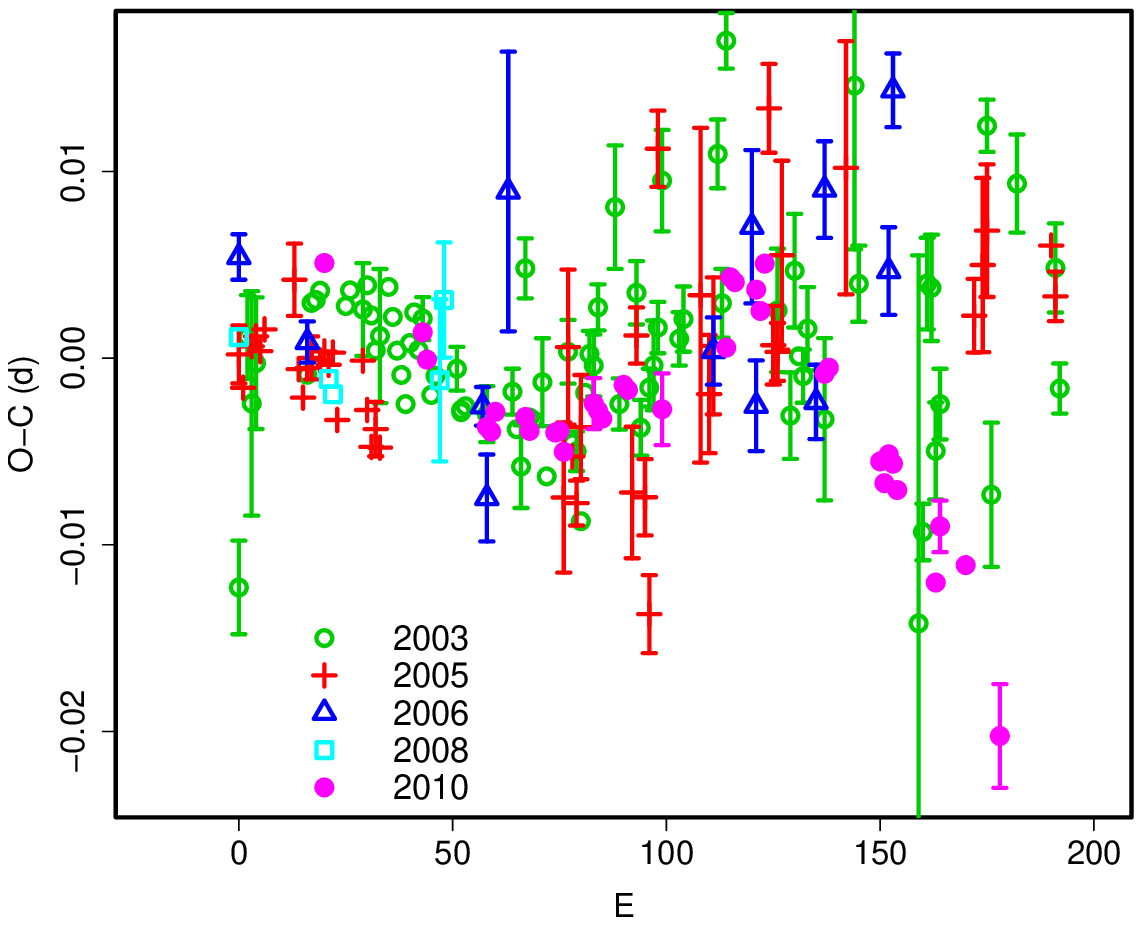}
  \end{center}
  \caption{Comparison of $O-C$ diagrams of GO Com between different
  superoutbursts.  A period of 0.06303 d was used to draw this figure.
  Approximate cycle counts ($E$) after the appearance of the
  superhumps were used.}
  \label{fig:gocomcomp2}
\end{figure}

   The 2010 superoutburst was particularly well observed and stages
B and C were very clearly defined (table \ref{tab:gocomoc2010}).
These observations have confirmed the finding in \citet{ima05gocom}
and \citet{Pdot}.
Figure \ref{fig:gocomcomp2} illustrates a comparison of $O-C$ diagrams
between different superoutbursts.  The 2010 observation better recorded
the stage C superhumps with higher accuracy, while the 2003 observation
well recorded the earlier part.  A combination of these superoutbursts
very well demonstrates the ``canonical'' period variation in
a short-$P_{\rm SH}$ system.  It is not clear whether the 2010 superoutburst
was preceded by a precursor outburst as in the 2003 one.

\begin{table}
\caption{Superhump maxima of GO Com (2010).}\label{tab:gocomoc2010}
\begin{center}
\begin{tabular}{ccccc}
\hline
$E$ & max\commenta & error & $O-C$\commentb & $N$\commentc \\
\hline
0 & 55286.1498 & 0.0006 & 0.0028 & 93 \\
23 & 55287.5958 & 0.0004 & 0.0005 & 61 \\
24 & 55287.6574 & 0.0004 & $-$0.0009 & 37 \\
38 & 55288.5362 & 0.0006 & $-$0.0036 & 41 \\
39 & 55288.5989 & 0.0003 & $-$0.0038 & 61 \\
40 & 55288.6630 & 0.0004 & $-$0.0027 & 55 \\
47 & 55289.1040 & 0.0002 & $-$0.0026 & 127 \\
48 & 55289.1663 & 0.0004 & $-$0.0032 & 127 \\
54 & 55289.5443 & 0.0004 & $-$0.0030 & 50 \\
55 & 55289.6075 & 0.0003 & $-$0.0028 & 66 \\
56 & 55289.6694 & 0.0005 & $-$0.0039 & 66 \\
63 & 55290.1132 & 0.0014 & $-$0.0008 & 40 \\
64 & 55290.1758 & 0.0004 & $-$0.0012 & 40 \\
65 & 55290.2384 & 0.0009 & $-$0.0015 & 40 \\
70 & 55290.5554 & 0.0005 & 0.0006 & 64 \\
71 & 55290.6181 & 0.0004 & 0.0004 & 61 \\
79 & 55291.1213 & 0.0019 & $-$0.0002 & 36 \\
94 & 55292.0701 & 0.0008 & 0.0041 & 135 \\
95 & 55292.1369 & 0.0006 & 0.0079 & 159 \\
96 & 55292.1997 & 0.0006 & 0.0077 & 96 \\
101 & 55292.5144 & 0.0005 & 0.0076 & 60 \\
102 & 55292.5763 & 0.0004 & 0.0065 & 64 \\
103 & 55292.6419 & 0.0005 & 0.0091 & 66 \\
117 & 55293.5184 & 0.0004 & 0.0041 & 60 \\
118 & 55293.5817 & 0.0003 & 0.0045 & 39 \\
130 & 55294.3331 & 0.0006 & 0.0002 & 66 \\
131 & 55294.3949 & 0.0008 & $-$0.0009 & 62 \\
132 & 55294.4595 & 0.0004 & 0.0007 & 77 \\
133 & 55294.5220 & 0.0005 & 0.0003 & 114 \\
134 & 55294.5837 & 0.0004 & $-$0.0011 & 132 \\
143 & 55295.1460 & 0.0009 & $-$0.0055 & 178 \\
144 & 55295.2120 & 0.0014 & $-$0.0024 & 38 \\
150 & 55295.5881 & 0.0005 & $-$0.0041 & 63 \\
158 & 55296.0832 & 0.0028 & $-$0.0128 & 28 \\
\hline
  \multicolumn{5}{l}{\commenta BJD$-$2400000.} \\
  \multicolumn{5}{l}{\commentb Against $max = 2455286.1470 + 0.062968 E$.} \\
  \multicolumn{5}{l}{\commentc Number of points used to determine the maximum.} \\
\end{tabular}
\end{center}
\end{table}

\subsection{TV Corvi}\label{obj:tvcrv}

\begin{figure}
  \begin{center}
%    \FigureFile(88mm,70mm){tvcrvcomp2.eps}
    \FigureFile(88mm,70mm){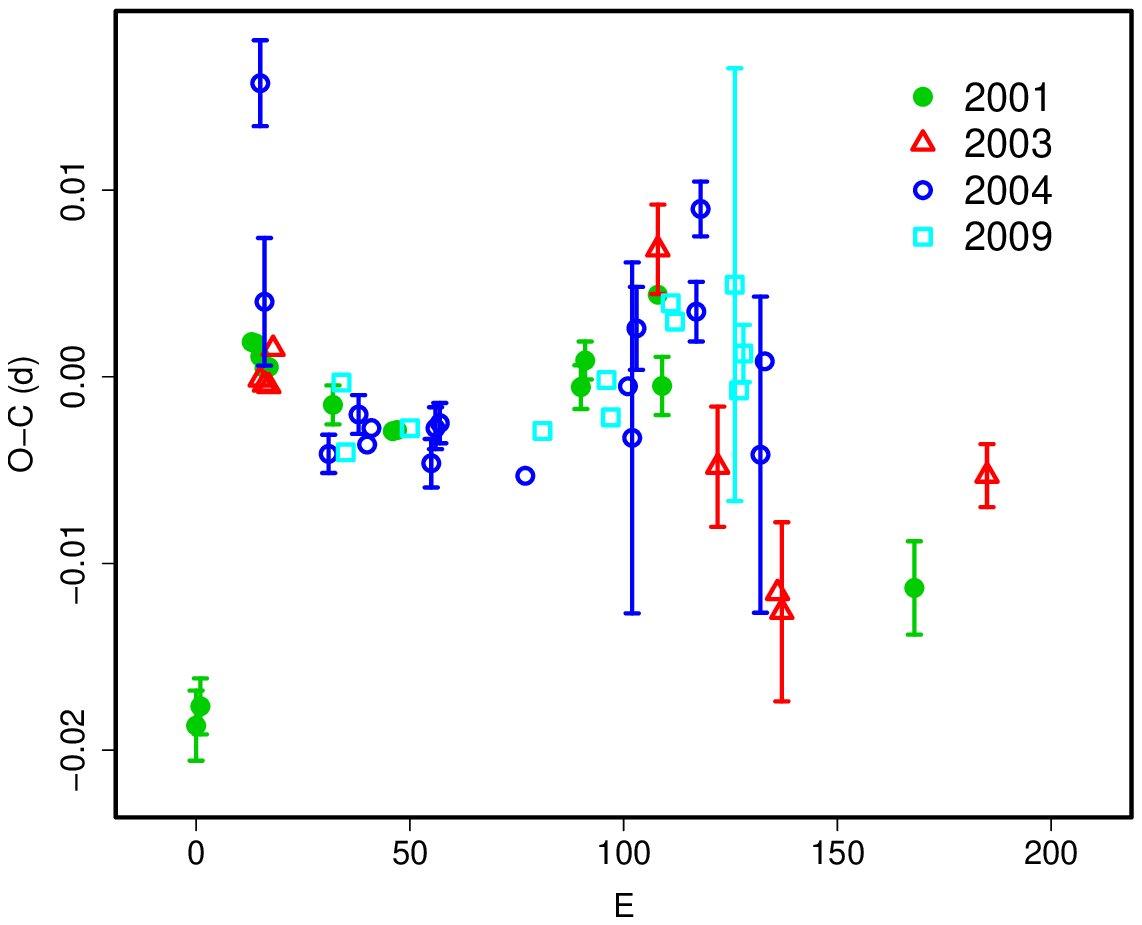}
  \end{center}
  \caption{Comparison of $O-C$ diagrams of TV Crv between different
  superoutbursts.  A period of 0.06500 d was used to draw this figure.
  Approximate cycle counts ($E$) after the appearance of the
  superhumps were used.}
  \label{fig:tvcrvcomp2}
\end{figure}

   We observed an superoutburst in 2009 December.  The times of
superhump maxima are listed in table \ref{tab:tvcrvoc2009}.
A combined $O-C$ diagram (figure \ref{fig:tvcrvcomp2}) strengthens
the assertion in \citet{Pdot} that the $O-C$ behavior is common
between different superoutbursts.

\begin{table}
\caption{Superhump maxima of TV Crv (2009).}\label{tab:tvcrvoc2009}
\begin{center}
\begin{tabular}{ccccc}
\hline
$E$ & max\commenta & error & $O-C$\commentb & $N$\commentc \\
\hline
0 & 55183.2495 & 0.0004 & 0.0028 & 125 \\
1 & 55183.3107 & 0.0003 & $-$0.0010 & 144 \\
16 & 55184.2870 & 0.0005 & $-$0.0005 & 108 \\
47 & 55186.3019 & 0.0006 & $-$0.0024 & 143 \\
62 & 55187.2796 & 0.0008 & $-$0.0005 & 142 \\
63 & 55187.3426 & 0.0007 & $-$0.0025 & 121 \\
77 & 55188.2587 & 0.0008 & 0.0028 & 143 \\
78 & 55188.3227 & 0.0006 & 0.0018 & 144 \\
92 & 55189.2347 & 0.0116 & 0.0030 & 114 \\
93 & 55189.2941 & 0.0009 & $-$0.0027 & 144 \\
94 & 55189.3610 & 0.0015 & $-$0.0008 & 82 \\
\hline
  \multicolumn{5}{l}{\commenta BJD$-$2400000.} \\
  \multicolumn{5}{l}{\commentb Against $max = 2455183.2467 + 0.065054 E$.} \\
  \multicolumn{5}{l}{\commentc Number of points used to determine the maximum.} \\
\end{tabular}
\end{center}
\end{table}

\subsection{V337 Cygni}\label{obj:v337cyg}

   We observed the 2010 superoutburst of this object.  The times of
superhump maxima are listed in table \ref{tab:v337cygoc2010}.
Although the period derivative was not well determined, we detected
a stage B--C transition.  The obtained parameters are listed in
table \ref{tab:perlist}.

\begin{table}
\caption{Superhump maxima of V337 Cyg (2010).}\label{tab:v337cygoc2010}
\begin{center}
\begin{tabular}{ccccc}
\hline
$E$ & max\commenta & error & $O-C$\commentb & $N$\commentc \\
\hline
0 & 55421.4715 & 0.0012 & $-$0.0046 & 118 \\
1 & 55421.5390 & 0.0007 & $-$0.0072 & 153 \\
29 & 55423.5087 & 0.0007 & $-$0.0009 & 75 \\
42 & 55424.4236 & 0.0007 & 0.0024 & 73 \\
43 & 55424.4944 & 0.0006 & 0.0032 & 72 \\
44 & 55424.5626 & 0.0007 & 0.0012 & 77 \\
45 & 55424.6340 & 0.0011 & 0.0025 & 54 \\
51 & 55425.0473 & 0.0078 & $-$0.0050 & 36 \\
52 & 55425.1288 & 0.0019 & 0.0064 & 63 \\
95 & 55428.1525 & 0.0013 & 0.0149 & 30 \\
126 & 55430.3163 & 0.0065 & 0.0050 & 119 \\
138 & 55431.1449 & 0.0010 & $-$0.0079 & 128 \\
139 & 55431.2128 & 0.0020 & $-$0.0101 & 147 \\
\hline
  \multicolumn{5}{l}{\commenta BJD$-$2400000.} \\
  \multicolumn{5}{l}{\commentb Against $max = 2455421.4761 + 0.070121 E$.} \\
  \multicolumn{5}{l}{\commentc Number of points used to determine the maximum.} \\
\end{tabular}
\end{center}
\end{table}

\subsection{V1113 Cygni}\label{obj:v1113cyg}

\begin{figure}
  \begin{center}
%    \FigureFile(88mm,70mm){v1113cygcomp.eps}
    \FigureFile(88mm,70mm){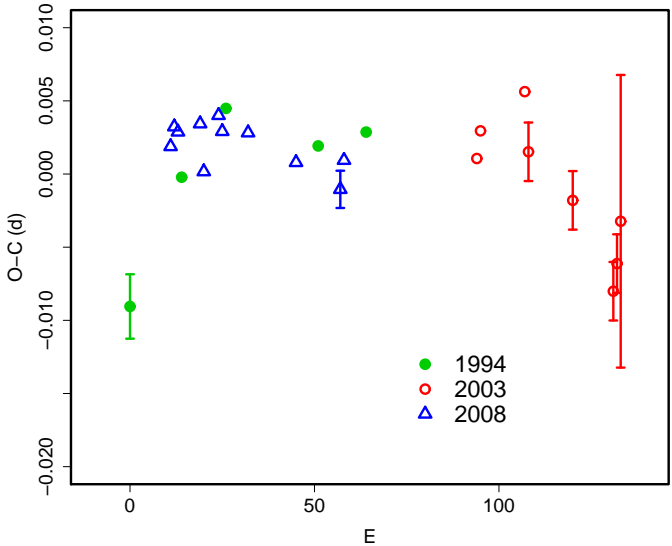}
  \end{center}
  \caption{Comparison of $O-C$ diagrams of V1113 Cyg between different
  superoutbursts.  A period of 0.07911 d was used to draw this figure.
  Approximate cycle counts ($E$) after the start of the superoutburst
  were used.}
  \label{fig:v1113cygcomp}
\end{figure}

   \citet{bak10v1113cyg} recently reported new observations of 2003
and 2005 superoutbursts, and claimed the presence of a large negative
period derivative.  Upon examination of their observations, it has become
evident that they observed the apparently late stage of a superoutburst
in 2003.  Using the typical duration ($\sim 11$ d) of superoutbursts
in this system, their observation probably started $\sim$ 94 cycles
after the start of the superoutburst, and they most likely caught
a stage B--C transition.  A combined $O-C$ diagram based on this
interpretation, supplemented by early observations reported in
\citet{Pdot}, is shown in figure \ref{fig:v1113cygcomp}.  As judged
from this figure, the evolution of the superhump period is not particularly
unusual in this system.

\subsection{V1454 Cygni}\label{obj:v1454cyg}

\begin{figure}
  \begin{center}
%    \FigureFile(88mm,110mm){v1454cygshpdm.eps}
    \FigureFile(88mm,110mm){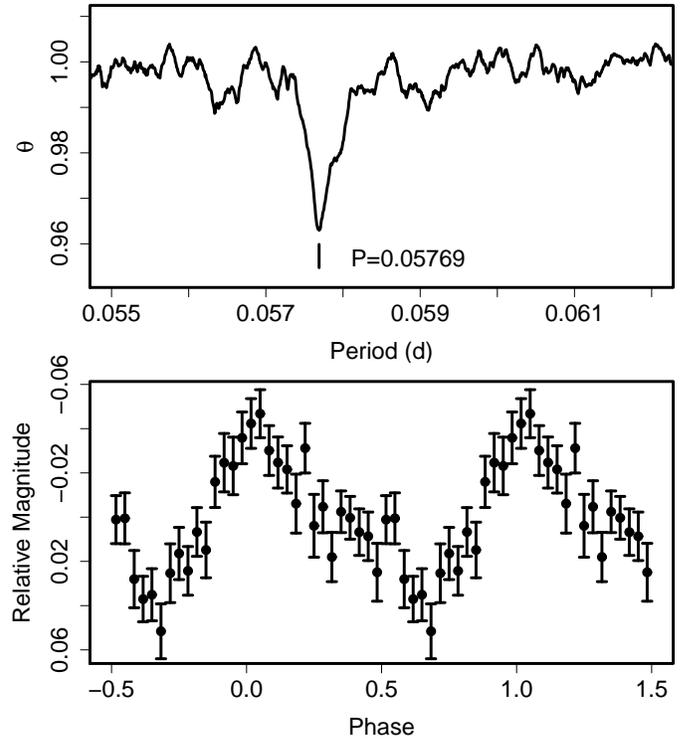}
  \end{center}
  \caption{Superhumps in V1454 Cyg (2009).
     (Upper): PDM analysis.
     (Lower): Phase-averaged profile.}
  \label{fig:v1454cygshpdm}
\end{figure}

   This object was only partly observed during the 2006 superoutburst
\citep{Pdot}.
The 2009 superoutburst was detected during its early stage
(I. Miller, baavss-alert 2020, vsnet-alert 11376).
A delay of $\geq$6.5 d in the full growth of ordinary superhumps was
recorded (vsnet-alert 11393, 11395).
The new observation now clarified the alias selection
[0.05769(2) d with the PDM method] and safely excluded the 0.0610 d
earlier reported (figure \ref{fig:v1454cygshpdm}).
The $P_{\rm dot}$ for stage B was $+10.3(3.5) \times 10^{-5}$,
fairly common for this short $P_{\rm SH}$.
Although there was likely a stage B--C transition after $E=95$,
we could not determine the period of stage C superhumps due to the
lack of observations.
An analysis of the earlier observation (BJD before 2455058) has
yielded a weak signal of 0.05777(2) d.  Although the extraction of times
of superhump maxima was difficult due to the low amplitudes, this period
suggests that this interval involved stage A evolution rather than
a manifestation of early superhumps.

\begin{table}
\caption{Superhump maxima of V1454 Cyg (2009).}\label{tab:v1454cygoc2009}
\begin{center}
\begin{tabular}{ccccc}
\hline
$E$ & max\commenta & error & $O-C$\commentb & $N$\commentc \\
\hline
0 & 55058.1434 & 0.0004 & 0.0055 & 122 \\
1 & 55058.2036 & 0.0013 & 0.0081 & 65 \\
39 & 55060.3868 & 0.0010 & 0.0001 & 57 \\
40 & 55060.4431 & 0.0024 & $-$0.0013 & 54 \\
41 & 55060.4996 & 0.0010 & $-$0.0025 & 56 \\
42 & 55060.5540 & 0.0035 & $-$0.0057 & 54 \\
50 & 55061.0219 & 0.0009 & 0.0009 & 109 \\
51 & 55061.0756 & 0.0019 & $-$0.0031 & 107 \\
52 & 55061.1361 & 0.0024 & $-$0.0003 & 121 \\
53 & 55061.1919 & 0.0020 & $-$0.0021 & 121 \\
54 & 55061.2405 & 0.0033 & $-$0.0112 & 115 \\
56 & 55061.3703 & 0.0016 & 0.0033 & 54 \\
57 & 55061.4258 & 0.0009 & 0.0012 & 52 \\
58 & 55061.4806 & 0.0009 & $-$0.0017 & 56 \\
59 & 55061.5380 & 0.0011 & $-$0.0020 & 57 \\
60 & 55061.5956 & 0.0042 & $-$0.0021 & 52 \\
75 & 55062.4672 & 0.0019 & 0.0046 & 57 \\
76 & 55062.5230 & 0.0027 & 0.0027 & 57 \\
86 & 55063.0867 & 0.0032 & $-$0.0101 & 114 \\
87 & 55063.1513 & 0.0064 & $-$0.0032 & 121 \\
88 & 55063.2097 & 0.0155 & $-$0.0024 & 86 \\
92 & 55063.4463 & 0.0024 & 0.0035 & 56 \\
93 & 55063.5010 & 0.0033 & 0.0005 & 57 \\
94 & 55063.5635 & 0.0023 & 0.0054 & 57 \\
95 & 55063.6254 & 0.0029 & 0.0095 & 30 \\
160 & 55067.3705 & 0.0019 & 0.0066 & 57 \\
161 & 55067.4330 & 0.0020 & 0.0115 & 54 \\
163 & 55067.5267 & 0.0037 & $-$0.0102 & 57 \\
164 & 55067.5890 & 0.0019 & $-$0.0055 & 54 \\
\hline
  \multicolumn{5}{l}{\commenta BJD$-$2400000.} \\
  \multicolumn{5}{l}{\commentb Against $max = 2455058.1379 + 0.057662 E$.} \\
  \multicolumn{5}{l}{\commentc Number of points used to determine the maximum.} \\
\end{tabular}
\end{center}
\end{table}

\subsection{AQ Eridani}\label{obj:aqeri}

\begin{figure}
  \begin{center}
%    \FigureFile(88mm,70mm){aqericomp2.eps}
    \FigureFile(88mm,70mm){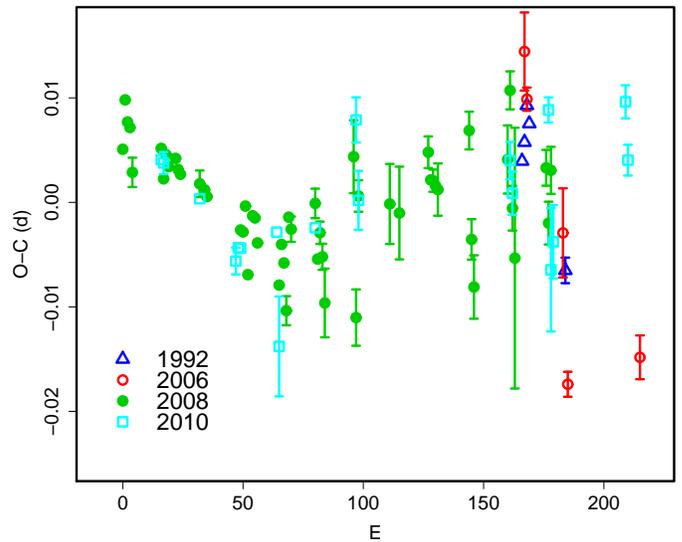}
  \end{center}
  \caption{Comparison of $O-C$ diagrams of AQ Eri between different
  superoutbursts.  A period of 0.06238 d was used to draw this figure.
  Approximate cycle counts ($E$) after the appearance of the
  superhumps were used.  Due to the lack of early observations, the
  $O-C$ diagrams for the 1992 and 2006 were shifted to better match
  the 2008 one.
  }
  \label{fig:aqericomp2}
\end{figure}

   We observed another superoutburst in 2010 (table \ref{tab:aqerioc2010}).
There was a relatively large scatter in the $O-C$ diagram after the
rapid fading due to its faintness.  We therefore did not attempt
to determine the period during stage C.  The mean $P_{\rm SH}$ appears
to confirm the previous period determination (cf. figure
\ref{fig:aqericomp2}).

\begin{table}
\caption{Superhump maxima of AQ Eri (2010).}\label{tab:aqerioc2010}
\begin{center}
\begin{tabular}{ccccc}
\hline
$E$ & max\commenta & error & $O-C$\commentb & $N$\commentc \\
\hline
0 & 55201.0094 & 0.0004 & 0.0067 & 54 \\
1 & 55201.0715 & 0.0011 & 0.0064 & 37 \\
16 & 55202.0038 & 0.0003 & 0.0025 & 102 \\
31 & 55202.9335 & 0.0013 & $-$0.0039 & 65 \\
32 & 55202.9972 & 0.0003 & $-$0.0027 & 223 \\
33 & 55203.0595 & 0.0004 & $-$0.0027 & 73 \\
48 & 55203.9967 & 0.0006 & $-$0.0016 & 90 \\
49 & 55204.0482 & 0.0048 & $-$0.0126 & 51 \\
64 & 55204.9952 & 0.0008 & $-$0.0017 & 100 \\
81 & 55206.0660 & 0.0022 & 0.0081 & 45 \\
82 & 55206.1207 & 0.0028 & 0.0004 & 37 \\
145 & 55210.0545 & 0.0018 & 0.0023 & 142 \\
146 & 55210.1137 & 0.0021 & $-$0.0008 & 146 \\
161 & 55211.0574 & 0.0012 & 0.0067 & 68 \\
162 & 55211.1045 & 0.0059 & $-$0.0086 & 65 \\
163 & 55211.1695 & 0.0035 & $-$0.0060 & 101 \\
193 & 55213.0543 & 0.0016 & 0.0065 & 61 \\
194 & 55213.1111 & 0.0015 & 0.0009 & 91 \\
\hline
  \multicolumn{5}{l}{\commenta BJD$-$2400000.} \\
  \multicolumn{5}{l}{\commentb Against $max = 2455201.0027 + 0.062410 E$.} \\
  \multicolumn{5}{l}{\commentc Number of points used to determine the maximum.} \\
\end{tabular}
\end{center}
\end{table}

\subsection{VX Fornacis}\label{obj:vxfor}

   VX For was discovered in 1990 as a probable dwarf nova showing Balmer,
He\textsc{I} and possibly He\textsc{II} emission lines
\citep{lil90vxforiauc}.  The large outburst amplitude and the outburst
lasting for more than 10 d \citep{lil99v1494aqliauc7327} were already
suggestive of an SU UMa-type superoutburst.
The object has been listed as a good candidate for a WZ Sge-type
dwarf nova \citep{kat01hvvir}.

\begin{figure}
  \begin{center}
%    \FigureFile(88mm,70mm){vxforlc.eps}
    \FigureFile(88mm,70mm){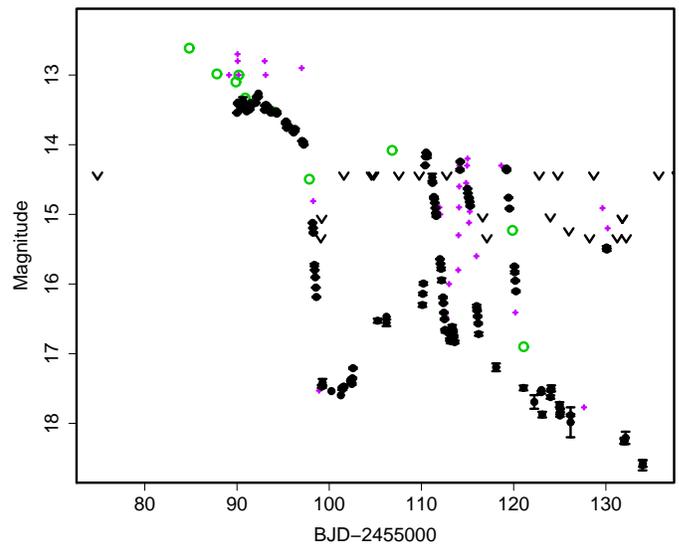}
  \end{center}
  \caption{Light curve of VX For (2009).
     The filled circles, open circles and small crosses represent CCD
     observations used here, ASAS-3 $V$ data and visual observations,
     respectively.}
  \label{fig:vxforlc}
\end{figure}

   The 2009 outburst, first-ever since the initial discovery, was
detected by R. Stubbings on 2009 Sep. 14 at a visual magnitude of
13.0 (vsnet-alert 11471).
Ordinary superhump were soon observed
following this outburst detection (vsnet-alert 11474).  Although the
early appearance of ordinary superhumps was initially considered as
unfavorable for the WZ Sge-type interpretation, a later retrospective
detection of earlier positive observation in the ASAS-3 data
($V$ = 12.61 on September 10) indicated that the early stage of
the outburst was missed when early superhumps were expected
(vsnet-alert 11492).  Using the empirical classification of
WZ Sge-type dwarf novae introduced in \citet{Pdot},
T. Kato suggested that the object is expected to undergo multiple
rebrightenings as in EG Cnc based on relatively small $P_{\rm dot}$
obtained from early observations and a relatively long $P_{\rm SH}$
(vsnet-alert 11492).  The object indeed underwent five rebrightenings 
(vsnet-alert 11521, 11526, 11536, 11577, 11602; figure \ref{fig:vxforlc}),
making it the first object predicted for its multiple rebrightening
in real-time.  The overall behavior is very similar to that of
ASAS J153616$-$0839.1 \citep{Pdot}.

\begin{figure}
  \begin{center}
%    \FigureFile(88mm,110mm){vxforshpdm.eps}
    \FigureFile(88mm,110mm){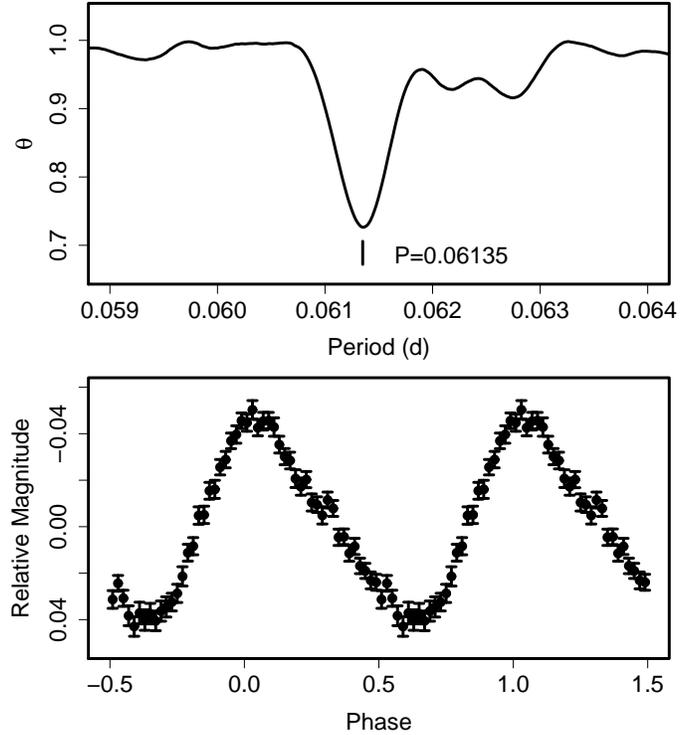}
  \end{center}
  \caption{Superhumps in VX For during the superoutburst plateau (2009).
     (Upper): PDM analysis.
     (Lower): Phase-averaged profile.}
  \label{fig:vxforshpdm}
\end{figure}

   The mean $P_{\rm SH}$ during the plateau phase was 0.061355(7) d
(PDM method, figure \ref{fig:vxforshpdm}).  The times of superhump maxima
during the plateau phase and fading stage are listed in table
\ref{tab:vxforoc2009}.

\begin{figure}
  \begin{center}
%    \FigureFile(88mm,110mm){vxforlatepdm.eps}
    \FigureFile(88mm,110mm){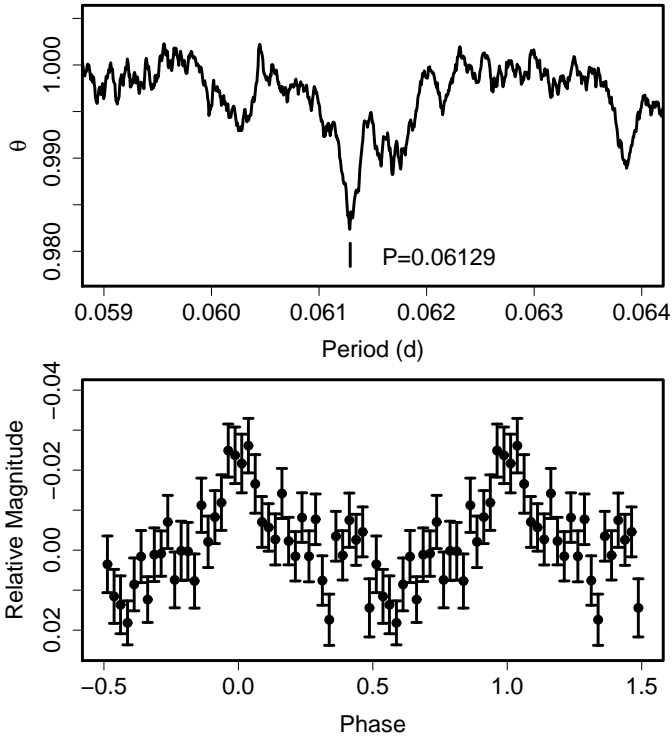}
  \end{center}
  \caption{Superhumps in VX For during the post-superoutburst and rebrightening
     phase (2009).
     (Upper): PDM analysis.
     (Lower): Phase-averaged profile.}
  \label{fig:vxforlatepdm}
\end{figure}

   The superhumps persisted during the post-superoutburst and rebrightening
phase.  We obtained $P_{\rm SH} = 0.06129(2)$ d, slightly short than
that of the superoutburst plateau phase (figure \ref{fig:vxforlatepdm}).
There was some indication of a signal around $P = 0.06156$ d (small
excess signal in figure \ref{fig:vxforlatepdm}) and transient appearance
of this period (cf. vsnet-alert 11513).  Although this signal needs to be
confirmed with better data, this might be analogous to long-period
late-stage superhumps in WZ Sge-type dwarf novae (\cite{kat08wzsgelateSH};
\cite{Pdot}).

\begin{table}
\caption{Superhump maxima of VX For (2009).}\label{tab:vxforoc2009}
\begin{center}
\begin{tabular}{ccccc}
\hline
$E$ & max\commenta & error & $O-C$\commentb & $N$\commentc \\
\hline
0 & 55090.0917 & 0.0002 & $-$0.0100 & 991 \\
6 & 55090.4665 & 0.0004 & $-$0.0026 & 258 \\
7 & 55090.5274 & 0.0004 & $-$0.0029 & 258 \\
8 & 55090.5873 & 0.0003 & $-$0.0042 & 230 \\
19 & 55091.2630 & 0.0007 & $-$0.0019 & 104 \\
20 & 55091.3216 & 0.0006 & $-$0.0046 & 104 \\
21 & 55091.3893 & 0.0004 & 0.0019 & 252 \\
22 & 55091.4483 & 0.0003 & $-$0.0003 & 229 \\
23 & 55091.5073 & 0.0015 & $-$0.0025 & 51 \\
24 & 55091.5683 & 0.0015 & $-$0.0028 & 116 \\
32 & 55092.0586 & 0.0010 & $-$0.0022 & 42 \\
33 & 55092.1223 & 0.0005 & 0.0003 & 68 \\
34 & 55092.1822 & 0.0009 & $-$0.0011 & 66 \\
35 & 55092.2450 & 0.0008 & 0.0005 & 84 \\
36 & 55092.3078 & 0.0007 & 0.0021 & 30 \\
48 & 55093.0420 & 0.0003 & 0.0016 & 349 \\
49 & 55093.0999 & 0.0003 & $-$0.0018 & 127 \\
50 & 55093.1617 & 0.0003 & $-$0.0012 & 159 \\
51 & 55093.2244 & 0.0006 & 0.0003 & 95 \\
55 & 55093.4687 & 0.0003 & $-$0.0003 & 223 \\
56 & 55093.5293 & 0.0003 & $-$0.0009 & 261 \\
57 & 55093.5909 & 0.0003 & $-$0.0006 & 261 \\
66 & 55094.1453 & 0.0015 & 0.0028 & 51 \\
67 & 55094.2044 & 0.0011 & 0.0007 & 41 \\
68 & 55094.2659 & 0.0008 & 0.0010 & 162 \\
69 & 55094.3264 & 0.0018 & 0.0002 & 68 \\
85 & 55095.3094 & 0.0012 & 0.0037 & 126 \\
87 & 55095.4308 & 0.0008 & 0.0026 & 258 \\
88 & 55095.4932 & 0.0006 & 0.0038 & 214 \\
89 & 55095.5545 & 0.0005 & 0.0039 & 213 \\
99 & 55096.1696 & 0.0013 & 0.0068 & 52 \\
100 & 55096.2289 & 0.0011 & 0.0049 & 46 \\
101 & 55096.2960 & 0.0020 & 0.0108 & 33 \\
115 & 55097.1400 & 0.0049 & $-$0.0024 & 24 \\
132 & 55098.1979 & 0.0017 & 0.0147 & 22 \\
133 & 55098.2561 & 0.0012 & 0.0117 & 32 \\
137 & 55098.4994 & 0.0014 & 0.0100 & 258 \\
149 & 55099.2215 & 0.0059 & $-$0.0025 & 23 \\
185 & 55101.4220 & 0.0094 & $-$0.0061 & 134 \\
186 & 55101.4843 & 0.0031 & $-$0.0050 & 134 \\
187 & 55101.5307 & 0.0027 & $-$0.0198 & 134 \\
201 & 55102.3908 & 0.0103 & $-$0.0169 & 125 \\
202 & 55102.4749 & 0.0167 & 0.0061 & 135 \\
203 & 55102.5323 & 0.0018 & 0.0022 & 135 \\
204 & 55102.5907 & 0.0018 & $-$0.0006 & 133 \\
\hline
  \multicolumn{5}{l}{\commenta BJD$-$2400000.} \\
  \multicolumn{5}{l}{\commentb Against $max = 2455090.1017 + 0.061224 E$.} \\
  \multicolumn{5}{l}{\commentc Number of points used to determine the maximum.} \\
\end{tabular}
\end{center}
\end{table}

\subsection{AW Geminorum}\label{obj:awgem}

\begin{figure}
  \begin{center}
%    \FigureFile(88mm,70mm){awgemcomp.eps}
    \FigureFile(88mm,70mm){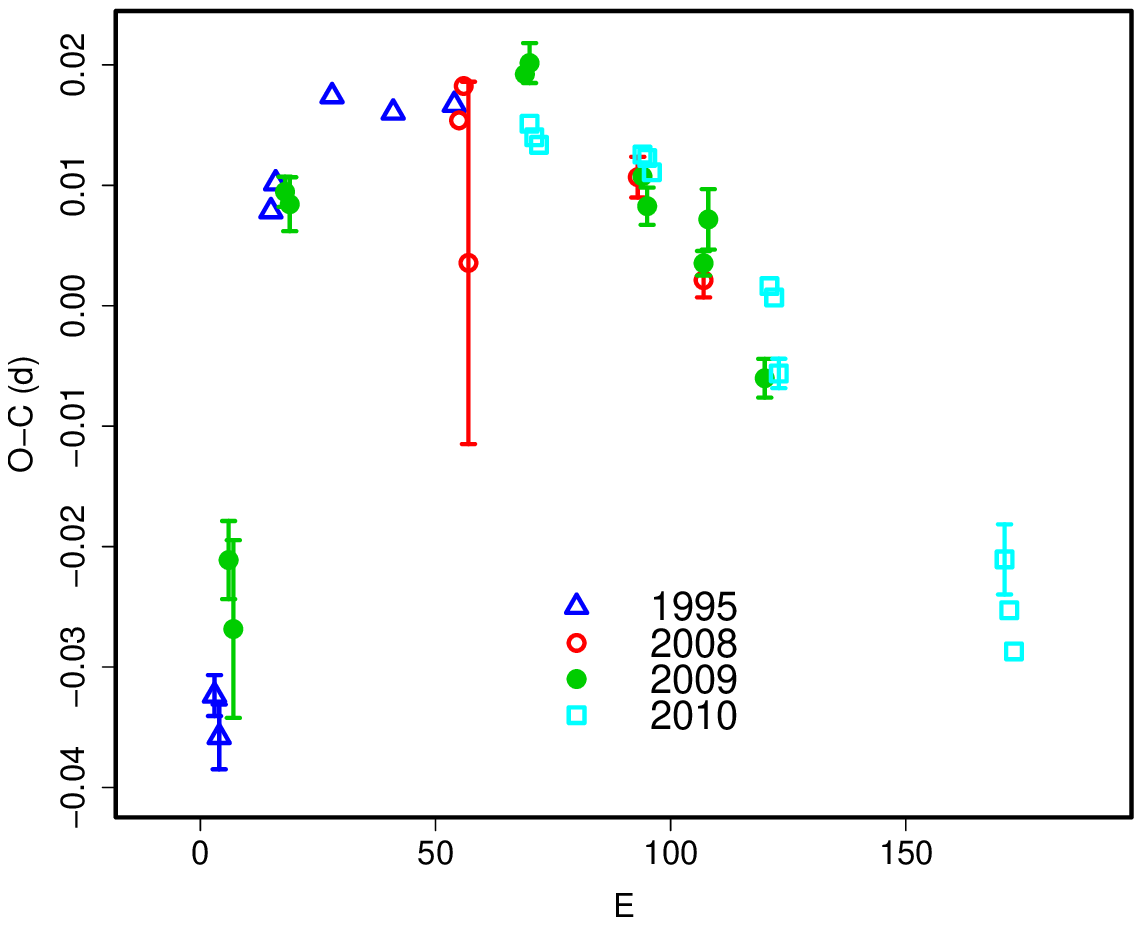}
  \end{center}
  \caption{Comparison of $O-C$ diagrams of AW Gem between different
  superoutbursts.  A period of 0.07915 d was used to draw this figure.
  Approximate cycle counts ($E$) after the start of the
  superoutburst were used.
  Since the start of 2008 and 2010 superoutburst were poorly constrained,
  we shifted the $O-C$ diagrams to best match other other superoutbursts.
  }
  \label{fig:awgemcomp}
\end{figure}

   The times of superhump maxima during the 2010 superoutburst is
listed in table \ref{tab:awgemoc2010}.  A stage B--C transition was
well recorded.  Although only parts of superoutbursts were observed
in individual years, the combined $O-C$ diagram
(figure \ref{fig:awgemcomp}) clearly demonstrates the universal
pattern of period evolution.

\begin{table}
\caption{Superhump maxima of AW Gem (2010).}\label{tab:awgemoc2010}
\begin{center}
\begin{tabular}{ccccc}
\hline
$E$ & max\commenta & error & $O-C$\commentb & $N$\commentc \\
\hline
0 & 55257.3860 & 0.0002 & $-$0.0033 & 72 \\
1 & 55257.4640 & 0.0003 & $-$0.0040 & 40 \\
2 & 55257.5425 & 0.0004 & $-$0.0042 & 69 \\
24 & 55259.2830 & 0.0004 & 0.0040 & 81 \\
25 & 55259.3619 & 0.0004 & 0.0041 & 80 \\
26 & 55259.4398 & 0.0005 & 0.0033 & 80 \\
51 & 55261.4091 & 0.0004 & 0.0041 & 69 \\
52 & 55261.4873 & 0.0005 & 0.0035 & 61 \\
53 & 55261.5602 & 0.0012 & $-$0.0024 & 37 \\
101 & 55265.3439 & 0.0029 & 0.0018 & 43 \\
102 & 55265.4189 & 0.0008 & $-$0.0020 & 78 \\
103 & 55265.4946 & 0.0006 & $-$0.0050 & 79 \\
\hline
  \multicolumn{5}{l}{\commenta BJD$-$2400000.} \\
  \multicolumn{5}{l}{\commentb Against $max = 2455257.3892 + 0.078742 E$.} \\
  \multicolumn{5}{l}{\commentc Number of points used to determine the maximum.} \\
\end{tabular}
\end{center}
\end{table}

\subsection{IR Geminorum}\label{obj:irgem}

   The times of superhump maxima during the 2010 superoutburst is
listed in table \ref{tab:irgemoc2010}.  Although the object was
observed on only two nights, the period is in good agreement with
earlier observations.  The observations was likely performed during
stage B.

\begin{table}
\caption{Superhump maxima of IR Gem (2010).}\label{tab:irgemoc2010}
\begin{center}
\begin{tabular}{ccccc}
\hline
$E$ & max\commenta & error & $O-C$\commentb & $N$\commentc \\
\hline
0 & 55265.4201 & 0.0004 & 0.0003 & 65 \\
1 & 55265.4903 & 0.0004 & $-$0.0003 & 67 \\
56 & 55269.3867 & 0.0003 & 0.0002 & 71 \\
57 & 55269.4573 & 0.0004 & $-$0.0000 & 73 \\
58 & 55269.5280 & 0.0007 & $-$0.0002 & 50 \\
\hline
  \multicolumn{5}{l}{\commenta BJD$-$2400000.} \\
  \multicolumn{5}{l}{\commentb Against $max = 2455265.4198 + 0.070834 E$.} \\
  \multicolumn{5}{l}{\commentc Number of points used to determine the maximum.} \\
\end{tabular}
\end{center}
\end{table}

\subsection{V592 Herculis}\label{obj:v592her}

   V592 Her was discovered as a possible fast nova in 1968 on Sonneberg
plates \citep{ric68v592her}.  \citet{ric91v592her} further discovered
a second outburst in 1986 on historical plates.  In 1998, another outburst
was recorded which led to the identification of this object as being
a WZ Sge-type dwarf nova (\cite{due98v592her}; \cite{kat02v592her}).
\citet{kat02v592her} first identified the true superhump period
of 0.05648(2) d, although their observation was not very sufficient
to determine its period variation.

\begin{figure}
  \begin{center}
%    \FigureFile(88mm,110mm){v592hershpdm.eps}
    \FigureFile(88mm,110mm){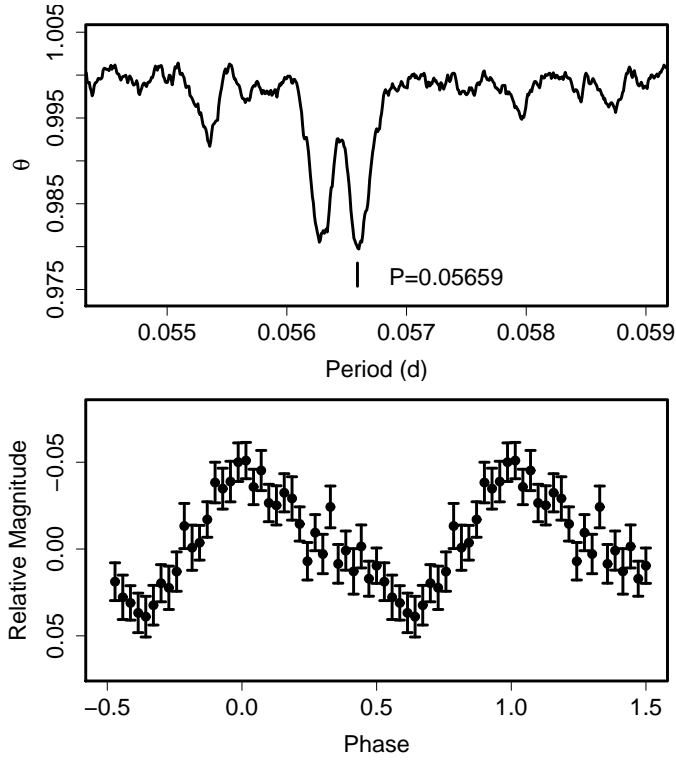}
  \end{center}
  \caption{Superhumps in V592 Her (2010).
     (Upper): PDM analysis.
     (Lower): Phase-averaged profile.}
  \label{fig:v592hershpdm}
\end{figure}

   The 2010 outburst of this object was detected by M. Reszelski at
a relatively faint magnitude of 14.16 (unfiltered CCD magnitude,
vsnet-obs 67929).  Subsequent observations detected growing superhumps
(vsnet-alert 12092, 12094, 12095).  This new outburst has confirmed
the selection of the superhump period in (\cite{kat02v592her};
figure \ref{fig:v592hershpdm}).

\begin{figure}
  \begin{center}
%    \FigureFile(88mm,90mm){v592heroc2010.eps}
    \FigureFile(88mm,90mm){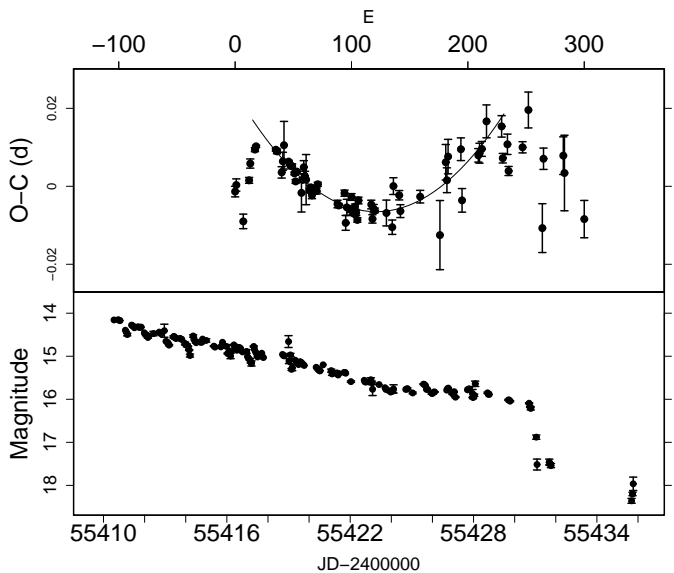}
  \end{center}
  \caption{$O-C$ of superhumps in V592 Her (2010).
  (Upper): $O-C$ diagram.  The $O-C$ values were against the mean period
  for the stage B ($35 \le E \le 216$, thin curve)
  (Lower): Light curve.  The outburst entered the rapid decline phase
  soon after the stage B--C transition.
  }
  \label{fig:v592heroc2010}
\end{figure}

   The times of superhump maxima are listed in table \ref{tab:v592heroc2010}.
All A--C stages for superhump evolution were clearly detected
(figure \ref{fig:v592heroc2010}).  We obtained a clearly positive
$P_{\rm dot}$ of $+7.4(0.6) \times 10^{-5}$ during stage B, which is
likely a more reliable value than the 1998 estimate thanks to the greatly
improved statistics.  As judged from this relatively large period derivative
and the presence of stage C evolution, this object is less likely an
extreme WZ Sge-type dwarf nova with a very small period variation
what was supposed from the 1998 data \citet{kat02v592her}.
The relatively short ($\sim$ 12 yr) recurrence time would qualify the
system as a WZ Sge-type dwarf nova similar to HV Vir \citep{ish03hvvir}.
The lack of repetitive rebrightenings would also support this classification.

\begin{figure}
  \begin{center}
%    \FigureFile(88mm,110mm){v592hereshpdm.eps}
    \FigureFile(88mm,110mm){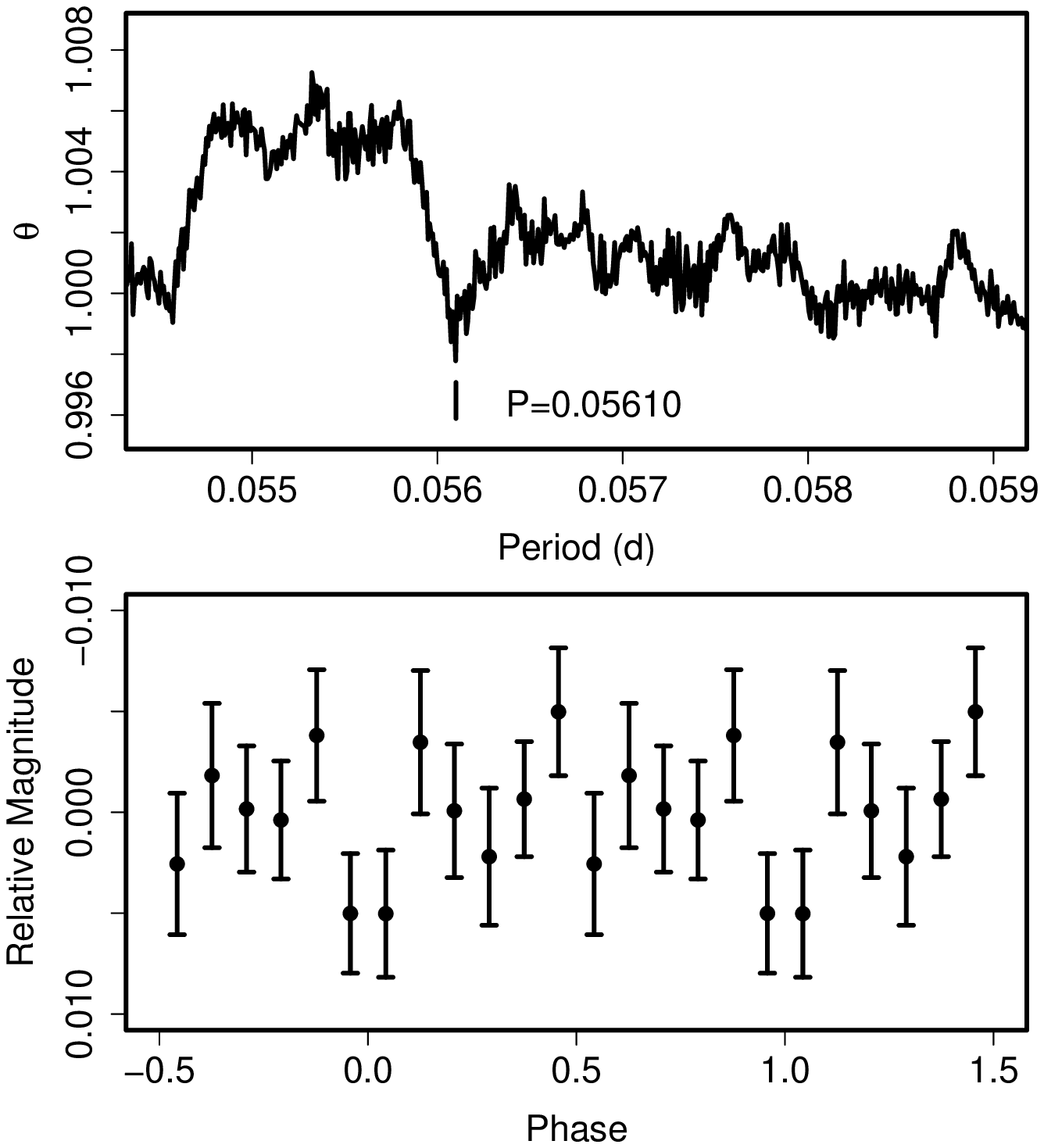}
  \end{center}
  \caption{Candidate early superhumps in V592 Her (2010).
     (Upper): PDM analysis.
     (Lower): Phase-averaged profile.}
  \label{fig:v592hereshpdm}
\end{figure}

   There was a weaker signal slightly shorter than the superhump period
(figure \ref{fig:v592hershpdm}), which might be attributed to the orbital
period.  Although the detection was not statistically significant,
a similar periodicity may have been present before the appearance of
ordinary superhumps (figure \ref{fig:v592hereshpdm}).  Since this
period agrees with the candidate period from radial-velocity study
(\cite{men02v592her}; see \cite{kat02v592her} for the alias selection),
we adopted it as a candidate for the orbital period, which corresponds
to a fractional superhump excess of 0.9 \%.

\begin{table}
\caption{Superhump maxima of V592 Her (2010).}\label{tab:v592heroc2010}
\begin{center}
\begin{tabular}{ccccc}
\hline
$E$ & max\commenta & error & $O-C$\commentb & $N$\commentc \\
\hline
0 & 55413.7374 & 0.0014 & $-$0.0009 & 53 \\
1 & 55413.7957 & 0.0016 & 0.0008 & 53 \\
7 & 55414.1260 & 0.0019 & $-$0.0086 & 82 \\
12 & 55414.4196 & 0.0008 & 0.0018 & 53 \\
13 & 55414.4805 & 0.0012 & 0.0061 & 54 \\
17 & 55414.7105 & 0.0005 & 0.0096 & 48 \\
18 & 55414.7679 & 0.0003 & 0.0105 & 53 \\
35 & 55415.7294 & 0.0004 & 0.0094 & 52 \\
36 & 55415.7856 & 0.0006 & 0.0089 & 38 \\
40 & 55416.0066 & 0.0015 & 0.0035 & 58 \\
41 & 55416.0661 & 0.0023 & 0.0063 & 79 \\
42 & 55416.1268 & 0.0061 & 0.0104 & 127 \\
46 & 55416.3490 & 0.0003 & 0.0062 & 33 \\
47 & 55416.4049 & 0.0002 & 0.0054 & 104 \\
48 & 55416.4612 & 0.0003 & 0.0051 & 71 \\
49 & 55416.5176 & 0.0006 & 0.0049 & 42 \\
51 & 55416.6290 & 0.0003 & 0.0031 & 62 \\
52 & 55416.6836 & 0.0005 & 0.0010 & 69 \\
53 & 55416.7427 & 0.0004 & 0.0035 & 52 \\
57 & 55416.9637 & 0.0049 & $-$0.0020 & 71 \\
58 & 55417.0236 & 0.0006 & 0.0013 & 175 \\
59 & 55417.0835 & 0.0017 & 0.0046 & 89 \\
60 & 55417.1378 & 0.0013 & 0.0022 & 82 \\
61 & 55417.1935 & 0.0064 & 0.0013 & 50 \\
64 & 55417.3606 & 0.0005 & $-$0.0015 & 89 \\
65 & 55417.4180 & 0.0005 & $-$0.0007 & 143 \\
66 & 55417.4725 & 0.0009 & $-$0.0027 & 55 \\
70 & 55417.7001 & 0.0008 & $-$0.0017 & 35 \\
71 & 55417.7584 & 0.0005 & 0.0000 & 51 \\
88 & 55418.7157 & 0.0010 & $-$0.0053 & 52 \\
89 & 55418.7719 & 0.0005 & $-$0.0057 & 51 \\
94 & 55419.0581 & 0.0008 & $-$0.0026 & 27 \\
95 & 55419.1071 & 0.0019 & $-$0.0102 & 68 \\
96 & 55419.1677 & 0.0021 & $-$0.0062 & 66 \\
99 & 55419.3370 & 0.0008 & $-$0.0068 & 61 \\
100 & 55419.3968 & 0.0009 & $-$0.0036 & 71 \\
101 & 55419.4498 & 0.0009 & $-$0.0072 & 75 \\
102 & 55419.5057 & 0.0004 & $-$0.0080 & 62 \\
103 & 55419.5641 & 0.0006 & $-$0.0061 & 38 \\
104 & 55419.6192 & 0.0007 & $-$0.0077 & 62 \\
105 & 55419.6739 & 0.0006 & $-$0.0096 & 72 \\
106 & 55419.7355 & 0.0009 & $-$0.0046 & 53 \\
117 & 55420.3571 & 0.0011 & $-$0.0058 & 62 \\
118 & 55420.4101 & 0.0011 & $-$0.0095 & 69 \\
119 & 55420.4693 & 0.0011 & $-$0.0069 & 20 \\
120 & 55420.5254 & 0.0006 & $-$0.0074 & 52 \\
130 & 55421.0909 & 0.0033 & $-$0.0081 & 62 \\
135 & 55421.3703 & 0.0018 & $-$0.0119 & 60 \\
136 & 55421.4374 & 0.0021 & $-$0.0013 & 68 \\
141 & 55421.7181 & 0.0012 & $-$0.0037 & 52 \\
142 & 55421.7706 & 0.0016 & $-$0.0078 & 41 \\
159 & 55422.7367 & 0.0016 & $-$0.0043 & 52 \\
176 & 55423.6891 & 0.0089 & $-$0.0144 & 42 \\
181 & 55423.9909 & 0.0046 & 0.0042 & 149 \\
182 & 55424.0429 & 0.0032 & $-$0.0004 & 113 \\
183 & 55424.1055 & 0.0044 & 0.0056 & 60 \\
194 & 55424.7301 & 0.0029 & 0.0073 & 52 \\
\hline
  \multicolumn{5}{l}{\commenta BJD$-$2400000.} \\
  \multicolumn{5}{l}{\commentb Against $max = 2455413.7383 + 0.056621 E$.} \\
  \multicolumn{5}{l}{\commentc Number of points used to determine the maximum.} \\
\end{tabular}
\end{center}
\end{table}

\addtocounter{table}{-1}
\begin{table}
\caption{Superhump maxima of V592 Her (2010). (continued)}
\begin{center}
\begin{tabular}{ccccc}
\hline
$E$ & max\commenta & error & $O-C$\commentb & $N$\commentc \\
\hline
195 & 55424.7736 & 0.0030 & $-$0.0058 & 42 \\
209 & 55425.5776 & 0.0017 & 0.0056 & 47 \\
210 & 55425.6348 & 0.0026 & 0.0061 & 59 \\
212 & 55425.7491 & 0.0019 & 0.0072 & 53 \\
216 & 55425.9826 & 0.0042 & 0.0142 & 162 \\
229 & 55426.7172 & 0.0027 & 0.0127 & 53 \\
230 & 55426.7657 & 0.0012 & 0.0046 & 45 \\
234 & 55426.9957 & 0.0026 & 0.0081 & 82 \\
235 & 55427.0454 & 0.0012 & 0.0012 & 113 \\
247 & 55427.7308 & 0.0014 & 0.0071 & 53 \\
252 & 55428.0234 & 0.0046 & 0.0166 & 62 \\
264 & 55428.6724 & 0.0063 & $-$0.0138 & 36 \\
265 & 55428.7468 & 0.0027 & 0.0039 & 52 \\
282 & 55429.7099 & 0.0048 & 0.0045 & 52 \\
283 & 55429.7621 & 0.0097 & 0.0000 & 52 \\
300 & 55430.7126 & 0.0048 & $-$0.0120 & 50 \\
\hline
  \multicolumn{5}{l}{\commenta BJD$-$2400000.} \\
  \multicolumn{5}{l}{\commentb Against $max = 2455413.7389 + 0.056620 E$.} \\
  \multicolumn{5}{l}{\commentc Number of points used to determine the maximum.} \\
\end{tabular}
\end{center}
\end{table}

\subsection{V660 Herculis}\label{obj:v660her}

   The times of superhump maxima during the 2009 superoutburst are listed
in table \ref{tab:v660heroc2009}.  Since the late stage of the superoutburst
was only observed, the recorded superhumps are likely stage C superhumps.

\begin{table}
\caption{Superhump maxima of V660 Her (2009).}\label{tab:v660heroc2009}
\begin{center}
\begin{tabular}{ccccc}
\hline
$E$ & max\commenta & error & $O-C$\commentb & $N$\commentc \\
\hline
0 & 55056.4789 & 0.0008 & $-$0.0032 & 30 \\
12 & 55057.4501 & 0.0007 & $-$0.0004 & 36 \\
25 & 55058.5061 & 0.0019 & 0.0067 & 9 \\
49 & 55060.4337 & 0.0049 & $-$0.0023 & 17 \\
61 & 55061.4035 & 0.0098 & $-$0.0008 & 23 \\
\hline
  \multicolumn{5}{l}{\commenta BJD$-$2400000.} \\
  \multicolumn{5}{l}{\commentb Against $max = 2455056.4821 + 0.080692 E$.} \\
  \multicolumn{5}{l}{\commentc Number of points used to determine the maximum.} \\
\end{tabular}
\end{center}
\end{table}

\subsection{V844 Herculis}\label{obj:v844her}

   Two further superoutbursts in 2009 February--March
(table \ref{tab:v844heroc2009}) and in 2010 April--May
(table \ref{tab:v844heroc2010}; see also vsnet-alert 11959 for the outburst
detection) were observed.  Both superoutbursts were
relatively faint and short ones.  During the former superoutburst
We obtained a clearly positive $P_{\rm dot}$ of $+9.5(1.7) \times 10^{-5}$
(likely for stage B).

\begin{figure}
  \begin{center}
%    \FigureFile(88mm,70mm){v844hercomp2.eps}
    \FigureFile(88mm,70mm){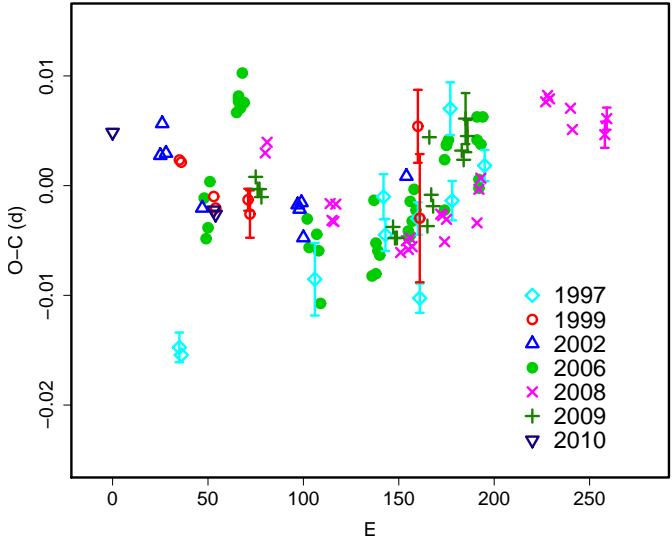}
  \end{center}
  \caption{Comparison of $O-C$ diagrams of V844 Her between different
  superoutbursts.  A period of 0.05590 d was used to draw this figure.
  Approximate cycle counts ($E$) after the start of the
  superoutburst were used.
  Since the start of the 2009 superoutburst was not well constrained,
  we shifted the $O-C$ diagram to best match the others.
  The initial superhump maximum of the 2010 superoutburst may have
  a one-cycle ambiguity in the cycle count, and the large $O-C$ may
  have been a result of the rapidly evolving stage A superhumps. 
  }
  \label{fig:v844hercomp2}
\end{figure}

   The object underwent yet another superoutburst
between them: 2009 October--November (vsnet-alert 11622).
There was also a normal outburst in 2009 June (vsnet-alert 11289, 11290).
The object was thus unusually active for this star (cf. \cite{kat00v844her};
\cite{oiz07v844her}).  As suggested in \citet{kat08wzsgelateSH} and
\citet{Pdot}, these superoutburst would provide an excellent opportunity
to study the dependence of period derivatives on the extent of
the superoutburst.  As already seen, the $P_{\rm dot}$ of the 2009
superoutburst is not significantly different from those of longer
superoutbursts.  Although the superhumps had already developed at the epoch
of initial observation ($\sim$ 1.5 d after the initial outburst detection),
we couldn't severely constrain the delay time in evolution of superhumps
since there was a three-day gap of observation before this detection.
The 2010 observation started $\sim$ 1 d after the rising phase of the
outburst, and superhumps were already present.  Although these superhumps
were possibly stage A superhumps (see figure \ref{fig:v844hercomp2}),
this observations seems to support the hypothesis in \citet{Pdot} that
the delay time in development of superhumps is shorter in smaller
superoutbursts.

\begin{table}
\caption{Superhump maxima of V844 Her (2009).}\label{tab:v844heroc2009}
\begin{center}
\begin{tabular}{ccccc}
\hline
$E$ & max\commenta & error & $O-C$\commentb & $N$\commentc \\
\hline
0 & 54887.4721 & 0.0003 & 0.0024 & 87 \\
1 & 54887.5268 & 0.0002 & 0.0012 & 104 \\
2 & 54887.5827 & 0.0002 & 0.0012 & 109 \\
3 & 54887.6379 & 0.0002 & 0.0005 & 125 \\
72 & 54891.4923 & 0.0003 & $-$0.0038 & 101 \\
73 & 54891.5472 & 0.0003 & $-$0.0049 & 108 \\
74 & 54891.6031 & 0.0004 & $-$0.0049 & 108 \\
90 & 54892.4986 & 0.0008 & $-$0.0042 & 105 \\
91 & 54892.5626 & 0.0008 & 0.0039 & 108 \\
92 & 54892.6132 & 0.0006 & $-$0.0014 & 107 \\
93 & 54892.6681 & 0.0009 & $-$0.0024 & 93 \\
108 & 54893.5116 & 0.0006 & 0.0023 & 106 \\
109 & 54893.5667 & 0.0005 & 0.0014 & 103 \\
110 & 54893.6264 & 0.0023 & 0.0052 & 60 \\
111 & 54893.6807 & 0.0015 & 0.0035 & 51 \\
\hline
  \multicolumn{5}{l}{\commenta BJD$-$2400000.} \\
  \multicolumn{5}{l}{\commentb Against $max = 2454887.4697 + 0.055923 E$.} \\
  \multicolumn{5}{l}{\commentc Number of points used to determine the maximum.} \\
\end{tabular}
\end{center}
\end{table}

\begin{table}
\caption{Superhump maxima of V844 Her (2010).}\label{tab:v844heroc2010}
\begin{center}
\begin{tabular}{ccccc}
\hline
$E$ & max\commenta & error & $O-C$\commentb & $N$\commentc \\
\hline
0 & 55316.1758 & 0.0002 & $-$0.0000 & 101 \\
53 & 55319.1315 & 0.0002 & 0.0001 & 134 \\
54 & 55319.1870 & 0.0002 & $-$0.0001 & 177 \\
\hline
  \multicolumn{5}{l}{\commenta BJD$-$2400000.} \\
  \multicolumn{5}{l}{\commentb Against $max = 2455316.1758 + 0.055764 E$.} \\
  \multicolumn{5}{l}{\commentc Number of points used to determine the maximum.} \\
\end{tabular}
\end{center}
\end{table}

\subsection{CT Hydrae}\label{obj:cthya}

   The times of superhump maxima during the 2010 superoutburst is
listed in table \ref{tab:cthyaoc2010}.  This observation first time
recorded a stage A--B transition and growth of superhumps.

\begin{table}
\caption{Superhump maxima of CT Hya (2010).}\label{tab:cthyaoc2010}
\begin{center}
\begin{tabular}{ccccc}
\hline
$E$ & max\commenta & error & $O-C$\commentb & $N$\commentc \\
\hline
0 & 55269.0068 & 0.0049 & $-$0.0021 & 91 \\
1 & 55269.0698 & 0.0024 & $-$0.0057 & 143 \\
14 & 55269.9458 & 0.0013 & 0.0047 & 86 \\
15 & 55270.0122 & 0.0005 & 0.0045 & 129 \\
90 & 55275.0002 & 0.0009 & $-$0.0014 & 142 \\
\hline
  \multicolumn{5}{l}{\commenta BJD$-$2400000.} \\
  \multicolumn{5}{l}{\commentb Against $max = 2455269.0089 + 0.066585 E$.} \\
  \multicolumn{5}{l}{\commentc Number of points used to determine the maximum.} \\
\end{tabular}
\end{center}
\end{table}

\subsection{V699 Ophiuchi}\label{obj:v699oph}

\begin{figure}
  \begin{center}
%    \FigureFile(88mm,70mm){v699ophcomp.eps}
    \FigureFile(88mm,70mm){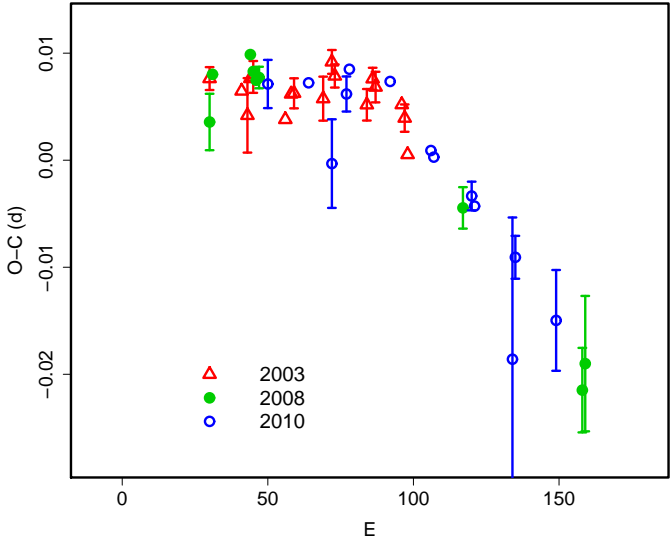}
  \end{center}
  \caption{Comparison of $O-C$ diagrams of V699 Oph between different
  superoutbursts.  A period of 0.07031 d was used to draw this figure.
  Since the start of the superoutbursts or appearance of superhumps
  were not well constrained, we shifted the $O-C$ diagrams
  to best fit each other.
  }
  \label{fig:v699ophcomp}
\end{figure}

   Our new observation of the 2010 superoutburst
(table \ref{tab:v699ophoc2010}) clearly caught a stage B--C transition.
A comparison of $O-C$ diagrams is shown in figure \ref{fig:v699ophcomp}.

\begin{table}
\caption{Superhump maxima of V699 Oph (2010).}\label{tab:v699ophoc2010}
\begin{center}
\begin{tabular}{ccccc}
\hline
$E$ & max\commenta & error & $O-C$\commentb & $N$\commentc \\
\hline
0 & 55362.5280 & 0.0023 & $-$0.0044 & 12 \\
14 & 55363.5125 & 0.0007 & $-$0.0008 & 77 \\
22 & 55364.0674 & 0.0041 & $-$0.0064 & 47 \\
27 & 55364.4254 & 0.0016 & 0.0014 & 41 \\
28 & 55364.4981 & 0.0007 & 0.0039 & 68 \\
42 & 55365.4813 & 0.0007 & 0.0063 & 65 \\
56 & 55366.4592 & 0.0008 & 0.0033 & 69 \\
57 & 55366.5288 & 0.0008 & 0.0029 & 60 \\
70 & 55367.4392 & 0.0013 & 0.0025 & 62 \\
71 & 55367.5086 & 0.0008 & 0.0018 & 62 \\
84 & 55368.4083 & 0.0132 & $-$0.0092 & 39 \\
85 & 55368.4882 & 0.0020 & 0.0005 & 67 \\
99 & 55369.4666 & 0.0047 & $-$0.0019 & 65 \\
\hline
  \multicolumn{5}{l}{\commenta BJD$-$2400000.} \\
  \multicolumn{5}{l}{\commentb Against $max = 2455362.5324 + 0.070061 E$.} \\
  \multicolumn{5}{l}{\commentc Number of points used to determine the maximum.} \\
\end{tabular}
\end{center}
\end{table}

\subsection{V1032 Ophiuchi}\label{obj:v1032oph}

   V1032 Oph was originally discovered as a candidate RR Lyr-type
variable star \citep{kin65v1032oph}.  Based on its bright UV emission,
\citet{wil09v1032oph} conducted a systematic observation which led to
a conclusion that the object is a likely SU UMa-type dwarf nova.

   The 2010 outburst was detected by E. Muyllaert (vsnet-alert 11898).
E. de Miguel and H. Maehara confirmed that this object is an eclipsing
SU UMa-type dwarf nova (vsnet-alert 11904, 11905).
The times of recorded eclipses, determined with the Kwee and van Woerden
(KW) method \citep{KWmethod}, are summarized in table \ref{tab:v1032ophecl}.
We obtained an ephemeris of

\begin{equation}
{\rm Min(BJD)} = 2455286.68267(17) + 0.0810564(13) E
\label{equ:v1032ophecl}.
\end{equation}

\begin{figure}
  \begin{center}
%    \FigureFile(88mm,110mm){v1032ophshpdm.eps}
    \FigureFile(88mm,110mm){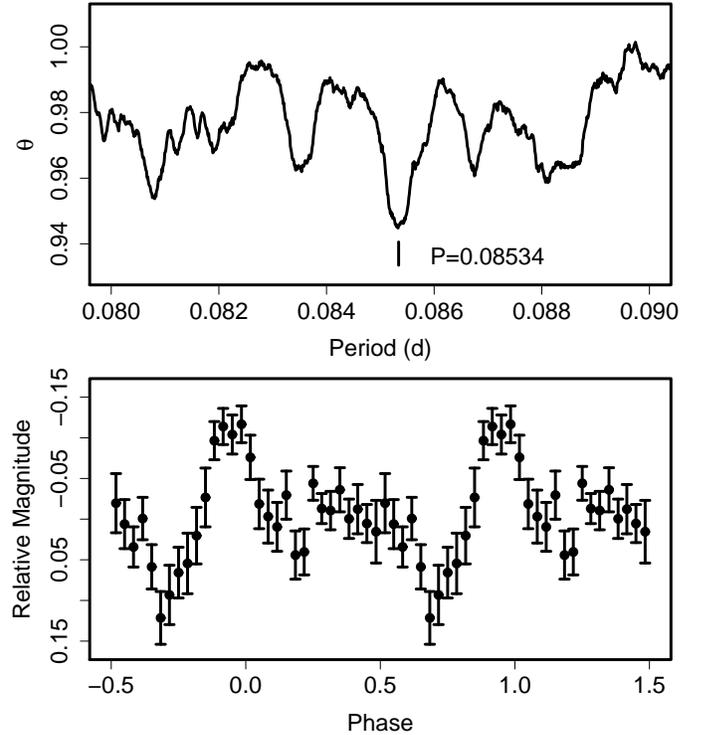}
  \end{center}
  \caption{Superhumps in V1032 Oph (2010).
     (Upper): PDM analysis.
     (Lower): Phase-averaged profile.}
  \label{fig:v1032ophshpdm}
\end{figure}

In the following analysis, we removed observations within 0.07
$P_{\rm orb}$ of eclipses.
The times of superhump maxima are listed in table \ref{tab:v1032ophoc2010}.
Due to the faintness of the object and overlapping eclipsing feature,
the scatter is relatively large.  The last maximum ($E=117$) may have been
a traditional late superhump with an $\sim$0.5 phase offset, when the
observation was performed after a brightness drop from the superoutburst.
A linear fit to the observed epochs for $E < 117$ yielded a mean period
of 0.08529(12) d.  This period is in agreement with 0.08534(5) d
determined with the PDM method (figure \ref{fig:v1032ophshpdm}).
We adopted the latter because of a smaller error.  Although the period
derivative could not be unambiguously determined, restricting to well-defined
maxima ($E=30, 47, 85, 86, 117$) and allowing a large period variation
(as in MN Dra), we can derive a global
$P_{\rm dot}$ of $-36(3) \times 10^{-5}$.  This value needs to be confirmed
by future observations.
The fractional superhump excess amounts to 5.3 \%, which is one of the
largest among SU UMa-type dwarf novae below the period gap
(cf. subsection \ref{obj:gzcnc}).

\begin{table}
\caption{Eclipse Minima of V1032 Oph.}\label{tab:v1032ophecl}
\begin{center}
\begin{tabular}{cccc}
\hline
$E$ & Minimum\commenta & error & $O-C$\commentb \\
\hline
0 & 55286.6829 & 0.0008 & 0.0002 \\
32 & 55289.2765 & 0.0006 & 0.0001 \\
49 & 55290.6544 & 0.0007 & -0.0000 \\
69 & 55292.2763 & 0.0010 & 0.0008 \\
74 & 55292.6809 & 0.0006 & 0.0001 \\
89 & 55293.8967 & 0.0005 & 0.0000 \\
90 & 55293.9775 & 0.0004 & -0.0002 \\
99 & 55294.7075 & 0.0006 & 0.0002 \\
102 & 55294.9505 & 0.0004 & 0.0001 \\
103 & 55295.0310 & 0.0006 & -0.0004 \\
111 & 55295.6801 & 0.0006 & 0.0001 \\
114 & 55295.9231 & 0.0004 & 0.0000 \\
115 & 55296.0041 & 0.0005 & -0.0001 \\
122 & 55296.5710 & 0.0005 & -0.0005 \\
123 & 55296.6517 & 0.0004 & -0.0009 \\
212 & 55303.8664 & 0.0011 & -0.0002 \\
213 & 55303.9478 & 0.0009 & 0.0001 \\
238 & 55305.9747 & 0.0006 & 0.0006 \\
296 & 55310.6754 & 0.0005 & 0.0001 \\
\hline
  \multicolumn{4}{l}{\commenta BJD$-$2400000.} \\
  \multicolumn{4}{l}{\commentb Against equation \ref{equ:v1032ophecl}.} \\
\end{tabular}
\end{center}
\end{table}

\begin{table}
\caption{Superhump maxima of V1032 Oph (2010).}\label{tab:v1032ophoc2010}
\begin{center}
\begin{tabular}{ccccc}
\hline
$E$ & max\commenta & error & $O-C$\commentb & $N$\commentc \\
\hline
0 & 55286.6842 & 0.0023 & 0.0062 & 74 \\
24 & 55288.7046 & 0.0018 & $-$0.0174 & 24 \\
30 & 55289.2294 & 0.0011 & $-$0.0036 & 167 \\
41 & 55290.1909 & 0.0040 & 0.0210 & 126 \\
43 & 55290.3086 & 0.0017 & $-$0.0316 & 73 \\
47 & 55290.6862 & 0.0006 & 0.0053 & 73 \\
65 & 55292.2223 & 0.0023 & 0.0083 & 147 \\
66 & 55292.3013 & 0.0072 & 0.0022 & 141 \\
85 & 55293.9201 & 0.0012 & 0.0028 & 34 \\
86 & 55294.0063 & 0.0015 & 0.0039 & 32 \\
94 & 55294.6949 & 0.0015 & 0.0110 & 65 \\
97 & 55294.9549 & 0.0018 & 0.0156 & 37 \\
98 & 55295.0222 & 0.0083 & $-$0.0023 & 26 \\
100 & 55295.2186 & 0.0035 & 0.0238 & 37 \\
106 & 55295.6961 & 0.0060 & $-$0.0097 & 60 \\
117 & 55296.6072 & 0.0006 & $-$0.0355 & 69 \\
\hline
  \multicolumn{5}{l}{\commenta BJD$-$2400000.} \\
  \multicolumn{5}{l}{\commentb Against $max = 2455286.6780 + 0.085169 E$.} \\
  \multicolumn{5}{l}{\commentc Number of points used to determine the maximum.} \\
\end{tabular}
\end{center}
\end{table}

\subsection{V2051 Ophiuchi}\label{obj:v2051oph}

   The times of superhump maxima during the 2010 superoutburst are listed
in table \ref{tab:v2051ophoc2010}.  Although the outburst was detected
during its relatively early stage (vsnet-alert 12049), the subsequent
observational coverage was rather insufficient.  The observed superhumps
were likely stage B superhumps.  We could not meaningfully determine
\citet{Pdot}.

\begin{table}
\caption{Superhump maxima of V2051 Oph (2010).}\label{tab:v2051ophoc2010}
\begin{center}
\begin{tabular}{ccccc}
\hline
$E$ & max\commenta & error & $O-C$\commentb & $N$\commentc \\
\hline
0 & 55382.9240 & 0.0003 & 0.0014 & 107 \\
1 & 55382.9876 & 0.0003 & 0.0008 & 104 \\
2 & 55383.0492 & 0.0002 & $-$0.0019 & 78 \\
3 & 55383.1165 & 0.0003 & 0.0012 & 94 \\
14 & 55383.8220 & 0.0014 & 0.0000 & 34 \\
17 & 55384.0142 & 0.0006 & $-$0.0004 & 92 \\
18 & 55384.0865 & 0.0005 & 0.0076 & 89 \\
34 & 55385.1019 & 0.0009 & $-$0.0048 & 71 \\
35 & 55385.1613 & 0.0027 & $-$0.0096 & 40 \\
65 & 55387.1037 & 0.0013 & 0.0057 & 37 \\
\hline
  \multicolumn{5}{l}{\commenta BJD$-$2400000.} \\
  \multicolumn{5}{l}{\commentb Against $max = 2455382.9226 + 0.064238 E$.} \\
  \multicolumn{5}{l}{\commentc Number of points used to determine the maximum.} \\
\end{tabular}
\end{center}
\end{table}

\subsection{EF Pegasi}\label{obj:efpeg}

   EF Peg is one of representatives of long-$P_{\rm SH}$ SU UMa-type
dwarf novae with infrequent outbursts.  \citet{Pdot} reported mildly
negative $P_{\rm dot}$ for the 1991 and 1997 superoutbursts.
The object underwent an outburst in 2009 December (vsnet-alert 11738,
baavss-alert 2177, vsnet-alert 11740), first time since its 2001
superoutburst.  The outburst was caught during its early stage and
the development of superhumps was recorded.

   The times of superhump maxima are listed in table \ref{tab:efpegoc2009}.
Due to the short visibility in the evening sky, the number of maxima
was relatively small.  The $O-C$ diagram, however, clearly shows the
presence of stage A during the evolutionary stage of superhumps
($E \le 11$), and subsequent phase of a slow period decrease.
We attributed the latter phase to a transition from stage B to C
and identified the periods in table \ref{tab:perlist}.

\begin{figure}
  \begin{center}
%    \FigureFile(88mm,70mm){efpegcomp2.eps}
    \FigureFile(88mm,70mm){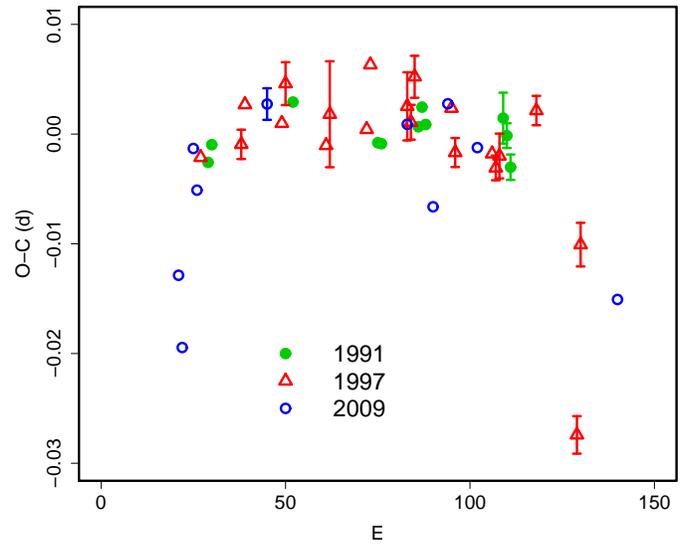}
  \end{center}
  \caption{Comparison of $O-C$ diagrams of EF Peg between different
  superoutbursts.  A period of 0.08705 d was used to draw this figure.
  Approximate cycle counts ($E$) after the start of the
  superoutburst were used.
  }
  \label{fig:efpegcomp2}
\end{figure}

   A comparison of $O-C$ variations between different superoutbursts
is presented in figure \ref{fig:efpegcomp2}.  Among long-$P_{\rm SH}$
systems, the $O-C$ variation looks similar to those of AX Cap and
SDSS J1627 (cf. \cite{Pdot}) with a discontinuous period variation
between stages B and C.  It would be noteworthy that all systems are
known to show only rare superoutbursts.

\begin{table}
\caption{Superhump maxima of EF Peg (2009).}\label{tab:efpegoc2009}
\begin{center}
\begin{tabular}{ccccc}
\hline
$E$ & max\commenta & error & $O-C$\commentb & $N$\commentc \\
\hline
0 & 55187.3273 & 0.0007 & $-$0.0296 & 229 \\
6 & 55187.8804 & 0.0008 & 0.0006 & 353 \\
7 & 55187.9609 & 0.0007 & $-$0.0061 & 521 \\
10 & 55188.2402 & 0.0005 & 0.0117 & 287 \\
11 & 55188.3234 & 0.0004 & 0.0078 & 277 \\
30 & 55189.9852 & 0.0014 & 0.0135 & 296 \\
68 & 55193.2913 & 0.0005 & 0.0073 & 268 \\
75 & 55193.8931 & 0.0008 & $-$0.0010 & 464 \\
79 & 55194.2507 & 0.0005 & 0.0080 & 80 \\
87 & 55194.9431 & 0.0008 & 0.0031 & 440 \\
125 & 55198.2372 & 0.0006 & $-$0.0151 & 77 \\
\hline
  \multicolumn{5}{l}{\commenta BJD$-$2400000.} \\
  \multicolumn{5}{l}{\commentb Against $max = 2455187.3569 + 0.087163 E$.} \\
  \multicolumn{5}{l}{\commentc Number of points used to determine the maximum.} \\
\end{tabular}
\end{center}
\end{table}

\subsection{V368 Pegasi}\label{obj:v368peg}

\begin{figure}
  \begin{center}
%    \FigureFile(88mm,70mm){v368pegcomp2.eps}
    \FigureFile(88mm,70mm){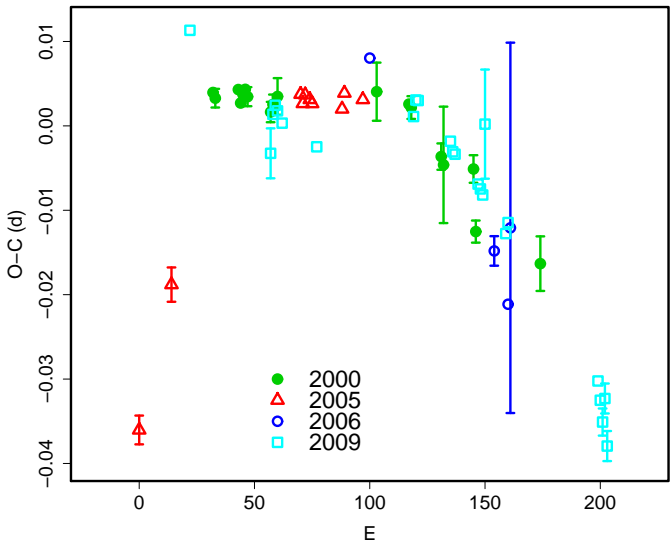}
  \end{center}
  \caption{Comparison of $O-C$ diagrams of V368 Peg between different
  superoutbursts.  A period of 0.07039 d was used to draw this figure.
  Approximate cycle counts ($E$) after the start of the
  superoutburst were used.
  }
  \label{fig:v368pegcomp2}
\end{figure}

   We also observed the 2009 superoutburst.  This outburst was one of the
brightest in recent years (cf. vsnet-alert 11507).
The times of superhump maxima are listed in table \ref{tab:v368pegoc2009},
which clearly shows a stage B--C transition.  The superhumps were not
very apparent on the first night (BJD 2455102), and it was likely that
the development of superhumps took more than 1 d.
The relatively large $P_{\rm dot}$ for stage B strongly depends on
$E = 0$, and may not be real.
A comparison of $O-C$ diagrams between different superoutbursts
is shown in figure \ref{fig:v368pegcomp2}.  Despite its brightness,
the behavior of the 2009 superoutburst was not strikingly different
from that of other superoutbursts.

\begin{table}
\caption{Superhump maxima of V368 Peg (2009).}\label{tab:v368pegoc2009}
\begin{center}
\begin{tabular}{ccccc}
\hline
$E$ & max\commenta & error & $O-C$\commentb & $N$\commentc \\
\hline
0 & 55102.8556 & 0.0009 & $-$0.0046 & 109 \\
35 & 55105.3047 & 0.0030 & $-$0.0110 & 31 \\
36 & 55105.3797 & 0.0003 & $-$0.0062 & 73 \\
37 & 55105.4512 & 0.0004 & $-$0.0048 & 69 \\
38 & 55105.5209 & 0.0005 & $-$0.0053 & 43 \\
40 & 55105.6602 & 0.0008 & $-$0.0063 & 24 \\
55 & 55106.7133 & 0.0003 & $-$0.0056 & 110 \\
97 & 55109.6732 & 0.0006 & 0.0078 & 153 \\
98 & 55109.7456 & 0.0005 & 0.0100 & 156 \\
99 & 55109.8159 & 0.0004 & 0.0102 & 148 \\
113 & 55110.7965 & 0.0006 & 0.0087 & 36 \\
114 & 55110.8657 & 0.0006 & 0.0076 & 33 \\
115 & 55110.9358 & 0.0008 & 0.0076 & 26 \\
125 & 55111.6361 & 0.0004 & 0.0064 & 74 \\
126 & 55111.7060 & 0.0004 & 0.0061 & 102 \\
127 & 55111.7756 & 0.0006 & 0.0056 & 56 \\
128 & 55111.8544 & 0.0065 & 0.0142 & 13 \\
137 & 55112.4750 & 0.0006 & 0.0033 & 57 \\
138 & 55112.5466 & 0.0007 & 0.0049 & 73 \\
177 & 55115.2731 & 0.0007 & $-$0.0048 & 49 \\
178 & 55115.3412 & 0.0009 & $-$0.0068 & 46 \\
179 & 55115.4090 & 0.0016 & $-$0.0092 & 50 \\
180 & 55115.4822 & 0.0018 & $-$0.0061 & 50 \\
181 & 55115.5469 & 0.0018 & $-$0.0115 & 49 \\
220 & 55118.2843 & 0.0009 & $-$0.0103 & 30 \\
\hline
  \multicolumn{5}{l}{\commenta BJD$-$2400000.} \\
  \multicolumn{5}{l}{\commentb Against $max = 2455102.8602 + 0.070156 E$.} \\
  \multicolumn{5}{l}{\commentc Number of points used to determine the maximum.} \\
\end{tabular}
\end{center}
\end{table}

\subsection{UV Persei}\label{obj:uvper}

\begin{figure}
  \begin{center}
%    \FigureFile(88mm,70mm){uvpercomp2.eps}
    \FigureFile(88mm,70mm){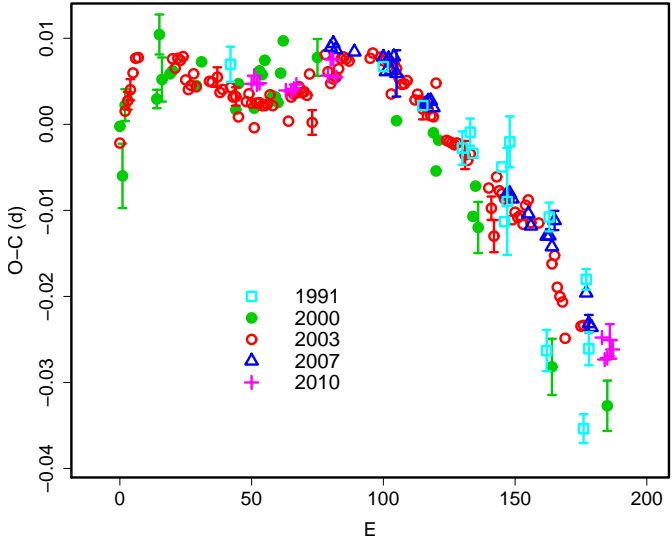}
  \end{center}
  \caption{Comparison of $O-C$ diagrams of UV Per between different
  superoutbursts.  A period of 0.06665 d was used to draw this figure.
  Approximate cycle counts ($E$) after the appearance of the
  superhumps were used.
  }
  \label{fig:uvpercomp2}
\end{figure}

   We observed the 2010 superoutburst during this middle and final
stages (table \ref{tab:uvperoc2010}).  Although a stage B--C transition
was recorded, the long gap in observation hindered precise determination
of periods.  We only list representative values for stage B.
Late-stage superhumps superimposed on the rapid fading from the
superoutburst plateau were clearly recorded as in the 1992 superoutburst.
A comparison of $O-C$ diagrams between different superoutbursts is
shown in figure \ref{fig:uvpercomp2}.  This figure is an improvement
of the corresponding one presented in \citet{Pdot}.

\begin{table}
\caption{Superhump maxima of UV Per (2010).}\label{tab:uvperoc2010}
\begin{center}
\begin{tabular}{ccccc}
\hline
$E$ & max\commenta & error & $O-C$\commentb & $N$\commentc \\
\hline
0 & 55203.4188 & 0.0003 & $-$0.0041 & 74 \\
1 & 55203.4864 & 0.0004 & $-$0.0029 & 74 \\
2 & 55203.5521 & 0.0003 & $-$0.0037 & 73 \\
3 & 55203.6189 & 0.0004 & $-$0.0033 & 62 \\
13 & 55204.2845 & 0.0004 & $-$0.0016 & 71 \\
15 & 55204.4177 & 0.0005 & $-$0.0012 & 40 \\
16 & 55204.4847 & 0.0004 & $-$0.0006 & 46 \\
17 & 55204.5518 & 0.0006 & 0.0001 & 34 \\
29 & 55205.3527 & 0.0005 & 0.0042 & 41 \\
30 & 55205.4212 & 0.0005 & 0.0063 & 35 \\
31 & 55205.4878 & 0.0006 & 0.0065 & 37 \\
32 & 55205.5524 & 0.0009 & 0.0047 & 28 \\
133 & 55212.2538 & 0.0004 & $-$0.0001 & 63 \\
134 & 55212.3179 & 0.0008 & $-$0.0024 & 67 \\
135 & 55212.3847 & 0.0009 & $-$0.0019 & 62 \\
136 & 55212.4535 & 0.0018 & 0.0004 & 48 \\
137 & 55212.5190 & 0.0011 & $-$0.0004 & 62 \\
\hline
  \multicolumn{5}{l}{\commenta BJD$-$2400000.} \\
  \multicolumn{5}{l}{\commentb Against $max = 2455203.4229 + 0.066398 E$.} \\
  \multicolumn{5}{l}{\commentc Number of points used to determine the maximum.} \\
\end{tabular}
\end{center}
\end{table}

\subsection{EI Piscium}\label{obj:eipsc}

   We observed the 2009 superoutburst first detected by ASAS-3
on June 8 at $V=12.46$.  Due to the poor seasonal condition,
we could only determine the mean superhump period of 0.04635(5) d
(with the PDM method), in agreement with previous measurements.
The times of superhump maxima are listed in table \ref{tab:eipscoc2009}.
The overall light curve of the outburst suggests that the main superoutburst
plateau lasted less than 10 d and experienced a rebrightening on
June 18.  The course of the outburst could have been similar to
the 2005 one (\cite{uem02j2329letter}; \cite{ski02j2329}).
The shortness of the superoutburst plateau in such short-$P_{\rm orb}$
systems with evolved secondaries would require a special explanation.

\begin{table}
\caption{Superhump maxima of EI Psc (2009).}\label{tab:eipscoc2009}
\begin{center}
\begin{tabular}{ccccc}
\hline
$E$ & max\commenta & error & $O-C$\commentb & $N$\commentc \\
\hline
0 & 54994.1841 & 0.0011 & 0.0006 & 44 \\
1 & 54994.2292 & 0.0005 & $-$0.0006 & 67 \\
10 & 54994.6467 & 0.0004 & 0.0001 & 29 \\
\hline
  \multicolumn{5}{l}{\commenta BJD$-$2400000.} \\
  \multicolumn{5}{l}{\commentb Against $max = 2454994.1836 + 0.046311 E$.} \\
  \multicolumn{5}{l}{\commentc Number of points used to determine the maximum.} \\
\end{tabular}
\end{center}
\end{table}

\subsection{EK Trianguli Australis}\label{obj:ektra}

   We observed the 2009 superoutburst.  A clear stae A--B transition
was recorded (table \ref{tab:ektraoc2009}).  Although the period variation
was alost absent during the supposed stage B ($E \ge 29$), this may
have been a result of fragmentary observations of stages B and C.

\begin{table}
\caption{Superhump maxima of EK TrA (2009).}\label{tab:ektraoc2009}
\begin{center}
\begin{tabular}{ccccc}
\hline
$E$ & max\commenta & error & $O-C$\commentb & $N$\commentc \\
\hline
0 & 55027.0310 & 0.0029 & $-$0.0009 & 74 \\
1 & 55027.0787 & 0.0026 & $-$0.0181 & 118 \\
29 & 55028.9255 & 0.0002 & 0.0101 & 159 \\
30 & 55028.9901 & 0.0003 & 0.0098 & 100 \\
76 & 55031.9722 & 0.0003 & 0.0043 & 180 \\
77 & 55032.0358 & 0.0003 & 0.0030 & 184 \\
137 & 55035.9273 & 0.0011 & $-$0.0023 & 108 \\
138 & 55035.9924 & 0.0006 & $-$0.0022 & 137 \\
139 & 55036.0559 & 0.0005 & $-$0.0036 & 139 \\
\hline
  \multicolumn{5}{l}{\commenta BJD$-$2400000.} \\
  \multicolumn{5}{l}{\commentb Against $max = 2455027.0319 + 0.064947 E$.} \\
  \multicolumn{5}{l}{\commentc Number of points used to determine the maximum.} \\
\end{tabular}
\end{center}
\end{table}

\subsection{SU Ursae Majoris}\label{obj:suuma}

   The 2010 January superoutburst was observed for its early and late
stages (table \ref{tab:suumaoc2010}).  Although we could only measure
the mean period of stage B, the value is in agreement with those obtained
during previous superoutbursts.  The humps observed for $E \ge 163$ may
not be genuine superhumps.

\begin{table}
\caption{Superhump maxima of SU UMa (2010).}\label{tab:suumaoc2010}
\begin{center}
\begin{tabular}{ccccc}
\hline
$E$ & max\commenta & error & $O-C$\commentb & $N$\commentc \\
\hline
0 & 55219.5098 & 0.0002 & $-$0.0003 & 132 \\
1 & 55219.5883 & 0.0002 & $-$0.0009 & 149 \\
2 & 55219.6689 & 0.0002 & 0.0005 & 150 \\
7 & 55220.0702 & 0.0014 & 0.0061 & 89 \\
11 & 55220.3809 & 0.0002 & 0.0002 & 145 \\
12 & 55220.4613 & 0.0002 & 0.0014 & 156 \\
13 & 55220.5392 & 0.0002 & 0.0002 & 153 \\
14 & 55220.6182 & 0.0004 & $-$0.0001 & 157 \\
15 & 55220.6987 & 0.0003 & 0.0013 & 122 \\
25 & 55221.4871 & 0.0004 & $-$0.0018 & 155 \\
26 & 55221.5668 & 0.0004 & $-$0.0012 & 151 \\
27 & 55221.6454 & 0.0004 & $-$0.0018 & 154 \\
50 & 55223.4634 & 0.0004 & $-$0.0044 & 96 \\
51 & 55223.5457 & 0.0004 & $-$0.0012 & 118 \\
163 & 55232.4126 & 0.0012 & 0.0006 & 75 \\
164 & 55232.4926 & 0.0012 & 0.0014 & 83 \\
\hline
  \multicolumn{5}{l}{\commenta BJD$-$2400000.} \\
  \multicolumn{5}{l}{\commentb Against $max = 2455219.5101 + 0.079153 E$.} \\
  \multicolumn{5}{l}{\commentc Number of points used to determine the maximum.} \\
\end{tabular}
\end{center}
\end{table}

\subsection{BC Ursae Majoris}\label{obj:bcuma}

\begin{figure}
  \begin{center}
%    \FigureFile(88mm,70mm){bcumacomp2.eps}
    \FigureFile(88mm,70mm){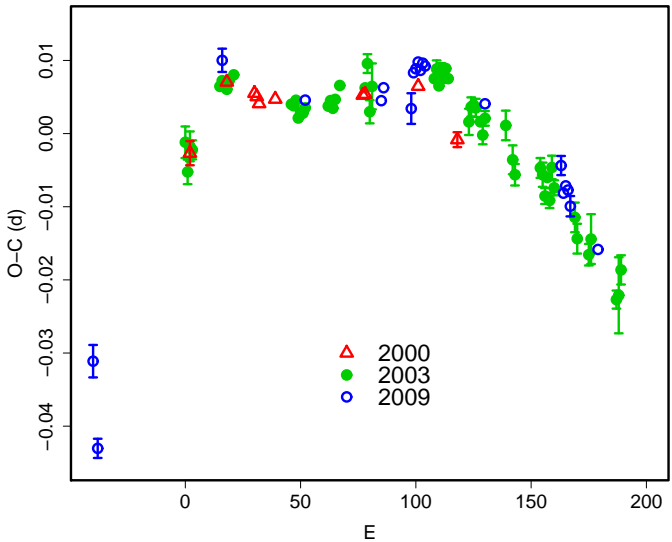}
  \end{center}
  \caption{Comparison of $O-C$ diagrams of BC UMa between different
  superoutbursts.  A period of 0.06455 d was used to draw this figure.
  Approximate cycle counts ($E$) after the appearance of the
  superhumps were used.
  }
  \label{fig:bcumacomp2}
\end{figure}

   BC UMa underwent a superoutburst in 2009 September--October
after a period of 6.7 yr (cf. vsnet-alert 11514).  The start of
the outburst was not well constrained dur to the poor visibility
in the morning sky.
Superhumps were observed despite unfavorable seasonal conditions
(cf. vsnet-alert 11540, 11550, 11559).
The times of superhump maxima are listed in table \ref{tab:bcumaoc2009}.
There was a clear stage B--C transition and the $P_{\rm dot}$ for stage B
($56 \le E \le 144$) was $+9.5(2.7) \times 10^{-5}$.
We attributed the interval $E \le 2$ to likely stage A superhumps
based on comparison with other superoutbursts
(figure \ref{fig:bcumacomp2}).  If this identification is correct,
the superhumps during this superoutburst appears to have taken
a longer time to fully develop.

\begin{table}
\caption{Superhump maxima of BC UMa (2009).}\label{tab:bcumaoc2009}
\begin{center}
\begin{tabular}{ccccc}
\hline
$E$ & max\commenta & error & $O-C$\commentb & $N$\commentc \\
\hline
0 & 55105.2681 & 0.0022 & $-$0.0189 & 70 \\
2 & 55105.3853 & 0.0013 & $-$0.0310 & 43 \\
56 & 55108.9240 & 0.0016 & 0.0181 & 182 \\
92 & 55111.2424 & 0.0004 & 0.0101 & 49 \\
125 & 55113.3725 & 0.0007 & 0.0076 & 45 \\
126 & 55113.4388 & 0.0004 & 0.0093 & 84 \\
138 & 55114.2105 & 0.0021 & 0.0056 & 40 \\
139 & 55114.2800 & 0.0004 & 0.0104 & 86 \\
140 & 55114.3451 & 0.0005 & 0.0109 & 72 \\
141 & 55114.4106 & 0.0005 & 0.0117 & 85 \\
142 & 55114.4740 & 0.0005 & 0.0105 & 86 \\
143 & 55114.5395 & 0.0005 & 0.0114 & 87 \\
144 & 55114.6037 & 0.0004 & 0.0109 & 87 \\
170 & 55116.2768 & 0.0005 & 0.0039 & 113 \\
203 & 55118.3985 & 0.0013 & $-$0.0070 & 67 \\
204 & 55118.4593 & 0.0007 & $-$0.0108 & 72 \\
205 & 55118.5248 & 0.0006 & $-$0.0099 & 73 \\
206 & 55118.5888 & 0.0005 & $-$0.0105 & 73 \\
207 & 55118.6512 & 0.0014 & $-$0.0128 & 45 \\
219 & 55119.4198 & 0.0007 & $-$0.0196 & 46 \\
\hline
  \multicolumn{5}{l}{\commenta BJD$-$2400000.} \\
  \multicolumn{5}{l}{\commentb Against $max = 2455105.2870 + 0.064623 E$.} \\
  \multicolumn{5}{l}{\commentc Number of points used to determine the maximum.} \\
\end{tabular}
\end{center}
\end{table}

\subsection{EL Ursae Majoris}\label{obj:eluma}

   Although EL UMa was discovered as an eruptive object relatively early
in the history \citep{pes87eluma}, only little had been known until recent
years.  \citet{kat01hvvir} listed this object among candidate WZ Sge-type
dwarf novae.  Based on the similarity of its quiescent SDSS color to
those of known WZ Sge-type dwarf novae, we started monitoring since 2008.
Another outburst at $r$ = 13.7 in 2003 April was found on an archival
image \citep{wil10newCVs}.

   On 2009 January 13, H. Maehara finally detected this object in
outburst at an unfiltered CCD magnitude of 17.4 (vsnet-alert 11771)
The object further brightened to a magnitude of 14.9 on January 16
(vsnet-alert 11772).  The object has been confirmed to be a hydrogen-rich
dwarf niva in outburst by spectroscopy (Takahashi and Kinugasa, private
communication).
Although the nature of this brightening was
unclear at the time, the detection of additional sequence of outbursts
(vsnet-alert 11789, 11808), led to an interpretation that they are
post-superoutburst rebrightenings of a WZ Sge-type superoutburst,
whose main superoutburst was missed (vsnet-alert 11795).
The detection of modulations attributable to superhumps seems to
strengthen this interpretation (vsnet-alert 11799).  We include this
object based on this interpretation.

\begin{figure}
  \begin{center}
%    \FigureFile(88mm,110mm){elumashpdm.eps}
    \FigureFile(88mm,110mm){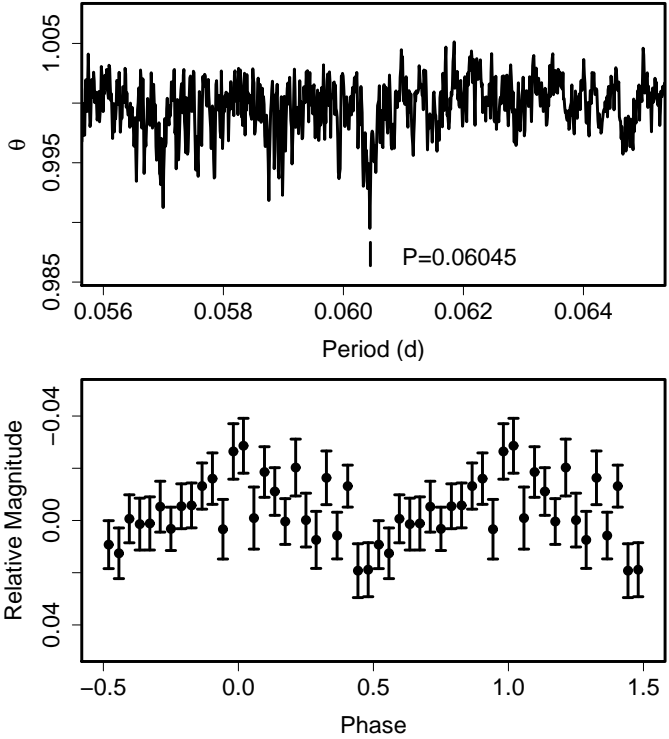}
  \end{center}
  \caption{Superhumps of EL UMa during the rebrightening phase (2010).
     (Upper): PDM analysis.
     (Lower): Phase-averaged profile.}
  \label{fig:elumashpdm}
\end{figure}

\begin{figure}
  \begin{center}
%    \FigureFile(88mm,110mm){elumalc.eps}
    \FigureFile(88mm,110mm){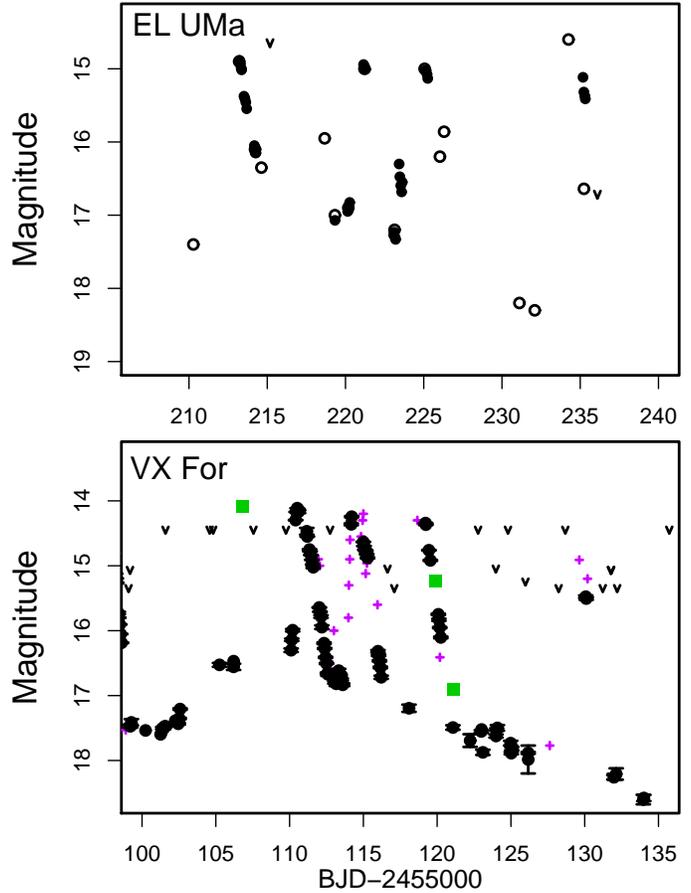}
  \end{center}
  \caption{Light curve of EL UMa during the rebrightening phase (2010).
     (Upper): EL UMa.  Open circles and "v'' signs are snapshot
     CCD observations and upper limits, respectively.
     (Lower): VX For (2009, cf. figure \ref{fig:vxforlc}).
     The overall feature is remarkably similar between these objects.}
  \label{fig:elumalc}
\end{figure}

   A period analysis of the rebrightening phase, after subtracting the
trends of outbursts (cf. \cite{kat09j0804}), has yielded a period of
0.06045(6) d (figure \ref{fig:elumashpdm}).  Although this period
needs to be confirmed by future observations, its relatively long
$P_{\rm SH}$ appears to be similar to that of VX For
(subsection \ref{obj:vxfor}) that underwent multiple rebrightenings
(figure \ref{fig:elumalc}).  There is a common tendency that the
quiescent interval preceding the last rebrightening is longer than
the other intervals between rebrightenings.

   This object is a good candidate for a CV passed the period minimum
in evolution (see a discussion in \cite{Pdot}).  Further radial-velocity
study in quiescence in encouraged in order to determine the orbital
period and the nature of the period during the rebrightening phase.

\subsection{IY Ursae Majoris}\label{obj:iyuma}

\begin{figure}
  \begin{center}
%    \FigureFile(88mm,70mm){iyumacomp2.eps}
    \FigureFile(88mm,70mm){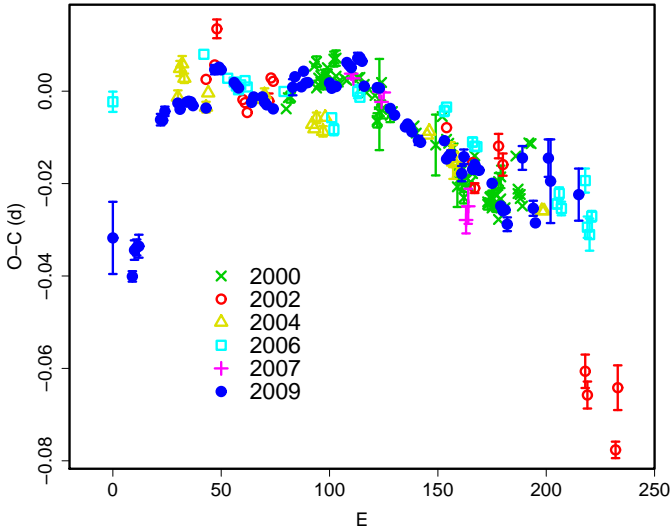}
  \end{center}
  \caption{Comparison of $O-C$ diagrams of IY UMa between different
  superoutbursts.  A period of 0.07616 d was used to draw this figure.
  Approximate cycle counts ($E$) after the start of the
  superoutburst were used.
  The figure is an improvement of the corresponding figure in \citet{Pdot}.
  The base period and symbols were modified for better visibility.
  }
  \label{fig:iyumacomp2}
\end{figure}

   Although the 2009 superoutburst of this object was reported in
\citet{Pdot}, we provide greatly improved results by combining newly
available observations.  The times of superhump maxima
(table \ref{tab:iyumaoc2009}) now clearly illustrate the presence of
all A--C stages, and a definitely positive
$P_{\rm dot}$ = $+15.1(2.3) \times 10^{-5}$ for stage B superhumps,
confirming the suggestion in \citet{Pdot} based on the combined
$O-C$ diagram (cf. figure \ref{fig:iyumacomp2}).
Although there was a signature of distinct stages during stage A,
we listed a mean period in table \ref{tab:perlist}.
There was a very clear signature of a stage B--C transition thanks
to the high quality of data.
Although the phases of superhump at late epochs ($E \ge 189$)
deviate from extrapolations of stage C superhumps, they can still
be interpreted as a continuation of stage C superhump, rather than
traditional late superhumps.

\begin{figure}
  \begin{center}
%    \FigureFile(88mm,110mm){iyumabeatoc.eps}
    \FigureFile(88mm,110mm){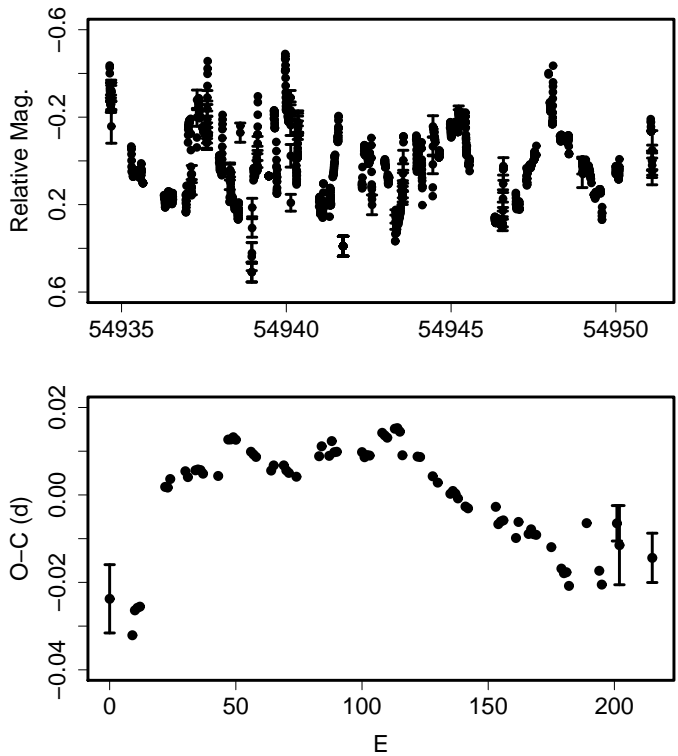}
  \end{center}
  \caption{Beat phenomenon in IY UMa (2009).
     (Upper): Residuals from the global trend (fitted by a third-order
     polynomial) of the outburst outside the eclipses.
     Each point represents an average of one superhump period.
     (Lower): $O-C$ of superhumps.}
  \label{fig:iyumabeatoc}
\end{figure}

   This object showed a strong ``textbook'' beat phenomenon between
superhumps and orbital modulations (figure \ref{fig:iyumabeatoc}).
The period of the beat phenomenon was shorter ($\sim$ 2.5 d) during
stage B, while it became longer ($\sim$ 3.0 d) during stage C.
These periods are in very good agreement with the expected beat
periods for stage B and C superhumps, 2.45 d and 3.08 d, respectively.
This can be understood if the variation of the beat period reflects
the variation of the angular velocity of the apsidal motion of
the elliptical accretion disk.  The close correlation between the
beat period and the superhump period suggests that the change in the
angular velocity of the global apsidal motion is more responsible
for the stage B--C transition rather than the appearance of a more
localized new component.

\begin{table}
\caption{Superhump maxima of IY UMa (2009).}\label{tab:iyumaoc2009}
\begin{center}
\begin{tabular}{ccccc}
\hline
$E$ & max\commenta & error & $O-C$\commentb & $N$\commentc \\
\hline
0 & 54934.6865 & 0.0078 & $-$0.0294 & 52 \\
9 & 54935.3637 & 0.0011 & $-$0.0372 & 77 \\
10 & 54935.4455 & 0.0021 & $-$0.0315 & 74 \\
11 & 54935.5222 & 0.0017 & $-$0.0309 & 64 \\
12 & 54935.5987 & 0.0025 & $-$0.0305 & 33 \\
22 & 54936.3877 & 0.0013 & $-$0.0026 & 105 \\
23 & 54936.4637 & 0.0010 & $-$0.0027 & 121 \\
24 & 54936.5418 & 0.0010 & $-$0.0007 & 82 \\
30 & 54937.0005 & 0.0003 & 0.0014 & 144 \\
31 & 54937.0753 & 0.0002 & 0.0001 & 201 \\
34 & 54937.3054 & 0.0002 & 0.0019 & 59 \\
35 & 54937.3817 & 0.0003 & 0.0021 & 68 \\
36 & 54937.4578 & 0.0003 & 0.0020 & 63 \\
37 & 54937.5331 & 0.0003 & 0.0012 & 66 \\
43 & 54937.9895 & 0.0007 & 0.0010 & 119 \\
47 & 54938.3025 & 0.0010 & 0.0096 & 66 \\
48 & 54938.3787 & 0.0009 & 0.0097 & 68 \\
49 & 54938.4553 & 0.0008 & 0.0102 & 67 \\
50 & 54938.5309 & 0.0007 & 0.0097 & 58 \\
56 & 54938.9851 & 0.0003 & 0.0073 & 123 \\
57 & 54939.0607 & 0.0002 & 0.0067 & 125 \\
58 & 54939.1363 & 0.0003 & 0.0062 & 82 \\
64 & 54939.5901 & 0.0007 & 0.0034 & 86 \\
65 & 54939.6675 & 0.0002 & 0.0047 & 122 \\
69 & 54939.9721 & 0.0003 & 0.0049 & 83 \\
70 & 54940.0471 & 0.0004 & 0.0038 & 123 \\
71 & 54940.1227 & 0.0007 & 0.0033 & 108 \\
74 & 54940.3503 & 0.0009 & 0.0026 & 61 \\
83 & 54941.0404 & 0.0017 & 0.0077 & 104 \\
84 & 54941.1189 & 0.0009 & 0.0101 & 97 \\
87 & 54941.3452 & 0.0003 & 0.0080 & 136 \\
88 & 54941.4247 & 0.0004 & 0.0115 & 133 \\
89 & 54941.4983 & 0.0004 & 0.0090 & 105 \\
90 & 54941.5746 & 0.0002 & 0.0091 & 130 \\
100 & 54942.3361 & 0.0002 & 0.0096 & 151 \\
101 & 54942.4111 & 0.0002 & 0.0085 & 254 \\
102 & 54942.4877 & 0.0003 & 0.0090 & 203 \\
103 & 54942.5638 & 0.0004 & 0.0090 & 134 \\
108 & 54942.9498 & 0.0008 & 0.0144 & 53 \\
109 & 54943.0254 & 0.0006 & 0.0139 & 88 \\
110 & 54943.1010 & 0.0004 & 0.0134 & 62 \\
113 & 54943.3315 & 0.0012 & 0.0156 & 68 \\
114 & 54943.4078 & 0.0008 & 0.0158 & 67 \\
115 & 54943.4832 & 0.0007 & 0.0151 & 62 \\
116 & 54943.5539 & 0.0008 & 0.0097 & 27 \\
122 & 54944.0106 & 0.0005 & 0.0098 & 86 \\
123 & 54944.0867 & 0.0005 & 0.0097 & 88 \\
128 & 54944.4631 & 0.0004 & 0.0056 & 51 \\
130 & 54944.6139 & 0.0008 & 0.0042 & 65 \\
135 & 54944.9922 & 0.0004 & 0.0020 & 177 \\
136 & 54945.0690 & 0.0004 & 0.0027 & 213 \\
\hline
  \multicolumn{5}{l}{\commenta BJD$-$2400000.} \\
  \multicolumn{5}{l}{\commentb Against $max = 2454934.7159 + 0.076106 E$.} \\
  \multicolumn{5}{l}{\commentc Number of points used to determine the maximum.} \\
\end{tabular}
\end{center}
\end{table}

\addtocounter{table}{-1}
\begin{table}
\caption{Superhump maxima of IY UMa (2009) (continued).}
\begin{center}
\begin{tabular}{ccccc}
\hline
$E$ & max\commenta & error & $O-C$\commentb & $N$\commentc \\
\hline
137 & 54945.1445 & 0.0006 & 0.0021 & 79 \\
138 & 54945.2196 & 0.0008 & 0.0011 & 79 \\
141 & 54945.4462 & 0.0010 & $-$0.0006 & 37 \\
142 & 54945.5220 & 0.0010 & $-$0.0010 & 55 \\
153 & 54946.3601 & 0.0009 & $-$0.0000 & 67 \\
154 & 54946.4323 & 0.0008 & $-$0.0040 & 66 \\
155 & 54946.5090 & 0.0006 & $-$0.0033 & 65 \\
156 & 54946.5854 & 0.0010 & $-$0.0030 & 30 \\
161 & 54946.9622 & 0.0016 & $-$0.0067 & 44 \\
162 & 54947.0420 & 0.0016 & $-$0.0030 & 61 \\
166 & 54947.3439 & 0.0007 & $-$0.0055 & 72 \\
167 & 54947.4212 & 0.0007 & $-$0.0044 & 156 \\
168 & 54947.4962 & 0.0005 & $-$0.0055 & 134 \\
169 & 54947.5722 & 0.0006 & $-$0.0056 & 103 \\
175 & 54948.0264 & 0.0008 & $-$0.0081 & 139 \\
179 & 54948.3261 & 0.0006 & $-$0.0128 & 67 \\
180 & 54948.4012 & 0.0005 & $-$0.0138 & 67 \\
181 & 54948.4776 & 0.0008 & $-$0.0135 & 64 \\
182 & 54948.5506 & 0.0015 & $-$0.0166 & 64 \\
189 & 54949.0981 & 0.0026 & $-$0.0018 & 191 \\
194 & 54949.4680 & 0.0016 & $-$0.0124 & 66 \\
195 & 54949.5410 & 0.0007 & $-$0.0156 & 67 \\
201 & 54950.0120 & 0.0041 & $-$0.0012 & 169 \\
202 & 54950.0831 & 0.0091 & $-$0.0062 & 149 \\
215 & 54951.0703 & 0.0056 & $-$0.0084 & 108 \\
\hline
  \multicolumn{5}{l}{\commenta BJD$-$2400000.} \\
  \multicolumn{5}{l}{\commentb Against $max = 2454934.7159 + 0.076106 E$.} \\
  \multicolumn{5}{l}{\commentc Number of points used to determine the maximum.} \\
\end{tabular}
\end{center}
\end{table}

\subsection{KS Ursae Majoris}\label{obj:ksuma}

   The times of superhump maxima during the 2010 superoutburst are
listed in table \ref{tab:ksumaoc2010}.  Since there was a large gap
in observations, we did not attempt to determine $P_{\rm dot}$
from these data.  The observation was probably recorded during
stages B and C.

\begin{table}
\caption{Superhump maxima of KS UMa (2010).}\label{tab:ksumaoc2010}
\begin{center}
\begin{tabular}{ccccc}
\hline
$E$ & max\commenta & error & $O-C$\commentb & $N$\commentc \\
\hline
0 & 55304.0436 & 0.0004 & $-$0.0009 & 92 \\
1 & 55304.1148 & 0.0003 & 0.0002 & 100 \\
2 & 55304.1837 & 0.0003 & $-$0.0011 & 99 \\
3 & 55304.2563 & 0.0006 & 0.0013 & 99 \\
99 & 55310.9915 & 0.0013 & 0.0034 & 99 \\
100 & 55311.0610 & 0.0011 & 0.0028 & 100 \\
101 & 55311.1262 & 0.0011 & $-$0.0021 & 96 \\
102 & 55311.1953 & 0.0011 & $-$0.0031 & 100 \\
103 & 55311.2691 & 0.0012 & 0.0005 & 93 \\
113 & 55311.9785 & 0.0015 & 0.0086 & 87 \\
114 & 55312.0382 & 0.0023 & $-$0.0019 & 95 \\
115 & 55312.1023 & 0.0018 & $-$0.0079 & 75 \\
\hline
  \multicolumn{5}{l}{\commenta BJD$-$2400000.} \\
  \multicolumn{5}{l}{\commentb Against $max = 2455304.0445 + 0.070136 E$.} \\
  \multicolumn{5}{l}{\commentc Number of points used to determine the maximum.} \\
\end{tabular}
\end{center}
\end{table}

\subsection{MR Ursae Majoris}\label{obj:mruma}

\begin{figure}
  \begin{center}
%    \FigureFile(88mm,70mm){mrumacomp2.eps}
    \FigureFile(88mm,70mm){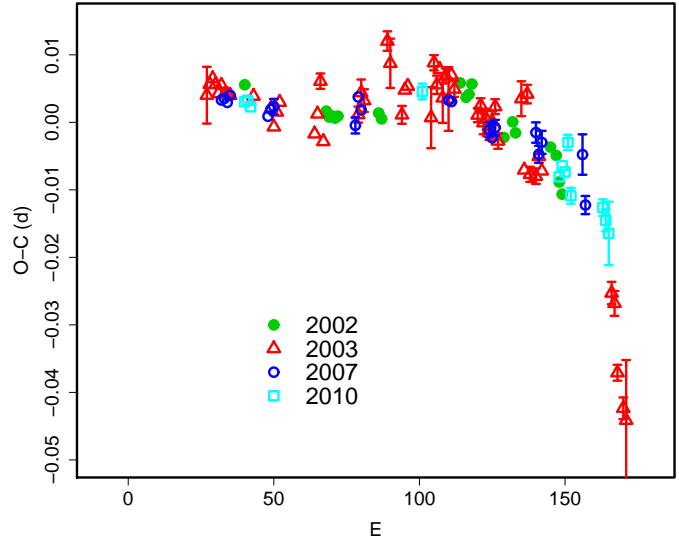}
  \end{center}
  \caption{Comparison of $O-C$ diagrams of MR UMa between different
  superoutbursts.  A period of 0.06512 d was used to draw this figure.
  Approximate cycle counts ($E$) after the start of the
  2007 superoutburst were used.  Since the starts of the other
  superoutbursts were not well constrained, we shifted the $O-C$ diagrams
  to best fit the 2007 one.
  }
  \label{fig:mrumacomp2}
\end{figure}

   The times of superhump maxima during the 2010 superoutburst are
listed in table \ref{tab:mrumaoc2010}.  Since there were relatively
large gaps in observations, we did not attempt to determine $P_{\rm dot}$
from these data.  The period of the presumable stage C superhumps
is listed in table \ref{tab:perlist}.
An updated comparison of $O-C$ diagrams between different superoutbursts
is given in figure \ref{fig:mrumacomp2}.

\begin{table}
\caption{Superhump maxima of MR UMa (2010).}\label{tab:mrumaoc2010}
\begin{center}
\begin{tabular}{ccccc}
\hline
$E$ & max\commenta & error & $O-C$\commentb & $N$\commentc \\
\hline
0 & 55303.9991 & 0.0002 & 0.0004 & 155 \\
1 & 55304.0645 & 0.0005 & 0.0007 & 107 \\
2 & 55304.1286 & 0.0005 & $-$0.0001 & 97 \\
61 & 55307.9730 & 0.0012 & 0.0074 & 42 \\
108 & 55311.0210 & 0.0009 & $-$0.0011 & 150 \\
109 & 55311.0877 & 0.0009 & 0.0007 & 157 \\
110 & 55311.1519 & 0.0006 & $-$0.0002 & 158 \\
111 & 55311.2214 & 0.0011 & 0.0043 & 113 \\
112 & 55311.2787 & 0.0012 & $-$0.0035 & 89 \\
123 & 55311.9932 & 0.0012 & $-$0.0043 & 138 \\
124 & 55312.0565 & 0.0016 & $-$0.0061 & 155 \\
125 & 55312.1196 & 0.0047 & $-$0.0079 & 71 \\
184 & 55315.9675 & 0.0006 & 0.0031 & 37 \\
185 & 55316.0361 & 0.0019 & 0.0067 & 44 \\
\hline
  \multicolumn{5}{l}{\commenta BJD$-$2400000.} \\
  \multicolumn{5}{l}{\commentb Against $max = 2455303.9987 + 0.065031 E$.} \\
  \multicolumn{5}{l}{\commentc Number of points used to determine the maximum.} \\
\end{tabular}
\end{center}
\end{table}

\subsection{TY Vulpeculae}\label{obj:tyvul}

   The 2010 superoutburst of this SU UMa-type dwarf nova was relatively
well-observed during its later stage.  The times of superhump maxima are
listed in table \ref{tab:tyvuloc2010}.  The period evolution now clearly
demonstrates the presence of stage B--C transition, whose existence was
suggested from the 2003 observation \citep{Pdot}.  The sudden change
in the superhump period at this transition favors the suggestion
in \citet{Pdot} that this object is analogous to AX Cap and SDSS J1627.

\begin{table}
\caption{Superhump maxima of TY Vul (2010).}\label{tab:tyvuloc2010}
\begin{center}
\begin{tabular}{ccccc}
\hline
$E$ & max\commenta & error & $O-C$\commentb & $N$\commentc \\
\hline
0 & 55368.7467 & 0.0008 & $-$0.0048 & 41 \\
1 & 55368.8262 & 0.0004 & $-$0.0055 & 39 \\
9 & 55369.4721 & 0.0007 & $-$0.0017 & 202 \\
10 & 55369.5530 & 0.0003 & $-$0.0010 & 188 \\
13 & 55369.7944 & 0.0009 & $-$0.0003 & 40 \\
14 & 55369.8748 & 0.0008 & $-$0.0002 & 39 \\
21 & 55370.4365 & 0.0007 & $-$0.0003 & 135 \\
22 & 55370.5183 & 0.0008 & 0.0012 & 160 \\
23 & 55370.5991 & 0.0008 & 0.0018 & 89 \\
25 & 55370.7518 & 0.0030 & $-$0.0060 & 20 \\
26 & 55370.8390 & 0.0007 & 0.0009 & 40 \\
27 & 55370.9180 & 0.0008 & $-$0.0004 & 32 \\
33 & 55371.3982 & 0.0090 & $-$0.0017 & 64 \\
34 & 55371.4837 & 0.0011 & 0.0036 & 158 \\
35 & 55371.5635 & 0.0008 & 0.0031 & 154 \\
37 & 55371.7267 & 0.0017 & 0.0059 & 37 \\
38 & 55371.8037 & 0.0033 & 0.0026 & 39 \\
39 & 55371.8884 & 0.0033 & 0.0070 & 22 \\
46 & 55372.4457 & 0.0004 & 0.0026 & 105 \\
47 & 55372.5236 & 0.0004 & 0.0002 & 163 \\
48 & 55372.6074 & 0.0005 & 0.0037 & 74 \\
58 & 55373.4079 & 0.0011 & 0.0017 & 51 \\
59 & 55373.4872 & 0.0005 & 0.0007 & 156 \\
60 & 55373.5673 & 0.0006 & 0.0005 & 146 \\
75 & 55374.7689 & 0.0014 & $-$0.0016 & 27 \\
76 & 55374.8473 & 0.0019 & $-$0.0035 & 27 \\
88 & 55375.8077 & 0.0021 & $-$0.0061 & 58 \\
89 & 55375.8919 & 0.0033 & $-$0.0022 & 109 \\
\hline
  \multicolumn{5}{l}{\commenta BJD$-$2400000.} \\
  \multicolumn{5}{l}{\commentb Against $max = 2455368.7515 + 0.080254 E$.} \\
  \multicolumn{5}{l}{\commentc Number of points used to determine the maximum.} \\
\end{tabular}
\end{center}
\end{table}

\subsection{1RXS J042332$+$745300}\label{obj:j0423}

   In \citet{Pdot}, we reported observations of the 2008 superoutburst
of this object (=HS 0417$+$7445, hereafter 1RXS J0423).
The 2010 superoutburst was again fortunately detected during its rising stage
and early stage evolution of superhumps was recorded.

   The times of superhump maxima are listed in table \ref{tab:j0423oc2010}.
A stage A--B and transition was clearly recorded.  The shorter superhump
period after $E=79$ probably corresponds to stage C superhumps.
Although the later half of the stage B and the stage B--C transition itself
were not observed, there was a possible indication of a positive
$P_{\rm dot}$ during the stage B.
The parameters are listed in table \ref{tab:perlist}.

\begin{figure}
  \begin{center}
%    \FigureFile(88mm,70mm){j0423occomp.eps}
    \FigureFile(88mm,70mm){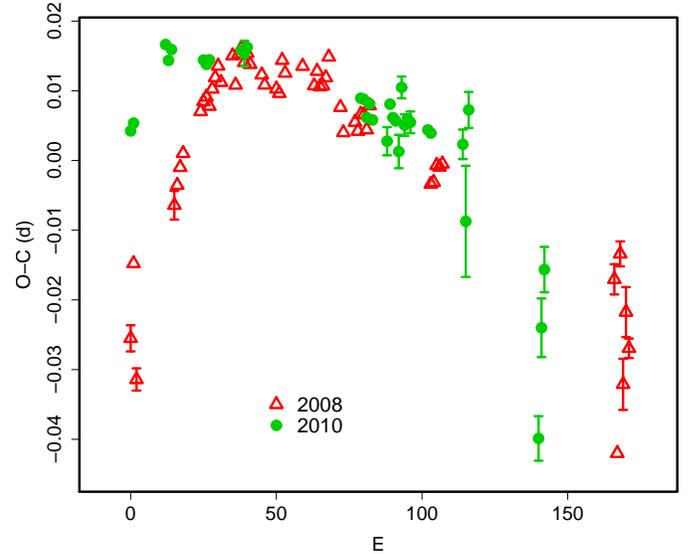}
  \end{center}
  \caption{Comparison of $O-C$ diagrams of 1RXS J0423 between different
  superoutbursts.  A period of 0.07845 d was used to draw this figure.
  Approximate cycle counts ($E$) after the start of the superoutburst
  were used.}
  \label{fig:j0423occomp}
\end{figure}

   Figure \ref{fig:j0423occomp} illustrates the comparison of $O-C$
diagrams between the 2008 and 2010 superoutbursts.  Although the later
parts of the $O-C$ diagrams were similar, there was a distinction during
the early stage.  This may have been a result of the presence of
a precursor outburst and early appearance of superhumps during the 2008
superoutburst (cf. \cite{Pdot}).  The $O-C$ evolution in the 2010
superoutburst resembled those of ordinary SU UMa-type dwarf novae
than in the 2008 superoutburst.

\begin{table}
\caption{Superhump maxima of 1RXS J0423 (2010).}\label{tab:j0423oc2010}
\begin{center}
\begin{tabular}{ccccc}
\hline
$E$ & max\commenta & error & $O-C$\commentb & $N$\commentc \\
\hline
0 & 55442.1696 & 0.0006 & $-$0.0155 & 124 \\
1 & 55442.2491 & 0.0005 & $-$0.0141 & 94 \\
12 & 55443.1233 & 0.0003 & $-$0.0007 & 114 \\
13 & 55443.1995 & 0.0004 & $-$0.0028 & 309 \\
14 & 55443.2795 & 0.0008 & $-$0.0010 & 108 \\
25 & 55444.1410 & 0.0004 & $-$0.0003 & 268 \\
26 & 55444.2188 & 0.0002 & $-$0.0007 & 470 \\
27 & 55444.2979 & 0.0003 & 0.0001 & 406 \\
38 & 55445.1623 & 0.0007 & 0.0037 & 67 \\
39 & 55445.2402 & 0.0018 & 0.0034 & 42 \\
40 & 55445.3196 & 0.0007 & 0.0045 & 73 \\
79 & 55448.3718 & 0.0009 & 0.0049 & 47 \\
80 & 55448.4501 & 0.0004 & 0.0050 & 79 \\
81 & 55448.5259 & 0.0004 & 0.0025 & 76 \\
82 & 55448.6063 & 0.0005 & 0.0047 & 79 \\
83 & 55448.6825 & 0.0010 & 0.0026 & 62 \\
88 & 55449.0717 & 0.0020 & 0.0005 & 157 \\
89 & 55449.1554 & 0.0008 & 0.0061 & 240 \\
90 & 55449.2320 & 0.0009 & 0.0043 & 475 \\
91 & 55449.3099 & 0.0005 & 0.0041 & 205 \\
92 & 55449.3840 & 0.0024 & $-$0.0001 & 49 \\
93 & 55449.4716 & 0.0016 & 0.0093 & 80 \\
94 & 55449.5447 & 0.0015 & 0.0041 & 78 \\
95 & 55449.6240 & 0.0009 & 0.0052 & 79 \\
96 & 55449.7020 & 0.0016 & 0.0048 & 42 \\
102 & 55450.1716 & 0.0009 & 0.0050 & 218 \\
103 & 55450.2496 & 0.0007 & 0.0047 & 197 \\
114 & 55451.1109 & 0.0021 & 0.0053 & 90 \\
115 & 55451.1783 & 0.0080 & $-$0.0056 & 114 \\
116 & 55451.2728 & 0.0026 & 0.0106 & 108 \\
140 & 55453.1084 & 0.0032 & $-$0.0318 & 77 \\
141 & 55453.2028 & 0.0042 & $-$0.0157 & 87 \\
142 & 55453.2896 & 0.0033 & $-$0.0071 & 85 \\
\hline
  \multicolumn{5}{l}{\commenta BJD$-$2400000.} \\
  \multicolumn{5}{l}{\commentb Against $max = 2455442.1850 + 0.078251 E$.} \\
  \multicolumn{5}{l}{\commentc Number of points used to determine the maximum.} \\
\end{tabular}
\end{center}
\end{table}

\subsection{1RXS J053234.9$+$624755}\label{obj:j0532}

   The period evolution of this SU UMa-type dwarf nova during the 2005
and 2008 superoutbursts has been described in \citet{ima09j0532} and
\citet{Pdot}.

\begin{figure}
  \begin{center}
%    \FigureFile(88mm,90mm){j05322009oc.eps}
    \FigureFile(88mm,90mm){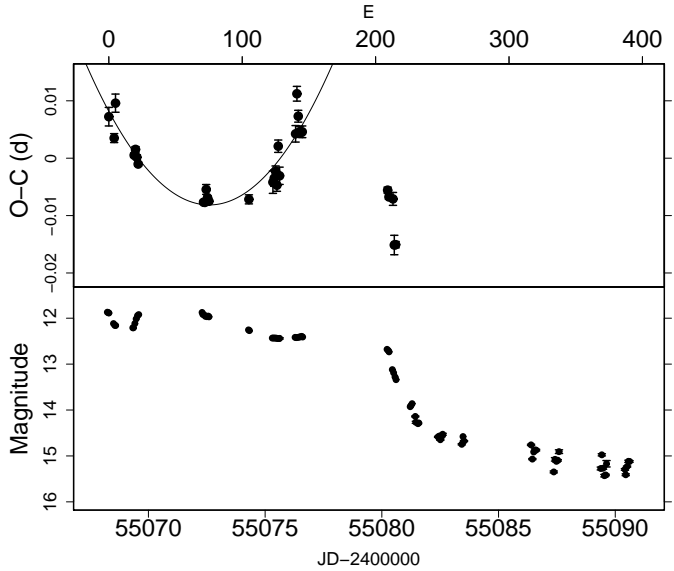}
  \end{center}
  \caption{$O-C$ of superhumps 1RXS J0532 (2009).
  (Upper): $O-C$ diagram.  The $O-C$ values were against the mean period
  for the stage B ($E \le 145$, thin curve)
  (Lower): Light curve.}
  \label{fig:j05322009oc}
\end{figure}

\begin{figure}
  \begin{center}
%    \FigureFile(88mm,110mm){j0532lateshpdm.eps}
    \FigureFile(88mm,110mm){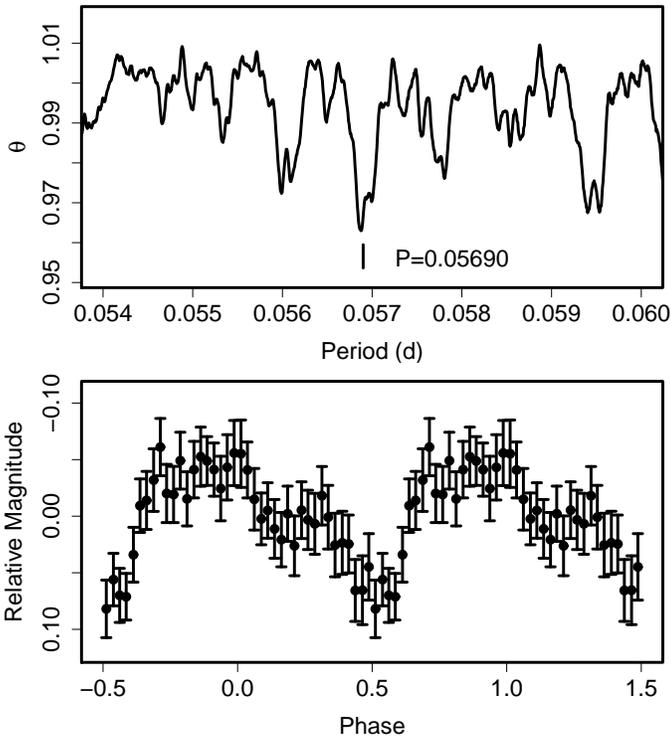}
  \end{center}
  \caption{Superhumps during the post-superoutburst stage in 1RXS J0532 (2009).
     (Upper): PDM analysis.
     (Lower): Phase-averaged profile.}
  \label{fig:j0532lateshpdm}
\end{figure}

   We also observed the 2009 superoutburst which was accompanied by
a precursor outburst, as in the 2005 one (figure \ref{fig:j05322009oc}).
The times of superhump maxima are listed in table \ref{tab:j0532oc2009}.
Superhumps were already present during the fading stage of the
rebrightening, and the period was smoothly decreasing as in the
2005 one \citet{ima09j0532}.  This indicates that the stage B, rather
than stage A, already started during the precursor outburst.
This behavior very well reproduced the features observed during
the 2005 superoutburst.
The observed $P_{\rm dot}$ for stage B ($E \le 145$) was
$+10.1(1.0) \times 10^{-5}$,
similar to the one observed in 2008, but is larger than in 2005.
The 2009 outburst was also well-observed during the post-superoutburst
stage.  Although times of individual superhumps were not sufficiently
measured due to strong flickering, the period analysis has yielded
a periodicity of 0.05690(2) d (BJD 2455081.2--2455090.6, figure
\ref{fig:j0532lateshpdm}), which is in agreement with the period of
stage C superhumps.  The signal from the orbital period, if present,
was still much smaller than the
superhump signal.  This analysis suggests the long endurance of
superhumps even after the termination of the superoutburst.

\begin{table}
\caption{Superhump maxima of 1RXS J0532 (2009).}\label{tab:j0532oc2009}
\begin{center}
\begin{tabular}{ccccc}
\hline
$E$ & max\commenta & error & $O-C$\commentb & $N$\commentc \\
\hline
0 & 55068.3087 & 0.0016 & 0.0047 & 106 \\
4 & 55068.5335 & 0.0008 & 0.0011 & 114 \\
5 & 55068.5967 & 0.0016 & 0.0072 & 72 \\
19 & 55069.3875 & 0.0004 & $-$0.0013 & 106 \\
20 & 55069.4457 & 0.0005 & $-$0.0003 & 114 \\
21 & 55069.5014 & 0.0004 & $-$0.0016 & 111 \\
22 & 55069.5574 & 0.0003 & $-$0.0027 & 113 \\
71 & 55072.3504 & 0.0003 & $-$0.0075 & 112 \\
72 & 55072.4075 & 0.0004 & $-$0.0075 & 114 \\
73 & 55072.4669 & 0.0009 & $-$0.0052 & 36 \\
74 & 55072.5226 & 0.0004 & $-$0.0066 & 72 \\
75 & 55072.5791 & 0.0003 & $-$0.0071 & 107 \\
105 & 55074.2935 & 0.0008 & $-$0.0056 & 108 \\
123 & 55075.3249 & 0.0019 & $-$0.0019 & 115 \\
124 & 55075.3828 & 0.0014 & $-$0.0012 & 102 \\
125 & 55075.4411 & 0.0009 & 0.0000 & 97 \\
126 & 55075.4958 & 0.0011 & $-$0.0023 & 108 \\
127 & 55075.5598 & 0.0011 & 0.0045 & 114 \\
128 & 55075.6117 & 0.0015 & $-$0.0006 & 77 \\
140 & 55076.3047 & 0.0014 & 0.0072 & 112 \\
141 & 55076.3688 & 0.0013 & 0.0142 & 114 \\
142 & 55076.4220 & 0.0010 & 0.0103 & 113 \\
143 & 55076.4764 & 0.0009 & 0.0076 & 112 \\
144 & 55076.5335 & 0.0006 & 0.0076 & 107 \\
145 & 55076.5907 & 0.0010 & 0.0077 & 110 \\
209 & 55080.2372 & 0.0005 & 0.0001 & 145 \\
210 & 55080.2932 & 0.0004 & $-$0.0011 & 172 \\
213 & 55080.4643 & 0.0011 & $-$0.0013 & 63 \\
214 & 55080.5134 & 0.0017 & $-$0.0093 & 54 \\
215 & 55080.5705 & 0.0006 & $-$0.0092 & 44 \\
\hline
  \multicolumn{5}{l}{\commenta BJD$-$2400000.} \\
  \multicolumn{5}{l}{\commentb Against $max = 2455068.3040 + 0.057097 E$.} \\
  \multicolumn{5}{l}{\commentc Number of points used to determine the maximum.} \\
\end{tabular}
\end{center}
\end{table}

\subsection{ASAS J224349$+$0809.5}\label{obj:asas2243}

\begin{figure}
  \begin{center}
%    \FigureFile(88mm,90mm){asas22432009oc.eps}
    \FigureFile(88mm,90mm){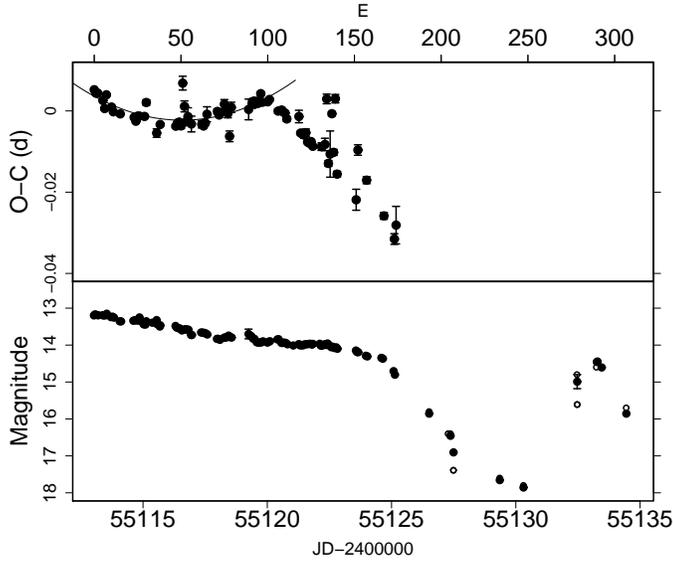}
  \end{center}
  \caption{$O-C$ of superhumps ASAS J2243 (2009).
  (Upper): $O-C$ diagram.  The $O-C$ values were against the mean period
  for the stage B ($E \le 101$, thin curve)
  (Lower): Light curve.  Open circles are snapshot unfiltered CCD
  observations.}
  \label{fig:asas22432009oc}
\end{figure}

   ASAS J224349$+$0809.5 (hereafter ASAS J2243) was selected as a dwarf nova
by P. Wils (cvnet-discussion 1320; \citet{she10asas2243}).
The 2009 outburst and the superhumps were detected by I. Miller
(cvnet-outburst 3358).  The object showed well-developed superhumps
and their time-evolution was intensively studied.
The times of superhump maxima are listed in table \ref{tab:asas2243oc2009}.
As also reported in \citet{she10asas2243}, a textbook stage B--C transition
was observed.
The $P_{\rm dot}$ during stage B was $+6.6(1.0) \times 10^{-5}$
($E \le 101$).\footnote{
   \citet{she10asas2243} reported $dP/dt$ of $+1.24(5) \times 10^{-3}$.
Our analysis of their timing data yielded $P_{\rm dot}$ of
$+9.4(0.7) \times 10^{-5}$ ($E \le 97$) by our definition.
}
The object also showed a post-superoutburst rebrightening
(figure \ref{fig:asas22432009oc}).

\begin{table}
\caption{Superhump maxima of ASAS J2243 (2009).}\label{tab:asas2243oc2009}
\begin{center}
\begin{tabular}{ccccc}
\hline
$E$ & max\commenta & error & $O-C$\commentb & $N$\commentc \\
\hline
0 & 55113.0454 & 0.0002 & 0.0005 & 132 \\
1 & 55113.1144 & 0.0003 & $-$0.0003 & 131 \\
2 & 55113.1842 & 0.0003 & $-$0.0001 & 132 \\
5 & 55113.3918 & 0.0003 & $-$0.0017 & 64 \\
6 & 55113.4596 & 0.0003 & $-$0.0036 & 75 \\
7 & 55113.5328 & 0.0003 & $-$0.0001 & 67 \\
10 & 55113.7392 & 0.0002 & $-$0.0028 & 182 \\
11 & 55113.8079 & 0.0002 & $-$0.0039 & 182 \\
15 & 55114.0866 & 0.0003 & $-$0.0040 & 134 \\
23 & 55114.6442 & 0.0002 & $-$0.0041 & 184 \\
24 & 55114.7131 & 0.0003 & $-$0.0049 & 180 \\
25 & 55114.7841 & 0.0003 & $-$0.0036 & 179 \\
26 & 55114.8539 & 0.0005 & $-$0.0035 & 104 \\
29 & 55115.0633 & 0.0004 & $-$0.0033 & 109 \\
30 & 55115.1365 & 0.0007 & 0.0003 & 64 \\
36 & 55115.5479 & 0.0010 & $-$0.0067 & 16 \\
38 & 55115.6896 & 0.0004 & $-$0.0044 & 55 \\
47 & 55116.3174 & 0.0004 & $-$0.0039 & 57 \\
48 & 55116.3881 & 0.0003 & $-$0.0030 & 83 \\
49 & 55116.4581 & 0.0005 & $-$0.0027 & 69 \\
50 & 55116.5269 & 0.0007 & $-$0.0036 & 31 \\
51 & 55116.6073 & 0.0017 & 0.0071 & 39 \\
52 & 55116.6714 & 0.0013 & 0.0014 & 49 \\
53 & 55116.7374 & 0.0005 & $-$0.0023 & 104 \\
54 & 55116.8085 & 0.0022 & $-$0.0009 & 69 \\
56 & 55116.9463 & 0.0019 & $-$0.0025 & 30 \\
62 & 55117.3650 & 0.0009 & $-$0.0020 & 80 \\
63 & 55117.4344 & 0.0005 & $-$0.0023 & 87 \\
64 & 55117.5049 & 0.0005 & $-$0.0015 & 76 \\
65 & 55117.5769 & 0.0018 & 0.0008 & 22 \\
71 & 55117.9965 & 0.0006 & 0.0020 & 196 \\
72 & 55118.0654 & 0.0005 & 0.0012 & 176 \\
75 & 55118.2775 & 0.0011 & 0.0042 & 26 \\
76 & 55118.3461 & 0.0009 & 0.0030 & 35 \\
77 & 55118.4148 & 0.0010 & 0.0021 & 58 \\
78 & 55118.4790 & 0.0013 & $-$0.0034 & 56 \\
79 & 55118.5559 & 0.0013 & 0.0038 & 57 \\
89 & 55119.2535 & 0.0025 & 0.0043 & 24 \\
91 & 55119.3946 & 0.0012 & 0.0059 & 56 \\
92 & 55119.4650 & 0.0007 & 0.0066 & 65 \\
93 & 55119.5342 & 0.0009 & 0.0060 & 32 \\
94 & 55119.6046 & 0.0007 & 0.0068 & 90 \\
95 & 55119.6740 & 0.0005 & 0.0065 & 81 \\
96 & 55119.7461 & 0.0005 & 0.0088 & 83 \\
97 & 55119.8138 & 0.0005 & 0.0068 & 79 \\
100 & 55120.0233 & 0.0004 & 0.0072 & 143 \\
101 & 55120.0937 & 0.0004 & 0.0079 & 149 \\
106 & 55120.4399 & 0.0005 & 0.0055 & 63 \\
108 & 55120.5797 & 0.0007 & 0.0060 & 73 \\
109 & 55120.6490 & 0.0004 & 0.0055 & 91 \\
110 & 55120.7186 & 0.0004 & 0.0054 & 88 \\
111 & 55120.7870 & 0.0006 & 0.0041 & 88 \\
118 & 55121.2762 & 0.0015 & 0.0053 & 113 \\
119 & 55121.3420 & 0.0004 & 0.0014 & 156 \\
120 & 55121.4116 & 0.0007 & 0.0013 & 182 \\
121 & 55121.4811 & 0.0008 & 0.0011 & 149 \\
122 & 55121.5515 & 0.0009 & 0.0017 & 92 \\
123 & 55121.6190 & 0.0003 & $-$0.0005 & 182 \\
\hline
  \multicolumn{5}{l}{\commenta BJD$-$2400000.} \\
  \multicolumn{5}{l}{\commentb Against $max = 2455113.0449 + 0.069712 E$.} \\
  \multicolumn{5}{l}{\commentc Number of points used to determine the maximum.} \\
\end{tabular}
\end{center}
\end{table}

\addtocounter{table}{-1}
\begin{table}
\caption{Superhump maxima of ASAS J2243 (2009) (continued).}
\begin{center}
\begin{tabular}{ccccc}
\hline
$E$ & max\commenta & error & $O-C$\commentb & $N$\commentc \\
\hline
124 & 55121.6885 & 0.0003 & $-$0.0006 & 182 \\
125 & 55121.7587 & 0.0005 & $-$0.0002 & 184 \\
126 & 55121.8274 & 0.0004 & $-$0.0012 & 130 \\
131 & 55122.1763 & 0.0009 & $-$0.0008 & 131 \\
133 & 55122.3165 & 0.0015 & $-$0.0001 & 24 \\
134 & 55122.3975 & 0.0012 & 0.0112 & 140 \\
135 & 55122.4514 & 0.0007 & $-$0.0046 & 220 \\
136 & 55122.5235 & 0.0057 & $-$0.0022 & 85 \\
137 & 55122.6033 & 0.0005 & 0.0079 & 178 \\
138 & 55122.6636 & 0.0007 & $-$0.0016 & 181 \\
139 & 55122.7466 & 0.0010 & 0.0117 & 178 \\
140 & 55122.7979 & 0.0007 & $-$0.0067 & 180 \\
151 & 55123.5594 & 0.0026 & $-$0.0120 & 94 \\
152 & 55123.6415 & 0.0013 & 0.0004 & 178 \\
157 & 55123.9831 & 0.0009 & $-$0.0066 & 147 \\
167 & 55124.6724 & 0.0008 & $-$0.0144 & 92 \\
173 & 55125.0856 & 0.0013 & $-$0.0195 & 149 \\
174 & 55125.1588 & 0.0046 & $-$0.0160 & 73 \\
\hline
  \multicolumn{5}{l}{\commenta BJD$-$2400000.} \\
  \multicolumn{5}{l}{\commentb Against $max = 2455113.0449 + 0.069712 E$.} \\
  \multicolumn{5}{l}{\commentc Number of points used to determine the maximum.} \\
\end{tabular}
\end{center}
\end{table}

\subsection{Lanning 420}\label{obj:lanning420}

\begin{figure}
  \begin{center}
%    \FigureFile(88mm,110mm){lanning420shpdm.eps}
    \FigureFile(88mm,110mm){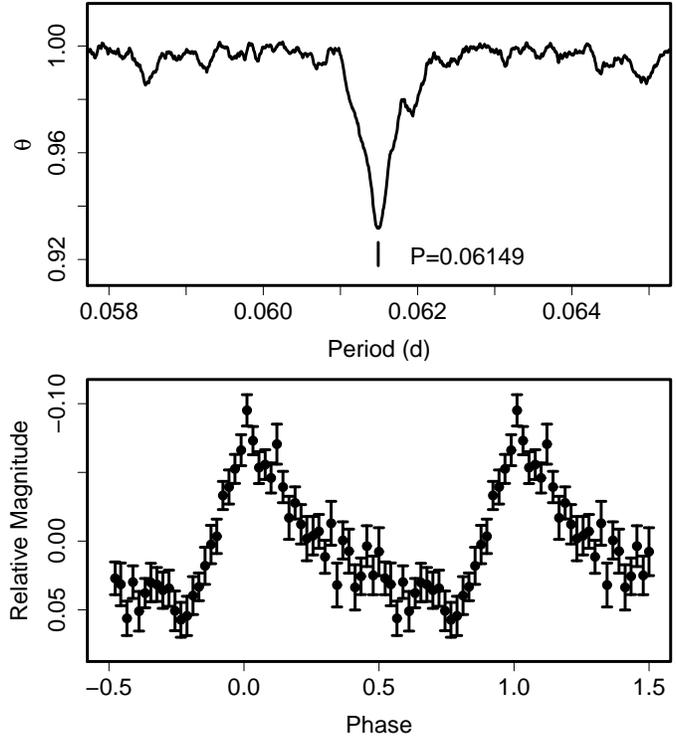}
  \end{center}
  \caption{Superhumps in Lanning 420 (2010).
     (Upper): PDM analysis.
     (Lower): Phase-averaged profile.}
  \label{fig:lanning420shpdm}
\end{figure}

   Lanning 420 was selected as a UV-bright transient object
\citep{lan00UVstar5}, which was considered to be a possible nova because
of the absence on the Digitized Sky Survey image.  \citet{bra08lanning386}
listed Lanning CVs and further investigated these objects.
S. Brady indeed detected an outburst in 2007 and another one in 2010,
which turned out to be a superoutburst (BAAVSS alert 2374).
Follow-up observation confirmed the presence of superhumps
(vsnet-alert 12131, 12132, 12138; figure \ref{fig:lanning420shpdm}).

   The times of superhump maxima are listed in table \ref{tab:lanning4202010}.
There was a distinct shortening of the superhump period around $E=70$,
which we interpreted as a stage B--C transition.
The last two epochs were measured during the early post-superoutburst stage.
We included these epochs because short-$P_{\rm SH}$ tend to show persistent
superhumps during this stage and the times of maxima were in good agreement
with extrapolated stage C superhumps.
The measured period derivative during the stage B had a relatively large
error because only the late stage of the stage B was observed.
Observations of the earlier stage are crucial to better characterize
the period variation in this system.

\begin{table}
\caption{Superhump maxima of Lanning 420 (2010).}\label{tab:lanning4202010}
\begin{center}
\begin{tabular}{ccccc}
\hline
$E$ & max\commenta & error & $O-C$\commentb & $N$\commentc \\
\hline
0 & 55438.5841 & 0.0005 & $-$0.0084 & 20 \\
1 & 55438.6469 & 0.0005 & $-$0.0070 & 26 \\
2 & 55438.7074 & 0.0005 & $-$0.0078 & 24 \\
3 & 55438.7691 & 0.0005 & $-$0.0075 & 25 \\
4 & 55438.8277 & 0.0018 & $-$0.0102 & 24 \\
14 & 55439.4478 & 0.0011 & $-$0.0033 & 51 \\
15 & 55439.5051 & 0.0018 & $-$0.0074 & 46 \\
16 & 55439.5672 & 0.0011 & $-$0.0066 & 89 \\
17 & 55439.6314 & 0.0012 & $-$0.0037 & 86 \\
18 & 55439.6895 & 0.0010 & $-$0.0069 & 88 \\
19 & 55439.7535 & 0.0008 & $-$0.0042 & 76 \\
20 & 55439.8132 & 0.0012 & $-$0.0059 & 46 \\
21 & 55439.8767 & 0.0009 & $-$0.0037 & 60 \\
22 & 55439.9369 & 0.0008 & $-$0.0048 & 56 \\
23 & 55439.9982 & 0.0012 & $-$0.0049 & 45 \\
24 & 55440.0589 & 0.0015 & $-$0.0055 & 125 \\
25 & 55440.1225 & 0.0030 & $-$0.0032 & 115 \\
31 & 55440.4990 & 0.0049 & 0.0054 & 20 \\
32 & 55440.5511 & 0.0022 & $-$0.0039 & 21 \\
33 & 55440.6082 & 0.0073 & $-$0.0081 & 19 \\
34 & 55440.6785 & 0.0020 & 0.0008 & 49 \\
35 & 55440.7379 & 0.0006 & $-$0.0010 & 56 \\
36 & 55440.7997 & 0.0009 & $-$0.0006 & 31 \\
37 & 55440.8611 & 0.0006 & $-$0.0005 & 45 \\
38 & 55440.9243 & 0.0007 & 0.0014 & 56 \\
39 & 55440.9852 & 0.0023 & 0.0009 & 163 \\
40 & 55441.0392 & 0.0054 & $-$0.0064 & 128 \\
41 & 55441.1128 & 0.0018 & 0.0059 & 253 \\
42 & 55441.1713 & 0.0019 & 0.0031 & 115 \\
46 & 55441.4149 & 0.0030 & 0.0014 & 15 \\
47 & 55441.4792 & 0.0028 & 0.0044 & 21 \\
48 & 55441.5345 & 0.0020 & $-$0.0017 & 25 \\
49 & 55441.5992 & 0.0038 & 0.0017 & 39 \\
50 & 55441.6608 & 0.0016 & 0.0019 & 48 \\
51 & 55441.7225 & 0.0008 & 0.0023 & 73 \\
52 & 55441.7874 & 0.0017 & 0.0059 & 60 \\
53 & 55441.8443 & 0.0020 & 0.0015 & 53 \\
54 & 55441.9112 & 0.0015 & 0.0071 & 56 \\
55 & 55441.9710 & 0.0019 & 0.0056 & 56 \\
62 & 55442.4036 & 0.0011 & 0.0088 & 21 \\
63 & 55442.4636 & 0.0010 & 0.0075 & 71 \\
64 & 55442.5285 & 0.0026 & 0.0111 & 21 \\
65 & 55442.5846 & 0.0016 & 0.0059 & 21 \\
66 & 55442.6469 & 0.0029 & 0.0068 & 21 \\
76 & 55443.2586 & 0.0018 & 0.0053 & 110 \\
78 & 55443.3918 & 0.0051 & 0.0159 & 35 \\
79 & 55443.4458 & 0.0014 & 0.0085 & 48 \\
80 & 55443.5039 & 0.0015 & 0.0053 & 46 \\
81 & 55443.5651 & 0.0017 & 0.0051 & 48 \\
82 & 55443.6274 & 0.0015 & 0.0061 & 45 \\
83 & 55443.6835 & 0.0083 & 0.0009 & 17 \\
92 & 55444.2430 & 0.0024 & 0.0085 & 122 \\
93 & 55444.3012 & 0.0021 & 0.0053 & 63 \\
94 & 55444.3635 & 0.0024 & 0.0064 & 15 \\
95 & 55444.4301 & 0.0130 & 0.0116 & 21 \\
96 & 55444.4863 & 0.0023 & 0.0065 & 20 \\
\hline
  \multicolumn{5}{l}{\commenta BJD$-$2400000.} \\
  \multicolumn{5}{l}{\commentb Against $max = 2455438.5926 + 0.061325 E$.} \\
  \multicolumn{5}{l}{\commentc Number of points used to determine the maximum.} \\
\end{tabular}
\end{center}
\end{table}

\addtocounter{table}{-1}
\begin{table}
\caption{Superhump maxima of Lanning 420 (2010) (continued).}
\begin{center}
\begin{tabular}{ccccc}
\hline
$E$ & max\commenta & error & $O-C$\commentb & $N$\commentc \\
\hline
97 & 55444.5442 & 0.0020 & 0.0031 & 15 \\
98 & 55444.6082 & 0.0026 & 0.0058 & 19 \\
99 & 55444.6637 & 0.0014 & $-$0.0001 & 19 \\
109 & 55445.2851 & 0.0155 & 0.0081 & 65 \\
116 & 55445.7083 & 0.0024 & 0.0020 & 17 \\
117 & 55445.7670 & 0.0032 & $-$0.0006 & 16 \\
118 & 55445.8309 & 0.0017 & 0.0019 & 16 \\
124 & 55446.1980 & 0.0028 & 0.0011 & 111 \\
125 & 55446.2583 & 0.0046 & 0.0001 & 80 \\
126 & 55446.3132 & 0.0059 & $-$0.0063 & 60 \\
174 & 55449.2331 & 0.0071 & $-$0.0301 & 117 \\
175 & 55449.2880 & 0.0028 & $-$0.0365 & 109 \\
\hline
  \multicolumn{5}{l}{\commenta BJD$-$2400000.} \\
  \multicolumn{5}{l}{\commentb Against $max = 2455438.5926 + 0.061325 E$.} \\
  \multicolumn{5}{l}{\commentc Number of points used to determine the maximum.} \\
\end{tabular}
\end{center}
\end{table}

\subsection{PG 0149$+$138}\label{obj:pg0149}

   This object (hereafter PG 0149) was originally discovered
as an eruptive object with strong UV excess, and was suspected to be
a supernova \citep{PGsurvey}.  \citet{szk02egcnchvvirHST} selected
this object during the course of the SDSS survey and classified it
as a CV (dwarf nova).  The object as been monitored as a dwarf nova
since then.  \citet{dil08SDSSCV} photometrically identified a
$P_{\rm orb}$ of 0.08242(3) d.

\begin{figure}
  \begin{center}
%    \FigureFile(88mm,110mm){pg0149shpdm.eps}
    \FigureFile(88mm,110mm){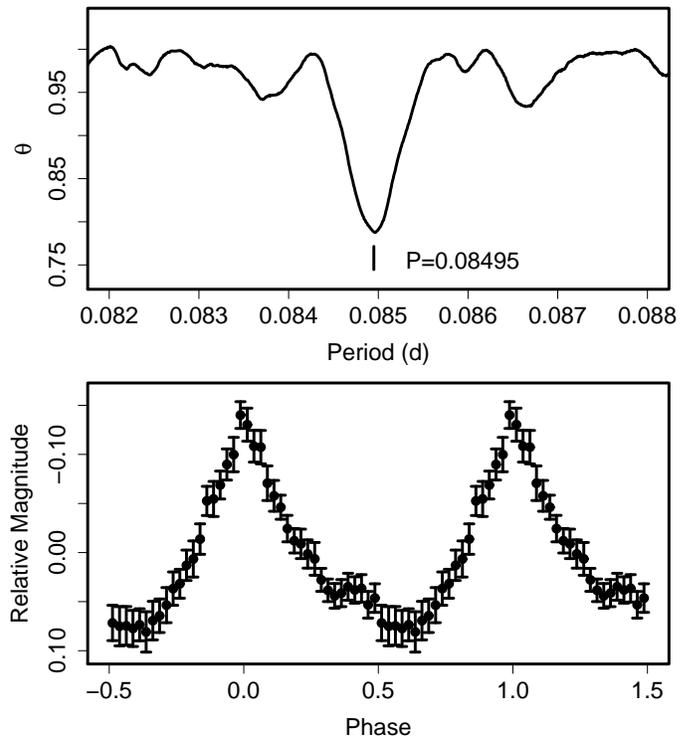}
  \end{center}
  \caption{Superhumps in PG 0149 (2009).
     (Upper): PDM analysis.
     (Lower): Phase-averaged profile.}
  \label{fig:pg0149shpdm}
\end{figure}

   The 2009 September outburst of this object was detected by the
the Catalina Real-time Transient Survey (CRTS, \cite{CRTS})\footnote{
   $<$http://nesssi.cacr.caltech.edu/catalina/$>$.
   For the information of the individual Catalina CVs, see
   $<$http://nesssi.cacr.caltech.edu/catalina/AllCV.html$>$.
}
(= CSS090911:015152$+$140047).  Superhumps were detected immediately
following the announcement (vsnet-alert 11465).
The mean $P_{\rm SH}$ with the PDM method was 0.08495(2) d
(figure \ref{fig:pg0149shpdm}).
The times of superhump maxima are summarized in table
\ref{tab:pg0149oc2009}.  There was a clear stage B--C transition with
a positive $P_{\rm dot}$ = $+14.6(2.5) \times 10^{-5}$ ($E \le 60$).
Such a positive $P_{\rm dot}$ is rare for SU UMa-type dwarf novae
with this $P_{\rm SH}$.  The behavior in the superhump period
resemble that of long-$P_{\rm SH}$ systems like QW Ser
(cf. \cite{nog04qwser}; \cite{Pdot}).

\begin{table}
\caption{Superhump maxima of PG 0149 (2009).}\label{tab:pg0149oc2009}
\begin{center}
\begin{tabular}{ccccc}
\hline
$E$ & max\commenta & error & $O-C$\commentb & $N$\commentc \\
\hline
0 & 55086.5129 & 0.0008 & $-$0.0038 & 44 \\
1 & 55086.5987 & 0.0003 & $-$0.0028 & 89 \\
11 & 55087.4448 & 0.0007 & $-$0.0055 & 57 \\
12 & 55087.5312 & 0.0005 & $-$0.0040 & 114 \\
13 & 55087.6173 & 0.0005 & $-$0.0029 & 89 \\
23 & 55088.4654 & 0.0006 & $-$0.0036 & 40 \\
24 & 55088.5484 & 0.0006 & $-$0.0054 & 83 \\
55 & 55091.1917 & 0.0006 & 0.0065 & 180 \\
59 & 55091.5320 & 0.0007 & 0.0073 & 77 \\
60 & 55091.6176 & 0.0008 & 0.0080 & 89 \\
70 & 55092.4661 & 0.0006 & 0.0076 & 91 \\
71 & 55092.5517 & 0.0007 & 0.0083 & 130 \\
72 & 55092.6371 & 0.0006 & 0.0089 & 111 \\
93 & 55094.4113 & 0.0014 & 0.0005 & 40 \\
94 & 55094.4977 & 0.0009 & 0.0020 & 45 \\
95 & 55094.5776 & 0.0009 & $-$0.0029 & 44 \\
103 & 55095.2556 & 0.0008 & $-$0.0039 & 146 \\
154 & 55099.5741 & 0.0013 & $-$0.0145 & 78 \\
\hline
  \multicolumn{5}{l}{\commenta BJD$-$2400000.} \\
  \multicolumn{5}{l}{\commentb Against $max = 2455086.5166 + 0.084883 E$.} \\
  \multicolumn{5}{l}{\commentc Number of points used to determine the maximum.} \\
\end{tabular}
\end{center}
\end{table}

\subsection{RX J1715.6$+$6856}\label{obj:j1715}

   This object (hereafter RX J1715) is a CV identified from the
ROSAT North Ecliptic Pole survey \citep{pre07ROSATCVs}.
\citet{pre07ROSATCVs} reported the detection of a 1.64(2) hr period
from radial-velocity study.  \citet{she10j1715} identified the dwarf-nova
type behavior and detected superhumps during a superoutburst in
2009 August.  We analyzed the combined data from \citet{she10j1715}
(from the AAVSO database) and our own observations.
As reported in \citet{she10j1715}, these observations covered the
final stage of the superoutburst.

\begin{figure}
  \begin{center}
%    \FigureFile(88mm,110mm){j1715shpdm.eps}
    \FigureFile(88mm,110mm){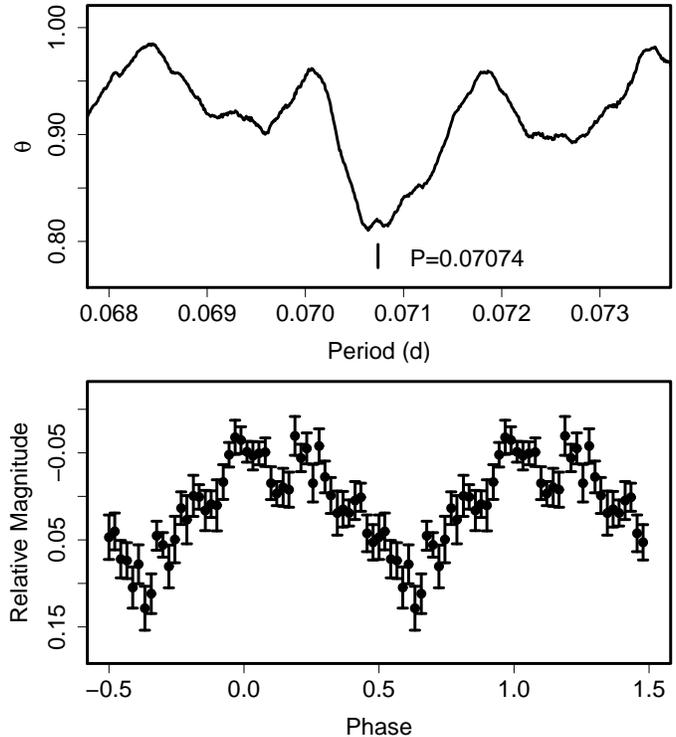}
  \end{center}
  \caption{Superhumps in RX J1715 (2009).
     (Upper): PDM analysis.
     (Lower): Phase-averaged profile.}
  \label{fig:j1715shpdm}
\end{figure}

   The times of maxima are listed in table \ref{tab:j1715oc2009}.
Since the $O-C$'s for maxima for $E \ge 63$ largely deviate from the
earlier trend, these humps are less likely a continuation of
superhumps observed during the plateau phase ($E \le 48$).
Restricting to $E \le 48$, we obtained a mean $P_{\rm SH}$
of 0.07074(4) d with the PDM method (figure \ref{fig:j1715shpdm}).
and is in agreement with the corresponding period of 0.07086(78) d
by \citet{she10j1715} within respective errors.
The $\epsilon$ against $P_{\rm orb}$ was 3.5 \%.
Since the observation covered the final part of the superoutburst,
these superhumps are likely stage C superhumps.

\begin{figure}
  \begin{center}
%    \FigureFile(88mm,110mm){j1715prof.eps}
    \FigureFile(88mm,110mm){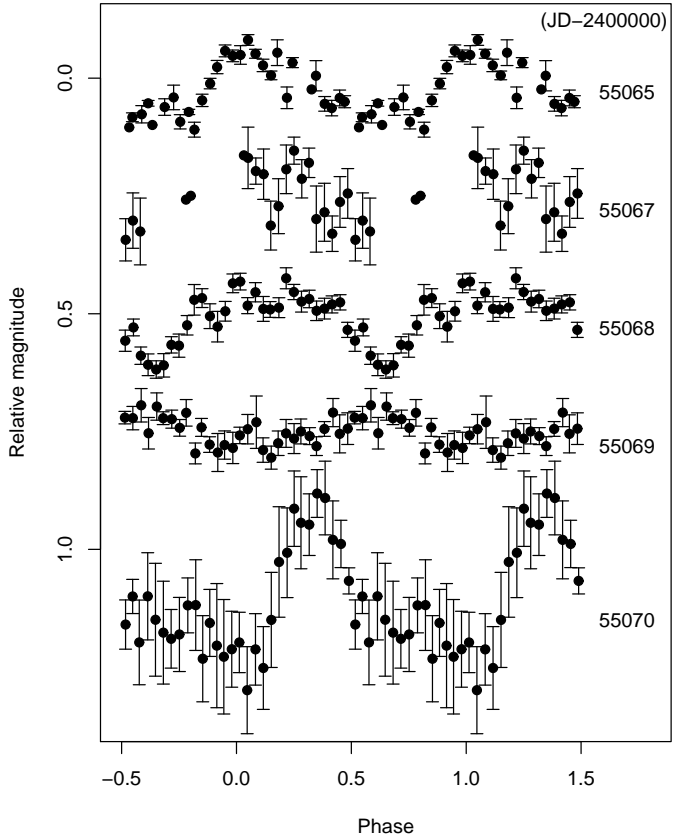}
  \end{center}
  \caption{Nightly variation of profiles of superhumps in RX J1715 (2009).
     The profiles were drawn against a period of 0.07074 d, determined
     from the superoutburst plateau.  Although this period well expresses
     the variations on the first three nights, there was a significant
     phase offset at later stages.}
  \label{fig:j1715prof}
\end{figure}

   Although \citet{she10j1715} interpreted observed humps for $E \ge 63$
(rapid fading stage and post-superoutburst stage) as orbital humps,
the reported period [0.06944(90) d] is somewhat different from the one
reported from radial-velocity study.
Our analysis also confirmed a significant phase offset during the
rapid fading stage (figure \ref{fig:j1715prof}).  We could not,
however, determine whether the newly arising signals were from
orbital humps or were from traditional late superhumps because
the amplitude on JD 2455069 was very small and it was difficult to
derive a meaningful period based on the last two nights of observations.
Future observations are needed to confirm the nature of these
large-amplitude humps.   Since stage C superhumps frequently endure
during the post-superoutburst stage in many well-observed systems
(cf. \cite{Pdot}), the early disappearance of superhumps in RX J1715
appears to be rather unique.  The case may be similar to DT Oct
(cf. \citet{Pdot}), which showed an early switch to traditional
late superhumps.  Although both relatively low outburst amplitudes
and the relatively strong X-ray emission are common to these objects,
the frequency of outbursts in RX J1715 appears to be smaller
\citep{she10j1715}.  A further detailed comparison between these
objects might be fruitful.

\begin{table}
\caption{Superhump maxima of RX J1715 (2009).}\label{tab:j1715oc2009}
\begin{center}
\begin{tabular}{ccccc}
\hline
$E$ & max\commenta & error & $O-C$\commentb & $N$\commentc \\
\hline
0 & 55065.3776 & 0.0007 & 0.0037 & 61 \\
1 & 55065.4528 & 0.0010 & 0.0078 & 87 \\
2 & 55065.5181 & 0.0008 & 0.0021 & 30 \\
23 & 55067.0079 & 0.0080 & $-$0.0000 & 86 \\
39 & 55068.1405 & 0.0021 & $-$0.0041 & 134 \\
42 & 55068.3635 & 0.0014 & 0.0058 & 62 \\
43 & 55068.4183 & 0.0014 & $-$0.0105 & 90 \\
44 & 55068.4953 & 0.0015 & $-$0.0045 & 101 \\
47 & 55068.7020 & 0.0021 & $-$0.0110 & 37 \\
48 & 55068.7722 & 0.0024 & $-$0.0118 & 27 \\
63 & 55069.8143 & 0.0045 & $-$0.0353 & 37 \\
63 & 55069.8756 & 0.0019 & 0.0259 & 38 \\
64 & 55069.9378 & 0.0010 & 0.0171 & 32 \\
75 & 55070.7114 & 0.0019 & 0.0092 & 27 \\
76 & 55070.7752 & 0.0016 & 0.0020 & 25 \\
77 & 55070.8431 & 0.0024 & $-$0.0012 & 26 \\
78 & 55070.9205 & 0.0012 & 0.0051 & 24 \\
\hline
  \multicolumn{5}{l}{\commenta BJD$-$2400000.} \\
  \multicolumn{5}{l}{\commentb Against $max = 2455065.3739 + 0.071044 E$.} \\
  \multicolumn{5}{l}{\commentc Number of points used to determine the maximum.} \\
\end{tabular}
\end{center}
\end{table}

\subsection{SDSS J012940.05$+$384210.4}\label{obj:j0129}

   This object (hereafter SDSS J0129) is an AM CVn-type CV selected
during the course of the SDSS \citep{and05SDSSamcvn}, who reported
broad He\textsc{I} emission lines.  The orbital period has not been
reported yet.

\begin{figure}
  \begin{center}
%    \FigureFile(88mm,110mm){j0129lc.eps}
    \FigureFile(88mm,110mm){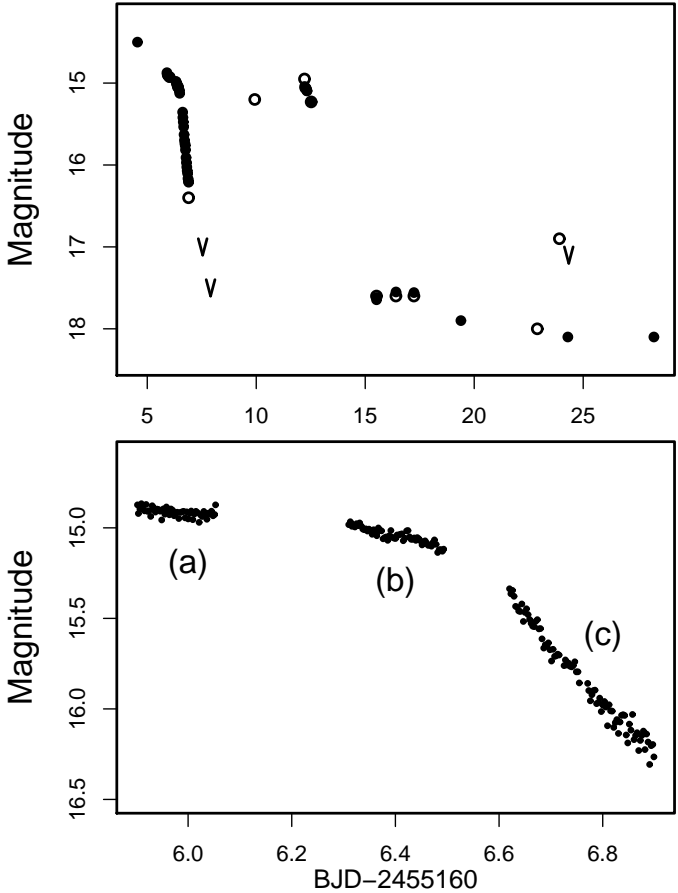}
  \end{center}
  \caption{Outburst of SDSS J0129 in 2009 November--December.
     (Upper): Overall outburst including the rebrightening phase.
     (Lower): Enlargement of the main superoutburst.  The segments
     (a)--(c) correspond to the intervals to draw figure \ref{fig:j0129shprof}.}
  \label{fig:j0129lc}
\end{figure}

   The object was reported in outburst on 2009 November 29 at a unfiltered
CCD magnitude of 14.5 (cvnet-outburst 3479, vsnet-alert 11702).
Two days after this detection, the object started to fade rapidly.
Immediately following this transient fading, the object experienced
rebrightenings on December 4 and 18 (vsnet-alert 11707, 11737).
The overall behavior of the outburst (figure \ref{fig:j0129lc}) was
extremely similar to the 2003 August superoutburst of V803 Cen
(\cite{kat04v803cen}; vsnet-alert 11709).

\begin{figure}
  \begin{center}
%    \FigureFile(88mm,110mm){j0129shpdm.eps}
    \FigureFile(88mm,110mm){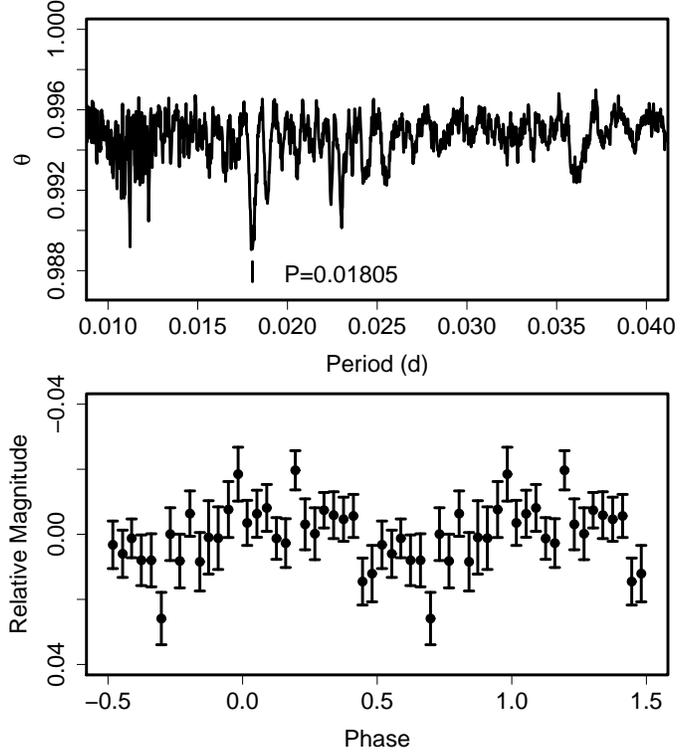}
  \end{center}
  \caption{Superhumps in SDSS J0129 (2009).
     (Upper): PDM analysis.  The signal around $P = 0.036$ is the
     first harmonic of the fundamental period.
     (Lower): Phase-averaged profile.}
  \label{fig:j0129shpdm}
\end{figure}

\begin{figure}
  \begin{center}
%    \FigureFile(88mm,110mm){j0129shprof.eps}
    \FigureFile(88mm,110mm){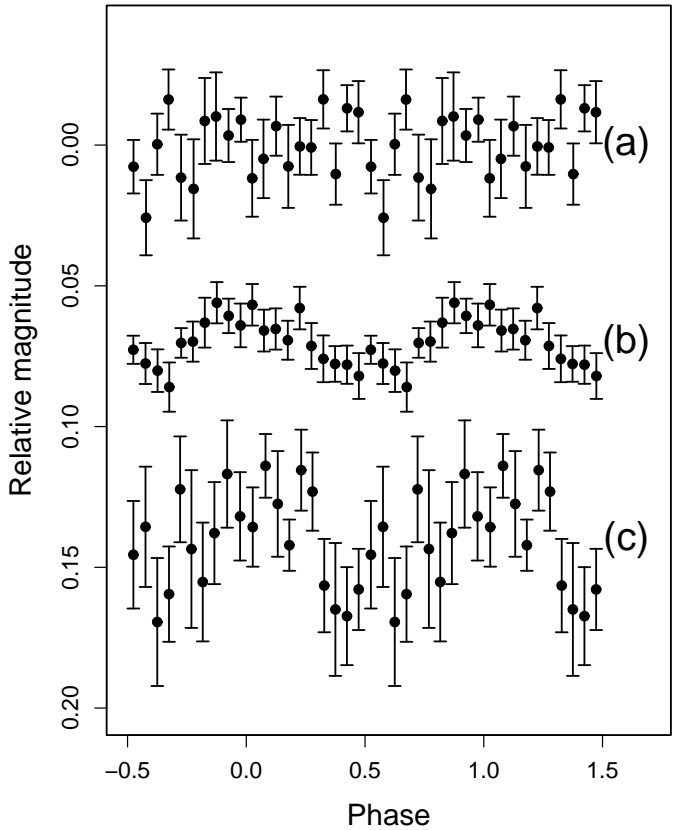}
  \end{center}
  \caption{Variation of superhump profiles during the fading stage of
  the superoutburst of SDSS J0129 (2009).  A mean period of 0.01805 d
  was used to draw this figure.  The segments (a)--(c) are
  shown in figure \ref{fig:j0129lc}.}
  \label{fig:j0129shprof}
\end{figure}

   The object developed hump signals during its terminal stage of
the plateau phase (figures \ref{fig:j0129shpdm}, \ref{fig:j0129shprof}).
Although these humps may reflect orbital modulations, we identified
them as superhumps based on extreme analogy with V803 Cen
\citep{kat04v803cen}.  The mean period determined with the PDM method
was 0.01805(10) d.  The period is also very similar to that of V803 Cen
($P$ = 0.018686(4) d = 1614.5(4) s, \cite{kat04v803cen}).

   The times of maxima during the fading branch from the superoutburst
are listed in table \ref{tab:j0129oc2009}.

\begin{table}
\caption{Superhump maxima of SDSS J0129 (2009).}\label{tab:j0129oc2009}
\begin{center}
\begin{tabular}{ccccc}
\hline
$E$ & max\commenta & error & $O-C$\commentb & $N$\commentc \\
\hline
0 & 55166.3297 & 0.0030 & $-$0.0020 & 40 \\
1 & 55166.3491 & 0.0037 & $-$0.0007 & 40 \\
2 & 55166.3686 & 0.0012 & 0.0007 & 41 \\
4 & 55166.4080 & 0.0013 & 0.0039 & 41 \\
5 & 55166.4231 & 0.0006 & 0.0009 & 52 \\
6 & 55166.4433 & 0.0013 & 0.0031 & 62 \\
7 & 55166.4567 & 0.0012 & $-$0.0016 & 61 \\
8 & 55166.4739 & 0.0015 & $-$0.0026 & 22 \\
18 & 55166.6542 & 0.0011 & $-$0.0032 & 17 \\
19 & 55166.6730 & 0.0011 & $-$0.0024 & 17 \\
20 & 55166.6947 & 0.0017 & 0.0012 & 17 \\
22 & 55166.7329 & 0.0019 & 0.0031 & 17 \\
23 & 55166.7458 & 0.0017 & $-$0.0020 & 17 \\
25 & 55166.7840 & 0.0010 & 0.0000 & 17 \\
26 & 55166.8034 & 0.0012 & 0.0013 & 17 \\
28 & 55166.8384 & 0.0010 & 0.0001 & 17 \\
29 & 55166.8547 & 0.0017 & $-$0.0016 & 16 \\
30 & 55166.8764 & 0.0008 & 0.0020 & 16 \\
\hline
  \multicolumn{5}{l}{\commenta BJD$-$2400000.} \\
  \multicolumn{5}{l}{\commentb Against $max = 2455166.3317 + 0.018090 E$.} \\
  \multicolumn{5}{l}{\commentc Number of points used to determine the maximum.} \\
\end{tabular}
\end{center}
\end{table}

\subsection{SDSS J031051.66$-$075500.3}\label{obj:j0310}

   In addition to the 2004 superoutburst of this object (hereafter
SDSS J0310), two superoutbursts were recorded in 2009.
Table \ref{tab:j0310oc2009b} gives the times of superhump maxima during
its second superoutburst in 2009 (2009 November, designated as 2009b
in table in order to avoid confusion with the 2009 superoutburst
described in \cite{Pdot}).
The mean $P_{\rm SH}$ during this superoutburst was 0.06786(3) d
(PDM method, alias selected based on the 2004 observation), which is
shorter than the period obtained during the 2004 superoutburst.
Although this difference may have resulted from different stages
observed in different outbursts, the difference appears to be larger
than those associated with typical stage B--C transitions \citep{Pdot}.
This needs to be clarified by future observations.
The shortest interval between superoutbursts was 284 d, which is
likely the supercycle of this object.

\begin{table}
\caption{Superhump maxima of SDSS J0310 (2009 November).}\label{tab:j0310oc2009b}
\begin{center}
\begin{tabular}{ccccc}
\hline
$E$ & max\commenta & error & $O-C$\commentb & $N$\commentc \\
\hline
0 & 55144.5155 & 0.0003 & 0.0028 & 67 \\
1 & 55144.5777 & 0.0003 & $-$0.0029 & 73 \\
31 & 55146.6161 & 0.0009 & 0.0001 & 61 \\
\hline
  \multicolumn{5}{l}{\commenta BJD$-$2400000.} \\
  \multicolumn{5}{l}{\commentb Against $max = 2455144.5127 + 0.067848 E$.} \\
  \multicolumn{5}{l}{\commentc Number of points used to determine the maximum.} \\
\end{tabular}
\end{center}
\end{table}

\subsection{SDSS J073208.11$+$413008.7}\label{obj:j0732}

\begin{figure}
  \begin{center}
%    \FigureFile(88mm,110mm){j0732shpdm.eps}
    \FigureFile(88mm,110mm){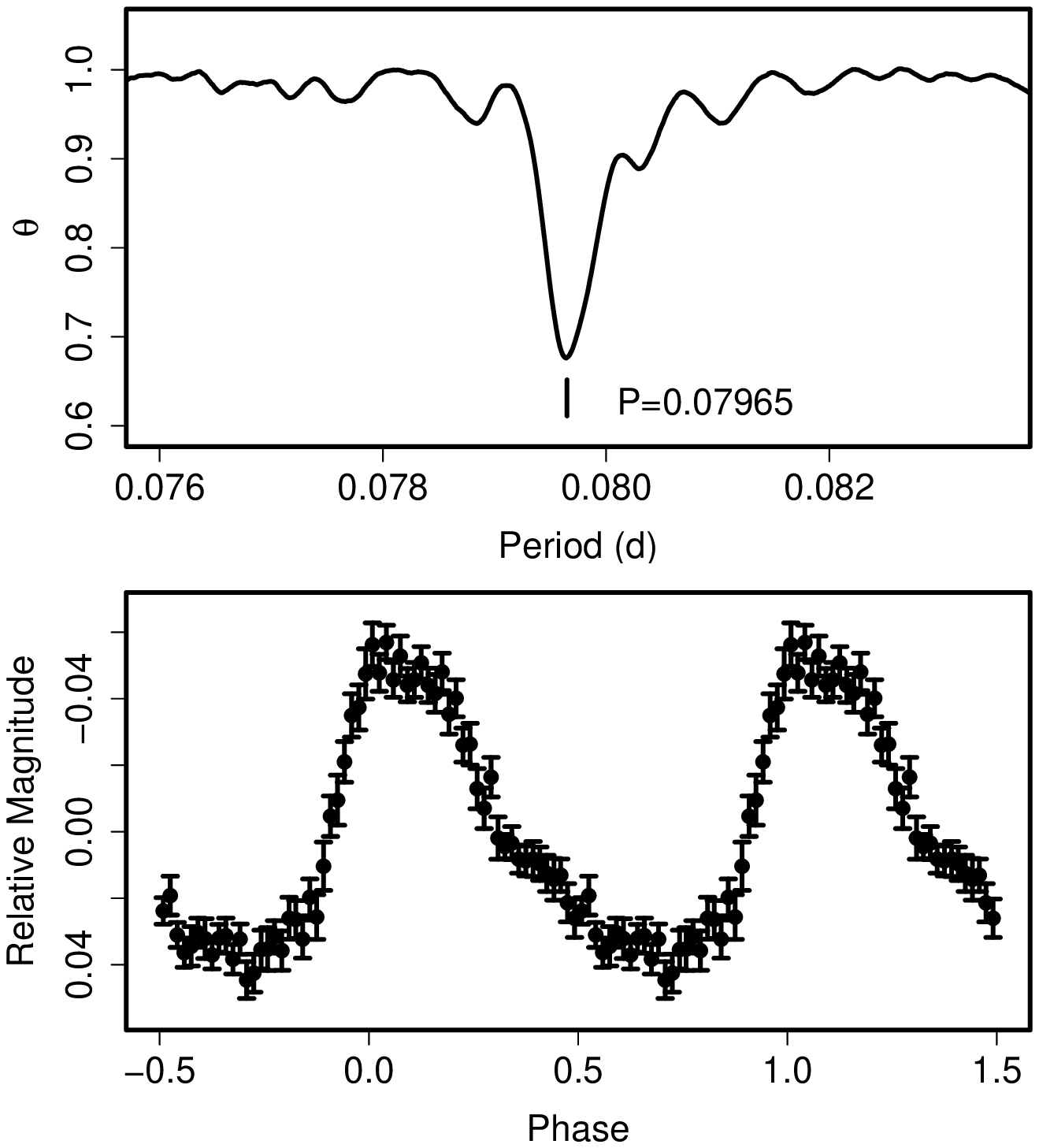}
  \end{center}
  \caption{Superhumps in SDSS J0732 (2010).
     (Upper): PDM analysis.
     (Lower): Phase-averaged profile.}
  \label{fig:j0732shpdm}
\end{figure}

   This object (hereafter SDSS J0732) was selected using SDSS and CRTS
data as a candidate dwarf nova by \citet{wil10newCVs}.\footnote{
   Although this object was not discovered during the regular course of
   the SDSS survey, we used SDSS designation for convention.
}
The object was detected in outburst by J. Shears on 2009 December 31 at
an unfiltered CCD magnitude of 16.2 (cvnet-outburst 3528; \citet{she10j0732}).
Subsequent observations confirmed the presence of superhumps
(cvnet-outburst 3535, figure \ref{fig:j0732shpdm}).

   The times of superhump maxima determined from AAVSO observations
are listed in table \ref{tab:j0732oc2010}.\footnote{
   We designated this outburst as SDSS J0732 (2010) because all
   time-resolved observation were undertaken in 2010.
}
The object showed a clear stage B--C transition around $E = 60$.
The $P_{\rm dot}$ for stage B was $+3.9(2.4) \times 10^{-5}$.\footnote{
   \citet{she10j0732} reported $dP/dt$ of $+2.81(9) \times 10^{-3}$.
They used a data set including our present data set.  We do not attempt
to choose a better $P_{\rm dot}$ between ours and theirs, since $P_{\rm dot}$
is dependent on the segment used, and the observed baseline for the
stage B was too short.
The $P_{\rm dot}$ determined from the timing data in \citet{she10j0732}
was $+22.5(6.5) \times 10^{-5}$ ($E \le 41$) while it was
$-1.7(3.3) \times 10^{-5}$ for $E \le 64$ (both values are by our definition
of $P_{\rm dot}$).
}
Relatively few objects with similar $P_{\rm SH}$, including
RZ Leo and QY Per, are known to show positive $P_{\rm dot}$
\citep{Pdot}.  Although the relatively large (5.2 mag) outburst amplitude
also might suggest a low mass-transfer rate as in RZ Leo and QY Per,
the outburst frequency estimated in \citet{she10j0732} is much higher
than in these objects.  An exact determination of $P_{\rm dot}$ during
the next superoutburst is desired.

\begin{table}
\caption{Superhump maxima of SDSS J0732 (2010).}\label{tab:j0732oc2010}
\begin{center}
\begin{tabular}{ccccc}
\hline
$E$ & max\commenta & error & $O-C$\commentb & $N$\commentc \\
\hline
0 & 55199.6270 & 0.0020 & $-$0.0080 & 42 \\
1 & 55199.7096 & 0.0007 & $-$0.0050 & 83 \\
2 & 55199.7887 & 0.0006 & $-$0.0054 & 76 \\
3 & 55199.8666 & 0.0006 & $-$0.0071 & 81 \\
4 & 55199.9462 & 0.0008 & $-$0.0071 & 82 \\
10 & 55200.4238 & 0.0018 & $-$0.0071 & 60 \\
14 & 55200.7446 & 0.0006 & $-$0.0047 & 82 \\
15 & 55200.8244 & 0.0007 & $-$0.0044 & 68 \\
16 & 55200.9040 & 0.0006 & $-$0.0044 & 82 \\
17 & 55200.9833 & 0.0011 & $-$0.0047 & 73 \\
26 & 55201.7046 & 0.0006 & 0.0003 & 82 \\
27 & 55201.7859 & 0.0007 & 0.0020 & 82 \\
28 & 55201.8656 & 0.0008 & 0.0020 & 74 \\
29 & 55201.9427 & 0.0010 & $-$0.0004 & 82 \\
30 & 55202.0233 & 0.0012 & 0.0005 & 83 \\
38 & 55202.6663 & 0.0010 & 0.0068 & 71 \\
39 & 55202.7456 & 0.0007 & 0.0065 & 82 \\
40 & 55202.8253 & 0.0006 & 0.0066 & 76 \\
41 & 55202.9044 & 0.0009 & 0.0061 & 82 \\
42 & 55202.9841 & 0.0011 & 0.0062 & 75 \\
59 & 55204.3438 & 0.0017 & 0.0129 & 88 \\
60 & 55204.4230 & 0.0011 & 0.0125 & 90 \\
75 & 55205.6122 & 0.0006 & 0.0077 & 73 \\
76 & 55205.6924 & 0.0007 & 0.0083 & 83 \\
77 & 55205.7727 & 0.0008 & 0.0090 & 82 \\
78 & 55205.8501 & 0.0007 & 0.0069 & 79 \\
79 & 55205.9295 & 0.0007 & 0.0067 & 82 \\
90 & 55206.8014 & 0.0006 & 0.0031 & 76 \\
91 & 55206.8798 & 0.0008 & 0.0019 & 82 \\
92 & 55206.9602 & 0.0010 & 0.0027 & 67 \\
114 & 55208.7037 & 0.0013 & $-$0.0049 & 82 \\
115 & 55208.7840 & 0.0011 & $-$0.0042 & 75 \\
116 & 55208.8664 & 0.0010 & $-$0.0014 & 83 \\
117 & 55208.9413 & 0.0011 & $-$0.0061 & 75 \\
118 & 55209.0165 & 0.0023 & $-$0.0104 & 82 \\
126 & 55209.6628 & 0.0019 & $-$0.0009 & 83 \\
127 & 55209.7402 & 0.0013 & $-$0.0031 & 76 \\
128 & 55209.8098 & 0.0014 & $-$0.0130 & 76 \\
129 & 55209.8958 & 0.0014 & $-$0.0067 & 82 \\
\hline
  \multicolumn{5}{l}{\commenta BJD$-$2400000.} \\
  \multicolumn{5}{l}{\commentb Against $max = 2455199.6349 + 0.079593 E$.} \\
  \multicolumn{5}{l}{\commentc Number of points used to determine the maximum.} \\
\end{tabular}
\end{center}
\end{table}

\subsection{SDSSp J083845.23$+$491055.5}\label{obj:j0838}

   This object (hereafter SDSS J0838) underwent another superoutburst
in 2010 in addition to the 2007 and 2009 ones discussed in \citet{Pdot}.
The 2010 superoutburst was notable in that it was preceded by a prominent
precursor outburst (vsnet-alert 11910).  Although only two superhump
maxima were measured, we listed them in table \ref{tab:j0838oc2010}.
The interval between maxima is in agreement with the period of
stage C superhumps in 2009.  Since the 2010 observation was undertaken
during the early stage of the superoutburst, the period is expected to be
close to the $P_{\rm SH}$ at the start of stage B.  As discussed in
\citet{Pdot}, this period is expected to be close to $P_{\rm SH}$ for
stage C superhumps.  The present finding is consistent with this
interpretation.  There was a slight indication of modulation
at a period around 0.0709 d and an amplitude of 0.07 mag.
Although the presence of this signal is not conclusive, the periodicity
might suggest a rare evolution of $P_{\rm SH}$ starting from $P_{\rm orb}$,
as recorded in QZ Vir \citet{kat97tleo}.  This needs to be confirmed
by future observations.

\begin{table}
\caption{Superhump maxima of SDSS J0838 (2010).}\label{tab:j0838oc2010}
\begin{center}
\begin{tabular}{cccc}
\hline
$E$ & max\commenta & error & $N$\commentb \\
\hline
0 & 55294.9551 & 0.0008 & 99 \\
1 & 55295.0289 & 0.0004 & 153 \\
\hline
  \multicolumn{4}{l}{\commenta BJD$-$2400000.} \\
  \multicolumn{4}{l}{\commentb Number of points used to determine the maximum.} \\
\end{tabular}
\end{center}
\end{table}

\subsection{SDSS J083931.35$+$282824.0}\label{obj:j0839}

   This object (hereafter SDSS J0839) was selected as a CV during the
course of SDSS \citet{szk05SDSSCV4}.  Although the spectrum was suggestive
of that of a dwarf nova, no outburst had been recorded.
On 2010 April 8, K. Itagaki detected an outbursting object which can be
identified with this CV (vsnet-alert 11911).  Subsequent observations
have confirmed the presence of superhumps (vsnet-alert 11916, 11921,
11926, 11930).

\begin{figure}
  \begin{center}
%    \FigureFile(88mm,110mm){j0839shpdm.eps}
    \FigureFile(88mm,110mm){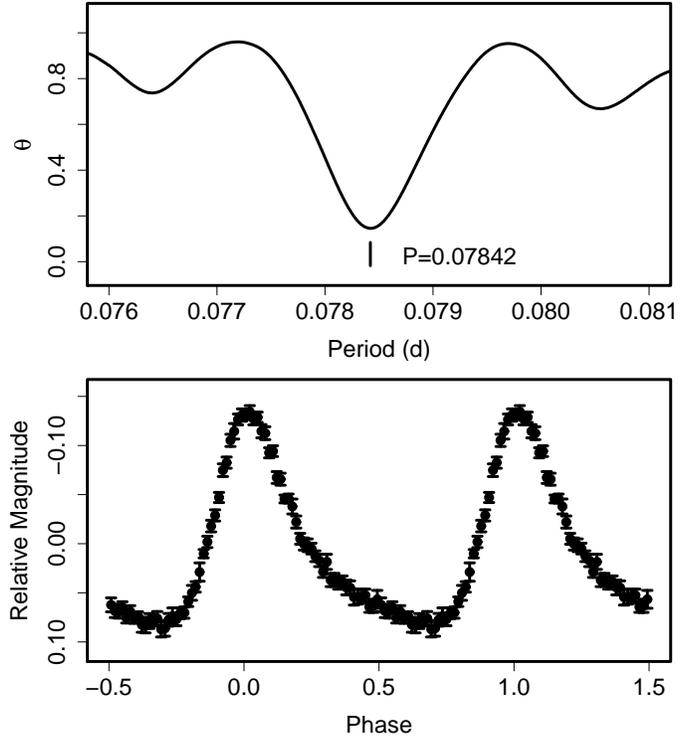}
  \end{center}
  \caption{Superhumps in SDSS J0839 (2010).
     (Upper): PDM analysis.
     (Lower): Phase-averaged profile.}
  \label{fig:j0839shpdm}
\end{figure}

   The times of superhump maxima are listed in table \ref{tab:j0839oc2010}.
There was an apparent break in the $O-C$ diagram around $E=15$, which
is likely a stage B--C transition.  The periods given in table
\ref{tab:perlist} are based on this interpretation.
The mean $P_{\rm SH}$ determined with the PDM method is 0.078423(7) d
(figure \ref{fig:j0839shpdm}).

\begin{table}
\caption{Superhump maxima of SDSS J0839 (2010).}\label{tab:j0839oc2010}
\begin{center}
\begin{tabular}{ccccc}
\hline
$E$ & max\commenta & error & $O-C$\commentb & $N$\commentc \\
\hline
0 & 55295.3531 & 0.0002 & $-$0.0003 & 190 \\
1 & 55295.4309 & 0.0002 & $-$0.0009 & 220 \\
2 & 55295.5099 & 0.0004 & $-$0.0003 & 83 \\
13 & 55296.3732 & 0.0004 & 0.0006 & 79 \\
14 & 55296.4518 & 0.0003 & 0.0007 & 78 \\
15 & 55296.5311 & 0.0007 & 0.0016 & 41 \\
26 & 55297.3919 & 0.0008 & $-$0.0000 & 78 \\
27 & 55297.4694 & 0.0005 & $-$0.0009 & 78 \\
38 & 55298.3319 & 0.0006 & $-$0.0008 & 128 \\
39 & 55298.4115 & 0.0003 & 0.0003 & 246 \\
\hline
  \multicolumn{5}{l}{\commenta BJD$-$2400000.} \\
  \multicolumn{5}{l}{\commentb Against $max = 2455295.3534 + 0.078403 E$.} \\
  \multicolumn{5}{l}{\commentc Number of points used to determine the maximum.} \\
\end{tabular}
\end{center}
\end{table}

\subsection{SDSS J090350.73$+$330036.1}\label{obj:j0903}

\begin{figure}
  \begin{center}
%    \FigureFile(88mm,110mm){j0903shpdm.eps}
    \FigureFile(88mm,110mm){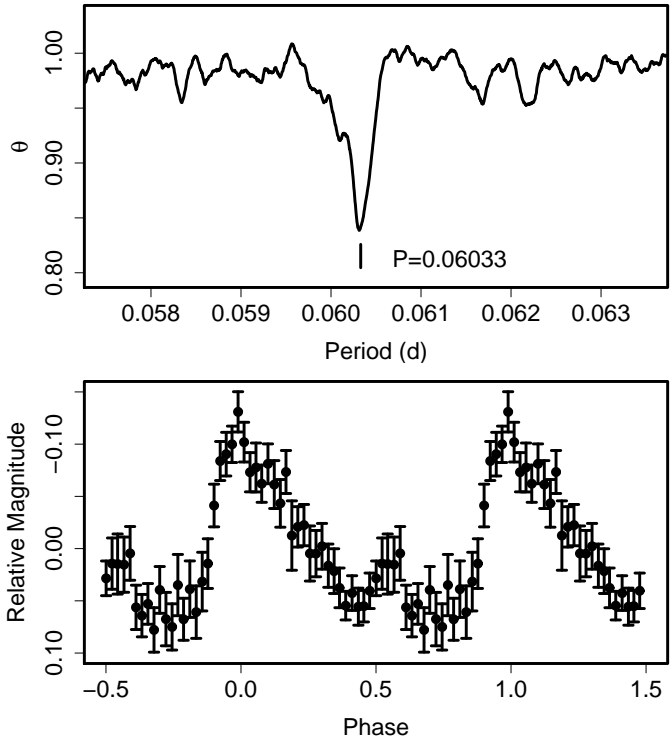}
  \end{center}
  \caption{Superhumps in SDSS J0903 (2010) outside the eclipses.
     (Upper): PDM analysis.
     (Lower): Phase-averaged profile.}
  \label{fig:j0903shpdm}
\end{figure}

   This object (hereafter SDSS J0903) is a CV selected
during the course of the SDSS \citep{szk05SDSSCV4}, who suspected
its eclipsing nature.  \citet{lit08eclCV} confirmed that this object
is a deeply eclipsing CV with a short orbital period of 0.059073543(9) d.
The 2010 outburst, the first-ever outburst reported in real-time,
was detected by CRTS (= CSS100522:090351$+$330036; cf. vsnet-alert 11994).
Subsequent observations have confirmed the eclipsing SU UMa-type nature
of this object (vsnet-alert 12006; figure \ref{fig:j0903shpdm}).

   The times of mid-eclipses were determined with the KW method
(table \ref{tab:j0903ecl}).
We obtained an updated orbital ephemeris (equation \ref{equ:j0903ecl})
using our data and times of eclipses in \citet{lit08eclCV}.

\begin{table}
\caption{Eclipse Minima of SDSS J0903.}\label{tab:j0903ecl}
\begin{center}
\begin{tabular}{ccccc}
\hline
$E$ & Minimum\commenta & error & $O-C$\commentb & Source\commentc \\
\hline
0 & 53800.394700 & 0.000006 & -0.00001 & 1 \\
2 & 53800.512854 & 0.000006 & -0.00000 & 1 \\
34 & 53802.403198 & 0.000006 & -0.00001 & 1 \\
35 & 53802.462279 & 0.000006 & -0.00000 & 1 \\
36 & 53802.521361 & 0.000006 & 0.00001 & 1 \\
37 & 53802.580442 & 0.000006 & 0.00001 & 1 \\
38 & 53802.639501 & 0.000006 & -0.00000 & 1 \\
50 & 53803.348386 & 0.000006 & 0.00000 & 1 \\
51 & 53803.407462 & 0.000006 & 0.00000 & 1 \\
52 & 53803.466529 & 0.000006 & -0.00000 & 1 \\
53 & 53803.525601 & 0.000006 & -0.00000 & 1 \\
26086 & 55341.38690 & 0.00030 & 0.00022 & 2 \\
26087 & 55341.44648 & 0.00036 & 0.00073 & 2 \\
26102 & 55342.33163 & 0.00038 & -0.00022 & 2 \\
26104 & 55342.44915 & 0.00038 & -0.00085 & 2 \\
26119 & 55343.33624 & 0.00039 & 0.00013 & 2 \\
26136 & 55344.34126 & 0.00034 & 0.00091 & 2 \\
26137 & 55344.39964 & 0.00039 & 0.00020 & 2 \\
26152 & 55345.28500 & 0.00030 & -0.00053 & 2 \\
26154 & 55345.40339 & 0.00046 & -0.00029 & 2 \\
26169 & 55346.29034 & 0.00029 & 0.00056 & 2 \\
26170 & 55346.34966 & 0.00032 & 0.00080 & 2 \\
26171 & 55346.40821 & 0.00045 & 0.00028 & 2 \\
26188 & 55347.41127 & 0.00056 & -0.00091 & 2 \\
26193 & 55347.70712 & 0.00027 & -0.00043 & 2 \\
26204 & 55348.35745 & 0.00037 & 0.00009 & 2 \\
26205 & 55348.41652 & 0.00035 & 0.00009 & 2 \\
26209 & 55348.65271 & 0.00030 & -0.00001 & 2 \\
26210 & 55348.71183 & 0.00036 & 0.00003 & 2 \\
26221 & 55349.36176 & 0.00037 & 0.00015 & 2 \\
26222 & 55349.42067 & 0.00040 & -0.00001 & 2 \\
26243 & 55350.66096 & 0.00034 & -0.00026 & 2 \\
26244 & 55350.71996 & 0.00044 & -0.00034 & 2 \\
26255 & 55351.36974 & 0.00028 & -0.00037 & 2 \\
26256 & 55351.42824 & 0.00041 & -0.00095 & 2 \\
26272 & 55352.37433 & 0.00072 & -0.00003 & 2 \\
\hline
  \multicolumn{5}{l}{\commenta BJD$-$2400000.} \\
  \multicolumn{5}{l}{\commentb Against equation \ref{equ:j0903ecl}.} \\
  \multicolumn{5}{l}{\commentc 1: \citet{lit08eclCV}, 2: this work.} \\
\end{tabular}
\end{center}
\end{table}

\begin{equation}
{\rm Min(BJD)} = 2453800.394708(2) + 0.059073525(4) E
\label{equ:j0903ecl}.
\end{equation}

In the following analysis, we removed observations within 0.08
$P_{\rm orb}$ of eclipses.

   The times of superhump maxima are listed in table \ref{tab:j0903oc2010}.
There was a clear stage B--C transition around $E=116$.  The $P_{\rm dot}$
for stage B was $+12.3(3.7) \times 10^{-5}$.  This clearly positive
$P_{\rm dot}$ in deeply eclipsing short-period SU UMa-type dwarf novae
confirmed the finding in XZ Eri (\cite{uem04xzeri}; \cite{Pdot}).
The fractional superhump excesses for stages B and C were
2.1 \% and 1.7 \%, respectively.  Other parameters are listed in
table \ref{tab:perlist}.

\begin{table}
\caption{Superhump maxma of SDSS J0903 (2010).}\label{tab:j0903oc2010}
\begin{center}
\begin{tabular}{ccccc}
\hline
$E$ & max\commenta & error & $O-C$\commentb & $N$\commentc \\
\hline
0 & 55340.4222 & 0.0009 & 0.0072 & 51 \\
31 & 55342.2882 & 0.0009 & 0.0033 & 26 \\
32 & 55342.3424 & 0.0007 & $-$0.0028 & 57 \\
33 & 55342.4003 & 0.0019 & $-$0.0052 & 31 \\
48 & 55343.3142 & 0.0013 & 0.0039 & 19 \\
49 & 55343.3624 & 0.0016 & $-$0.0083 & 55 \\
50 & 55343.4267 & 0.0017 & $-$0.0043 & 54 \\
65 & 55344.3326 & 0.0008 & $-$0.0032 & 57 \\
81 & 55345.2974 & 0.0008 & $-$0.0035 & 57 \\
82 & 55345.3554 & 0.0047 & $-$0.0058 & 48 \\
83 & 55345.4122 & 0.0040 & $-$0.0093 & 51 \\
98 & 55346.3301 & 0.0017 & 0.0038 & 83 \\
99 & 55346.3923 & 0.0017 & 0.0056 & 47 \\
114 & 55347.3027 & 0.0090 & 0.0112 & 74 \\
116 & 55347.4195 & 0.0024 & 0.0075 & 51 \\
121 & 55347.7186 & 0.0012 & 0.0050 & 52 \\
132 & 55348.3800 & 0.0009 & 0.0028 & 54 \\
133 & 55348.4419 & 0.0017 & 0.0044 & 29 \\
137 & 55348.6809 & 0.0007 & 0.0021 & 62 \\
149 & 55349.4016 & 0.0010 & $-$0.0010 & 50 \\
170 & 55350.6692 & 0.0019 & $-$0.0002 & 49 \\
182 & 55351.3798 & 0.0022 & $-$0.0133 & 29 \\
\hline
  \multicolumn{5}{l}{\commenta BJD$-$2400000.} \\
  \multicolumn{5}{l}{\commentb Against $max = 2455340.4150 + 0.060320 E$.} \\
  \multicolumn{5}{l}{\commentc Number of points used to determine the maximum.} \\
\end{tabular}
\end{center}
\end{table}

\subsection{SDSS J115207.00$+$404947.8}\label{obj:j1152}

   This object (hereafter SDSS J1152) is a CV selected
during the course of the SDSS \citep{szk07SDSSCV6}, who suspected
its eclipsing nature.  \citet{sou08CVperiod} established that this
CV is deeply eclipsing and determined its period to be 0.06770(28) d.

\begin{figure}
  \begin{center}
%    \FigureFile(88mm,110mm){j1152shpdm.eps}
    \FigureFile(88mm,110mm){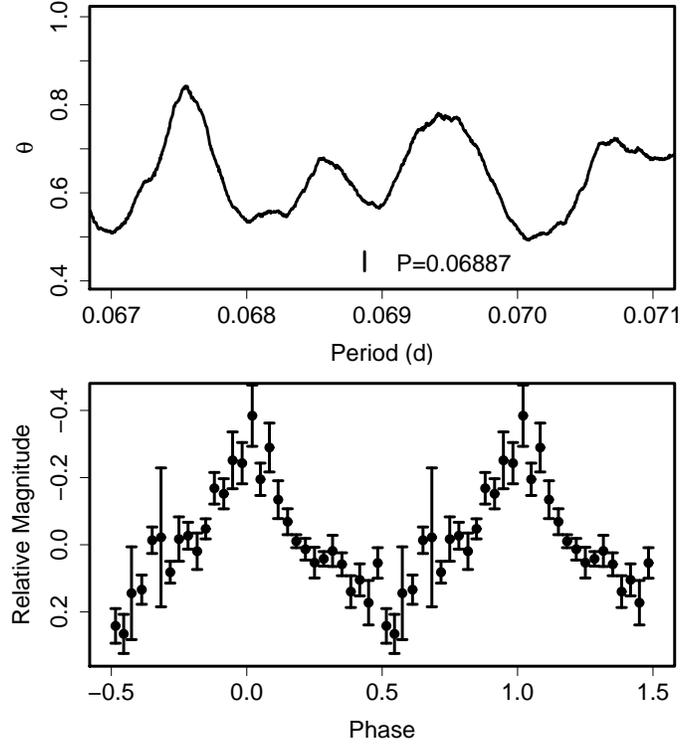}
  \end{center}
  \caption{Superhumps in SDSS J1152 (2009).
     (Upper): PDM analysis.  The alias selection was based on $P_{\rm orb}$.
     The best period was determined by Bayesian modeling.
     (Lower): Phase-averaged profile.}
  \label{fig:j1152shpdm}
\end{figure}

   The 2009 June outburst was detected by E. Muyllaert at a CCD
magnitude of 16.4 on June 9 (cvnet-outburst 3158).  Superhumps and
eclipses were soon detected (vsnet-alert 11288), establishing the
SU UMa-type nature of this object.  The times of superhump maxima
determined observations outside the eclipses
are listed in table \ref{tab:j1152oc2009}.  The intervening
clouds made the epoch of $E$ = 39 rather uncertain.
A period analysis with the PDM method has yielded a $P_{\rm SH}$ of
0.0689(1) d (figure \ref{fig:j1152shpdm}).
Since the signal is broad due to the poor phase coverage, we used
a Bayesian modeling of the light curve using the template superhump
light curve (see Appendix).  The best period by this method is 0.06887(4) d,
corresponding to $\epsilon$ of 1.7 \%.  We adopted this period
in the figure and table.
Although the basic nature of the object is well-established,
both orbital and superhump periods need to be refined by further
observations.

\begin{table}
\caption{Superhump maxima of SDSS J1152 (2009).}\label{tab:j1152oc2009}
\begin{center}
\begin{tabular}{ccccc}
\hline
$E$ & max\commenta & error & $O-C$\commentb & $N$\commentc \\
\hline
0 & 54994.0621 & 0.0005 & 0.0056 & 125 \\
39 & 54996.7376 & 0.0068 & $-$0.0135 & 9 \\
67 & 54998.6919 & 0.0021 & 0.0062 & 18 \\
68 & 54998.7564 & 0.0021 & 0.0016 & 28 \\
\hline
  \multicolumn{5}{l}{\commenta BJD$-$2400000.} \\
  \multicolumn{5}{l}{\commentb Against $max = 2454994.0565 + 0.069092 E$.} \\
  \multicolumn{5}{l}{\commentc Number of points used to determine the maximum.} \\
\end{tabular}
\end{center}
\end{table}

\subsection{SDSS J125023.85$+$665525.5}\label{obj:j1250}

   This object (hereafter J1250) is a CV selected
during the course of the SDSS \citep{szk03SDSSCV2}, who suggested
a high inclination CV.  \citet{dil08SDSSCV} confirmed that this is
indeed a deeply eclipsing CV with a very short orbital period of
0.058735687(4) d.

   Shortly before \citet{dil08SDSSCV} becomes available, the object was
found to be in outburst in 2008 January (S. Brady, cvnet-discussion 1104),
who successfully detected eclipses and reported a period of 0.059 d.
Based on these and 2009 observations, we have determined eclipse times
and updated the ephemeris.

   Both the 2008 and 2009 outbursts were superoutbursts, and the times
of superhump maxima are listed in tables \ref{tab:j1250oc2008}
and \ref{tab:j1250oc2009} determined from observations outside the eclipses.

\begin{figure}
  \begin{center}
%    \FigureFile(88mm,110mm){j1250shpdm.eps}
    \FigureFile(88mm,110mm){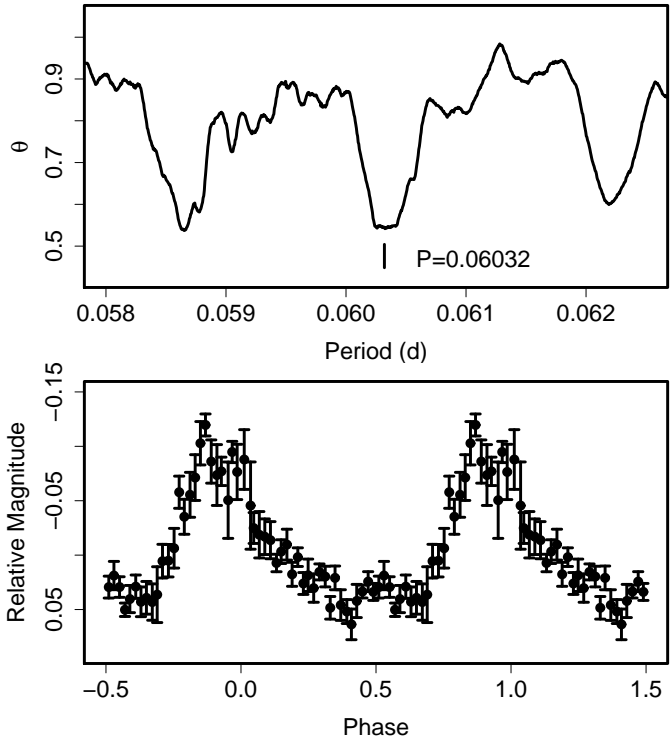}
  \end{center}
  \caption{Superhumps in SDSS J1250 (2008).
     (Upper): PDM analysis.  The selection of the alias is based on
     the orbital period.
     (Lower): Phase-averaged profile.}
  \label{fig:j1250shpdm}
\end{figure}

   A PDM analysis of the 2008 observations has yielded a mean $P_{\rm SH}$
of 0.06032(5) d (figure \ref{fig:j1250shpdm}).
The $P_{\rm dot}$ for $E \le 83$, apparently stage B,
was $+9.4(2.1) \times 10^{-5}$.  The period apparently decreased after
this, suggesting a transition to stage C.
The $\epsilon$ of 2.7 \% for the mean $P_{\rm SH}$ is relatively large
for this $P_{\rm orb}$.  Using the $\epsilon$--$q$ relation in \citet{Pdot},
this $\epsilon$ corresponds to $q$ = 0.14.  Assuming a moderate white-dwarf
mass, this $q$ would place the object around the upper (massive secondary)
boundary defined by SDSS J0903 and SDSS J1507 in figure 2 of
\citet{lit08eclCV} rather than around the period bounce.
This identification appears to be favored by the relatively large
$P_{\rm dot}$ and rather frequent outbursts (the interval of two
superoutburst was $\sim$ 650 d).
Further determination of system parameters of this object will improve
our knowledge of CV evolution near the period minimum.

\begin{table}
\caption{Superhump maxima of SDSS J1250 (2008).}\label{tab:j1250oc2008}
\begin{center}
\begin{tabular}{ccccc}
\hline
$E$ & max\commenta & error & $O-C$\commentb & $N$\commentc \\
\hline
0 & 54491.9263 & 0.0010 & 0.0039 & 6 \\
44 & 54494.5762 & 0.0018 & $-$0.0002 & 18 \\
45 & 54494.6331 & 0.0012 & $-$0.0037 & 41 \\
46 & 54494.6926 & 0.0014 & $-$0.0044 & 39 \\
47 & 54494.7553 & 0.0008 & $-$0.0020 & 42 \\
48 & 54494.8165 & 0.0007 & $-$0.0011 & 41 \\
77 & 54496.5690 & 0.0022 & 0.0021 & 17 \\
78 & 54496.6287 & 0.0011 & 0.0015 & 24 \\
79 & 54496.6906 & 0.0015 & 0.0031 & 24 \\
80 & 54496.7482 & 0.0011 & 0.0003 & 22 \\
81 & 54496.8082 & 0.0009 & $-$0.0000 & 24 \\
82 & 54496.8696 & 0.0013 & 0.0011 & 24 \\
83 & 54496.9293 & 0.0010 & 0.0005 & 20 \\
146 & 54500.7289 & 0.0033 & 0.0000 & 22 \\
147 & 54500.7882 & 0.0029 & $-$0.0010 & 19 \\
\hline
  \multicolumn{5}{l}{\commenta BJD$-$2400000.} \\
  \multicolumn{5}{l}{\commentb Against $max = 2454491.9224 + 0.060319 E$.} \\
  \multicolumn{5}{l}{\commentc Number of points used to determine the maximum.} \\
\end{tabular}
\end{center}
\end{table}

\begin{table}
\caption{Superhump maxima of SDSS J1250 (2009).}\label{tab:j1250oc2009}
\begin{center}
\begin{tabular}{ccccc}
\hline
$E$ & max\commenta & error & $N$\commentb \\
\hline
0 & 55144.2715 & 0.0015 & 77 \\
1 & 55144.3338 & 0.0004 & 54 \\
\hline
  \multicolumn{5}{l}{\commenta BJD$-$2400000.} \\
  \multicolumn{5}{l}{\commentb Number of points used to determine the maximum.} \\
\end{tabular}
\end{center}
\end{table}

\subsection{SDSS J150240.98$+$333423.9}\label{obj:j1502}

   This object (hereafter SDSS J1502) is an eclipsing CV selected
during the course of the SDSS \citep{szk06SDSSCV5}.
\citet{lit08eclCV} performed high-speed photometry in quiescence and
obtained orbital parameters [$P_{\rm dot}$ = 0.05890961(5) d and
estimated $q$ = 0.109(3)].  \citet{lit08eclCV} attributed this CV
to a CV before the ``period bounce''.

\begin{figure}
  \begin{center}
%    \FigureFile(88mm,110mm){j1502shpdm.eps}
    \FigureFile(88mm,110mm){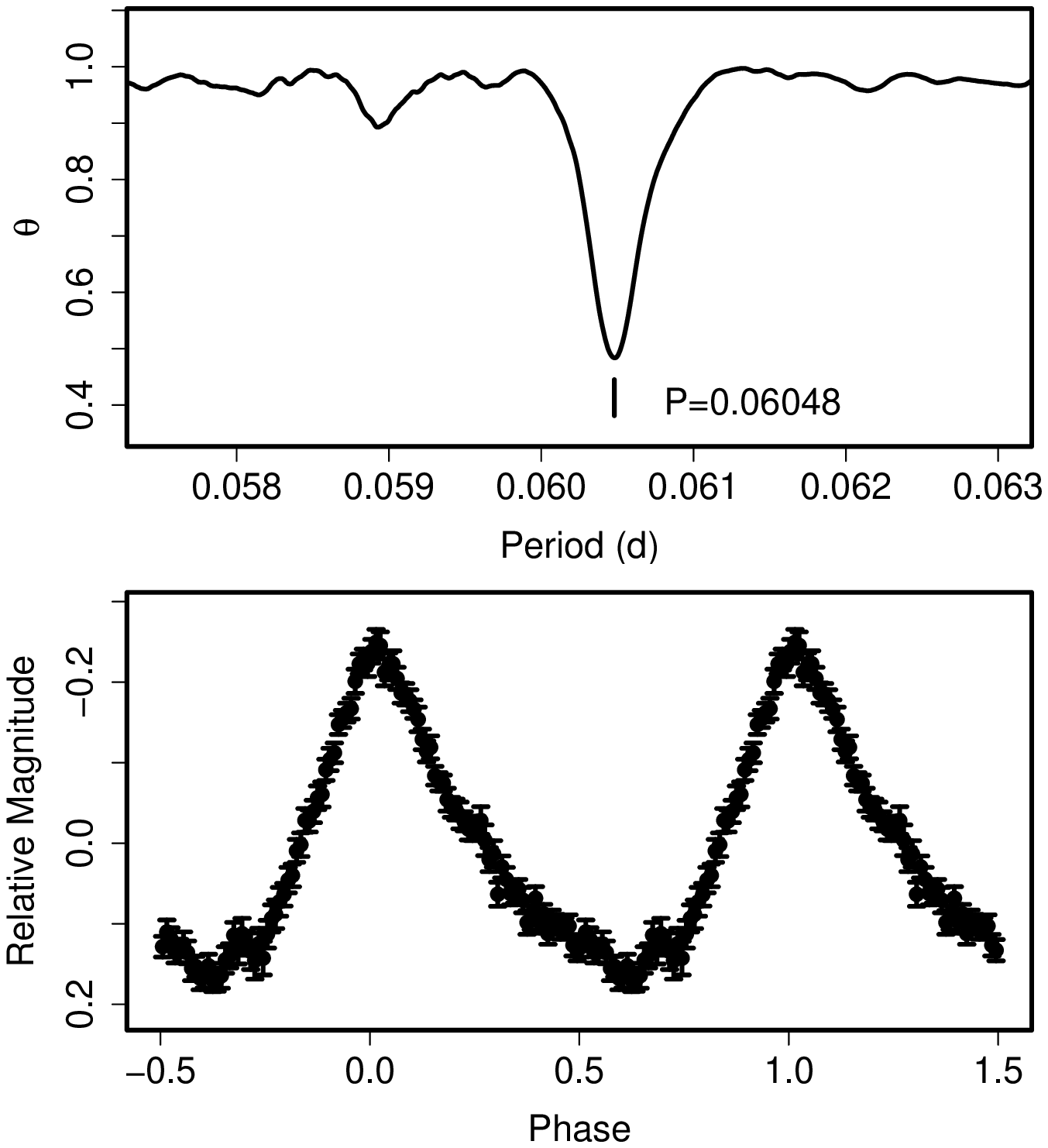}
  \end{center}
  \caption{Superhumps in SDSS J1502 (2009).
     (Upper): PDM analysis.
     (Lower): Phase-averaged profile.}
  \label{fig:j1502shpdm}
\end{figure}

   The 2009 July superoutburst of this object was detected by
J. Shears (baavss-alert 1997).  Subsequent observations detected
superhumps (cf. vsnet-alert 11332; figure \ref{fig:j1502shpdm}).
\citet{she10j1502} also reported an analysis of the same superoutburst
using the slightly different data set from ours.

   The times of mid-eclipses were determined with the KW method
(table \ref{tab:j1502ecl}).
We obtained an updated orbital ephemeris (equation \ref{equ:j1502ecl})
using our data and times of eclipses in \citet{lit08eclCV} and
\citet{she10j1502}.  We only used times of eclipses in \citet{she10j1502}
which were not covered by our observations.  We also disregarded $E=855$
eclipse due to its large $O-C$.

\begin{table}
\caption{Eclipse Minima of SDSS J1502.}\label{tab:j1502ecl}
\begin{center}
\begin{tabular}{ccccc}
\hline
$E$ & Minimum\commenta & error & $O-C$\commentb & Source\commentc \\
\hline
0 & 53799.640618 & 0.000004 & 0.00013 & 1 \\
2 & 53799.758414 & 0.000007 & 0.00011 & 1 \\
17 & 53800.642070 & 0.000006 & 0.00012 & 1 \\
18 & 53800.700966 & 0.000006 & 0.00011 & 1 \\
19 & 53800.759901 & 0.000006 & 0.00014 & 1 \\
52 & 53802.703911 & 0.000002 & 0.00013 & 1 \\
68 & 53803.646461 & 0.000003 & 0.00013 & 1 \\
69 & 53803.705371 & 0.000006 & 0.00013 & 1 \\
70 & 53803.764277 & 0.000003 & 0.00013 & 1 \\
854 & 53849.94908 & 0.00035 & -0.00009 & 2 \\
866 & 53850.65615 & 0.00012 & 0.00006 & 2 \\
867 & 53850.71498 & 0.00011 & -0.00002 & 2 \\
868 & 53850.77384 & 0.00021 & -0.00007 & 2 \\
869 & 53850.83284 & 0.00012 & 0.00002 & 2 \\
870 & 53850.89165 & 0.00014 & -0.00008 & 2 \\
871 & 53850.95066 & 0.00013 & 0.00003 & 2 \\
883 & 53851.65760 & 0.00014 & 0.00005 & 2 \\
884 & 53851.71630 & 0.00014 & -0.00016 & 2 \\
886 & 53851.83418 & 0.00016 & -0.00010 & 2 \\
887 & 53851.89334 & 0.00014 & 0.00015 & 2 \\
888 & 53851.95250 & 0.00014 & 0.00040 & 2 \\
900 & 53852.65899 & 0.00018 & -0.00002 & 2 \\
901 & 53852.71800 & 0.00015 & 0.00008 & 2 \\
902 & 53852.77661 & 0.00011 & -0.00022 & 2 \\
903 & 53852.83568 & 0.00014 & -0.00006 & 2 \\
1308 & 53876.69411 & 0.00004 & 0.00004 & 2 \\
6454 & 54179.84201 & 0.00009 & -0.00021 & 2 \\
6455 & 54179.90009 & 0.00009 & -0.00104 & 2 \\
6457 & 54180.01899 & 0.00011 & 0.00004 & 2 \\
13787 & 54611.82514 & 0.00012 & -0.00025 & 2 \\
13788 & 54611.88411 & 0.00012 & -0.00019 & 2 \\
13789 & 54611.94325 & 0.00017 & 0.00004 & 2 \\
13802 & 54612.70876 & 0.00019 & -0.00027 & 2 \\
13803 & 54612.76795 & 0.00012 & 0.00001 & 2 \\
13804 & 54612.82684 & 0.00018 & -0.00001 & 2 \\
13805 & 54612.88550 & 0.00012 & -0.00026 & 2 \\
13806 & 54612.94456 & 0.00027 & -0.00011 & 2 \\
20758 & 55022.48413 & 0.00005 & 0.00081 & 2 \\
20760 & 55022.60121 & 0.00032 & 0.00007 & 3 \\
20761 & 55022.66052 & 0.00036 & 0.00047 & 3 \\
20762 & 55022.71932 & 0.00026 & 0.00036 & 3 \\
20763 & 55022.77857 & 0.00031 & 0.00070 & 2 \\
20777 & 55023.60263 & 0.00029 & 0.00003 & 3 \\
20778 & 55023.66133 & 0.00028 & -0.00018 & 2 \\
20779 & 55023.72066 & 0.00021 & 0.00024 & 3 \\
20780 & 55023.77951 & 0.00018 & 0.00018 & 3 \\
20781 & 55023.83814 & 0.00024 & -0.00010 & 3 \\
20782 & 55023.89720 & 0.00040 & 0.00005 & 3 \\
20794 & 55024.60306 & 0.00050 & -0.00101 & 2 \\
20796 & 55024.72187 & 0.00026 & -0.00001 & 3 \\
20797 & 55024.78077 & 0.00022 & -0.00002 & 3 \\
20798 & 55024.83956 & 0.00028 & -0.00014 & 3 \\
20807 & 55025.37023 & 0.00027 & 0.00034 & 3 \\
\hline
  \multicolumn{5}{l}{\commenta BJD$-$2400000.} \\
  \multicolumn{5}{l}{\commentb Against equation \ref{equ:j1502ecl}.} \\
  \multicolumn{5}{l}{\commentc 1: \citet{lit08eclCV}, 2: \citet{she10j1502},} \\
  \multicolumn{5}{l}{3: this work.} \\
\end{tabular}
\end{center}
\end{table}

\addtocounter{table}{-1}
\begin{table}
\caption{Eclipse Minima of SDSS J1502 (continued).}
\begin{center}
\begin{tabular}{ccccc}
\hline
$E$ & Minimum\commenta & error & $O-C$\commentb & Source\commentc \\
\hline
20808 & 55025.42910 & 0.00030 & 0.00030 & 3 \\
20811 & 55025.60573 & 0.00042 & 0.00020 & 3 \\
20812 & 55025.66472 & 0.00014 & 0.00028 & 3 \\
20813 & 55025.72362 & 0.00016 & 0.00027 & 3 \\
20814 & 55025.78256 & 0.00017 & 0.00030 & 3 \\
20815 & 55025.84150 & 0.00023 & 0.00034 & 3 \\
20824 & 55026.37130 & 0.00026 & -0.00005 & 3 \\
20825 & 55026.43013 & 0.00033 & -0.00013 & 3 \\
20828 & 55026.60726 & 0.00011 & 0.00027 & 2 \\
20829 & 55026.66588 & 0.00029 & -0.00002 & 3 \\
20830 & 55026.72440 & 0.00022 & -0.00041 & 3 \\
20831 & 55026.78290 & 0.00033 & -0.00082 & 3 \\
20832 & 55026.84210 & 0.00100 & -0.00053 & 3 \\
20835 & 55027.01943 & 0.00032 & 0.00008 & 3 \\
20836 & 55027.07836 & 0.00028 & 0.00010 & 3 \\
20837 & 55027.13703 & 0.00053 & -0.00014 & 3 \\
20841 & 55027.37299 & 0.00026 & 0.00018 & 3 \\
20845 & 55027.60852 & 0.00038 & 0.00007 & 3 \\
20846 & 55027.66772 & 0.00040 & 0.00036 & 3 \\
20847 & 55027.72659 & 0.00021 & 0.00032 & 3 \\
20848 & 55027.78561 & 0.00022 & 0.00043 & 2 \\
20849 & 55027.84470 & 0.00051 & 0.00061 & 3 \\
20858 & 55028.37421 & 0.00027 & -0.00006 & 3 \\
20859 & 55028.43318 & 0.00019 & -0.00000 & 3 \\
20860 & 55028.49206 & 0.00023 & -0.00003 & 3 \\
20861 & 55028.55083 & 0.00026 & -0.00017 & 3 \\
20864 & 55028.72779 & 0.00037 & 0.00006 & 3 \\
20881 & 55029.72935 & 0.00035 & 0.00016 & 3 \\
20882 & 55029.78814 & 0.00036 & 0.00004 & 3 \\
20883 & 55029.84720 & 0.00033 & 0.00019 & 3 \\
20894 & 55030.49541 & 0.00105 & 0.00040 & 2 \\
20895 & 55030.55404 & 0.00024 & 0.00012 & 2 \\
20898 & 55030.73046 & 0.00014 & -0.00019 & 2 \\
20898 & 55030.73050 & 0.00033 & -0.00015 & 3 \\
20899 & 55030.78951 & 0.00034 & -0.00005 & 3 \\
20900 & 55030.84853 & 0.00035 & 0.00006 & 3 \\
20915 & 55031.73159 & 0.00076 & -0.00052 & 3 \\
20916 & 55031.79037 & 0.00043 & -0.00065 & 2 \\
20917 & 55031.84916 & 0.00138 & -0.00077 & 3 \\
20928 & 55032.49776 & 0.00068 & -0.00017 & 3 \\
20929 & 55032.55681 & 0.00079 & -0.00003 & 3 \\
20932 & 55032.73351 & 0.00064 & -0.00006 & 3 \\
20934 & 55032.85137 & 0.00065 & -0.00002 & 3 \\
20949 & 55033.73470 & 0.00075 & -0.00033 & 3 \\
20950 & 55033.79383 & 0.00077 & -0.00011 & 3 \\
20966 & 55034.73608 & 0.00118 & -0.00041 & 2 \\
20968 & 55034.85449 & 0.00133 & 0.00018 & 3 \\
21031 & 55038.56538 & 0.00065 & -0.00023 & 3 \\
21758 & 55081.39289 & 0.00168 & 0.00009 & 2 \\
\hline
  \multicolumn{5}{l}{\commenta BJD$-$2400000.} \\
  \multicolumn{5}{l}{\commentb Against equation \ref{equ:j1502ecl}.} \\
  \multicolumn{5}{l}{\commentc 1: \citet{lit08eclCV}, 2: \citet{she10j1502},} \\
  \multicolumn{5}{l}{3: this work.} \\
\end{tabular}
\end{center}
\end{table}

\begin{equation}
{\rm Min(BJD)} = 2453799.64048(6) + 0.058909473(3) E
\label{equ:j1502ecl}.
\end{equation}

In the following analysis, we removed observations within 0.08
$P_{\rm orb}$ of eclipses.

The times of superhump maxima are listed in table \ref{tab:j1502oc2009}.
There was a likely stage B--C transition around $E = 100$.
The values given in table \ref{tab:perlist} were estimated following
this interpretation.  The $P_{\rm dot}$ for stage B was
$+3.7(1.5) \times 10^{-5}$, which may have been underestimated because
the earliest part of the outburst was not sufficiently observed.
Although \citet{she10j1502} detected a pattern of $O-C$ variation
basically to ours, we used our values based on larger set of data.\footnote{
   Although their $dP/dt$ was reported to be $+2.8(1.0) \times 10^{-4}$,
our analysis of their timing data for $E \le 89$ yielded a
$P_{\rm dot}$ of $+12.1(1.3) \times 10^{-5}$ by our definition.
}
The maxima for $E \ge 152$ may not be superhumps,
probably strongly affected by orbital modulations, and are excluded
from this period analysis.

\begin{table}
\caption{Superhump maxima of SDSS J1502 (2009).}\label{tab:j1502oc2009}
\begin{center}
\begin{tabular}{ccccc}
\hline
$E$ & max\commenta & error & $O-C$\commentb & $N$\commentc \\
\hline
0 & 55022.6300 & 0.0004 & 0.0039 & 92 \\
1 & 55022.6895 & 0.0003 & 0.0029 & 78 \\
16 & 55023.5921 & 0.0004 & $-$0.0009 & 69 \\
17 & 55023.6541 & 0.0008 & 0.0007 & 9 \\
18 & 55023.7122 & 0.0004 & $-$0.0016 & 145 \\
19 & 55023.7736 & 0.0003 & $-$0.0007 & 179 \\
20 & 55023.8343 & 0.0004 & $-$0.0004 & 147 \\
21 & 55023.8946 & 0.0012 & $-$0.0005 & 74 \\
34 & 55024.6790 & 0.0005 & $-$0.0017 & 73 \\
35 & 55024.7401 & 0.0002 & $-$0.0011 & 95 \\
36 & 55024.8015 & 0.0003 & $-$0.0002 & 95 \\
46 & 55025.4060 & 0.0003 & 0.0000 & 91 \\
47 & 55025.4652 & 0.0003 & $-$0.0012 & 91 \\
49 & 55025.5904 & 0.0004 & 0.0031 & 29 \\
50 & 55025.6449 & 0.0002 & $-$0.0028 & 154 \\
51 & 55025.7061 & 0.0001 & $-$0.0020 & 279 \\
52 & 55025.7648 & 0.0001 & $-$0.0038 & 230 \\
53 & 55025.8260 & 0.0003 & $-$0.0029 & 187 \\
61 & 55026.3178 & 0.0011 & 0.0054 & 70 \\
62 & 55026.3694 & 0.0005 & $-$0.0035 & 110 \\
63 & 55026.4387 & 0.0005 & 0.0054 & 108 \\
66 & 55026.6156 & 0.0014 & 0.0010 & 24 \\
67 & 55026.6726 & 0.0004 & $-$0.0024 & 127 \\
68 & 55026.7367 & 0.0005 & 0.0012 & 115 \\
69 & 55026.7966 & 0.0006 & 0.0007 & 111 \\
73 & 55027.0410 & 0.0006 & 0.0033 & 91 \\
74 & 55027.1001 & 0.0003 & 0.0020 & 109 \\
79 & 55027.4019 & 0.0003 & 0.0017 & 94 \\
82 & 55027.5840 & 0.0005 & 0.0025 & 52 \\
83 & 55027.6446 & 0.0005 & 0.0027 & 71 \\
84 & 55027.7070 & 0.0005 & 0.0046 & 75 \\
86 & 55027.8266 & 0.0003 & 0.0033 & 75 \\
89 & 55028.0090 & 0.0006 & 0.0045 & 54 \\
90 & 55028.0678 & 0.0003 & 0.0028 & 131 \\
91 & 55028.1284 & 0.0007 & 0.0030 & 85 \\
95 & 55028.3691 & 0.0003 & 0.0020 & 105 \\
96 & 55028.4293 & 0.0003 & 0.0017 & 135 \\
97 & 55028.4904 & 0.0004 & 0.0024 & 157 \\
98 & 55028.5513 & 0.0003 & 0.0029 & 101 \\
101 & 55028.7293 & 0.0006 & $-$0.0005 & 59 \\
118 & 55029.7516 & 0.0007 & $-$0.0055 & 88 \\
119 & 55029.8127 & 0.0006 & $-$0.0048 & 92 \\
120 & 55029.8733 & 0.0006 & $-$0.0046 & 81 \\
134 & 55030.7147 & 0.0005 & $-$0.0093 & 76 \\
135 & 55030.7763 & 0.0005 & $-$0.0081 & 87 \\
136 & 55030.8392 & 0.0007 & $-$0.0057 & 88 \\
152 & 55031.7872 & 0.0023 & $-$0.0246 & 34 \\
153 & 55031.8347 & 0.0023 & $-$0.0375 & 34 \\
163 & 55032.4860 & 0.0016 & 0.0095 & 47 \\
164 & 55032.5491 & 0.0017 & 0.0121 & 46 \\
167 & 55032.7254 & 0.0010 & 0.0072 & 48 \\
168 & 55032.7880 & 0.0008 & 0.0093 & 46 \\
169 & 55032.8425 & 0.0011 & 0.0034 & 49 \\
183 & 55033.6874 & 0.0037 & 0.0023 & 31 \\
184 & 55033.7398 & 0.0070 & $-$0.0058 & 48 \\
185 & 55033.7996 & 0.0020 & $-$0.0065 & 49 \\
186 & 55033.8721 & 0.0038 & 0.0056 & 27 \\
200 & 55034.7195 & 0.0013 & 0.0070 & 47 \\
201 & 55034.7818 & 0.0016 & 0.0088 & 49 \\
202 & 55034.8427 & 0.0026 & 0.0093 & 49 \\
\hline
  \multicolumn{5}{l}{\commenta BJD$-$2400000.} \\
  \multicolumn{5}{l}{\commentb Against $max = 2455022.6261 + 0.060432 E$.} \\
  \multicolumn{5}{l}{\commentc Number of points used to determine the maximum.} \\
\end{tabular}
\end{center}
\end{table}

\subsection{SDSS J161027.61$+$090738.4}\label{obj:j1610}

\begin{figure}
  \begin{center}
%    \FigureFile(88mm,110mm){j1610shpdm.eps}
    \FigureFile(88mm,110mm){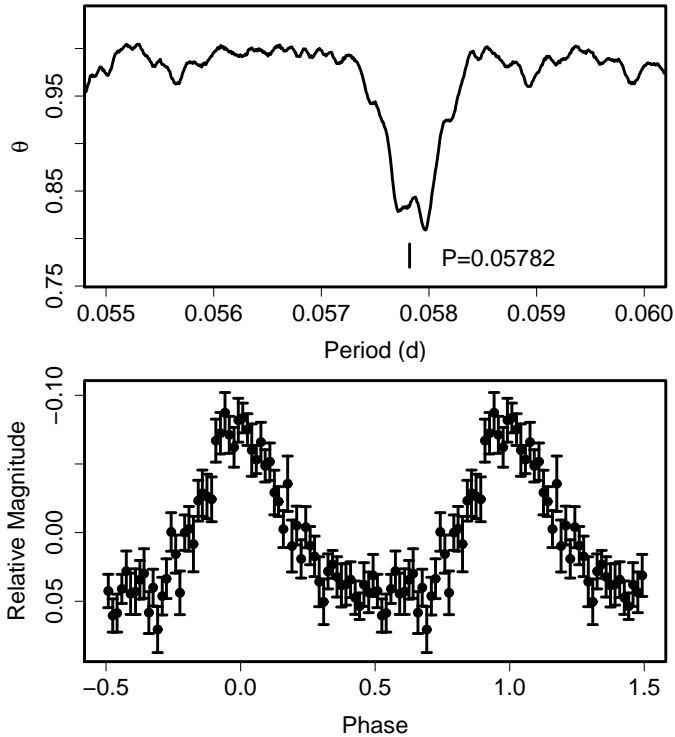}
  \end{center}
  \caption{Superhumps in SDSS J1610 (2009).
     (Upper): PDM analysis.
     (Lower): Phase-averaged profile.}
  \label{fig:j1610shpdm}
\end{figure}

   This object (hereafter SDSS J1610) was initially selected as
a dwarf nova by \citet{wil10newCVs}.  The 2009 July outburst was detected
by the CRTS (= CSS090727:161028+090739).
Double-peaked strong emission lines, characteristic to a relatively high
inclination dwarf nova, in SDSS spectrum was announced (vsnet-alert 11350).
A remarkable growth of superhumps was observed four days after the
outburst detection (vsnet-alert 11366), suggesting a substantial
delay in the growth of superhumps.  Further observations clarified
the potential WZ Sge-type characters (vsnet-alert 11367, 11368, 11381;
figure \ref{fig:j1610shpdm}.

   The times of ordinary superhumps are listed in table
\ref{tab:j1610oc2009}.  There was a clear stage A--B transition
around $E=33$.  The $P_{\rm dot}$ for the well-observed segment of
stage B was $+6.4(1.2) \times 10^{-5}$ ($33 \le E \le 137$).
Although an extrapolation to $E=269$ has yielded a $P_{\rm dot}$
of $+3.2(0.3) \times 10^{-5}$, the identification of superhumps
was slightly ambiguous due to the faintness.  We thus adopted the
former value for the $P_{\rm dot}$ of this object.

\begin{figure}
  \begin{center}
%    \FigureFile(88mm,110mm){j1610eshpdm.eps}
    \FigureFile(88mm,110mm){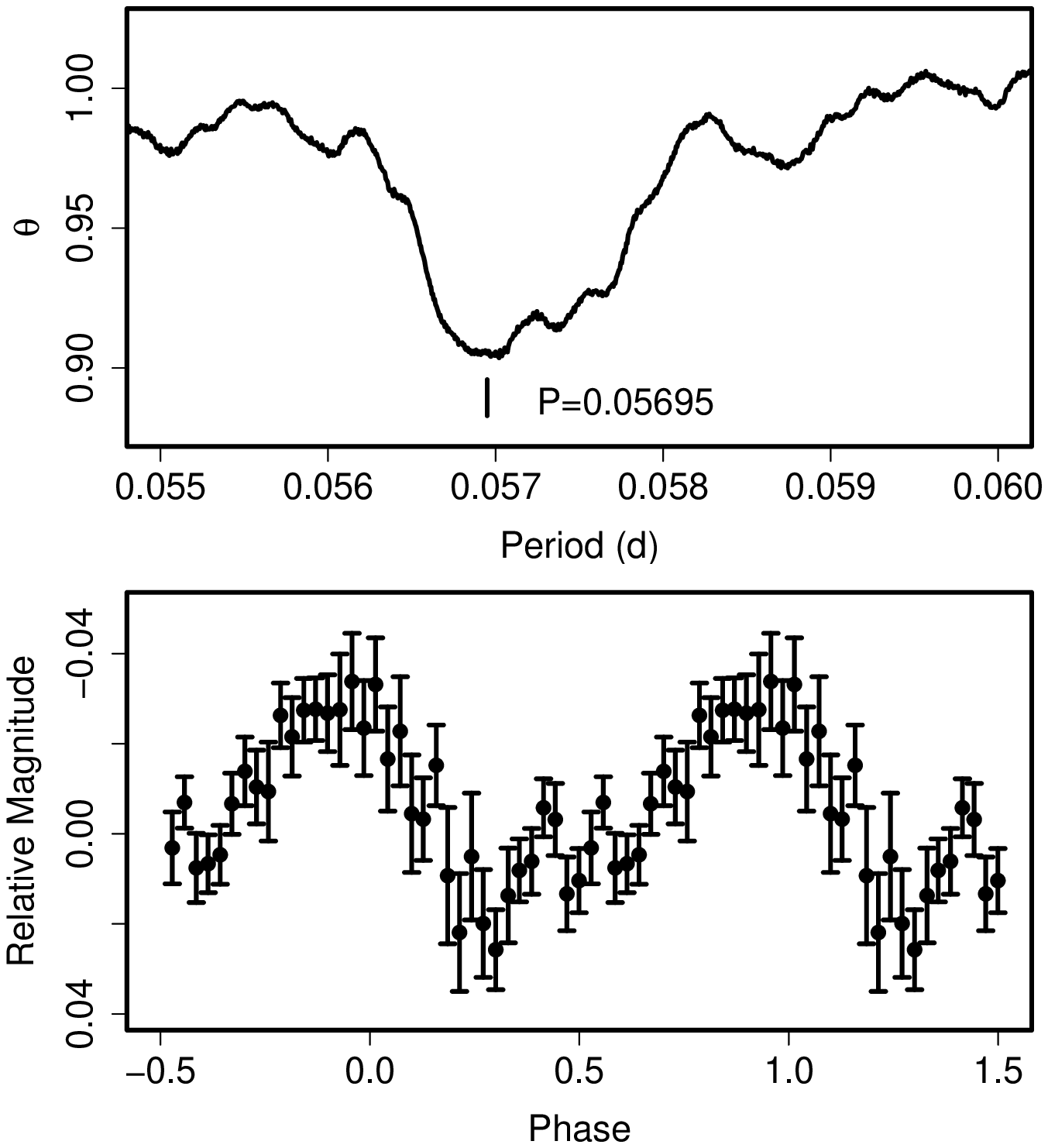}
  \end{center}
  \caption{Early superhumps in SDSS J1610 (2009).
     (Upper): PDM analysis.
     (Lower): Phase-averaged profile.}
  \label{fig:j1610eshpdm}
\end{figure}

   An analysis of the light curve before the growth of ordinary
superhumps detected early superhumps (figure \ref{fig:j1610eshpdm})
with a period of 0.05687(1) d (PDM and Bayesian methods),
confirming both the
WZ Sge-type nature and a relatively high inclination.
The resultant $\epsilon$ of 1.6 \% is relatively large among
WZ Sge-type dwarf novae (cf. \cite{Pdot}), consistent with
a relatively large $P_{\rm dot}$.

\begin{table}
\caption{Superhump maxima of SDSS J1610 (2009).}\label{tab:j1610oc2009}
\begin{center}
\begin{tabular}{ccccc}
\hline
$E$ & max\commenta & error & $O-C$\commentb & $N$\commentc \\
\hline
0 & 55042.4471 & 0.0006 & $-$0.0220 & 120 \\
10 & 55043.0435 & 0.0016 & $-$0.0053 & 99 \\
11 & 55043.1014 & 0.0011 & $-$0.0053 & 184 \\
16 & 55043.3959 & 0.0012 & $-$0.0007 & 29 \\
17 & 55043.4515 & 0.0016 & $-$0.0031 & 28 \\
26 & 55043.9811 & 0.0003 & 0.0048 & 61 \\
27 & 55044.0393 & 0.0002 & 0.0049 & 60 \\
28 & 55044.0986 & 0.0004 & 0.0063 & 60 \\
33 & 55044.3907 & 0.0003 & 0.0085 & 23 \\
34 & 55044.4468 & 0.0006 & 0.0066 & 26 \\
35 & 55044.5061 & 0.0013 & 0.0080 & 17 \\
43 & 55044.9678 & 0.0004 & 0.0059 & 58 \\
44 & 55045.0245 & 0.0006 & 0.0045 & 56 \\
45 & 55045.0839 & 0.0004 & 0.0060 & 46 \\
51 & 55045.4319 & 0.0005 & 0.0061 & 114 \\
52 & 55045.4864 & 0.0005 & 0.0026 & 76 \\
61 & 55046.0067 & 0.0006 & 0.0012 & 49 \\
62 & 55046.0641 & 0.0011 & 0.0006 & 23 \\
68 & 55046.4086 & 0.0009 & $-$0.0028 & 29 \\
68 & 55046.4086 & 0.0010 & $-$0.0027 & 30 \\
78 & 55046.9894 & 0.0012 & $-$0.0017 & 181 \\
79 & 55047.0467 & 0.0012 & $-$0.0023 & 167 \\
80 & 55047.1042 & 0.0008 & $-$0.0029 & 60 \\
85 & 55047.3919 & 0.0008 & $-$0.0050 & 31 \\
86 & 55047.4519 & 0.0009 & $-$0.0029 & 27 \\
102 & 55048.3792 & 0.0009 & $-$0.0033 & 22 \\
103 & 55048.4370 & 0.0010 & $-$0.0034 & 28 \\
137 & 55050.4060 & 0.0028 & $-$0.0055 & 28 \\
268 & 55058.0072 & 0.0025 & 0.0009 & 121 \\
269 & 55058.0659 & 0.0035 & 0.0017 & 123 \\
\hline
  \multicolumn{5}{l}{\commenta BJD$-$2400000.} \\
  \multicolumn{5}{l}{\commentb Against $max = 2455042.4690 + 0.057975 E$.} \\
  \multicolumn{5}{l}{\commentc Number of points used to determine the maximum.} \\
\end{tabular}
\end{center}
\end{table}

\subsection{SDSS J162520.29$+$120308.7}\label{obj:j1625}

   This object was selected as a CV candidate by \citet{wil10newCVs}.
Although the object was not recognized as a CV during the course of
the SDSS, we employed the SDSS designation in this paper
(hereafter SDSS J1625).

\begin{figure}
  \begin{center}
%    \FigureFile(88mm,110mm){j1625shpdm.eps}
    \FigureFile(88mm,110mm){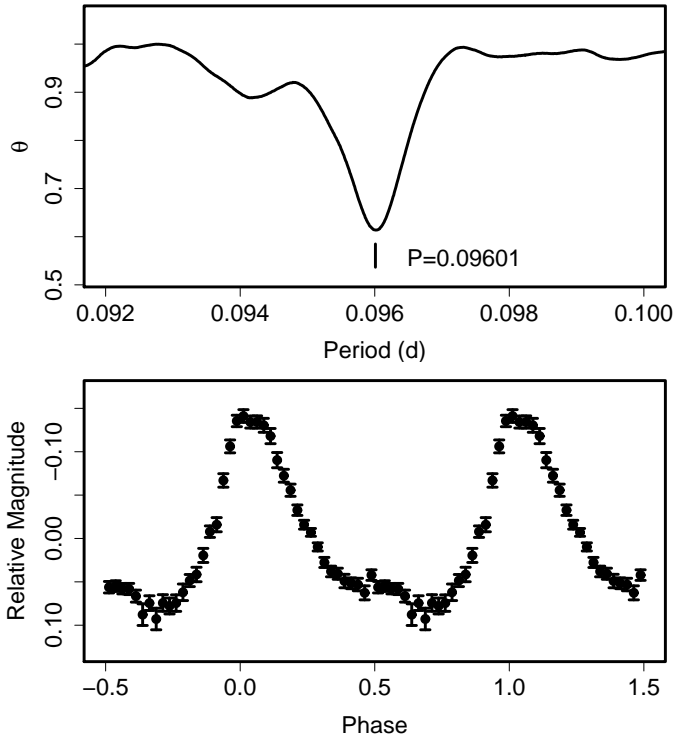}
  \end{center}
  \caption{Superhumps in SDSS J1625 (2010) during stage B.
     (Upper): PDM analysis.
     (Lower): Phase-averaged profile.}
  \label{fig:j1625shpdm}
\end{figure}

   The 2010 outburst of this object was detected by the CRTS
(= CSS100705:162520$+$120309), a discussion on the SDSS spectrum
and the earlier discovery of the object by \citet{wil10newCVs} can be found
in vsnet-alert 12052, 12053.  Short-term variations were detected
soon after this outburst detection (vsnet-alert 12054, 12059), which later
turned out to be developing superhumps (vsnet-alert 12061, 12062).
Fully developed superhumps and the course of period evolution were
subsequently observed (vsnet-alert 12064, 12065, 12066, 12068, 12071;
figure \ref{fig:j1625shpdm}).
The object entered the rapid decline phase 4--5 d after the development
of superhumps, which was unexpectedly early (vsnet-alert 12079).
The object was also unusual in its rebrightening phenomenon soon after
the rapid decline (vsnet-alert 12087).

\begin{figure}
  \begin{center}
%    \FigureFile(88mm,90mm){j1625oc2010.eps}
    \FigureFile(88mm,90mm){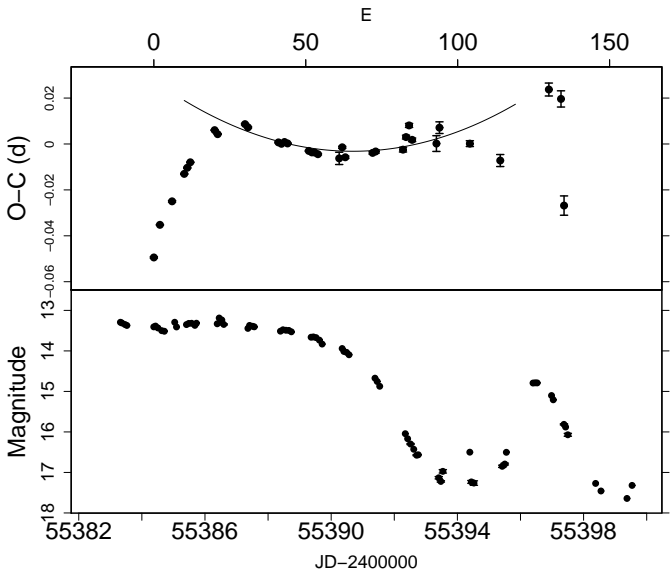}
  \end{center}
  \caption{$O-C$ of superhumps in SDSS J1625 (2010).
  (Upper): $O-C$ diagram.  The $O-C$ values were against the mean period
  for the stage B ($30 \le E \le 104$, thin curve)
  (Lower): Light curve.  The outburst entered the rapid decline phase
  during stage B, and was followed by a rebrightening.
  }
  \label{fig:j1625oc2010}
\end{figure}

   The times of superhump maxima are listed in table \ref{tab:j1625oc2010}.
The epochs for $E \le 21$ correspond to the growing stage of superhumps
(stage A).  The epochs for $E \ge 130$ were maxima after the rebrightening
and these humps may not be true superhumps.  We also excluded $E=114$
(just prior to the rebrightening) in determining period variation
due to its low signal.  The $P_{\rm dot}$ for $30 \ge E \ge 104$ was
$+14.9(3.7) \times 10^{-5}$, which is unusually large for this long
$P_{\rm SH}$ = 0.09605(5) d (mean period based on timing analysis).
There was also little indication of
a stage B--C transition during the superoutburst plateau.
These unusual evolution may be related to
the unexpectedly early fading during the superoutburst plateau
(see figure \ref{fig:j1625oc2010}).
This object appears to add another variety of superhump evolution in
long-$P_{\rm SH}$ systems (cf. \cite{Pdot}, subsection 4.10).

\begin{table}
\caption{Superhump maxima of SDSS J1625 (2010).}\label{tab:j1625oc2010}
\begin{center}
\begin{tabular}{ccccc}
\hline
$E$ & max\commenta & error & $O-C$\commentb & $N$\commentc \\
\hline
0 & 55384.4489 & 0.0003 & $-$0.0363 & 363 \\
2 & 55384.6553 & 0.0002 & $-$0.0224 & 176 \\
6 & 55385.0497 & 0.0003 & $-$0.0128 & 161 \\
10 & 55385.4459 & 0.0002 & $-$0.0015 & 587 \\
11 & 55385.5446 & 0.0002 & 0.0010 & 365 \\
12 & 55385.6431 & 0.0001 & 0.0032 & 179 \\
20 & 55386.4255 & 0.0001 & 0.0159 & 732 \\
21 & 55386.5198 & 0.0001 & 0.0140 & 662 \\
30 & 55387.3886 & 0.0003 & 0.0169 & 265 \\
31 & 55387.4833 & 0.0003 & 0.0153 & 289 \\
41 & 55388.4373 & 0.0002 & 0.0071 & 328 \\
42 & 55388.5327 & 0.0002 & 0.0064 & 291 \\
43 & 55388.6296 & 0.0002 & 0.0070 & 194 \\
44 & 55388.7250 & 0.0002 & 0.0062 & 158 \\
51 & 55389.3941 & 0.0003 & 0.0018 & 382 \\
52 & 55389.4895 & 0.0003 & 0.0010 & 430 \\
53 & 55389.5855 & 0.0002 & 0.0007 & 189 \\
54 & 55389.6808 & 0.0003 & $-$0.0001 & 200 \\
61 & 55390.3514 & 0.0027 & $-$0.0031 & 135 \\
62 & 55390.4523 & 0.0004 & 0.0016 & 338 \\
63 & 55390.5440 & 0.0006 & $-$0.0029 & 144 \\
72 & 55391.4104 & 0.0005 & $-$0.0025 & 193 \\
73 & 55391.5071 & 0.0007 & $-$0.0021 & 170 \\
82 & 55392.3723 & 0.0010 & $-$0.0027 & 102 \\
83 & 55392.4739 & 0.0008 & 0.0026 & 130 \\
84 & 55392.5750 & 0.0009 & 0.0075 & 144 \\
85 & 55392.6648 & 0.0008 & 0.0010 & 190 \\
93 & 55393.4316 & 0.0034 & $-$0.0019 & 153 \\
94 & 55393.5346 & 0.0025 & 0.0049 & 52 \\
104 & 55394.4882 & 0.0013 & $-$0.0037 & 43 \\
114 & 55395.4413 & 0.0026 & $-$0.0127 & 49 \\
130 & 55397.0091 & 0.0028 & 0.0156 & 200 \\
134 & 55397.3893 & 0.0035 & 0.0108 & 94 \\
135 & 55397.4389 & 0.0042 & $-$0.0358 & 149 \\
\hline
  \multicolumn{5}{l}{\commenta BJD$-$2400000.} \\
  \multicolumn{5}{l}{\commentb Against $max = 2455384.4852 + 0.096218 E$.} \\
  \multicolumn{5}{l}{\commentc Number of points used to determine the maximum.} \\
\end{tabular}
\end{center}
\end{table}

\subsection{SDSS J163722.21$-$001957.1}\label{obj:j1637}

   This object (hereafter SDSS J1637) is a CV selected
during the course of the SDSS \citep{szk02SDSSCVs}.
\citet{szk02SDSSCVs} reported the presence of high and low states
and suggested the dwarf nova-type classification.

\begin{figure}
  \begin{center}
%    \FigureFile(88mm,110mm){j1637shpdm.eps}
    \FigureFile(88mm,110mm){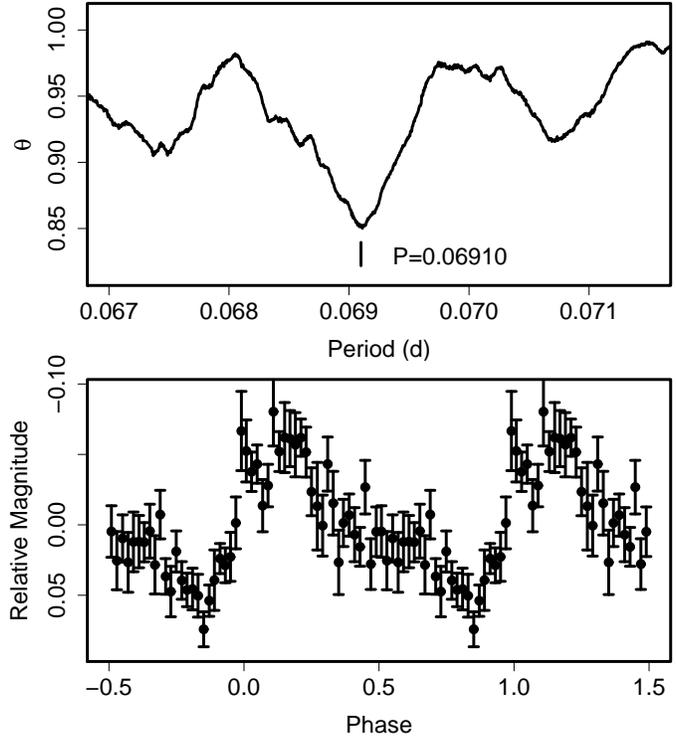}
  \end{center}
  \caption{Superhumps in SDSS J1637 (2004).
     (Upper): PDM analysis.
     (Lower): Phase-averaged profile.}
  \label{fig:j1637shpdm}
\end{figure}

   The 2004 outburst of this object was detected by R. Stubbings
at a visual magnitude of 15.0 on 2004 March 27 (vsnet-outburst 6212).
The presence of superhumps was soon confirmed (vsnet-alert 8084, 8086, 8088).
The mean superhump period determined from the observation was
0.06910(4) d (PDM method, figure \ref{fig:j1637shpdm}),
and the times of superhump maxima are listed in table \ref{tab:j1637oc2004}.
Since the outburst entered its rapid decline stage $\sim$ 4 d after
the initial detection, these superhumps were very likely stage C
superhumps.
Using the orbital period of 0.06739(1) d determined by \citet{sou08CVperiod},
we obtained an $\epsilon$ of 2.5 \%.

\begin{table}
\caption{Superhump maxima of SDSS J1637 (2004).}\label{tab:j1637oc2004}
\begin{center}
\begin{tabular}{ccccc}
\hline
$E$ & max\commenta & error & $O-C$\commentb & $N$\commentc \\
\hline
0 & 53093.2412 & 0.0016 & $-$0.0023 & 145 \\
1 & 53093.3132 & 0.0012 & 0.0004 & 76 \\
2 & 53093.3832 & 0.0012 & 0.0014 & 37 \\
4 & 53093.5206 & 0.0015 & 0.0004 & 18 \\
5 & 53093.5896 & 0.0008 & 0.0002 & 155 \\
29 & 53095.2464 & 0.0027 & $-$0.0030 & 36 \\
30 & 53095.3206 & 0.0025 & 0.0020 & 38 \\
31 & 53095.3890 & 0.0022 & 0.0013 & 30 \\
43 & 53096.2189 & 0.0034 & 0.0012 & 87 \\
44 & 53096.2853 & 0.0028 & $-$0.0016 & 147 \\
\hline
  \multicolumn{5}{l}{\commenta BJD$-$2400000.} \\
  \multicolumn{5}{l}{\commentb Against $max = 2453093.2435 + 0.069167 E$.} \\
  \multicolumn{5}{l}{\commentc Number of points used to determine the maximum.} \\
\end{tabular}
\end{center}
\end{table}

\subsection{SDSS J165359.06$+$201010.4}\label{obj:j1653}

   This object (hereafter SDSS J1653) is a CV selected
during the course of the SDSS \citep{szk06SDSSCV5}, who detected superhumps
with a period of 1.58 hr during one of its superoutburst.

   The 2010 superoutburst was detected by the CRTS (cf. vsnet-alert 11936).
The times of superhump maxima are listed in table \ref{tab:j1653oc2010}.
It is evident from these data that these observations recorded a
stage B--C transition.  The derived periods for each stage are listed
in table \ref{tab:perlist}.  Although the observation only covered the
late part of the superoutburst, the evolution of superhump period appears
to be typical.  As judged from these periods, \citet{szk06SDSSCV5}
appears to have recorded stage B superhumps.

\begin{table}
\caption{Superhump maxima of SDSS J1653 (2010).}\label{tab:j1653oc2010}
\begin{center}
\begin{tabular}{ccccc}
\hline
$E$ & max\commenta & error & $O-C$\commentb & $N$\commentc \\
\hline
0 & 55304.1037 & 0.0009 & $-$0.0007 & 67 \\
1 & 55304.1673 & 0.0007 & $-$0.0022 & 59 \\
2 & 55304.2313 & 0.0004 & $-$0.0032 & 66 \\
3 & 55304.2970 & 0.0005 & $-$0.0025 & 66 \\
5 & 55304.4272 & 0.0003 & $-$0.0024 & 73 \\
6 & 55304.4913 & 0.0004 & $-$0.0033 & 67 \\
7 & 55304.5579 & 0.0004 & $-$0.0018 & 72 \\
50 & 55307.3670 & 0.0017 & 0.0110 & 55 \\
51 & 55307.4263 & 0.0014 & 0.0052 & 65 \\
52 & 55307.4906 & 0.0010 & 0.0046 & 69 \\
93 & 55310.1499 & 0.0009 & $-$0.0025 & 45 \\
94 & 55310.2197 & 0.0015 & 0.0023 & 40 \\
215 & 55318.0817 & 0.0041 & $-$0.0045 & 22 \\
\hline
  \multicolumn{5}{l}{\commenta BJD$-$2400000.} \\
  \multicolumn{5}{l}{\commentb Against $max = 2455304.1044 + 0.065032 E$.} \\
  \multicolumn{5}{l}{\commentc Number of points used to determine the maximum.} \\
\end{tabular}
\end{center}
\end{table}

\subsection{SDSS J204817.85$-$061044.8}\label{obj:j2048}

   This object (hereafter SDSS J2048) is a CV selected
during the course of the SDSS \citep{szk03SDSSCV2}, who reported the
detection of a white dwarf in the spectrum, suggesting a dwarf nova
with a low mass-transfer rate.

\begin{figure}
  \begin{center}
%    \FigureFile(88mm,110mm){j2048shpdm.eps}
    \FigureFile(88mm,110mm){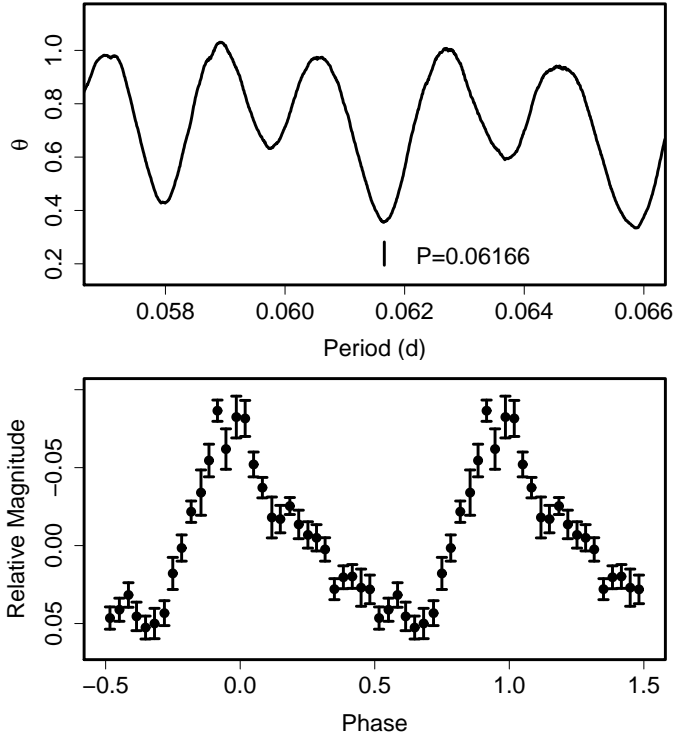}
  \end{center}
  \caption{Superhumps in SDSS J2048 (2009).
     (Upper): PDM analysis.  The selection of the alias was based on
     the known $P_{\rm orb}$.
     (Lower): Phase-averaged profile.}
  \label{fig:j2048shpdm}
\end{figure}

   The outburst in 2009 October was detected by E. Muyllaert
(cvnet-outburst 3367).  Subsequent observations by I. Miller detected
superhumps (cvnet-outburst 3383).
\citet{wou10CVperiod} observed the object in quiescence and obtained
an orbital period of 0.060597(2) d.
We identified the superhump period of 0.06166(2) d based om this
$P_{\rm orb}$ (figure \ref{fig:j2048shpdm}.
The times of superhump maxima are listed in table \ref{tab:j2048oc2009}.
The $\epsilon$ was 1.8 \%.

\begin{table}
\caption{Superhump maxima of SDSS J2048 (2009).}\label{tab:j2048oc2009}
\begin{center}
\begin{tabular}{ccccc}
\hline
$E$ & max\commenta & error & $O-C$\commentb & $N$\commentc \\
\hline
0 & 55119.4241 & 0.0008 & $-$0.0013 & 45 \\
16 & 55120.4152 & 0.0012 & 0.0029 & 32 \\
31 & 55121.3347 & 0.0007 & $-$0.0029 & 63 \\
32 & 55121.4006 & 0.0011 & 0.0014 & 56 \\
\hline
  \multicolumn{5}{l}{\commenta BJD$-$2400000.} \\
  \multicolumn{5}{l}{\commentb Against $max = 2455119.4254 + 0.061683 E$.} \\
  \multicolumn{5}{l}{\commentc Number of points used to determine the maximum.} \\
\end{tabular}
\end{center}
\end{table}

\subsection{OT J040659.8$+$005244}\label{obj:j0406}

   This object (hereafter OT J0406) was discovered by K. Itagaki
in 2008 \citep{yam08j0406cbet1463}.  The 2010 outburst, detected by the
CRTS, is the second known superoutburst of this object.  The outburst
was apparently detected in its late stage.  The times of superhump maxima
are listed in table \ref{tab:j0406oc2010}.  We attribute these superhumps
to stage C superhumps.  The mean period with the PDM method
was 0.07996(3) d, close to that (0.07992 d) recorded during the 2008
superoutburst \citep{Pdot}.  We adopted this period in table \ref{tab:perlist}.

\begin{table}
\caption{Superhump maxima of J0406 (2010).}\label{tab:j0406oc2010}
\begin{center}
\begin{tabular}{ccccc}
\hline
$E$ & max\commenta & error & $O-C$\commentb & $N$\commentc \\
\hline
0 & 55246.9899 & 0.0009 & $-$0.0011 & 126 \\
12 & 55247.9497 & 0.0010 & $-$0.0000 & 132 \\
25 & 55248.9899 & 0.0011 & 0.0015 & 205 \\
49 & 55250.9069 & 0.0030 & 0.0010 & 156 \\
62 & 55251.9432 & 0.0027 & $-$0.0014 & 207 \\
\hline
  \multicolumn{5}{l}{\commenta BJD$-$2400000.} \\
  \multicolumn{5}{l}{\commentb Against $max = 2455246.9909 + 0.079898 E$.} \\
  \multicolumn{5}{l}{\commentc Number of points used to determine the maximum.} \\
\end{tabular}
\end{center}
\end{table}

\subsection{OT J050617.4$+$354738}\label{obj:j0506}

\begin{figure}
  \begin{center}
%    \FigureFile(88mm,110mm){j0506shpdm.eps}
    \FigureFile(88mm,110mm){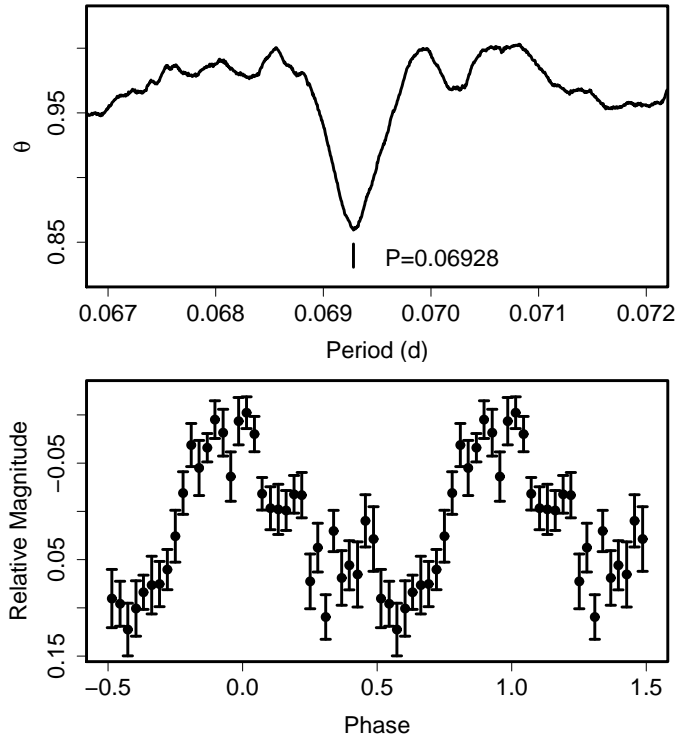}
  \end{center}
  \caption{Superhumps in OT J0506 (2009).
     (Upper): PDM analysis.
     (Lower): Phase-averaged profile.}
  \label{fig:j0506shpdm}
\end{figure}

   This object (=USNO$-$B1.0 1257$-$0089884, hereafter OT J0506)
is a dwarf nova discovered by \citet{kry10j0506} (see also
vsnet-alert 11686 for the initial announcement).
The object was rising at the time of the discovery and superhumps
were subsequently detected (vsnet-alert 11688; \cite{kry10j0506}).
Since the discovery announcement was made sufficiently early, the
early stages of the outburst was well observed.
The times of superhump maxima are listed in table \ref{tab:j0506oc2009}.
The mean superhump period excluding the initial night was 0.06928(2) d
(PDM method, figure \ref{fig:j0506shpdm}).

\begin{table}
\caption{Superhump maxima of OT J0506 (2009).}\label{tab:j0506oc2009}
\begin{center}
\begin{tabular}{ccccc}
\hline
$E$ & max\commenta & error & $O-C$\commentb & $N$\commentc \\
\hline
0 & 55160.5113 & 0.0034 & $-$0.0010 & 24 \\
1 & 55160.5777 & 0.0008 & $-$0.0039 & 48 \\
2 & 55160.6485 & 0.0007 & $-$0.0024 & 48 \\
23 & 55162.1113 & 0.0006 & 0.0045 & 130 \\
24 & 55162.1810 & 0.0007 & 0.0049 & 149 \\
52 & 55164.1160 & 0.0010 & $-$0.0012 & 102 \\
53 & 55164.1854 & 0.0011 & $-$0.0011 & 140 \\
56 & 55164.3920 & 0.0009 & $-$0.0025 & 12 \\
81 & 55166.1293 & 0.0023 & 0.0016 & 146 \\
82 & 55166.2028 & 0.0099 & 0.0058 & 92 \\
85 & 55166.4085 & 0.0029 & 0.0035 & 7 \\
88 & 55166.6129 & 0.0018 & $-$0.0001 & 13 \\
96 & 55167.1688 & 0.0020 & 0.0013 & 157 \\
97 & 55167.2387 & 0.0020 & 0.0018 & 14 \\
98 & 55167.3056 & 0.0017 & $-$0.0006 & 12 \\
99 & 55167.3755 & 0.0011 & $-$0.0000 & 14 \\
100 & 55167.4428 & 0.0015 & $-$0.0021 & 12 \\
101 & 55167.5116 & 0.0016 & $-$0.0026 & 13 \\
102 & 55167.5774 & 0.0022 & $-$0.0061 & 12 \\
\hline
  \multicolumn{5}{l}{\commenta BJD$-$2400000.} \\
  \multicolumn{5}{l}{\commentb Against $max = 2455160.5123 + 0.069326 E$.} \\
  \multicolumn{5}{l}{\commentc Number of points used to determine the maximum.} \\
\end{tabular}
\end{center}
\end{table}

\subsection{OT J102637.0$+$475426}\label{obj:j1026}

   This object (hereafter OT J1026) is a dwarf nova discovered by K. Itagaki
\citep{yam09j1026cbet1644}.  Although \citet{Pdot} reported on late-stage
observations of the 2009 superoutburst, the observational coverage was
short.  We observed the 2010 superoutburst, detected by I. Miller
(baavss-alert 2245).  We first time succeeded in recording both stages
B and C (table \ref{tab:j1026oc2010}).  The resultant period is in
disagreement with the 2009 result.  Since the present coverage is much
better than the 2009 observation, the present values are more reliable.
A reanalysis of the 2009 data could not yield a continuous
$O-C$ diagram, suggesting a discontinuous phase jump (or appearance of
stronger secondary superhump peaks) in the final stage of the 2009
superoutburst.

\begin{table}
\caption{Superhump maxima of OT J1026 (2010).}\label{tab:j1026oc2010}
\begin{center}
\begin{tabular}{ccccc}
\hline
$E$ & max\commenta & error & $O-C$\commentb & $N$\commentc \\
\hline
0 & 55270.5082 & 0.0002 & $-$0.0012 & 67 \\
1 & 55270.5783 & 0.0003 & 0.0002 & 75 \\
2 & 55270.6456 & 0.0003 & $-$0.0011 & 75 \\
26 & 55272.2894 & 0.0009 & $-$0.0048 & 30 \\
27 & 55272.3616 & 0.0005 & $-$0.0013 & 76 \\
28 & 55272.4296 & 0.0005 & $-$0.0020 & 58 \\
29 & 55272.5006 & 0.0006 & 0.0004 & 68 \\
43 & 55273.4636 & 0.0004 & 0.0024 & 60 \\
44 & 55273.5311 & 0.0008 & 0.0012 & 36 \\
57 & 55274.4268 & 0.0039 & 0.0045 & 47 \\
58 & 55274.4972 & 0.0021 & 0.0063 & 67 \\
59 & 55274.5603 & 0.0012 & 0.0007 & 72 \\
113 & 55278.2660 & 0.0013 & $-$0.0004 & 53 \\
114 & 55278.3322 & 0.0012 & $-$0.0029 & 76 \\
115 & 55278.4048 & 0.0011 & 0.0010 & 69 \\
116 & 55278.4718 & 0.0008 & $-$0.0006 & 76 \\
117 & 55278.5393 & 0.0008 & $-$0.0018 & 61 \\
118 & 55278.6091 & 0.0007 & $-$0.0006 & 72 \\
\hline
  \multicolumn{5}{l}{\commenta BJD$-$2400000.} \\
  \multicolumn{5}{l}{\commentb Against $max = 2455270.5094 + 0.068646 E$.} \\
  \multicolumn{5}{l}{\commentc Number of points used to determine the maximum.} \\
\end{tabular}
\end{center}
\end{table}

\subsection{OT J104411.4$+$211307}\label{obj:j1044}

   This object (= CSS100217:104411$+$211307, hereafter OT J1044) was
discovered by the CRTS on 2010 February 17.  The large outburst amplitude
was already suggestive of a WZ Sge-type like outburst (vsnet-alert 11817).
Subsequent observation soon established the presence of early superhumps
(vsnet-alert 11820, 11828, 11829; figure \ref{fig:j1044shpdm}).
The object later developed ordinary superhumps (vsnet-alert 11836, 11837).
There was a single post-superoutburst rebrightening (vsnet-alert 11885).

   The times of ordinary superhumps are listed in table \ref{tab:j1044oc2010}.
It was most likely that we only sufficiently observed stage A ($E \le 41$)
and stage C ($E \ge 195$).  Although the $O-C$ analysis suggests a mean
period of $\sim$0.0605 d during stage B, we did not adopt this value in
table \ref{tab:perlist}.  The superhump period during stage C was, however,
reliably determined (figure \ref{fig:j1044shpdm}).  The period [0.06024(4) d]
suggests an $\epsilon$ of 1.9 \% against the period [0.05909 (1) d]
of early superhumps.
The waveform of superhumps became double-humped during the rapid
fading stage ($E \ge 246$), we only listed maxima having the same phases
as in $E < 246$.  The last two points ($E=376$, $E=377$) were obtained
during the fading stage of a rebrightening.  Although these humps may
have been different from earlier superhumps, we included them because
WZ Sge-type dwarf novae are known to exhibit long-enduring superhumps
even during the rebrightening phase.

   All the observed features, including the outburst amplitude of $\sim$6 mag,
the existence of early superhumps, relatively large $\epsilon$,
the likely existence of stage C and the single post-superoutburst
rebrightening resemble those of large-amplitude borderline WZ Sge-type
dwarf novae such as BC UMa and RZ Leo (objects showing
``type-C'' WZ Sge-type outbursts in \cite{Pdot}).

\begin{figure}
  \begin{center}
%    \FigureFile(88mm,110mm){j1044eshpdm.eps}
    \FigureFile(88mm,110mm){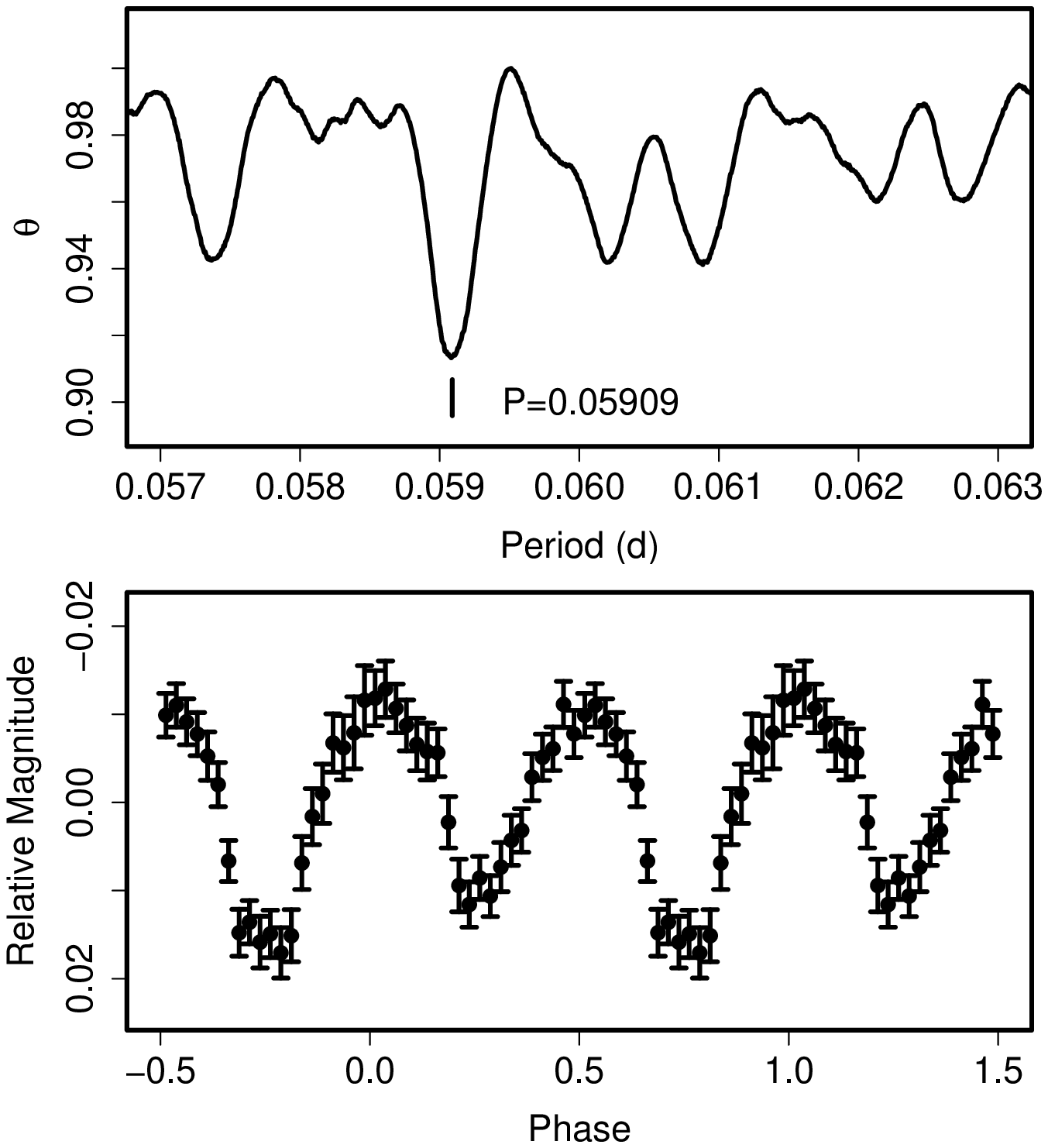}
  \end{center}
  \caption{Early superhumps in OT J1044 (2010).
     (Upper): PDM analysis.
     (Lower): Phase-averaged profile.}
  \label{fig:j1044eshpdm}
\end{figure}

\begin{figure}
  \begin{center}
%    \FigureFile(88mm,110mm){j1044shpdm.eps}
    \FigureFile(88mm,110mm){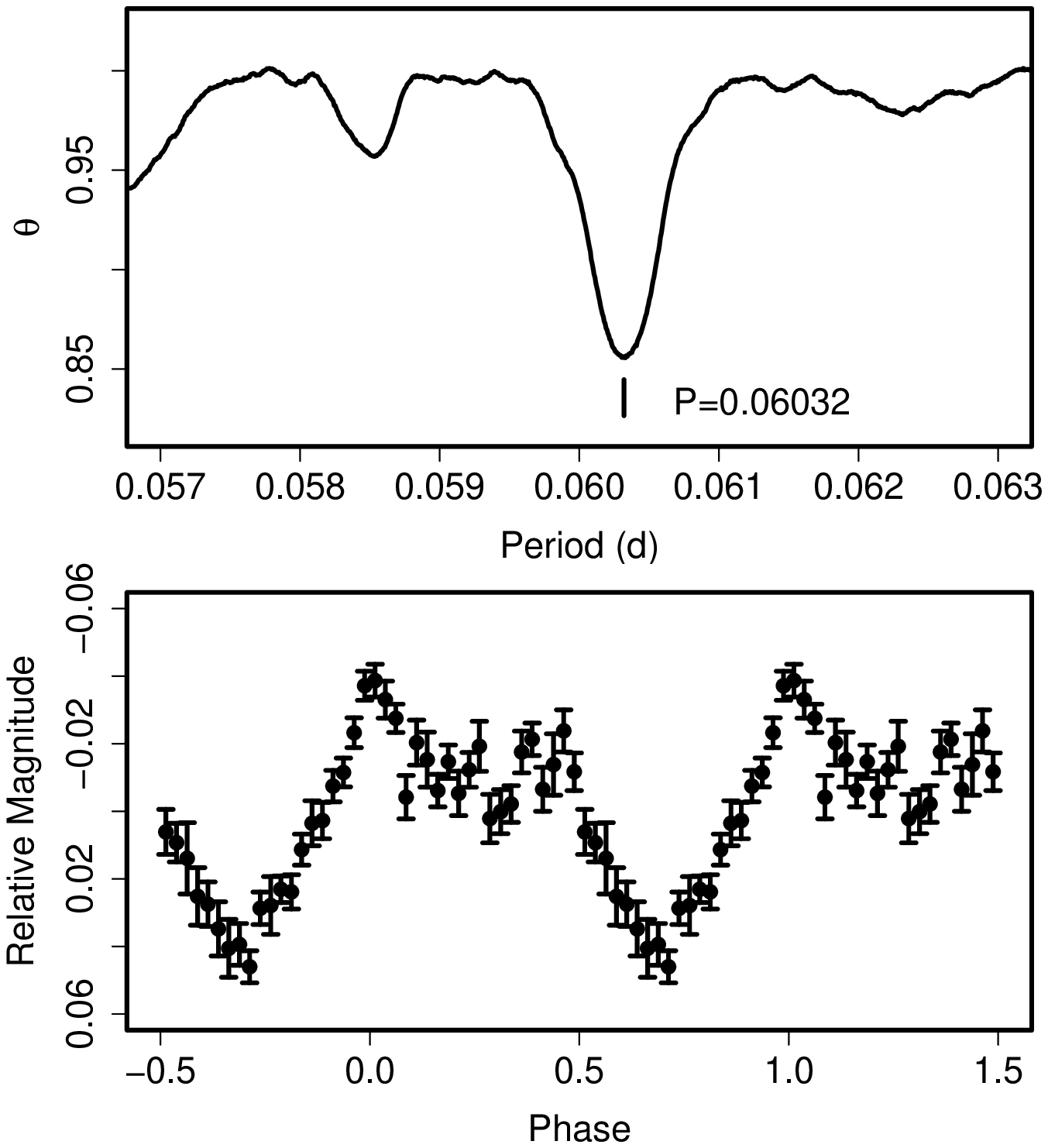}
  \end{center}
  \caption{Ordinary superhumps in OT J1044 during stage C (2010).
     (Upper): PDM analysis.
     (Lower): Phase-averaged profile.}
  \label{fig:j1044shpdm}
\end{figure}

\begin{table}
\caption{Superhump maxima of OT J1044 (2019).}\label{tab:j1044oc2010}
\begin{center}
\begin{tabular}{ccccc}
\hline
$E$ & max\commenta & error & $O-C$\commentb & $N$\commentc \\
\hline
0 & 55250.5223 & 0.0010 & $-$0.0081 & 65 \\
1 & 55250.5847 & 0.0008 & $-$0.0062 & 49 \\
2 & 55250.6440 & 0.0013 & $-$0.0074 & 25 \\
8 & 55251.0082 & 0.0030 & $-$0.0065 & 60 \\
9 & 55251.0696 & 0.0004 & $-$0.0056 & 95 \\
10 & 55251.1309 & 0.0003 & $-$0.0049 & 228 \\
11 & 55251.1920 & 0.0002 & $-$0.0044 & 229 \\
24 & 55251.9822 & 0.0004 & $-$0.0011 & 101 \\
25 & 55252.0459 & 0.0009 & 0.0019 & 100 \\
26 & 55252.1050 & 0.0003 & 0.0005 & 312 \\
27 & 55252.1643 & 0.0002 & $-$0.0008 & 302 \\
28 & 55252.2251 & 0.0004 & $-$0.0004 & 218 \\
41 & 55253.0074 & 0.0019 & $-$0.0052 & 61 \\
92 & 55256.1063 & 0.0017 & 0.0060 & 105 \\
113 & 55257.3823 & 0.0041 & 0.0106 & 32 \\
114 & 55257.4322 & 0.0019 & $-$0.0000 & 61 \\
115 & 55257.4891 & 0.0011 & $-$0.0037 & 67 \\
116 & 55257.5505 & 0.0013 & $-$0.0028 & 65 \\
195 & 55262.3421 & 0.0014 & 0.0059 & 21 \\
196 & 55262.4117 & 0.0011 & 0.0149 & 87 \\
197 & 55262.4689 & 0.0009 & 0.0116 & 85 \\
198 & 55262.5285 & 0.0008 & 0.0106 & 89 \\
199 & 55262.5863 & 0.0010 & 0.0079 & 96 \\
200 & 55262.6474 & 0.0011 & 0.0084 & 76 \\
212 & 55263.3763 & 0.0028 & 0.0109 & 30 \\
213 & 55263.4288 & 0.0010 & 0.0028 & 92 \\
214 & 55263.4911 & 0.0010 & 0.0046 & 93 \\
215 & 55263.5524 & 0.0017 & 0.0053 & 94 \\
216 & 55263.6158 & 0.0012 & 0.0082 & 97 \\
217 & 55263.6732 & 0.0017 & 0.0051 & 49 \\
227 & 55264.2828 & 0.0014 & 0.0093 & 62 \\
228 & 55264.3384 & 0.0012 & 0.0043 & 76 \\
229 & 55264.3995 & 0.0016 & 0.0048 & 86 \\
230 & 55264.4557 & 0.0015 & 0.0005 & 96 \\
231 & 55264.5205 & 0.0016 & 0.0047 & 82 \\
232 & 55264.5774 & 0.0011 & 0.0011 & 93 \\
233 & 55264.6271 & 0.0026 & $-$0.0097 & 86 \\
244 & 55265.2995 & 0.0022 & $-$0.0033 & 78 \\
245 & 55265.3571 & 0.0012 & $-$0.0062 & 91 \\
246 & 55265.4192 & 0.0027 & $-$0.0047 & 97 \\
247 & 55265.4772 & 0.0013 & $-$0.0072 & 92 \\
248 & 55265.5405 & 0.0026 & $-$0.0045 & 100 \\
249 & 55265.6008 & 0.0043 & $-$0.0047 & 85 \\
376 & 55273.2741 & 0.0018 & $-$0.0204 & 63 \\
377 & 55273.3328 & 0.0023 & $-$0.0222 & 63 \\
\hline
  \multicolumn{5}{l}{\commenta BJD$-$2400000.} \\
  \multicolumn{5}{l}{\commentb Against $max = 2455250.5304 + 0.060543 E$.} \\
  \multicolumn{5}{l}{\commentc Number of points used to determine the maximum.} \\
\end{tabular}
\end{center}
\end{table}

\subsection{OT J112253.3$-$111037}\label{obj:j1122}

\begin{figure}
  \begin{center}
%    \FigureFile(88mm,110mm){j1122shpdm.eps}
    \FigureFile(88mm,110mm){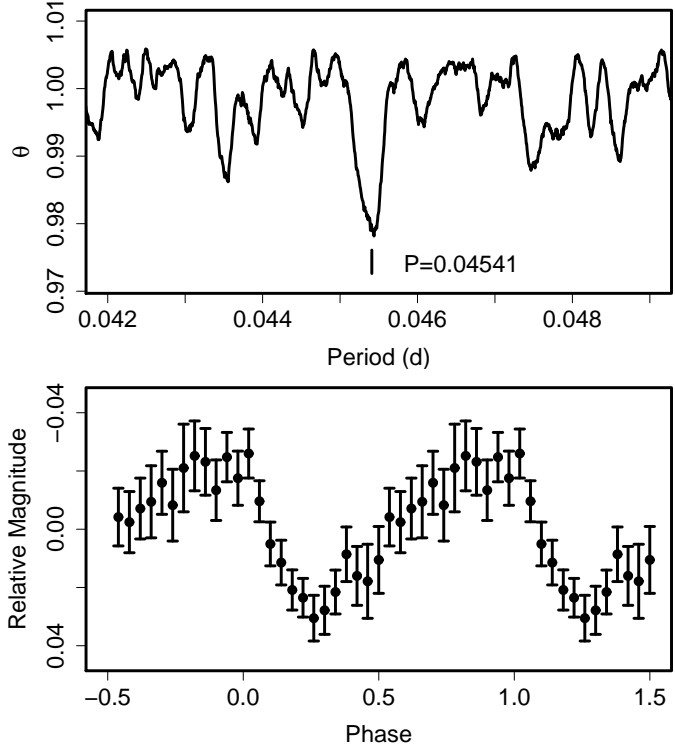}
  \end{center}
  \caption{Superhumps in OT J1122 (2010).
     (Upper): PDM analysis.
     (Lower): Phase-averaged profile.}
  \label{fig:j1122shpdm}
\end{figure}

   This object (= CSS100603:112253$-$111037, hereafter OT J1122) is a
transient discovered by the CRTS.  The object was soon suspected to be
a large-amplitude dwarf nova (vsnet-alert 12020).  H. Maehara detected
short-period superhumps (vsnet-alert 12025; figure \ref{fig:j1122shpdm})
indicating that the object is an unusual short-$P_{\rm orb}$ CV with
an evolved secondary.
J. Greaves pointed out that this object is included in the public
SDSS DR7 archive.  The spectrum indicates that the object is hydrogen-rich,
rather than an AM CVn-type helium CV (vsnet-alert 12025), although the
strength of helium emission lines suggests an helium enrichment
(vsnet-alert 12026).

   The times of superhump maxima are listed in table \ref{tab:j1122oc2010}.
There is an indication of a systematic period decrease after $E=96$.
We attributed this to a stage B--C transition and give the derived
parameters based in this interpretation in table \ref{tab:perlist}.

   More detailed analysis and discussion will be presented in
Maehara et al., in preparation.

\begin{table}
\caption{Superhump maxima of OT J1122 (2010).}\label{tab:j1122oc2010}
\begin{center}
\begin{tabular}{ccccc}
\hline
$E$ & max\commenta & error & $O-C$\commentb & $N$\commentc \\
\hline
0 & 55350.9981 & 0.0034 & $-$0.0001 & 95 \\
1 & 55351.0405 & 0.0033 & $-$0.0032 & 89 \\
36 & 55352.6341 & 0.0007 & 0.0019 & 54 \\
37 & 55352.6784 & 0.0007 & 0.0007 & 83 \\
44 & 55352.9942 & 0.0015 & $-$0.0012 & 96 \\
45 & 55353.0346 & 0.0033 & $-$0.0061 & 95 \\
49 & 55353.2263 & 0.0004 & 0.0040 & 148 \\
52 & 55353.3564 & 0.0022 & $-$0.0021 & 45 \\
72 & 55354.2685 & 0.0006 & 0.0022 & 189 \\
73 & 55354.3118 & 0.0012 & 0.0001 & 189 \\
93 & 55355.2223 & 0.0007 & 0.0029 & 188 \\
94 & 55355.2656 & 0.0007 & 0.0007 & 219 \\
95 & 55355.3096 & 0.0012 & $-$0.0006 & 211 \\
96 & 55355.3646 & 0.0048 & 0.0089 & 48 \\
132 & 55356.9831 & 0.0030 & $-$0.0066 & 95 \\
137 & 55357.2187 & 0.0012 & 0.0020 & 83 \\
138 & 55357.2635 & 0.0012 & 0.0014 & 94 \\
139 & 55357.3070 & 0.0011 & $-$0.0005 & 101 \\
154 & 55357.9839 & 0.0046 & $-$0.0044 & 42 \\
\hline
  \multicolumn{5}{l}{\commenta BJD$-$2400000.} \\
  \multicolumn{5}{l}{\commentb Against $max = 2455350.9982 + 0.045390 E$.} \\
  \multicolumn{5}{l}{\commentc Number of points used to determine the maximum.} \\
\end{tabular}
\end{center}
\end{table}

\subsection{OT J144011.0$+$494734}\label{obj:j1440}

   We reported on the detection of superhumps and period variation
in this object (= CSS090530:144011$+$494734, hereafter OT J1440)
in \citet{Pdot}.  We present here a re-analysis after incorporating
the data in \citet{boy10j1440}.  The times of superhump maxima are
listed in table \ref{tab:j1440oc2009}.

\begin{figure}
  \begin{center}
%    \FigureFile(88mm,90mm){j14402009oc.eps}
    \FigureFile(88mm,90mm){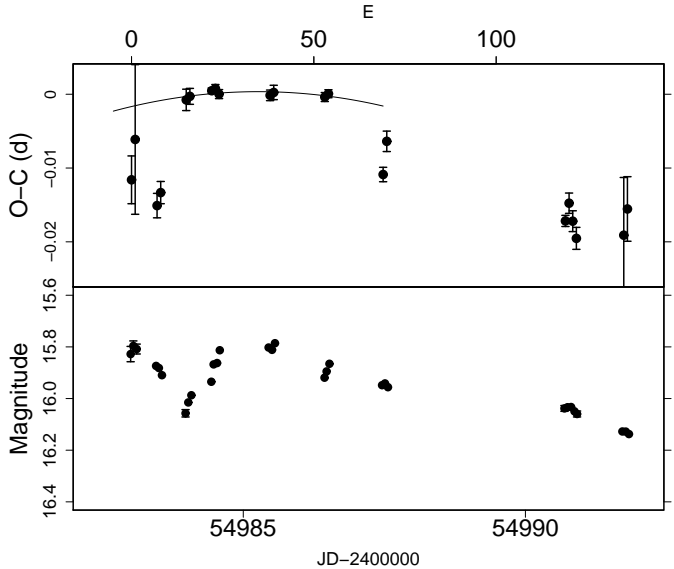}
  \end{center}
  \caption{$O-C$ of superhumps OT J1440 (2009).
  (Upper): $O-C$ diagram.  The $O-C$ values were against the mean period
  for the stage B ($15 \le E \le 54$, thin curve)
  (Lower): Light curve.  There was a temporary fading before
  the superhumps fully grew.}
  \label{fig:j14402009oc}
\end{figure}

   It has now become evident that $E < 10$ corresponds to stage A,
when the superhumps were indeed still in development and the
object once started to fade temporarily (figure \ref{fig:j14402009oc}).
The period break reported in \citet{Pdot} and \citet{boy10j1440}
after $E = 54$ this is most likely a stage B--C transition.
It is unusual for such a short-$P_{\rm SH}$ system to show
a stage B--C transition at such an early stage.  The apparent lack
of distinct positive $P_{\rm dot}$ during the stage B is also
unusual.  The development of the superhump period the presence of
a likely precursor (or a stagnation) in the light curve are similar to
those of BZ UMa, which has been proposed to critically exhibit
the SU UMa-type phenomenon \citet{Pdot}.  Since BZ UMa is known
to show many normal outbursts in addition to a very rare superoutburst
(\cite{jur94bzuma}; \cite{rin90bzuma}), a further systematic observation
of OT J1440 is encouraged to explore the similarity between these
objects and potential cause of the unusual period behavior.

\begin{table}
\caption{Superhump maxima of OT J1440 (2009).}\label{tab:j1440oc2009}
\begin{center}
\begin{tabular}{ccccc}
\hline
$E$ & max\commenta & error & $O-C$\commentb & $N$\commentc \\
\hline
0 & 54983.0245 & 0.0032 & $-$0.0095 & 69 \\
1 & 54983.0946 & 0.0101 & $-$0.0039 & 64 \\
7 & 54983.4733 & 0.0017 & $-$0.0123 & 68 \\
8 & 54983.5397 & 0.0015 & $-$0.0104 & 68 \\
15 & 54984.0046 & 0.0014 & 0.0029 & 191 \\
16 & 54984.0697 & 0.0011 & 0.0034 & 223 \\
22 & 54984.4581 & 0.0004 & 0.0048 & 164 \\
23 & 54984.5231 & 0.0005 & 0.0052 & 168 \\
24 & 54984.5869 & 0.0006 & 0.0046 & 74 \\
38 & 54985.4914 & 0.0007 & 0.0059 & 106 \\
39 & 54985.5564 & 0.0010 & 0.0063 & 105 \\
53 & 54986.4604 & 0.0006 & 0.0072 & 71 \\
54 & 54986.5254 & 0.0005 & 0.0077 & 64 \\
69 & 54987.4837 & 0.0010 & $-$0.0017 & 110 \\
70 & 54987.5528 & 0.0014 & 0.0029 & 125 \\
119 & 54990.7082 & 0.0008 & $-$0.0028 & 47 \\
120 & 54990.7752 & 0.0014 & $-$0.0003 & 102 \\
121 & 54990.8374 & 0.0014 & $-$0.0026 & 169 \\
122 & 54990.8997 & 0.0015 & $-$0.0049 & 143 \\
135 & 54991.7401 & 0.0078 & $-$0.0031 & 98 \\
136 & 54991.8083 & 0.0044 & 0.0006 & 119 \\
\hline
  \multicolumn{5}{l}{\commenta BJD$-$2400000.} \\
  \multicolumn{5}{l}{\commentb Against $max = 2454983.0341 + 0.064512 E$.} \\
  \multicolumn{5}{l}{\commentc Number of points used to determine the maximum.} \\
\end{tabular}
\end{center}
\end{table}

\subsection{OT J163120.9$+$103134}\label{obj:j1631}

   This object (= CSS080505:163121$+$103134, hereafter OT J1631) underwent
another superoutburst in 2010 (I. Miller, baavss-alert 2268).
The times of superhump maxima are listed in table \ref{tab:j1631oc2010}.
Although there seems to have a change in the period between $E=0$
and $E=14$, the change was apparently too large to be attributed to
a stage A--B or B--C transition.  It is likely that observation at $E=0$
was insufficient to derive a reliable maximum.  Disregarding observations
before BJD 2455307, we obtained a period of 0.06395(2) d with the PDM
method.

\begin{table}
\caption{Superhump maxima of OT J1631 (2010).}\label{tab:j1631oc2010}
\begin{center}
\begin{tabular}{ccccc}
\hline
$E$ & max\commenta & error & $O-C$\commentb & $N$\commentc \\
\hline
0 & 55306.6220 & 0.0008 & $-$0.0103 & 42 \\
14 & 55307.5341 & 0.0009 & 0.0047 & 68 \\
15 & 55307.6010 & 0.0010 & 0.0074 & 69 \\
55 & 55310.1595 & 0.0018 & 0.0026 & 133 \\
71 & 55311.1836 & 0.0020 & 0.0014 & 138 \\
72 & 55311.2404 & 0.0012 & $-$0.0059 & 108 \\
\hline
  \multicolumn{5}{l}{\commenta BJD$-$2400000.} \\
  \multicolumn{5}{l}{\commentb Against $max = 2455306.6323 + 0.064082 E$.} \\
  \multicolumn{5}{l}{\commentc Number of points used to determine the maximum.} \\
\end{tabular}
\end{center}
\end{table}

\subsection{OT J170343.6$+$090835}\label{obj:j1703}

\begin{figure}
  \begin{center}
%    \FigureFile(88mm,110mm){j1703shpdm.eps}
    \FigureFile(88mm,110mm){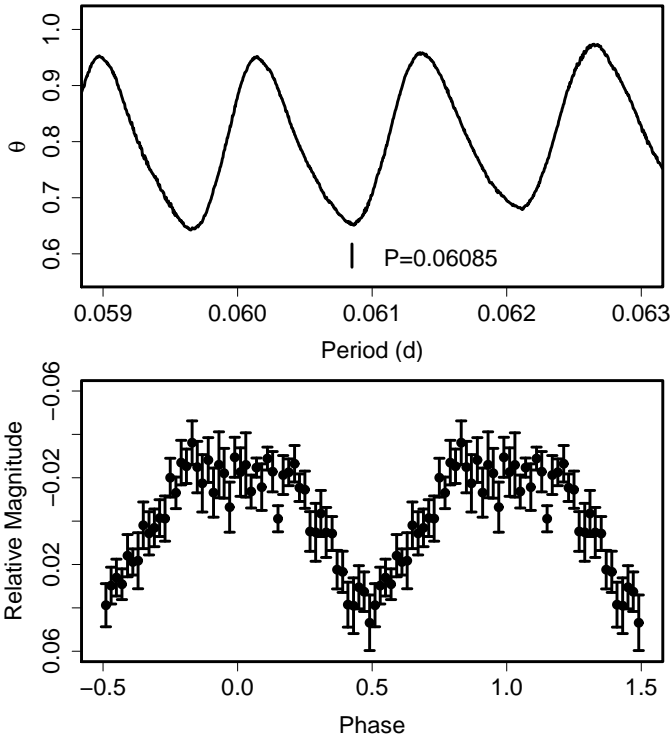}
  \end{center}
  \caption{Superhumps in OT J1703 (2009).
     (Upper): PDM analysis.
     (Lower): Phase-averaged profile.}
  \label{fig:j1703shpdm}
\end{figure}

   This object (= CSS090622:170344$+$090835, hereafter OT J1703) was
discovered by the CRTS on 2009 June 22.  The detection of superhumps
(vsnet-alert 11297, 11298) led to a classification as an SU UMa-type
dwarf nova.
Although there still remain possibilities of aliases, we have adopted
$P_{\rm SH}$ = 0.06085(2) d (figure \ref{fig:j1703shpdm}) based on
the best period determined from
the single-night observation (vsnet-alert 11298).
The times of superhump maxima are listed in table \ref{tab:j1703oc2009}.
The outburst appears to have been caught when the amplitude of
superhumps decayed (the initial observation was performed $\sim$ 5 d
after the CRTS detection).  According to CRTS observations, this
object were further detected in outburst on 2009 July 22 and 29.
Since there were no previous CRTS detections before 2009 June,
there may have been an enhanced outburst activity (either normal
outbursts or post-superoutburst rebrightenings) in 2009 July.

\begin{table}
\caption{OT J1703 (2009).}\label{tab:j1703oc2009}
\begin{center}
\begin{tabular}{ccccc}
\hline
$E$ & max\commenta & error & $O-C$\commentb & $N$\commentc \\
\hline
0 & 55009.4145 & 0.0021 & 0.0011 & 94 \\
1 & 55009.4716 & 0.0012 & $-$0.0026 & 100 \\
2 & 55009.5366 & 0.0014 & 0.0016 & 103 \\
51 & 55012.5157 & 0.0004 & $-$0.0000 & 58 \\
\hline
  \multicolumn{5}{l}{\commenta BJD$-$2400000.} \\
  \multicolumn{5}{l}{\commentb Against $max = 2455009.4134 + 0.060830 E$.} \\
  \multicolumn{5}{l}{\commentc Number of points used to determine the maximum.} \\
\end{tabular}
\end{center}
\end{table}

\subsection{OT J182142.8$+$212154}\label{obj:j1821}

\begin{figure}
  \begin{center}
%    \FigureFile(88mm,110mm){j1821shpdm.eps}
    \FigureFile(88mm,110mm){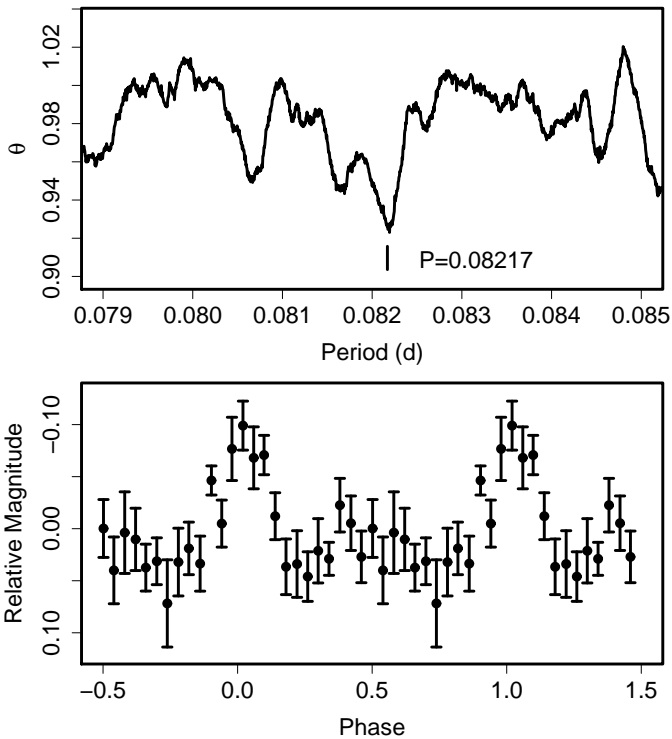}
  \end{center}
  \caption{Superhumps in OT J1821 (2010).
     (Upper): PDM analysis.
     (Lower): Phase-averaged profile.}
  \label{fig:j1821shpdm}
\end{figure}

   This object (hereafter OT J1821) was discovered by K. Itagaki
(vsnet-alert 11952).  The SDSS color suggested that the object is a dwarf
nova in outburst.  Superhumps were subsequently detected (vsnet-alert 11956;
figure \ref{fig:j1821shpdm}).
The times of superhump maxima are listed in table \ref{tab:j1821oc2010}.
These observations covered the final part of the superoutburst, and
the recorded superhumps were likely stage C superhumps.  The lack of
variation in the period is compatible with this interpretation.

\begin{table}
\caption{Superhump maxima of OT J1821 (2010).}\label{tab:j1821oc2010}
\begin{center}
\begin{tabular}{ccccc}
\hline
$E$ & max\commenta & error & $O-C$\commentb & $N$\commentc \\
\hline
0 & 55312.1638 & 0.0008 & $-$0.0034 & 80 \\
1 & 55312.2544 & 0.0039 & 0.0052 & 50 \\
18 & 55313.6433 & 0.0018 & $-$0.0016 & 53 \\
48 & 55316.1072 & 0.0011 & $-$0.0005 & 46 \\
49 & 55316.1883 & 0.0021 & $-$0.0014 & 59 \\
74 & 55318.2444 & 0.0055 & 0.0023 & 118 \\
85 & 55319.1458 & 0.0119 & 0.0007 & 19 \\
86 & 55319.2259 & 0.0284 & $-$0.0013 & 31 \\
\hline
  \multicolumn{5}{l}{\commenta BJD$-$2400000.} \\
  \multicolumn{5}{l}{\commentb Against $max = 2455312.1671 + 0.082094 E$.} \\
  \multicolumn{5}{l}{\commentc Number of points used to determine the maximum.} \\
\end{tabular}
\end{center}
\end{table}

\subsection{OT J213806.6$+$261957}\label{obj:j2138}

   This transient (hereafter OT J2138) was independently discovered by
D.-A. Yi \citep{yam10j2138cbet2273} and S. Kaneko \citep{nak10j2138cbet2275}.
\citet{ara10j2138cbet2275} spectroscopically confirmed that this is
an outbursting dwarf nova (see also vsnet-alert 11971).
The spectrum by \citet{gra10j2138cbet2275} indicated the presence of
highly excited emission lines, suggesting that this is a WZ Sge-type
outburst [cf. vsnet-alert 11974, see also vsnet-alert 11987 (K. Kinugasa,
Gunma Astronomical Observatory) and follow-up spectroscopy by
\cite{tov10j2138cbet2283}].

\begin{figure}
  \begin{center}
%    \FigureFile(88mm,110mm){j2138eshpdm.eps}
    \FigureFile(88mm,110mm){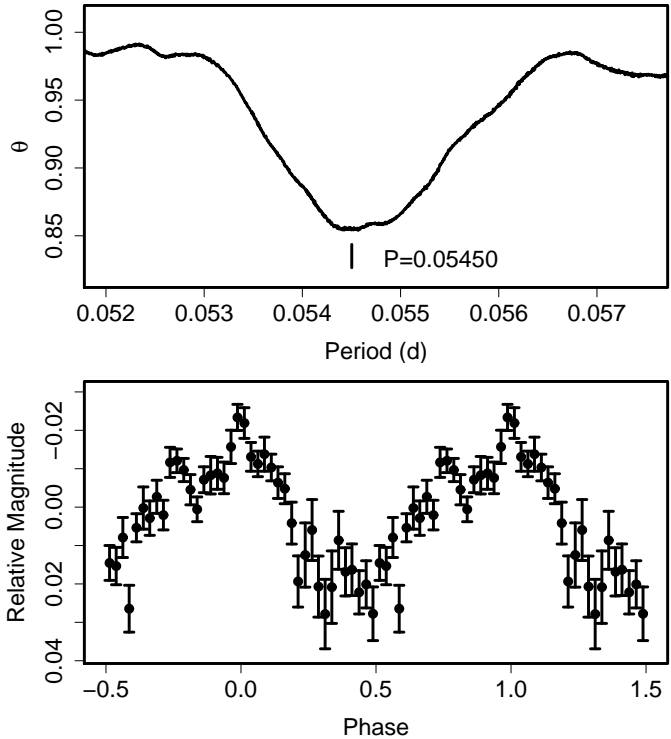}
  \end{center}
  \caption{Early superhumps in OT J2138 (2010) before BJD 2455327.5.
     (Upper): PDM analysis.
     (Lower): Phase-averaged profile.}
  \label{fig:j2138eshpdm}
\end{figure}

\begin{figure}
  \begin{center}
%    \FigureFile(88mm,110mm){j2138shpdm.eps}
    \FigureFile(88mm,110mm){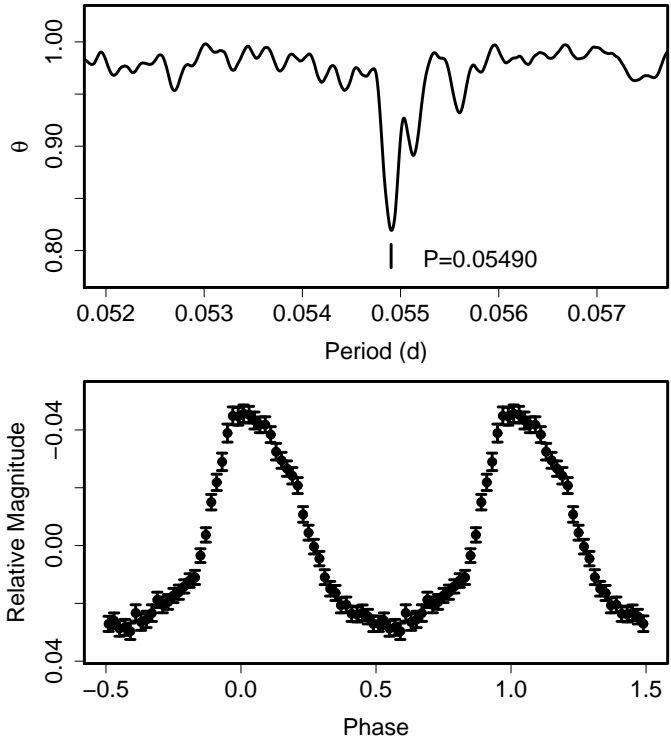}
  \end{center}
  \caption{Superhumps in OT J2138 (2010) during the plateau phase.
     (Upper): PDM analysis.
     (Lower): Phase-averaged profile.}
  \label{fig:j2138shpdm}
\end{figure}

   The earliest indication of short-period modulations (likely
early superhumps) was detected by G. Masi (vsnet-alert 11977,
see also 11978, 11985; figure \ref{fig:j2138eshpdm}).
Ordinary superhumps subsequently appeared
(vsnet-alert 11990; figure \ref{fig:j2138shpdm}),
strengthening the identification as a WZ Sge-type
dwarf nova.  \citet{hud10j2138atel2619} reported a detection of another
outburst in 1942, which is the only known outburst other than the
present one.

\begin{figure*}
  \begin{center}
%    \FigureFile(160mm,160mm){j2138humpall.eps}
    \FigureFile(160mm,160mm){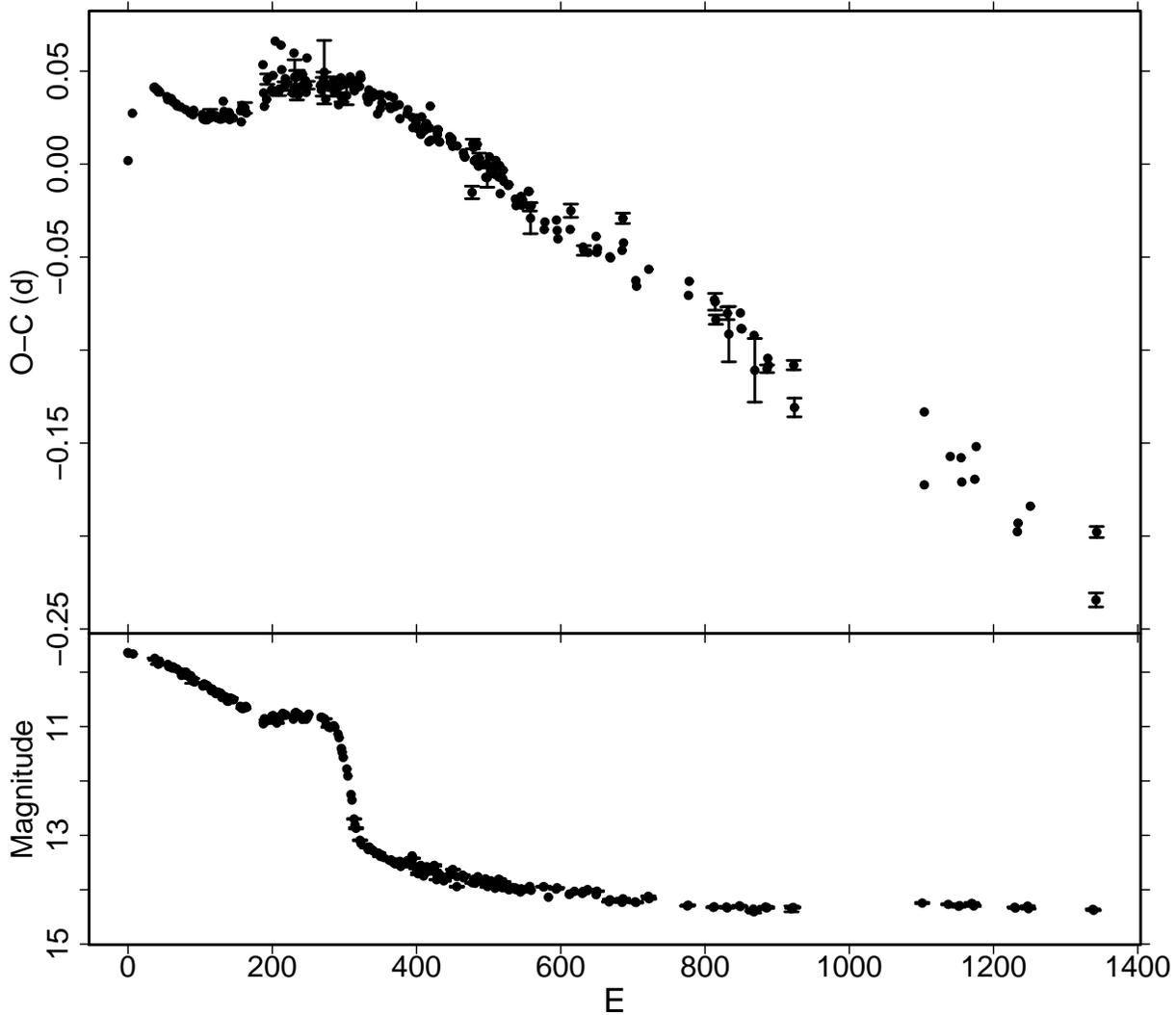}
  \end{center}
  \caption{$O-C$ variation in OT J2138 (2010).  (Upper) $O-C$.
  (Lower) Light curve.
  The earliest stage of the superoutburst is not shown in this figure.
  }
  \label{fig:j2138humpall}
\end{figure*}

\begin{figure}
  \begin{center}
%    \FigureFile(88mm,110mm){j2138latepdm.eps}
    \FigureFile(88mm,110mm){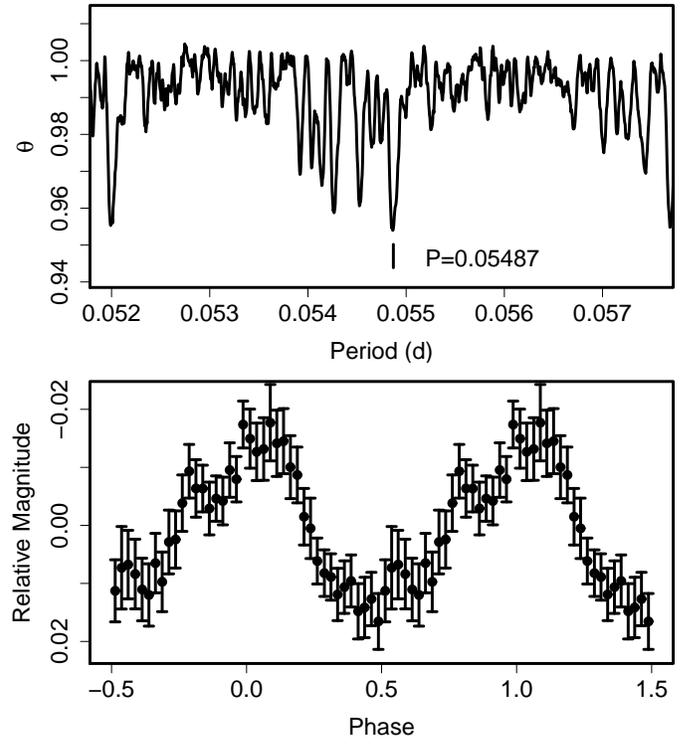}
  \end{center}
  \caption{Superhumps in OT J2138 (2010) during the late post-superoutburst
     stage.
     (Upper): PDM analysis.
     (Lower): Phase-averaged profile.}
  \label{fig:j2138latepdm}
\end{figure}

   The times of maxima for ordinary superhumps are listed in
table \ref{tab:j2138oc2010}.  Although there was an indication of double
maxima during the late plateau phase, we did not attempt to distinguish
the different peaks, and listed most prominent ones.
The overall $O-C$ diagram (figure \ref{fig:j2138humpall}) bears a high
degree of similarity to that of ASAS J0025 in 2004 \citet{Pdot}.\footnote{
  OT J2138 is a companion to a close visual double \citep{yam10j2138cbet2273},
  which dominates when the variable becomes fainter.  Since all the
  observers measured combined magnitudes, the contribution from the visual
  companion is responsible for the apparently low outburst amplitude
  as in figure \ref{fig:j2138humpall} and the low amplitudes of
  post-superoutburst superhumps.
}
The main difference is that OT J2138 did not fade from the plateau
before entering stage C.  Combined with the apparent presence of multiple
hump peaks, this stage in OT J2138 appears to phenomenologically correspond
to the phase between the rapid fading from the plateau and rebrightening
in ASAS J0025.  The apparent absence of a rebrightening
in OT J2138 may be explained if this object somehow succeeded in
maintaining the plateau phase (or avoiding the quenching of the outbursting
state) when ASAS J0025 once returned to quiescence
before the rebrightening.  The predominance of type-D superoutbursts
(superoutbursts without a rebrightening, see \cite{Pdot}) in very
short-$P_{\rm orb}$ systems may be understood as this kind of variety.
In table \ref{tab:perlist}, we listed intervals exhibiting best-defined
maxima and omitted a phase with multiple peaks.  The post-superoutburst
stage with $E > 521$ showed a slightly shorter superhump period
and the signal was persistent even during the very late post-superoutburst
stage ($P$ = 0.05487(1) d, figure \ref{fig:j2138latepdm}).
There was no indication of a longer late-stage superhumps as observed
in GW Lib, V455 And, EG Cnc, WZ Sge and several newly discovered
WZ Sge-type dwarf novae (\cite{kat08wzsgelateSH}; \cite{Pdot}).

   More detailed analysis and discussion will be presented in
Maehara et al., in preparation.

\begin{table}
\caption{Superhump maxima of OT J2138 (2010).}\label{tab:j2138oc2010}
\begin{center}
\begin{tabular}{ccccc}
\hline
$E$ & max\commenta & error & $O-C$\commentb & $N$\commentc \\
\hline
0 & 55330.8319 & 0.0006 & 0.0019 & 89 \\
6 & 55331.1881 & 0.0013 & 0.0274 & 309 \\
36 & 55332.8555 & 0.0002 & 0.0412 & 76 \\
37 & 55332.9109 & 0.0002 & 0.0414 & 83 \\
38 & 55332.9650 & 0.0002 & 0.0404 & 116 \\
41 & 55333.1300 & 0.0005 & 0.0401 & 125 \\
42 & 55333.1838 & 0.0002 & 0.0388 & 662 \\
43 & 55333.2391 & 0.0002 & 0.0390 & 832 \\
44 & 55333.2940 & 0.0002 & 0.0387 & 503 \\
54 & 55333.8428 & 0.0002 & 0.0363 & 128 \\
55 & 55333.8963 & 0.0002 & 0.0347 & 149 \\
56 & 55333.9519 & 0.0002 & 0.0351 & 106 \\
60 & 55334.1724 & 0.0004 & 0.0352 & 145 \\
61 & 55334.2265 & 0.0003 & 0.0342 & 477 \\
62 & 55334.2810 & 0.0004 & 0.0335 & 427 \\
66 & 55334.5003 & 0.0009 & 0.0324 & 35 \\
67 & 55334.5554 & 0.0004 & 0.0324 & 53 \\
68 & 55334.6094 & 0.0003 & 0.0313 & 53 \\
73 & 55334.8845 & 0.0002 & 0.0307 & 152 \\
80 & 55335.2689 & 0.0002 & 0.0293 & 110 \\
86 & 55335.5978 & 0.0005 & 0.0275 & 151 \\
87 & 55335.6535 & 0.0005 & 0.0281 & 117 \\
90 & 55335.8173 & 0.0003 & 0.0265 & 102 \\
91 & 55335.8749 & 0.0004 & 0.0290 & 130 \\
103 & 55336.5340 & 0.0014 & 0.0267 & 52 \\
104 & 55336.5871 & 0.0008 & 0.0246 & 134 \\
105 & 55336.6417 & 0.0009 & 0.0241 & 171 \\
108 & 55336.8067 & 0.0003 & 0.0238 & 153 \\
109 & 55336.8625 & 0.0002 & 0.0244 & 244 \\
110 & 55336.9171 & 0.0002 & 0.0239 & 145 \\
114 & 55337.1408 & 0.0024 & 0.0271 & 56 \\
115 & 55337.1938 & 0.0005 & 0.0250 & 97 \\
116 & 55337.2511 & 0.0002 & 0.0272 & 387 \\
122 & 55337.5806 & 0.0007 & 0.0260 & 179 \\
123 & 55337.6346 & 0.0007 & 0.0249 & 182 \\
127 & 55337.8545 & 0.0002 & 0.0242 & 104 \\
128 & 55337.9101 & 0.0002 & 0.0247 & 106 \\
129 & 55337.9647 & 0.0004 & 0.0242 & 146 \\
132 & 55338.1396 & 0.0018 & 0.0338 & 131 \\
133 & 55338.1893 & 0.0005 & 0.0284 & 500 \\
134 & 55338.2408 & 0.0004 & 0.0248 & 578 \\
137 & 55338.4063 & 0.0006 & 0.0248 & 92 \\
138 & 55338.4633 & 0.0003 & 0.0267 & 113 \\
139 & 55338.5180 & 0.0005 & 0.0263 & 163 \\
140 & 55338.5746 & 0.0008 & 0.0278 & 141 \\
141 & 55338.6259 & 0.0008 & 0.0239 & 159 \\
145 & 55338.8472 & 0.0004 & 0.0248 & 89 \\
146 & 55338.9023 & 0.0003 & 0.0247 & 74 \\
156 & 55339.4574 & 0.0005 & 0.0287 & 175 \\
157 & 55339.5065 & 0.0005 & 0.0227 & 208 \\
158 & 55339.5705 & 0.0017 & 0.0315 & 85 \\
159 & 55339.6238 & 0.0017 & 0.0297 & 76 \\
162 & 55339.7897 & 0.0029 & 0.0302 & 51 \\
163 & 55339.8426 & 0.0006 & 0.0280 & 82 \\
164 & 55339.8972 & 0.0007 & 0.0275 & 82 \\
187 & 55341.1909 & 0.0016 & 0.0535 & 82 \\
\hline
  \multicolumn{5}{l}{\commenta BJD$-$2400000.} \\
  \multicolumn{5}{l}{\commentb Against $max = 2455330.8300 + 0.055120 E$.} \\
  \multicolumn{5}{l}{\commentc Number of points used to determine the maximum.} \\
\end{tabular}
\end{center}
\end{table}

\addtocounter{table}{-1}
\begin{table}
\caption{Superhump maxima of OT J2138 (2010) (continued).}
\begin{center}
\begin{tabular}{ccccc}
\hline
$E$ & max\commenta & error & $O-C$\commentb & $N$\commentc \\
\hline
188 & 55341.2308 & 0.0011 & 0.0383 & 264 \\
189 & 55341.2788 & 0.0010 & 0.0311 & 292 \\
192 & 55341.4479 & 0.0015 & 0.0348 & 78 \\
193 & 55341.5139 & 0.0028 & 0.0457 & 75 \\
199 & 55341.8388 & 0.0019 & 0.0400 & 118 \\
200 & 55341.8928 & 0.0009 & 0.0388 & 119 \\
201 & 55341.9568 & 0.0011 & 0.0477 & 158 \\
204 & 55342.1406 & 0.0011 & 0.0662 & 44 \\
209 & 55342.3890 & 0.0021 & 0.0390 & 46 \\
210 & 55342.4457 & 0.0012 & 0.0405 & 113 \\
212 & 55342.5794 & 0.0020 & 0.0640 & 101 \\
213 & 55342.6214 & 0.0016 & 0.0508 & 84 \\
217 & 55342.8330 & 0.0023 & 0.0419 & 169 \\
218 & 55342.8921 & 0.0011 & 0.0460 & 174 \\
228 & 55343.4363 & 0.0005 & 0.0389 & 116 \\
229 & 55343.4904 & 0.0008 & 0.0379 & 115 \\
230 & 55343.5673 & 0.0012 & 0.0597 & 89 \\
231 & 55343.6096 & 0.0092 & 0.0469 & 99 \\
233 & 55343.7196 & 0.0036 & 0.0466 & 51 \\
234 & 55343.7661 & 0.0036 & 0.0381 & 62 \\
235 & 55343.8313 & 0.0023 & 0.0481 & 95 \\
236 & 55343.8757 & 0.0011 & 0.0374 & 116 \\
241 & 55344.1558 & 0.0013 & 0.0419 & 90 \\
242 & 55344.2173 & 0.0009 & 0.0483 & 103 \\
246 & 55344.4345 & 0.0012 & 0.0449 & 107 \\
247 & 55344.4832 & 0.0011 & 0.0386 & 107 \\
248 & 55344.5569 & 0.0009 & 0.0571 & 88 \\
249 & 55344.5973 & 0.0021 & 0.0424 & 57 \\
267 & 55345.5896 & 0.0014 & 0.0425 & 81 \\
268 & 55345.6420 & 0.0013 & 0.0399 & 83 \\
270 & 55345.7540 & 0.0050 & 0.0416 & 60 \\
271 & 55345.8132 & 0.0038 & 0.0457 & 59 \\
272 & 55345.8721 & 0.0171 & 0.0495 & 44 \\
274 & 55345.9678 & 0.0007 & 0.0350 & 83 \\
278 & 55346.1947 & 0.0004 & 0.0413 & 398 \\
279 & 55346.2512 & 0.0004 & 0.0427 & 508 \\
284 & 55346.5280 & 0.0011 & 0.0439 & 46 \\
285 & 55346.5840 & 0.0022 & 0.0448 & 130 \\
286 & 55346.6330 & 0.0022 & 0.0387 & 116 \\
290 & 55346.8566 & 0.0019 & 0.0418 & 104 \\
291 & 55346.9111 & 0.0011 & 0.0411 & 104 \\
292 & 55346.9569 & 0.0019 & 0.0319 & 165 \\
295 & 55347.1368 & 0.0010 & 0.0464 & 177 \\
296 & 55347.1814 & 0.0025 & 0.0359 & 198 \\
297 & 55347.2432 & 0.0016 & 0.0425 & 80 \\
302 & 55347.5203 & 0.0019 & 0.0440 & 47 \\
303 & 55347.5682 & 0.0049 & 0.0368 & 84 \\
304 & 55347.6304 & 0.0009 & 0.0440 & 113 \\
308 & 55347.8539 & 0.0005 & 0.0469 & 103 \\
309 & 55347.9080 & 0.0006 & 0.0459 & 104 \\
310 & 55347.9612 & 0.0008 & 0.0440 & 67 \\
313 & 55348.1229 & 0.0014 & 0.0403 & 98 \\
314 & 55348.1771 & 0.0002 & 0.0394 & 2499 \\
315 & 55348.2376 & 0.0018 & 0.0448 & 76 \\
321 & 55348.5653 & 0.0005 & 0.0418 & 76 \\
\hline
  \multicolumn{5}{l}{\commenta BJD$-$2400000.} \\
  \multicolumn{5}{l}{\commentb Against $max = 2455330.8300 + 0.055120 E$.} \\
  \multicolumn{5}{l}{\commentc Number of points used to determine the maximum.} \\
\end{tabular}
\end{center}
\end{table}

\addtocounter{table}{-1}
\begin{table}
\caption{Superhump maxima of OT J2138 (2010) (continued).}
\begin{center}
\begin{tabular}{ccccc}
\hline
$E$ & max\commenta & error & $O-C$\commentb & $N$\commentc \\
\hline
322 & 55348.6266 & 0.0011 & 0.0480 & 101 \\
323 & 55348.6800 & 0.0008 & 0.0462 & 45 \\
331 & 55349.1111 & 0.0008 & 0.0364 & 119 \\
332 & 55349.1650 & 0.0007 & 0.0352 & 264 \\
333 & 55349.2184 & 0.0008 & 0.0334 & 358 \\
334 & 55349.2800 & 0.0004 & 0.0399 & 396 \\
339 & 55349.5518 & 0.0004 & 0.0362 & 72 \\
340 & 55349.6091 & 0.0017 & 0.0383 & 82 \\
341 & 55349.6629 & 0.0007 & 0.0370 & 50 \\
346 & 55349.9285 & 0.0013 & 0.0270 & 49 \\
347 & 55349.9937 & 0.0006 & 0.0371 & 53 \\
349 & 55350.1040 & 0.0011 & 0.0371 & 142 \\
350 & 55350.1519 & 0.0008 & 0.0299 & 262 \\
351 & 55350.2145 & 0.0012 & 0.0374 & 363 \\
352 & 55350.2652 & 0.0005 & 0.0329 & 325 \\
362 & 55350.8202 & 0.0011 & 0.0367 & 53 \\
363 & 55350.8695 & 0.0005 & 0.0309 & 73 \\
364 & 55350.9235 & 0.0007 & 0.0299 & 73 \\
368 & 55351.1500 & 0.0012 & 0.0358 & 291 \\
369 & 55351.2004 & 0.0004 & 0.0311 & 374 \\
370 & 55351.2553 & 0.0004 & 0.0309 & 383 \\
376 & 55351.5870 & 0.0013 & 0.0319 & 57 \\
377 & 55351.6347 & 0.0004 & 0.0244 & 57 \\
387 & 55352.1884 & 0.0006 & 0.0270 & 47 \\
388 & 55352.2459 & 0.0008 & 0.0293 & 236 \\
394 & 55352.5725 & 0.0009 & 0.0252 & 58 \\
395 & 55352.6220 & 0.0009 & 0.0196 & 59 \\
399 & 55352.8451 & 0.0006 & 0.0223 & 61 \\
400 & 55352.9027 & 0.0005 & 0.0247 & 73 \\
401 & 55352.9523 & 0.0006 & 0.0191 & 151 \\
406 & 55353.2247 & 0.0008 & 0.0160 & 240 \\
407 & 55353.2892 & 0.0011 & 0.0254 & 175 \\
410 & 55353.4490 & 0.0002 & 0.0198 & 184 \\
411 & 55353.5024 & 0.0006 & 0.0181 & 172 \\
412 & 55353.5589 & 0.0006 & 0.0195 & 104 \\
413 & 55353.6161 & 0.0015 & 0.0216 & 58 \\
414 & 55353.6714 & 0.0007 & 0.0218 & 41 \\
417 & 55353.8271 & 0.0016 & 0.0120 & 59 \\
418 & 55353.8895 & 0.0013 & 0.0194 & 73 \\
419 & 55353.9565 & 0.0010 & 0.0312 & 121 \\
420 & 55353.9933 & 0.0008 & 0.0129 & 56 \\
428 & 55354.4397 & 0.0005 & 0.0183 & 120 \\
429 & 55354.4925 & 0.0004 & 0.0160 & 123 \\
430 & 55354.5502 & 0.0008 & 0.0186 & 80 \\
431 & 55354.5990 & 0.0003 & 0.0123 & 58 \\
432 & 55354.6537 & 0.0018 & 0.0119 & 58 \\
446 & 55355.4283 & 0.0004 & 0.0148 & 123 \\
447 & 55355.4806 & 0.0006 & 0.0119 & 122 \\
448 & 55355.5368 & 0.0010 & 0.0130 & 72 \\
449 & 55355.5925 & 0.0008 & 0.0136 & 56 \\
450 & 55355.6435 & 0.0004 & 0.0095 & 56 \\
456 & 55355.9745 & 0.0008 & 0.0098 & 52 \\
465 & 55356.4668 & 0.0003 & 0.0060 & 155 \\
466 & 55356.5203 & 0.0004 & 0.0044 & 169 \\
467 & 55356.5748 & 0.0010 & 0.0037 & 25 \\
477 & 55357.1070 & 0.0034 & $-$0.0152 & 54 \\
\hline
  \multicolumn{5}{l}{\commenta BJD$-$2400000.} \\
  \multicolumn{5}{l}{\commentb Against $max = 2455330.8300 + 0.055120 E$.} \\
  \multicolumn{5}{l}{\commentc Number of points used to determine the maximum.} \\
\end{tabular}
\end{center}
\end{table}

\addtocounter{table}{-1}
\begin{table}
\caption{Superhump maxima of OT J2138 (2010) (continued).}
\begin{center}
\begin{tabular}{ccccc}
\hline
$E$ & max\commenta & error & $O-C$\commentb & $N$\commentc \\
\hline
478 & 55357.1882 & 0.0026 & 0.0108 & 256 \\
479 & 55357.2414 & 0.0009 & 0.0089 & 272 \\
480 & 55357.2894 & 0.0009 & 0.0018 & 119 \\
482 & 55357.4003 & 0.0016 & 0.0024 & 61 \\
483 & 55357.4555 & 0.0010 & 0.0025 & 62 \\
484 & 55357.5187 & 0.0014 & 0.0107 & 63 \\
485 & 55357.5628 & 0.0011 & $-$0.0004 & 52 \\
486 & 55357.6172 & 0.0004 & $-$0.0011 & 54 \\
487 & 55357.6766 & 0.0027 & 0.0032 & 29 \\
496 & 55358.1624 & 0.0008 & $-$0.0071 & 259 \\
497 & 55358.2249 & 0.0011 & 0.0003 & 301 \\
498 & 55358.2725 & 0.0052 & $-$0.0072 & 269 \\
501 & 55358.4491 & 0.0007 & 0.0040 & 59 \\
502 & 55358.4989 & 0.0006 & $-$0.0013 & 61 \\
503 & 55358.5534 & 0.0006 & $-$0.0020 & 46 \\
504 & 55358.6081 & 0.0017 & $-$0.0024 & 54 \\
505 & 55358.6608 & 0.0010 & $-$0.0048 & 45 \\
508 & 55358.8288 & 0.0004 & $-$0.0022 & 59 \\
509 & 55358.8805 & 0.0006 & $-$0.0055 & 73 \\
510 & 55358.9432 & 0.0006 & 0.0020 & 73 \\
514 & 55359.1547 & 0.0013 & $-$0.0070 & 177 \\
515 & 55359.2161 & 0.0009 & $-$0.0007 & 222 \\
516 & 55359.2561 & 0.0006 & $-$0.0159 & 224 \\
519 & 55359.4297 & 0.0011 & $-$0.0075 & 60 \\
520 & 55359.4892 & 0.0004 & $-$0.0033 & 61 \\
521 & 55359.5382 & 0.0007 & $-$0.0093 & 35 \\
527 & 55359.8668 & 0.0004 & $-$0.0115 & 73 \\
528 & 55359.9224 & 0.0019 & $-$0.0110 & 73 \\
537 & 55360.4107 & 0.0006 & $-$0.0187 & 61 \\
538 & 55360.4622 & 0.0019 & $-$0.0223 & 61 \\
545 & 55360.8531 & 0.0004 & $-$0.0173 & 73 \\
546 & 55360.9034 & 0.0005 & $-$0.0221 & 73 \\
547 & 55360.9613 & 0.0011 & $-$0.0193 & 49 \\
555 & 55361.4070 & 0.0009 & $-$0.0146 & 53 \\
556 & 55361.4620 & 0.0008 & $-$0.0147 & 62 \\
557 & 55361.5089 & 0.0022 & $-$0.0229 & 54 \\
558 & 55361.5579 & 0.0083 & $-$0.0291 & 60 \\
559 & 55361.6198 & 0.0009 & $-$0.0223 & 56 \\
577 & 55362.5991 & 0.0005 & $-$0.0352 & 56 \\
578 & 55362.6583 & 0.0006 & $-$0.0311 & 42 \\
594 & 55363.5412 & 0.0005 & $-$0.0301 & 39 \\
595 & 55363.5908 & 0.0020 & $-$0.0356 & 56 \\
596 & 55363.6413 & 0.0010 & $-$0.0402 & 54 \\
613 & 55364.5834 & 0.0011 & $-$0.0351 & 55 \\
614 & 55364.6486 & 0.0036 & $-$0.0251 & 52 \\
631 & 55365.5662 & 0.0010 & $-$0.0446 & 56 \\
632 & 55365.6194 & 0.0026 & $-$0.0464 & 54 \\
638 & 55365.9491 & 0.0005 & $-$0.0475 & 85 \\
649 & 55366.5640 & 0.0007 & $-$0.0389 & 42 \\
650 & 55366.6105 & 0.0009 & $-$0.0475 & 55 \\
651 & 55366.6678 & 0.0011 & $-$0.0453 & 29 \\
668 & 55367.6003 & 0.0006 & $-$0.0498 & 52 \\
669 & 55367.6548 & 0.0011 & $-$0.0505 & 46 \\
685 & 55368.5408 & 0.0009 & $-$0.0464 & 36 \\
686 & 55368.6132 & 0.0028 & $-$0.0291 & 53 \\
687 & 55368.6551 & 0.0008 & $-$0.0423 & 47 \\
\hline
  \multicolumn{5}{l}{\commenta BJD$-$2400000.} \\
  \multicolumn{5}{l}{\commentb Against $max = 2455330.8300 + 0.055120 E$.} \\
  \multicolumn{5}{l}{\commentc Number of points used to determine the maximum.} \\
\end{tabular}
\end{center}
\end{table}

\addtocounter{table}{-1}
\begin{table}
\caption{Superhump maxima of OT J2138 (2010) (continued).}
\begin{center}
\begin{tabular}{ccccc}
\hline
$E$ & max\commenta & error & $O-C$\commentb & $N$\commentc \\
\hline
704 & 55369.5719 & 0.0015 & $-$0.0626 & 55 \\
705 & 55369.6239 & 0.0012 & $-$0.0657 & 52 \\
722 & 55370.5701 & 0.0011 & $-$0.0566 & 50 \\
777 & 55373.5876 & 0.0012 & $-$0.0706 & 47 \\
778 & 55373.6503 & 0.0012 & $-$0.0630 & 50 \\
813 & 55375.5696 & 0.0014 & $-$0.0729 & 55 \\
814 & 55375.6237 & 0.0045 & $-$0.0740 & 56 \\
815 & 55375.6692 & 0.0025 & $-$0.0836 & 35 \\
831 & 55376.5545 & 0.0034 & $-$0.0802 & 49 \\
833 & 55376.6535 & 0.0149 & $-$0.0914 & 53 \\
849 & 55377.5468 & 0.0006 & $-$0.0800 & 37 \\
850 & 55377.5936 & 0.0013 & $-$0.0884 & 56 \\
851 & 55377.6485 & 0.0013 & $-$0.0886 & 54 \\
868 & 55378.5821 & 0.0014 & $-$0.0921 & 57 \\
869 & 55378.6184 & 0.0171 & $-$0.1109 & 56 \\
886 & 55379.5563 & 0.0020 & $-$0.1100 & 45 \\
887 & 55379.6170 & 0.0012 & $-$0.1044 & 46 \\
888 & 55379.6683 & 0.0015 & $-$0.1082 & 39 \\
923 & 55381.5977 & 0.0025 & $-$0.1081 & 55 \\
924 & 55381.6300 & 0.0050 & $-$0.1309 & 54 \\
1104 & 55391.5100 & 0.0004 & $-$0.1725 & 49 \\
1104 & 55391.5492 & 0.0018 & $-$0.1333 & 58 \\
1140 & 55393.5096 & 0.0004 & $-$0.1572 & 92 \\
1155 & 55394.3357 & 0.0019 & $-$0.1579 & 53 \\
1156 & 55394.3778 & 0.0007 & $-$0.1710 & 71 \\
1174 & 55395.3714 & 0.0011 & $-$0.1695 & 82 \\
1176 & 55395.4992 & 0.0019 & $-$0.1519 & 54 \\
1233 & 55398.5953 & 0.0007 & $-$0.1977 & 54 \\
1234 & 55398.6551 & 0.0006 & $-$0.1930 & 43 \\
1251 & 55399.6012 & 0.0011 & $-$0.1839 & 53 \\
1342 & 55404.5667 & 0.0038 & $-$0.2344 & 42 \\
1343 & 55404.6583 & 0.0030 & $-$0.1978 & 42 \\
\hline
  \multicolumn{5}{l}{\commenta BJD$-$2400000.} \\
  \multicolumn{5}{l}{\commentb Against $max = 2455330.8300 + 0.055120 E$.} \\
  \multicolumn{5}{l}{\commentc Number of points used to determine the maximum.} \\
\end{tabular}
\end{center}
\end{table}

\subsection{OT J215815.3$+$094709}\label{obj:j2158}

\begin{figure}
  \begin{center}
%    \FigureFile(88mm,110mm){j2158shpdm.eps}
    \FigureFile(88mm,110mm){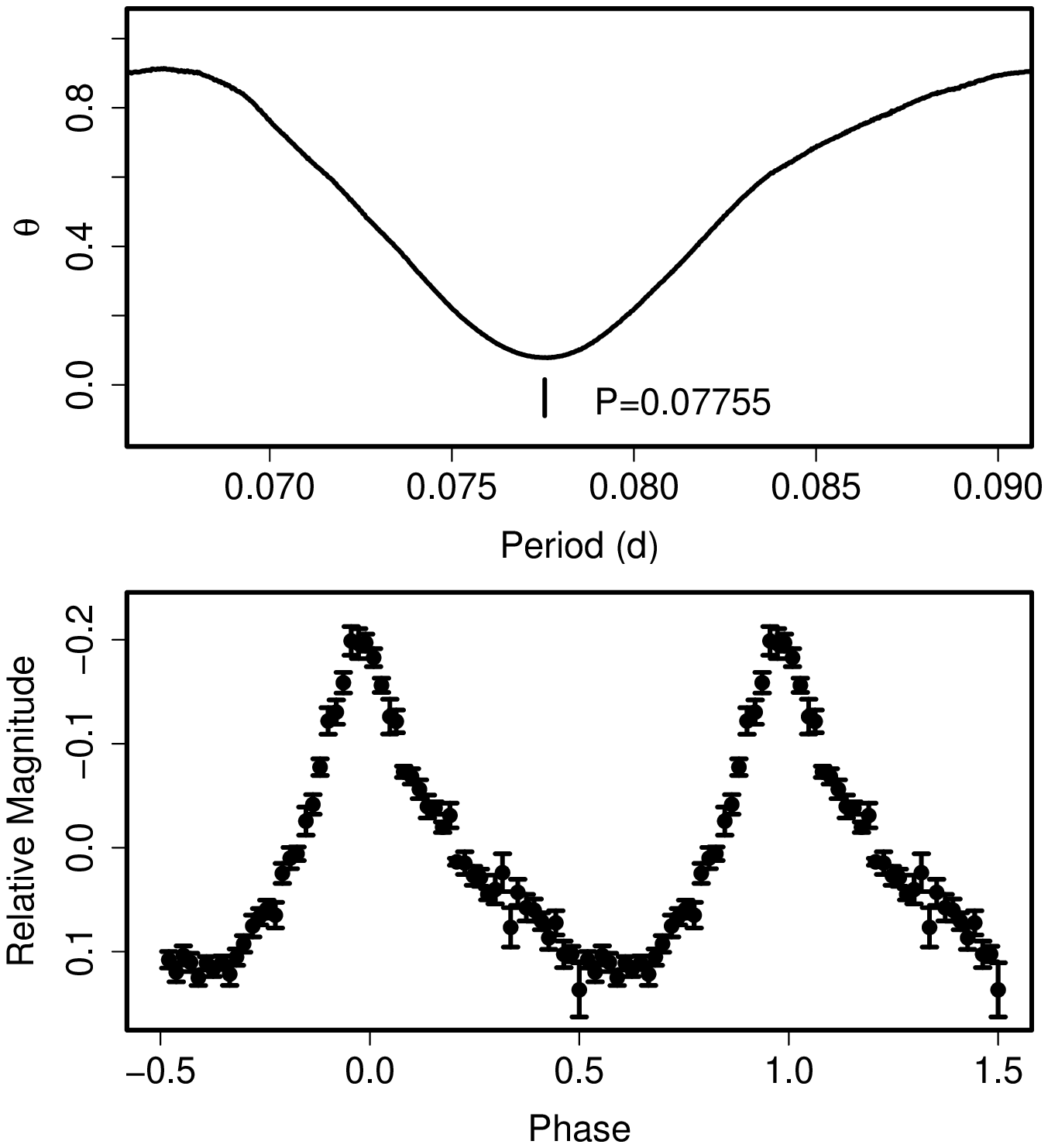}
  \end{center}
  \caption{Superhumps in OT J2158 (2010).
     (Upper): PDM analysis.
     (Lower): Phase-averaged profile.}
  \label{fig:j2158shpdm}
\end{figure}

   This object (= CSS100615:215815$+$094709, hereafter OT J2158) was
discovered by the CRTS in 2010 June.  Astrokolkhoz team immediately
detected superhumps (vsnet-outburst 11306).
The times of superhump maxima are listed in table \ref{tab:j2158oc2010}.
The mean period determined with the PDM method was 0.07755(9) d.
The large amplitudes (0.35 mag, figure \ref{fig:j2158shpdm}) and
waveform suggest that these superhumps
are stage B superhumps at their early evolution.

\begin{table}
\caption{Superhump maxima of OT J2158 (2010).}\label{tab:j2158oc2010}
\begin{center}
\begin{tabular}{ccccc}
\hline
$E$ & max\commenta & error & $O-C$\commentb & $N$\commentc \\
\hline
0 & 55366.5092 & 0.0008 & 0.0001 & 25 \\
1 & 55366.5868 & 0.0003 & 0.0000 & 50 \\
2 & 55366.6644 & 0.0005 & $-$0.0000 & 56 \\
4 & 55366.8190 & 0.0003 & $-$0.0007 & 108 \\
5 & 55366.8979 & 0.0005 & 0.0005 & 74 \\
\hline
  \multicolumn{5}{l}{\commenta BJD$-$2400000.} \\
  \multicolumn{5}{l}{\commentb Against $max = 2455366.5091 + 0.077655 E$.} \\
  \multicolumn{5}{l}{\commentc Number of points used to determine the maximum.} \\
\end{tabular}
\end{center}
\end{table}

\subsection{OT J223003.0$-$145835}\label{obj:j2230}

\begin{figure}
  \begin{center}
%    \FigureFile(88mm,110mm){j2230eshpdm.eps}
    \FigureFile(88mm,110mm){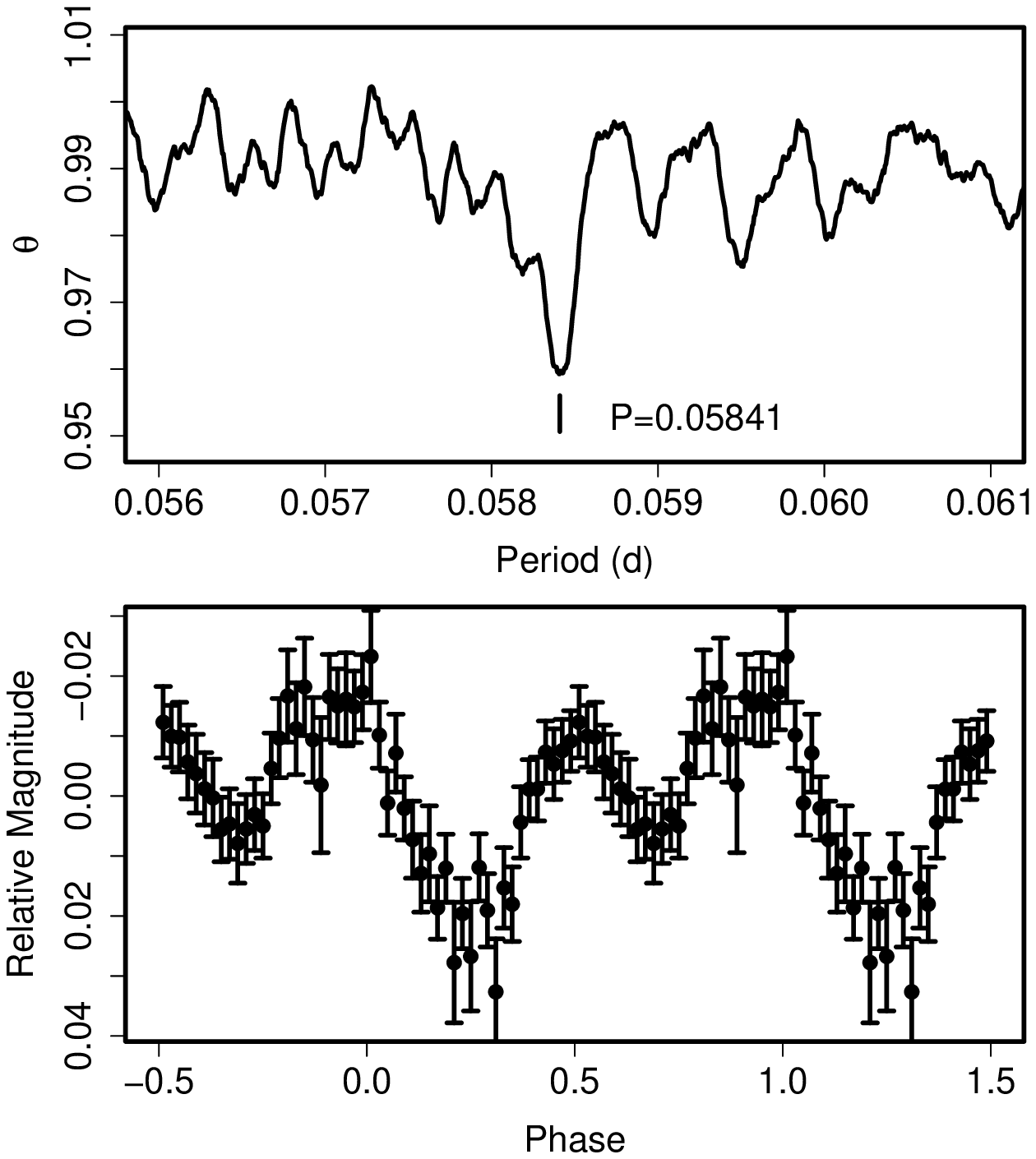}
  \end{center}
  \caption{Early superhumps in OT J2230 (2009).
     (Upper): PDM analysis.
     (Lower): Phase-averaged profile.}
  \label{fig:j2230eshpdm}
\end{figure}

   This object (= CSS090727:223003$-$145835, hereafter OT J2230) was
discovered by the CRTS in 2009 July.  The large outburst amplitude
exceeding 6 mag was immediately noted (vsnet-alert 11351).  Superhump-like
modulations were soon reported (vsnet-alert 11353).  It became evident
by later observations that these modulations were early superhumps
of a WZ Sge-type dwarf nova rather than ordinary superhumps
(vsnet-alert 11365, 11388).
We obtained a mean period of 0.05841(1) d with the PDM method
(figure \ref{fig:j2230eshpdm}).  Doubly humped modulations
characteristic to early superhumps (cf. \cite{kat02wzsgeESH})
are clearly seen.  Due to the faintness of the object, the appearance
of ordinary superhumps was not recorded.  The outburst lasted for
at least 18 d.

\subsection{OT J234440.5$-$001206}\label{obj:j2344}

\begin{figure}
  \begin{center}
%    \FigureFile(88mm,110mm){j2344shpdm.eps}
    \FigureFile(88mm,110mm){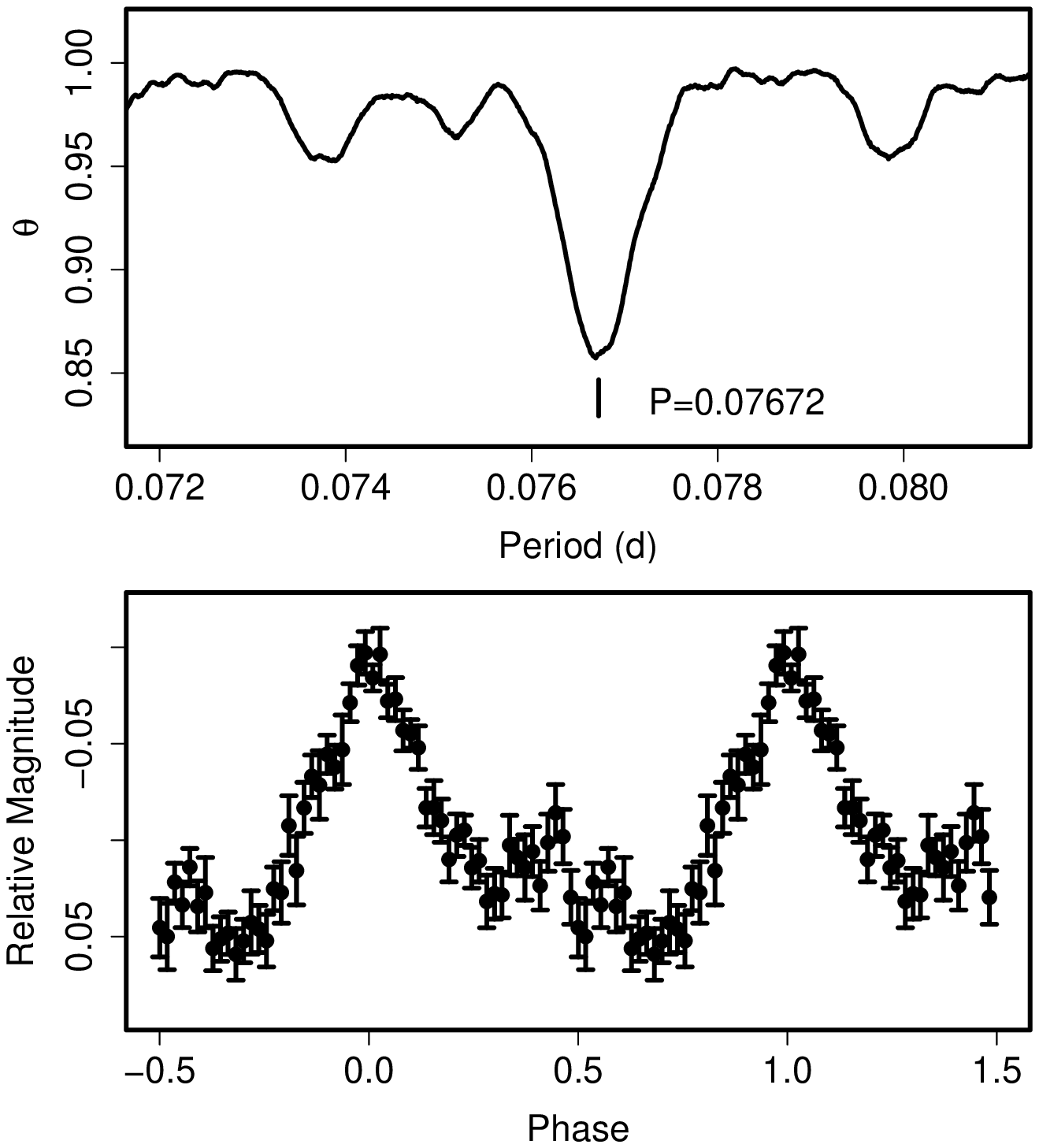}
  \end{center}
  \caption{Superhumps in OT J2344 (2010).
     (Upper): PDM analysis.
     (Lower): Phase-averaged profile.}
  \label{fig:j2344shpdm}
\end{figure}

   This object (= MLS100904:234441$-$001206, hereafter OT J2344) was
discovered by the CRTS in 2010 September during the course of
the Mount Lemmon survey.  P. Wils notified this event (cvnet-outburst
3828).  Immediately following this announcement, superhumps were
clearly detected (vsnet-alert 12143, 12144; figure \ref{fig:j2344shpdm}).

   The times of superhump maxima are listed in table \ref{tab:j2344oc2010}.
These superhumps were likely stage C superhumps since there was little
evidence of period variation and the object started a rapid fading 4 d
after the start of the observation.

\begin{table}
\caption{Superhump maxima of OT J2344 (2010).}\label{tab:j2344oc2010}
\begin{center}
\begin{tabular}{ccccc}
\hline
$E$ & max\commenta & error & $O-C$\commentb & $N$\commentc \\
\hline
0 & 55444.4661 & 0.0009 & 0.0024 & 38 \\
1 & 55444.5425 & 0.0005 & 0.0020 & 77 \\
2 & 55444.6163 & 0.0005 & $-$0.0008 & 104 \\
9 & 55445.1507 & 0.0066 & $-$0.0035 & 76 \\
12 & 55445.3861 & 0.0010 & 0.0019 & 38 \\
13 & 55445.4615 & 0.0012 & 0.0005 & 39 \\
14 & 55445.5387 & 0.0005 & 0.0010 & 39 \\
15 & 55445.6111 & 0.0011 & $-$0.0033 & 39 \\
26 & 55446.4576 & 0.0005 & $-$0.0006 & 170 \\
27 & 55446.5316 & 0.0007 & $-$0.0034 & 195 \\
28 & 55446.6107 & 0.0011 & $-$0.0009 & 149 \\
38 & 55447.3778 & 0.0009 & $-$0.0009 & 172 \\
39 & 55447.4547 & 0.0009 & $-$0.0008 & 173 \\
40 & 55447.5373 & 0.0020 & 0.0052 & 112 \\
48 & 55448.1513 & 0.0081 & 0.0054 & 111 \\
51 & 55448.3756 & 0.0011 & $-$0.0005 & 172 \\
52 & 55448.4519 & 0.0014 & $-$0.0009 & 172 \\
53 & 55448.5283 & 0.0018 & $-$0.0011 & 173 \\
60 & 55449.0629 & 0.0026 & $-$0.0035 & 40 \\
61 & 55449.1448 & 0.0042 & 0.0017 & 21 \\
\hline
  \multicolumn{5}{l}{\commenta BJD$-$2400000.} \\
  \multicolumn{5}{l}{\commentb Against $max = 2455444.4637 + 0.076711 E$.} \\
  \multicolumn{5}{l}{\commentc Number of points used to determine the maximum.} \\
\end{tabular}
\end{center}
\end{table}

\section{Discussion and summary}

\subsection{Period Derivatives during Stage B}

   The new data for SU UMa-type dwarf novae have improved the statistics
presented in \citet{Pdot}.  Although the presentations in \citet{Pdot}
covered a wide range of relations, we restrict only to key figures
after combining the samples with \citet{Pdot} in this paper.

\begin{figure*}
  \begin{center}
%    \FigureFile(120mm,80mm){pdotpsh2.eps}
    \FigureFile(120mm,80mm){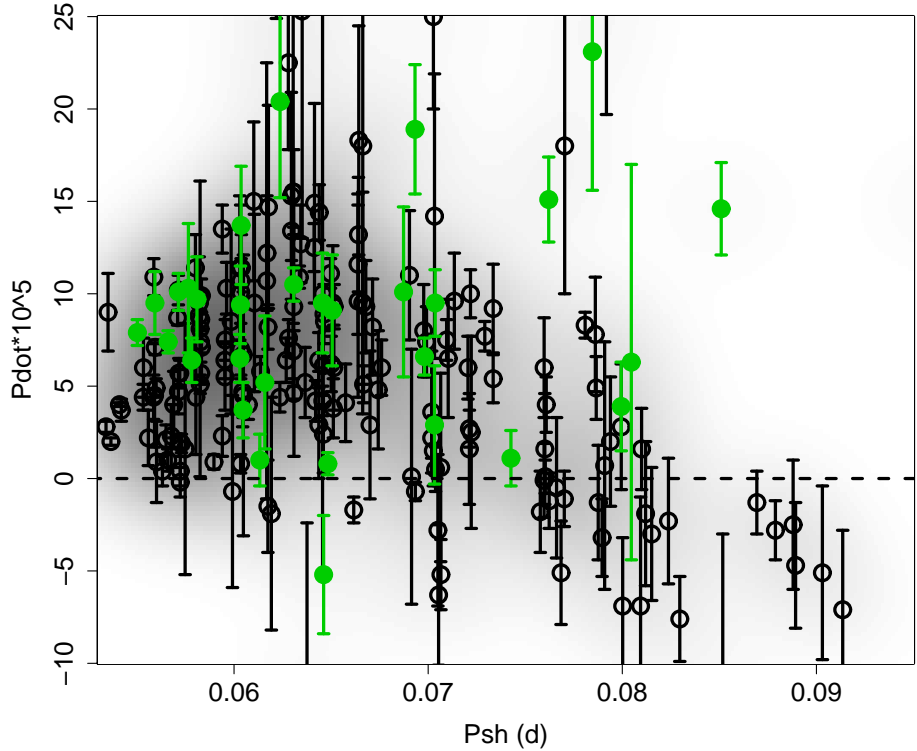}
  \end{center}
  \caption{$P_{\rm dot}$ for stage B versus $P_{\rm SH}$.
  Open and filled circles represent samples in \citet{Pdot} and
  this paper, respectively.
  An estimated density map is overlaid in gray scale.
  }
  \label{fig:pdotpsh2}
\end{figure*}

   Figure \ref{fig:pdotpsh2} represents the relation between $P_{\rm SH}$
and $P_{\rm dot}$ during stage B.
The enlarged figure (corresponding to figure 10 in \cite{Pdot})
is only shown here.  The estimated density map is overlaid for
a better visualization.  The new data generally confirmed the predominance
of positive period derivatives during stage B in systems with superhump
periods shorter than 0.07 d, in agreement with the tendency reported
in \citet{Pdot}.

\subsection{Period Derivatives in Long-$P_{\rm SH}$ Systems}

   In figure \ref{fig:pdotpsh2}, there appears to be a systematic difference
in $P_{\rm dot}$ for systems with $P_{\rm SH}$ longer than 0.075 d.
We have selected samples for $0.075 < P_{\rm SH} {\rm (d)} < 0.085$ from
\citet{Pdot} and this paper.  After using the same criterion for samples
described in \citet{Pdot} and rejecting poorly determined
(nominal error of $P_{\rm dot}$ exceeding $7 \times 10^{-5}$) superoutbursts,
we applied Student's $t$-test.  The test indicated that these samples are
different at a significance level of 0.09.  This test suggests that the
properties of samples are different between these two data sets.
It is already apparent all well-determined $P_{\rm dot}$'s in this paper
for this $P_{\rm SH}$ region has positive $P_{\rm dot}$'s, while many
systems had negative $P_{\rm dot}$'s in \citet{Pdot}.

   This might have been caused by a small number of long-$P_{\rm SH}$ samples
in this paper, since most of newly discovered SU UMa-type dwarf novae
have short $P_{\rm SH}$.  There would also be a possibility, however, that
most of long-$P_{\rm SH}$ with frequent superoutbursts were already discovered
at the time of \citet{Pdot} and that long-$P_{\rm SH}$ systems in this papers
were heavily biased toward systems with infrequent superoutbursts.
This possibility would be supported considering that all systems with
supercycles longer than 1000 d (V1251 Cyg, RZ Leo, QY Per) have large
positive $P_{\rm dot}$ (see also a discussion in \cite{Pdot} subsection 4.10)
while systems with $P_{\rm dot} < -4 \times 10^{-5}$ are restricted to
systems with supercycles shorter than 400 d (DH Aql, V1316 Cyg, CU Vel).
Although a direct correlation coefficient (0.20) between $\log$ supercycle
and $P_{\rm dot}$ is not significant, these results seem to strengthen the
idea in \citet{Pdot} that the degree of period variation is more diverse
in long-$P_{\rm SH}$ systems.  There is, however, a case of IY UMa, with
frequent superoutbursts, which showed a positive $P_{\rm dot}$ in this
new sample.  It is not yet clear whether some of long-$P_{\rm SH}$ have
different $P_{\rm dot}$'s between different superoutbursts or whether
this was a mere consequence of poor sampling in the past.
Future studies in these viewpoints would be fruitful.

\subsection{Difference Between Different Superoutbursts}

   The apparent lack of systematic difference between different
superoutbursts of the same system is one of the main conclusions
in \citet{Pdot}.  We further examined this issue using new materials.
The new data strengthen this conclusions, including well-observed
superoutbursts preceded by precursor outbursts (e.g. PU CMa and
1RXS J0532.  It would be noteworthy 1RXS J0532 almost completely
reproduced the behavior in 2005, which has been reported to be rather
unusual \citet{ima09j0532}.  This suggests that peculiarities found
in certain systems are specific to these systems, rather than a chance
occurrence.  This might suggest the existence of parameters specific
to systems in addition to general parameters such as $q$ or $P_{\rm orb}$.

   In V844 Her, our observations give additional support to the hypothesis
that the delay time in development of superhumps is shorter in significantly
smaller superoutbursts.

\subsection{Beat phenomenon and period variation}

   In the 2009 superoutburst of IY UMa, we recorded a strong
beat phenomenon (in the total luminosity) between superhumps and
orbital variation.
The beat period largely varied between stage B and C, and were in
very good agreement with the beat periods expected from superhump
periods during the corresponding stages.
The close correlation between the beat period and the superhump period
suggests that the change in the angular velocity of the global apsidal
motion is more responsible for the stage B--C transition rather than
the appearance of a more localized new component with a different period
(or angular velocity), such as a bright spot on the disk.
Future research focusing on this relation between the beat period and
superhump period will be an important key in clarifying the nature of
period variations of superhumps.

\subsection{Behavior in WZ Sge-Type Systems}

   The new sample includes three new WZ Sge-type objects with established
early superhumps, SDSS J1610, OT J1044 and OT J2230, and one WZ Sge-type
object with likely early superhumps (OT J2138).  All objects have
$P_{\rm SH}$ shorter than 0.061 d.  We also suggest that two systems,
VX For and EL UMa, are WZ Sge-type dwarf novae with multiple rebrightenings.
Although the stage with early superhumps was not observed in VX For,
the observed $P_{\rm dot}$ for the first time predicted for its multiple
rebrightening in real-time.  This success strengthens the relationship
between $P_{\rm dot}$ and type of rebrightenings in WZ Sge-type dwarf novae
proposed in \citet{Pdot}.  Although EL UMa was only observed for its
supposed rebrightening phase, all the known properties are strongly suggestive
of a WZ Sge-type system with multiple rebrightenings.
The $O-C$ analysis of OT J2138 and its comparison to ASAS J0025 suggest
an interpretation that the frequent absence of rebrightenings in
very short-$P_{\rm orb}$ objects can be a result of sustained superoutburst
plateau when usual SU UMa-type dwarf novae return to quiescence preceding
a rebrightening.  Although this phenomenon may be somehow related to
a small binary separation resulting stronger tidal torque and thereby
stronger tidal dissipation to maintain the outbursting state,
the exact mechanism should await further clarification.

\vskip 3mm

The authors are grateful to observers of VSNET Collaboration and
VSOLJ observers who supplied vital data.
We acknowledge with thanks the variable star
observations from the AAVSO International Database contributed by
observers worldwide and used in this research.
This work is deeply indebted to outburst detections and announcement
by a number of variable star observers worldwide, including participants of
CVNET, BAA VSS alert and AVSON networks. 
We are grateful to D. Boyd and his collaborators for making the data
for OT J1440 available.
The CCD operation of the Bronberg Observatory is partly sponsored by
the Center for Backyard Astrophysics.
The CCD operation by Peter Nelson is on loan from the AAVSO,
funded by the Curry Foundation.
We are grateful to the Catalina Real-time Transient Survey
team for making their real-time
detection of transient objects available to the public.
We are also grateful to H. Takahashi and K. Kinugasa for making unpublished
spectra of EL UMa available to us.
We are grateful to the anonymous referee for pointing out the apparent
difference of distribution of $P_{\rm dot}$ between this paper and \citet{Pdot}.

\appendix
\section{Bayesian Applications to Period Analysis}

   We explored three applications of Bayesian statistics in period
analysis employed in this paper.
Bayesian statistics (see e.g. \cite{BayesReview}) provide
a framework in estimating a posterior probability density function (PDF)
of model parameters from a combination of the observed data,
a likelihood function defined by the model, and a prior PDF of
the model parameters.
According to the Bayes' theorem, the posterior PDF of the model
parameters $\theta$ is

\begin{equation}
Pr(\theta|D) = \frac{Pr(D|\theta)\pi(\theta)}{Pr(D)}
  \,\propto\, Pr(D|\theta)\pi(\theta),
\end{equation}

where $\theta$ is the model parameter, $D$ is the data,
$Pr(\theta|D)$ is the posterior probability of the model,
$Pr(D|\theta) \equiv \mathcal{L}(\theta)$ is the likelihood,
$\pi(\theta)$ is the prior probability and

\begin{equation}
Pr(D) = \int Pr(D|\theta) d\theta
\end{equation}

is a normalization factor, also known as Bayesian evidence.

   We usually employ Markov-Chain Monte Carlo (MCMC) method
in order to obtain the PDF.  
The Metropolis-Hastings algorithm (\cite{Metropolis}; \cite{Hastings}),
one of the best known of MCMC implementation, is as follows:
starting with $\theta^{(1)}$, and at
the $n$-th step of MCMC, we obtain
$p(\theta^{(n)}) = Pr(D|\theta^{(n)})\pi(\theta^{(n)})$.
We then move to a new point $\theta^{(n+1)}$ following a proposal
density distribution $q(\theta^{(n)},\theta^{(n+1)})$.
The new point is accepted with a probability

\begin{equation}
\alpha(\theta^{(n)},\theta^{(n+1)}) =
\min \biggl\{1, \frac{
    p(\theta^{(n+1)}) q(\theta^{(n+1)},\theta^{(n)})}
    {p(\theta^{(n)}) q(\theta^{(n)},\theta^{(n+1)})
  } \biggr\} .
\end{equation}
Multivariate Gaussian movements are frequently used to cancel out
$q(\theta^{(n+1)},\theta^{(n)})$ and $q(\theta^{(n)},\theta^{(n+1)})$.
If the proposal is rejected, $\theta^{(n+1)} = \theta^{(n)}$.\footnote{
   The description in \citet{uem10blazar}, dealing with a similar
   class of problems, is incorrect in this step.
}
After discarding initial `burn-in' steps, we sample the remaining
steps of the Markov chain to obtain the PDF.\footnote{
   Sampling with constant (e.g. 3 to 100) spacings is formally used
   in order to avoid correlations between steps.  In our analysis,
   we took all remaining steps because our main interest is in means
   and dispersions of parameters of the PDF, which are little
   affected by correlation between steps.
}
Typical numbers of steps are $10^4$--$10^7$ for the entire chain and
$10^2$--$10^5$ for the `burn-in' steps.

\subsection{$O-C$ Analysis of SU UMa-Type Dwarf Novae}

   As reviewed in \citet{Pdot}, the $O-C$ diagrams of superhump maxima
in SU UMa-type dwarf novae generally follow discrete stages: most
frequently a segment with a positive $P_{\rm dot}$ and a later
segment with a discontinuously shorter constant period
(stages B and C in \citet{Pdot}).
Determining the parameters for periods and $P_{\rm dot}$ and the time
of transition is not a trivial task for the classical statistics.
\citet{uem10j0557} dealt with this problem using the Bayesian
statistics and the MCMC method.  We formulate this problem for
a wider usage.

   In this problem, $D = \{t_{obs}(E_i)\}$ are the observed $O-C$'s
(or directly observed maxima)
for the epochs $\{E_i\}$, and the parameter space is $\theta = \{a,b,c,T,p\}$
defined by the model ($t_{model}$)

\begin{eqnarray}
t_{early}(E_i) = & aE_i^2+bE_i+c & (0 \le E \le T) \nonumber \\
t_{late}(E_i)  = & pE_i+q        & (E > T) \nonumber\\
t_{early}(T)   = & t_{late}(T) , &         \nonumber \\
\end{eqnarray}

where $T$ is the epoch of the transition (cf. \cite{uem10j0557}).

   Assuming that $\epsilon_i = t_{obs}(E_i) - t_{model}(E_i)$
follows a normal distribution
$N(0,\sigma_i^2)$, the likelihood function can be written as

\begin{equation}
\mathcal{L}(\theta) = \prod_i \frac{1}{\sqrt{2\pi\sigma_i^2}}
\exp{\biggl[-\frac{\{t_{obs}(E_i)-t_{model}(E_i)\}^2}{2\sigma_i^2} \biggr]} .
\end{equation}

\begin{figure}
  \begin{center}
%    \FigureFile(88mm,110mm){ocsamp1.eps}
    \FigureFile(88mm,110mm){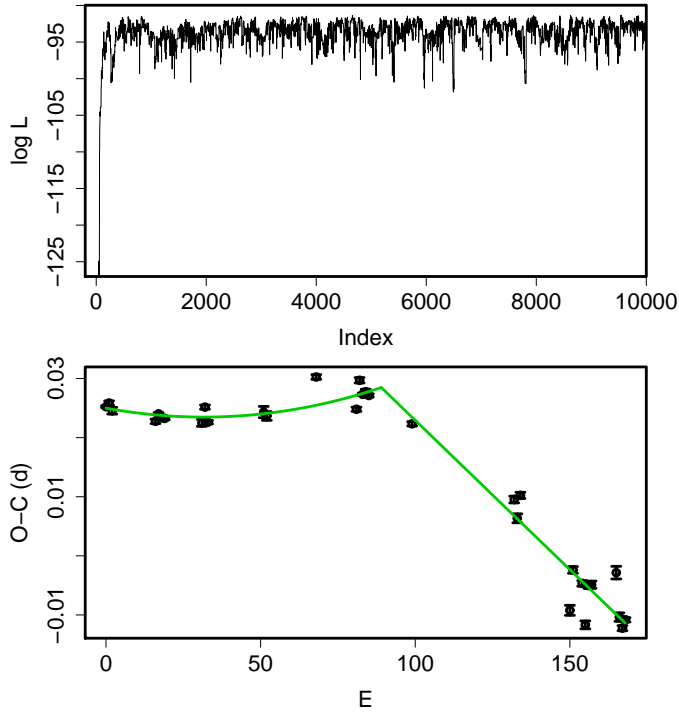}
  \end{center}
  \caption{MCMC analysis of $O-C$ diagram (1).
     (Upper): Log-likelihood value of the first 10000 chains of
     a MCMC sampling of $O-C$'s for superhump timings of QZ Vir in 2008.
     (Lower): Best-fit model using posterior means derived from the PDF.}
  \label{fig:ocsamp1}
\end{figure}

\begin{figure}
  \begin{center}
%    \FigureFile(88mm,110mm){ocsamp2.eps}
    \FigureFile(88mm,110mm){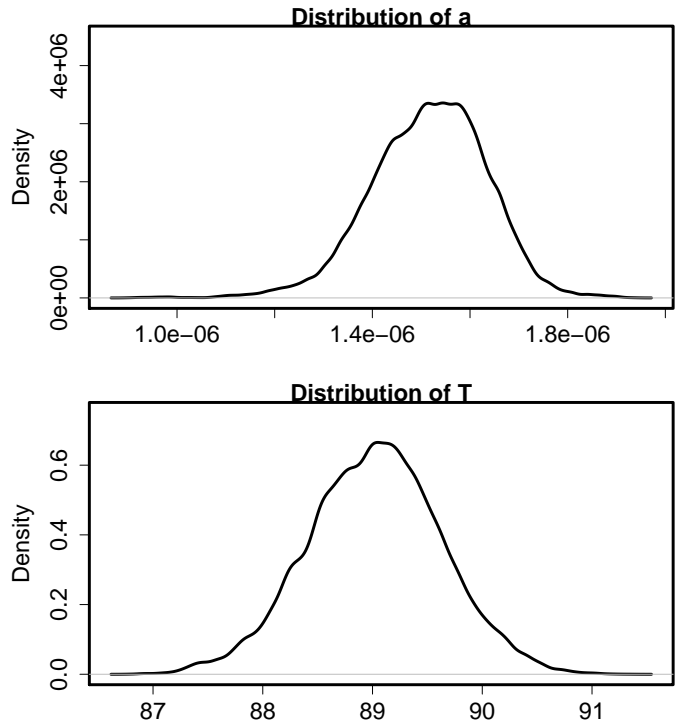}
  \end{center}
  \caption{MCMC analysis of $O-C$ diagram (2).
     (Upper): Distribution of $a$.
     (Lower): Distribution of $T$.}
  \label{fig:ocsamp2}
\end{figure}

   Using this likelihood, or its combination with priors, we can
obtain the PDF with the MCMC method.
Figure \ref{fig:ocsamp1} is a sample application to superhump timings
in QZ Vir in 2004 \citep{ohs10qzvir}.  The initial $T$ was chosen as 85
and other initial parameters were determined by linear least-squares
fitting.  The $\sigma$'s of the multivariate Gaussian distribution for
the proposal density were 0.04 times 1-$\sigma$ errors of the initial
fitting.  The total length of the chain was $10^5$ and the first 5000
samples were discarded for obtaining the PDF.
The mean parameters and standard errors were
$a=1.52(12) \times 10^{-6}$, $b= -9.6(1.1) \times 10^{-5}$,
$c=2.50(2) \times 10^{-2}$,
$T=89.0(6)$ and $p= -5.05(6) \times 10^{-4}$.
The posterior means of two parameters $a$ and $T$ are shown in figure
\ref{fig:ocsamp2}.

\begin{figure}
  \begin{center}
%    \FigureFile(88mm,70mm){ocsampprior.eps}
    \FigureFile(88mm,70mm){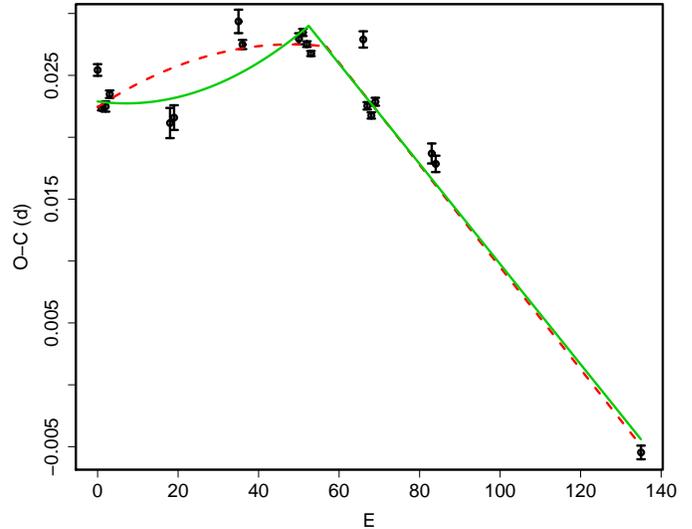}
  \end{center}
  \caption{Effect of priors for observations with a gaps.
     Dashed and solid curves represent Bayesian estimates
     for unrestricted model and a model with a prior, respectively.
     (See text for details).}
  \label{fig:ocsampprior}
\end{figure}

   We can also introduce priors such as a form of

\begin{equation}
\pi(a) = \frac{1}{\sqrt{2\pi\sigma_a^2}}
\exp{\biggl\{-\frac{(a-a_0)^2}{2\sigma_a^2} \biggr\}} ,
\end{equation}

when expected values of some parameters are empirically known.
This application of priors would be useful when $P_{\rm dot}$
of a certain object is known from observation of other superoutbursts
while a particular observation has a significant gap to determine
parameters by itself.
Figure \ref{fig:ocsampprior} demonstrates the effect of priors for
observations with a gaps.
The data are times of superhumps of QZ Vir during the
2007 superoutburst.  An unconstrained model gives a negative
$P_{\rm dot}$ of $a = -2.3(6) \times 10^{-6}$ and a break
at $T = 57.0(9)$ (dashed curve).
Incorporating prior knowledge that this object has a positive
$P_{\rm dot}$, i.e. setting a prior with
$a_0 = 4 \times 10^{-6}$ and $\sigma_a = 4 \times 10^{-7}$,
we get a reasonable fit (solid curve) with a break at $T = 48(4)$.
Since this example is for a demonstration purpose of Bayesian approach,
we did not use these values in the main text.
Although a proper way of using priors in such a problem need to be
further investigated, this formulation would provide a way of
analyzing a badly sampled observations when we have firm knowledge in
choosing the prior.

\subsection{Application to Period Analysis}

   We consider a Bayesian extension of parameter fitting of observed
light curves with a cyclic function.
In this case, $D = \{y_{obs}(t_i)\}$ is the observations at $\{t_i\}$, and
the parameter space is $\theta = \{f,a,b,c\}$
defined by the model

\begin{equation}
y_{model} = a F(2\pi f (t_i - t_0) - b) + c ,
\end{equation}
where $f$ is frequency and $t_0$ is an arbitrary constant defining the
zero phase.  $F$ is an arbitrary cyclic function having
a period of 1.

   Assuming that $\epsilon_i = y_{obs}(t_i) - y_{model}(t_i)$
follows a normal distribution
$N(0,\sigma_i^2)$, the likelihood function can be written as

\begin{equation}
\mathcal{L}(\theta) = \prod_i \frac{1}{\sqrt{2\pi\sigma_i^2}}
\exp{\biggl[-\frac{\{y_{obs}(t_i)-y_{model}(t_i)\}^2}{2\sigma_i^2} \biggr]} .
\end{equation}
 
   In practice, we can choose a sine function or a template function
for $F$.  The parameter $\sigma_i$  can be either individually set
or determined as a constant form $\sigma_i = \sigma$ from the best fit.

\begin{figure}
  \begin{center}
%    \FigureFile(88mm,70mm){v1454cygmcmc.eps}
    \FigureFile(88mm,70mm){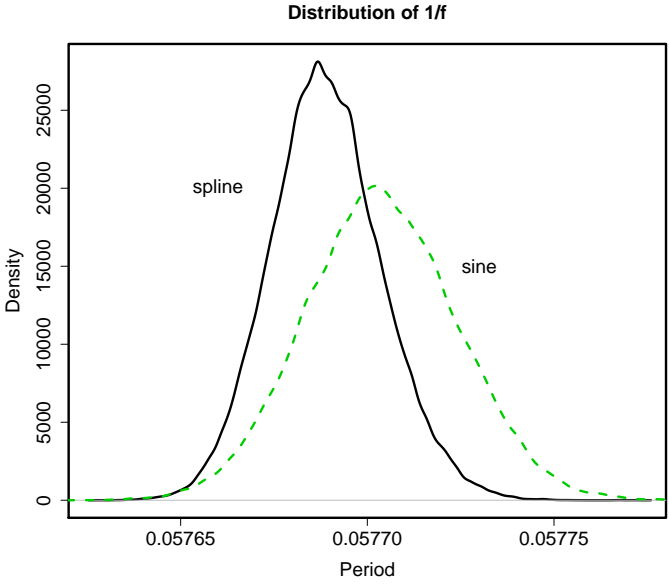}
  \end{center}
  \caption{Application to period analysis.
    The solid and dashed curves represent results of modeling by
    spline-interpolated templat profile of superhumps \citet{Pdot}
    sine function, respectively.  The used data were for V1454 Cyg (2009),
    the same as in figure \ref{fig:v1454cygshpdm}.
    }
  \label{fig:v1454cygmcmc}
\end{figure}

   Although a directed application of the original MCMC to solve the aliasing
problem is limited due to the slow mixing between widely separate peaks,
it is well suited for sampling the detailed structure of individual peaks.
Figure \ref{fig:v1454cygmcmc} illustrates the dependence on the model
functions.  Although a spline-interpolated model light curve usually gives
a smaller standard error for parameter estimates, the difference is not
usually very big.  In the present application to V1454 Cyg, spline and
sine fits give $1/f = 0.057689(15)$ and $0.057704(20)$, respectively.
These analysis has confirmed the results, both period and error estimate,
of the PDM analysis.  This method would be advantageous when the model
description becomes more complex than what the original PDM method describes.

\subsection{Bayesian Extension to PDM}

   The PDM method evaluates dispersions of the observed data against
phase-binned averages.  This prescription sometimes produces spurious
scatters in the resultant theta diagram when the number of data
(and data in each bin) is small.

   Instead of getting phase-averaged light curves (at given trial periods)
using a small number of discrete bins, we introduce a Bayesian approach
for obtaining continuous phase-averaged light curves.
We model light curves by a set of parameters
$\theta \equiv \{\theta(j)\} \equiv \{y_{model}(\phi_j)\}$
(at a trial period) for phases
$\{\phi_j\} (j=1,\dots,N$), where $N$ is the number of phase
bins, which are large ($\sim 100$) enough to describe continuous functions.

   The likelihood function is
\begin{equation}
\mathcal{L}(\theta) = \prod_i \frac{1}{\sqrt{2\pi\sigma_i^2}}
\exp{\biggl[-\frac{\{y_{obs}(\phi_i)-\bar{y}_{obs}-y_{model}(\phi_i)\}^2}{2\sigma_i^2} \biggr]} ,
\end{equation}
where $y_{model}(\phi_i)$ is linearly interpolated from $\phi_j$
and $\phi_{j+1}$ containing $\phi_i$ in the interval
$[\phi_j, \phi_{j+1})$.

   Following a standard technique in Bayesian analysis,
we then express the condition of smoothness of $\theta$ by
introducing a prior function assuming that second order
differences of $\{\theta(j)\}$ follow a normal distribution
$N(0,\sigma_s^2)$ \footnote{
   This distribution can be confirmed by differentiating well-sampled
   mean profiles of superhumps.
}
\begin{equation}
\pi(\theta) = \prod_j \frac{1}{\sqrt{2\pi\sigma_s^2}}
\exp{\biggl[-\frac{\{\theta(j-1)-2\theta(j)+\theta(j+1)\}^2}{2\sigma_s^2} \biggr]} ,
\end{equation}
where $\theta(0) = \theta(N-1)$ and $\theta(N+1) = \theta(1)$
reflecting the cyclic condition.

   In order to obtain maximum a posteriori (MAP) estimates,
we solve
\begin{equation}
\frac{\partial{\log({\mathcal{L}(\theta)\pi(\theta))}}}{\partial{\theta_j}} = 0
\end{equation}
and
\begin{equation}
\sum_j{\theta_j} = 0.
\end{equation}
These equations reduce to a set of linear, first-order, equations for
$\theta_j$.

   Although the smoothing parameter $\sigma_s$ can be estimated from
individual fits by different trial periods, we adopt a constant
$\sigma_s$ for all periods.  The value is estimated from the best fit,
i.e. from the period the giving smallest dispersion.
Using this smooth phase-average light curve, we can then estimate
overall variance as prescribed as in \citet{PDM}.

\begin{figure}
  \begin{center}
%    \FigureFile(88mm,110mm){pdmcomp.eps}
    \FigureFile(88mm,110mm){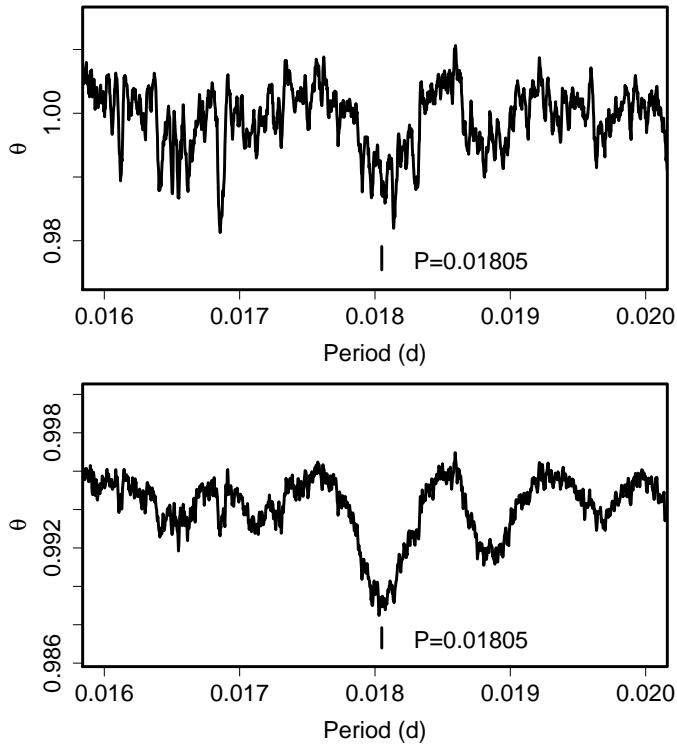}
  \end{center}
  \caption{Comparison of PDM and Bayesian-extended PDM.
  The used data were SDSS J0129 (subsection \ref{obj:j0129}).
  (Upper) PDM. (Lower) Bayesian-extended PDM.}
  \label{fig:pdmcomp}
\end{figure}

   This extension of the PDM method is particularly useful when the
number of observations is relatively small and the signal of variations
is comparable to the noise (figure \ref{fig:pdmcomp}).

\end{document}